# Biased AI improves human performance but reduces perceived helpfulness

Shiyang Lai[1,2], Jiwoong Choi[2†], Junsol Kim[1,2†], Nadav Kunievsky[2†], Yujin Potter[3†], James Evans[1,2,4*]

[1]Department of Sociology, University of Chicago; Chicago, 60637, USA

[2]Knowledge Lab, University of Chicago; Chicago, 60637, USA

[3]Department of Electrical Engineering and Computer Sciences, UC Berkeley; Berkeley, 94720, USA

[4]Santa Fe Institute; Santa Fe, NM 87505, USA

[*]Corresponding author. Email: jevans@uchicago.edu

[†]These authors contributed equally to this work and are alphabetically ordered.

**Artificial intelligence (AI) increasingly shapes how people think[1–3], engage[4–6], and evaluate information[7,8]. To minimize risk, most current systems are designed to present as ideologically neutral with standardized output[9–11]. Yet growing evidence suggests that these principles suppress cognitive engagement[12–14], impair human decision-making[15,16], and erode societal diversity[17,18]. Here we test the opposite approach by deliberately injecting bias into AI assistants. In three randomized experiments with 5,000 participants, biased AI improved human performance relative to default and neutral AI in tasks ranging from misinformation evaluation and financial investment to graduate education. These gains carried a subjective cost. Participants systematically undervalued AI they believed to be biased and inflated the helpfulness of AI they believed to be neutral, regardless of the systems' actual behavior. Interacting with two AIs whose biases flanked the participant's own perspective preserved the performance gains while limiting the subjective cost and one-sided influence. Our findings reveal the strategic value of intentional bias in AI design. Rather than performing a single fair, reliable, and authoritative voice, AI that speaks from specific viewpoints[19–22] triggers cognitive agency and elevates human-AI performance in judgment, decision-making, and problem-solving.**

Generative artificial intelligence (AI) systems have become increasingly integrated into human decision-making and everyday tasks[23–30]. In the pursuit of trustworthy AI systems, industry has aggressively pursued the production of unbiased, "fair" AI systems[31–34]. Despite a wide range of approaches to understand and evaluate algorithmic fairness, most efforts to debias Large Language Model (LLM)-based AI seek to remove systematic deviations from neutrality that reflect specific cultural, political, or ideological perspectives[35]. This motivation has led to extensive development of bias benchmarks and approaches that filter biased model responses through instruction fine-tuning[36], representation engineering[37], and reinforcement learning from human feedback (RLHF)[38]. Despite these bias-reduction attempts, systematic bias persists in ostensibly neutral AI systems[16,36,39], raising concerns about whether complete neutrality is attainable at all[31,40–43].

Beyond the feasibility of bias mitigation approaches, critics have argued that by treating fairness as an intrinsic property of AI models, dominant principles of "debiasing" overlook the dynamic and reciprocal nature of human-AI interaction[5,44,45]. Others have argued that formally "fair" models do not guarantee fair collective outcomes once deployed[46–48]. When AI systems appear overly irenic and sanitized, human users may be seduced to uncritically trust AI outputs without engagement[49–53], forfeiting the judgment required to discriminate between model outputs and catch errors[12,49]. Scholars have argued that this AI manifestation of automation bias could foster "metacognitive laziness" and "moral deskilling"[13,14,53,54], introduce accountability ambiguities[55,56], and contribute to cultural convergence and the erosion of innovation[57–61].

The principle that cognitive frictions can counteract automation bias, though rooted in a long research tradition[62,63], has only recently been applied to human-AI interaction, where an emerging line of work argues that deliberately slowing down or disrupting automatic acceptance of AI outputs can reactivate independent judgment[64,65]. We argue that intentionally biased AI[1] may produce precisely this kind of productive friction. Specifically, bias may lower the credibility of AI outputs, and the resulting grounded skepticism may disrupt passive consensus and stimulate constructive provocation, disagreement, and critical scrutiny, which could ultimately preserve or elevate overall human-AI collaborative performance. Existing literature from the social sciences also demonstrates how deliberately introducing strategic biases can enhance decision-making by reactivating human critical thinking. Kunda's motivated reasoning framework argues that activating accuracy motivations (i.e., the urge to find a correct answer) or directional motivations (i.e., the urge to defend a preferred answer) tends to increase cognitive effort[66], suggesting that purposely biased AI may heighten humans' engagement by motivating them to challenge competing views from AI[67,68]. Mercier and Sperber's argumentative theory of reasoning suggests that overtly biased AIs may be experienced as interlocutors that invite rebuttal and critical scrutiny, preventing the overly compliant, "sycophantic" drift of AI assistants[69,70] and, in turn, improving users' reasoning.

[1] Throughout this paper, "biased AI" refers to AI systems whose outputs reflect specific cultural, political, moral, or ideological positions. That is, AI with a stance that departs from neutrality. This is distinct from "bias" in the fairness-harms sense (e.g., racial, gender, or demographic discrimination), which is not what we study or advocate.

If perceived bias and disagreement stimulates more active human reasoning, then institutionalizing disagreement across multiple AI systems may further enhance the collaborative potential of human-AI interaction by giving them access to more perspectives against which they can calibrate their own engaged judgment. In this way, we extend the discussion on biased AI-assistant design by investigating the situation in which a single user collaborates with two AIs. This question is increasingly important in practice, as recent research suggests that AI users are increasingly relying on not one, but multiple AI models to generate competing opinions or configure more complex AI agent institutions[71], such as actor-critic architectures, where one agent proposes and another critiques[72–74]. Team process research demonstrates how perspective diversity and well-managed dissent lead to superior collective human outcomes[75,76], which may likewise benefit human users exposed to combinations of biased AI assistants. Moreover, as classic microsociological theory suggests, when two members of a triad are in conflict, the third can turn their disagreement to its own advantage[77,78]. We posit that humans working with multiple, distinct AI agents may occupy this structurally advantageous position, leveraging the tension between opposing perspectives to exercise independent judgment and triangulate toward more accurate evaluations.

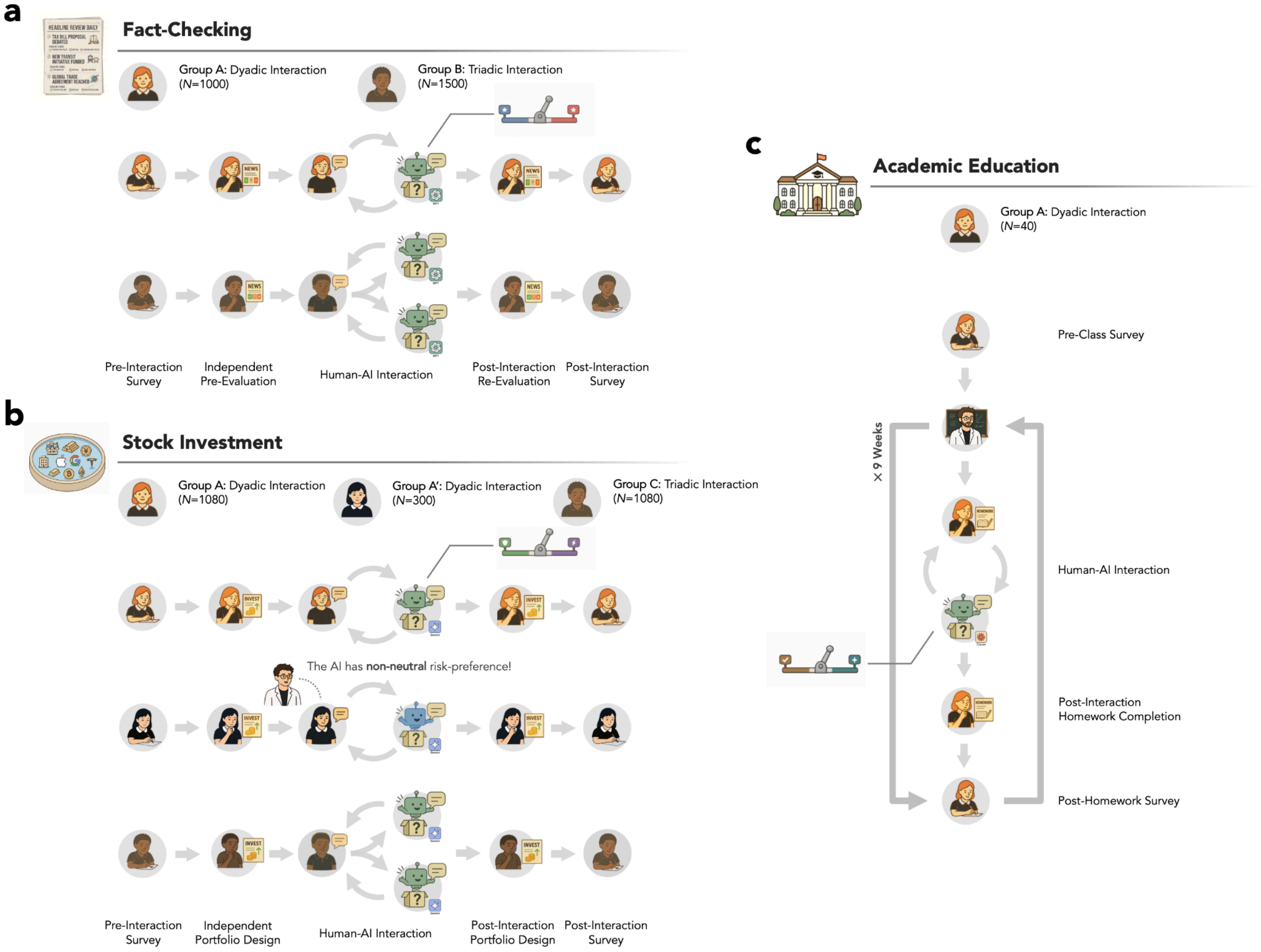

**Fig. 1 | Experimental designs across three consequential human–AI interaction settings. a,** Fact-checking experiment. Participants evaluated the veracity of news content before and after conversing with a single AI or two AIs whose political stances were randomized. **b,** Stock-investment experiment. Participants designed an investment portfolio before and after interacting with a single AI with a randomized risk preference, a single default AI that participants were falsely told had a steered risk preference, or two AIs whose risk preferences were jointly configured relative to the participant's own (see Methods). **c,** Academic-education field study. Over nine weeks, students attended weekly classes and completed coding and memo assignments in a class while interacting with an AI tutor; the tutor's working style was experimentally varied in two randomly selected weeks (see Methods).

To test these hypotheses, we conducted three randomized experiments, including two large-scale online experiments and a nine-week field experiment, spanning markedly different domains and task types: misinformation evaluation, stock investment, and graduate education (Fig. 1, a–c). We chose these domains for three reasons. First, each admits a salient axis along which AI bias can be characterized and experimentally steered: political stance, risk preference, and working style, respectively. Biases along these axes represent systematic deviations from neutrality that AI engineers often seek to remove in their debiasing efforts. Second, each is a high-stakes, real-world setting in which AI systems are increasingly deployed[79] and in which bias is especially consequential. Third, each is a demanding test bed: prior work suggests that default frontier LLMs offer limited benefit for human fact-checking[15], portfolio personalization[80], and active learning[81]. Across the three experiments, we enrolled 5,000 participants—4,960 online and 40 graduate students in the field experiment. In the fact-checking and investment experiments, each online participant assessed the veracity of news headlines or designed a stock portfolio for a given investment horizon, aided by one or two GPT-4o or Gemini 3.1 Pro assistants instructed to hold randomized political stances or risk preferences, yielding 9,960 human–AI conversations. In the education experiment, students completed weekly coding and research-memo assignments with a Claude Code assistant whose working style was experimentally varied in two randomly selected weeks, yielding 799 student–AI sessions.

We pre-registered the designs, hypotheses, and analysis plans for all experiments (Methods). Online participants were recruited from CloudResearch Connect. Study 1 examined fact-checking (Fig. 1a). We recruited 2,500 U.S. participants, balanced by sex and roughly representative of national partisan affiliation[82,83], randomly assigned to converse with one AI assistant (dyadic; $N = 1{,}000$) or two (triadic; $N = 1{,}500$) whose political stances were randomized across six levels, from extremely Republican-leaning to extremely Democrat-leaning. Each participant independently rated the veracity of three news headlines drawn from an 18-headline pool curated from post-2024 PolitiFact and Snopes fact-checks, conversed with the assistant(s), and then re-rated the same headlines. Primary outcomes were post-interaction accuracy (weighted by confidence), conversation length, and perceived improvement (Supplementary Information §1.8).

Study 2 examined stock investment (Fig. 1b). We recruited 2,460 U.S. participants across three waves, screened for a minimum level of stock-investment knowledge and balanced by sex. Participants were randomly assigned to one assistant with a randomized risk preference (dyadic; $N = 1{,}080$), one default assistant falsely described as risk-biased (dyadic; $N = 300$), or two assistants in default, opposing, or balanced configurations (triadic; $N = 1{,}080$; Methods). Before and after AI interaction, each participant independently constructed a portfolio for a 14-day investment horizon from a fixed menu of 25 stocks spanning a range of risk levels. Primary outcomes were post-interaction risk adjusted returns (active $M^2$), number of reply turns, and confidence in portfolio performance; active $M^2$ was computed from realized market prices over the 14 days following each wave, compared relative to the S&P 500 ETF (exchange traded fund). We also evaluated portfolios under counterfactual market regimes (SI §2.4).

Study 3 examined graduate education in a nine-week field experiment (Fig. 1c). Forty students who freely enrolled in MACS 37005 (AI Agents for Social Science & Society) at the University of Chicago completed weekly coding and research-memo assignments with a Claude Code assistant, yielding 312 student-week observations; the assistant's working style was default for seven weeks and randomized to novelty-biased, reliability-biased, or neutral in two randomly selected weeks. Primary outcomes were assignment grades, number of reply turns, and perceived helpfulness; each assignment was graded independently by two of four course teaching assistants and an AI grader (Gemini 3.1 Pro) who were all blind to the treatment assignment, and we analyzed average grades (SI §3.3).

**Biased AI improves or preserves human performance**

To assess how dialogue with a biased AI influenced objective performance, we regressed post-interaction performance on treatment condition, adapting the specification to each design (Methods). Across all three experiments, biased AI assistance matched or exceeded the default assistant's contribution to human performance. Interestingly, the domains differed in the condition under which the benefit most strongly appeared. In the fact-checking experiment (Fig. 2a), where the task has a discoverable ground truth, the benefit was largely direction-general: three of the four biased assistants improved veracity evaluations relative to the default—the somewhat Republican-leaning assistant by 9.5% (95% CI [2.9%, 16.6%]; $\Delta = 0.058$, Hedges' $g = 0.23$, adj. $p = 0.001$), the extremely Republican-leaning assistant by 9.1% ([2.2%, 16.5%]; $\Delta = 0.055$, $g = 0.22$, adj. $p = 0.002$) and the extremely Democrat-leaning assistant by 6.0% ([−0.8%, 13.2%]; $\Delta = 0.036$, $g = 0.15$, adj. $p = 0.037$)—with only the somewhat Democrat-leaning assistant showing no detectable effect (adj. $p = 0.777$). These benefits persisted, modestly attenuated, after controlling for the assistants' own judgment accuracy (SI Table S14). To probe the mechanism, we classified each exchange by whether the participant moved toward or away from the assistant's judgment, and whether that movement helped or hurt (Extended Data Fig. 1a; SI §1.5). Partisan assistants steered participants toward truth chiefly by reshaping persuasion rather than backfire: relative to the default, they reduced the odds of negative persuasion by 41% (OR = 0.59, [0.42, 0.84], $p = 0.003$) and raised the odds of positive persuasion by 40% (OR = 1.40, [1.02, 1.93], $p = 0.036$), leaving backfire rates unchanged.

In the stock-investment experiment (Fig. 2b), biased AI was beneficial, or at least preserved decision quality relative to the default AI, depending on whether its bias aligned with the market regime in the weeks following the experiment—for example, whether the market was risk-on or risk-off. In the relatively neutral market of Wave 1 (8–15 May 2026), biased conditions did not show significant difference from the default (all adj. $p \geq 0.078$). In the defensive-favouring market of Wave 2 (23–25 May 2026) and Wave 3 (8–10 June 2026), the regime-aligned bias paid off: an extremely risk-averse AI added 0.85% of financial performance (active $M^2$) in Wave 2 ([0.32%, 1.38%], $g = 0.47$, adj. $p = 0.007$) and 2.01% in Wave 3 ([1.27%, 2.75%], $g = 0.98$, adj. $p < 0.001$; somewhat risk-averse, 1.09%, $g = 0.53$, adj. $p = 0.003$). These gains came from what the assistants recommended. Accounting for each assistant's own realized portfolio performance absorbs the arm differences completely, and does so through the systematic shift the persona induces rather than through idiosyncratic variation in quality of advice (SI §2.7). As in the fact-checking experiment, these gains took the form of a shift in persuasion: relative to the default, risk-averse assistants raised the odds of positive persuasion (OR = 2.696, [1.565, 4.645], $p < 0.001$) and lowered those of negative persuasion (OR = 0.378, [0.234, 0.613], $p < 0.001$). Importantly, market-misaligned bias was costless: risk-seeking advice changed neither positive (OR = 1.106, $p = 0.724$) nor negative persuasion (OR = 1.055, $p = 0.818$), and left realized performance no different from the default (all adj. $p \geq 0.360$). This held even though risk-seeking AIs consistently produced the least mean-variance-efficient portfolios (GRS $F$ vs. default, [0.041, 0.042], adj. $p < 0.001$) and, in Wave 2, recommendations 1.12% to 1.58% worse than the default's on realized active return (adj. $p \leq 0.010$): participants realized only a fraction of this gap (0.38% to 0.41%). If we simulate counterfactual risk-on or bull-market returns (a market that gained about 4% over two weeks, rather than the falling one participants met), identical Wave-2 and Wave-3 portfolios instead place the risk-seeking arms above the default (Δ = 0.96–1.23%, $g$ = 0.72–0.93, all adj. $p < 0.001$; SI §2.6).

We note that the benefit conferred from biased AI required biased content, not merely the belief that an AI assistant was biased. Labelling a default assistant as risk-averse or risk-seeking, without actually altering its risk preference, left performance unchanged (all $p \geq 0.480$; Fig. 2e). Beyond portfolio performance, risk biased assistants also dampened the well-documented AI homogenization effect[84,85]: default assistance sharply reduced cross-participant portfolio diversity (Δ = −0.175, [−0.207, −0.142], adj. $p < 0.001$), whereas risk-averse assistance offset 36% of that homogenization (difference = 0.062, adj. $p = 0.006$) and risk-seeking assistance 91% (difference = 0.160, adj. $p < 0.001$), leaving diversity statistically indistinguishable from the pre-interaction baseline (−0.015, [−0.041, 0.009], adj. $p = 0.117$; Fig. 2d).

The classroom experiment (Fig. 2c; 12 January–14 March 2026; 312 student-weeks) extended this pattern to a field setting, showing that intentionally biased AI can improve classroom performance compared to default AI. Coding grades were highest under novelty-biased assistance, 6.1% above the default ([0.1%, 12.1%]; Δ = 0.541, $g = 0.48$, adj. $p = 0.093$), with a smaller, insignificant difference under reliability bias (Δ = 0.328, adj. $p = 0.229$). Memo grades rose 1.8% under biased assistance pooled across directions ([−0.2%, 3.8%]; Δ = 0.157, $g = 0.169$, $p = 0.080$), with no detectable difference between biased directions (Δ = 0.010, $p = 0.963$). These are intention-to-treat estimates: assistants were assigned by week but consulted voluntarily. The advantage concentrates in the more demanding weeks. Re-weighting each student-week by the volume of material in that week's assignment template—a property of the assignment fixed before any assistant was assigned, and therefore unaffected by treatment—raises the novelty-biased coding advantage to Δ = 0.663 ($g = 0.58$, $p = 0.014$). All template-derived weighting variants place it between Δ = 0.599 and 0.743 (all $p < 0.03$). The memo effect is unchanged under the same reweighting. Even on that conservative basis, all biased arms were directionally positive and none underperformed the default. This provides convergent field evidence for the controlled findings, from a deployment powered for large effects (minimum detectable $g = 0.67$ against an observed $g = 0.48$).

These comparisons leave unexamined whether the assistant's bias agreed or disagreed with the person it was advising. We therefore reclassified each interaction by the assistant's bias relative to the participant's own pre-interaction position (Fig. 2f): echo-chamber bias, leaning toward the human subject's position, versus opposition bias, leaning away. We found no consistent direction effect, suggesting that the benefit of bias robustly transcended any particular bias. Opposition bias assisted most in fact-checking, regime-aligned echo-chamber bias assisted most in investment, and neither pairing was detectably better in the classroom. In every experiment, however, none of the biased conditions fell below the default assistant (Extended Data Fig. 4; SI §5.1).

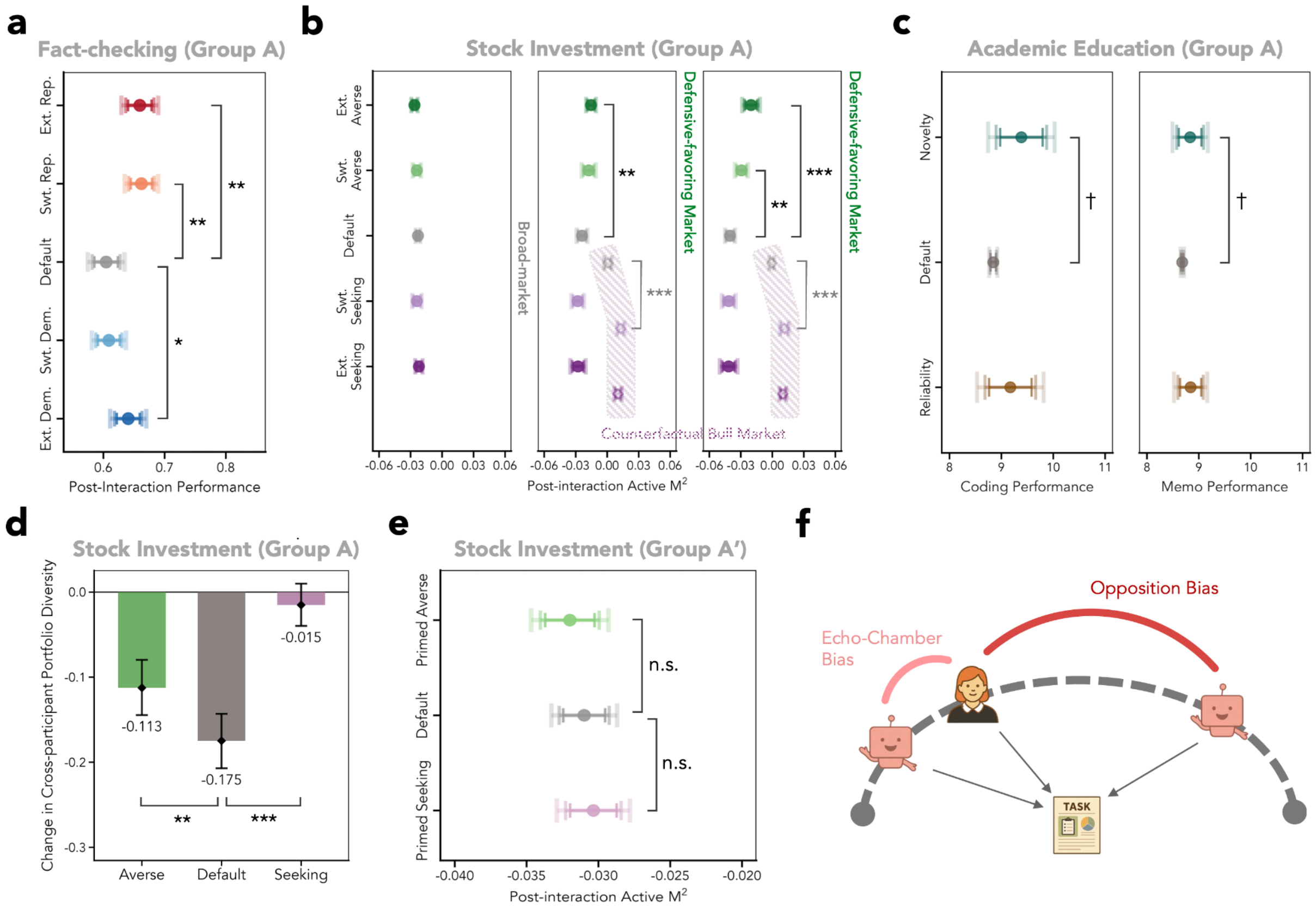

**Fig. 2 | Biased AI improves or preserves human performance. a,** Fact-checking performance by AI political stance (Ext = "extreme"; Swt = "somewhat"). **b,** Financial performance (Active $M^2$) by AI risk preference across the three waves; semi-transparent estimates show a bull-market counterfactual. **c,** Assignment performance by AI tutor working style. **d,** Change in cross-participant portfolio diversity (Ext and Swt conditions pooled). **e,** Financial performance for participants falsely primed that a default AI was biased (Group A′ in Fig. 1b). **f,** Bias direction defined relative to the participant: echo-chamber (toward the participant's position) versus opposition (away from it). Points and bars denote means; nested intervals (**a**–**c**, **e**) denote 90%, 95% and 99% confidence intervals, other error bars 95% confidence intervals. All tests two-sided: *$p < 0.05$; **$p < 0.01$; ***$p < 0.001$; †$p < 0.1$; *n.s.*, not significant.

## Scrutiny, not deference, drives the benefit of biased AI assistance

Across domains, users disliked biased AI and believed it worsened their decisions, even though biased AI actually improved, or at least preserved, their performance. In fact-checking (Fig. 3a), veracity evaluations improved with bias magnitude irrespective of direction (strong versus default: Δ = 0.047, [0.012, 0.082], adj. $p = 0.003$, $g = 0.190$), and participants wrote more to the assistant (Δ = 7.246 words, [2.234, 12.258], adj. $p = 0.002$, $g = 0.296$). Yet they perceived reduced benefit as the bias grew (Δ = 0.553, [0.118, 0.987], adj. $p = 0.036$, $g = 0.305$), increasingly construing the assistant as an influencing agent rather than an assistive tool (Fig. 3f) and growing less willing to recommend it (Δ = 0.779, [0.224, 1.308], adj. $p = 0.006$, $g = 0.429$).

In the investment experiment, engaging with a risk-seeking assistant with varied bias magnitude performed no worse than the default even though they gave poor portfolio advice, relative to the future market. We found that such assistants drew participants into more conversation (66.9% versus 57.6%; Δ = 0.094, [0.005, 0.182], $g = 0.19$, $p = 0.040$) and left them less confident in their revised portfolios (Δ = −0.284, [−0.568, −0.013], $g = -0.23$, $p = 0.037$). As in the fact-checking experiment, participants were also more likely to construe these AI assistants as influencing agents rather than assistive tools compared to the default (Fig. 3g). By contrast, assistants that avoided risk offered participants better advice during a bear market (poorly performing market returns), but participants did not engage with them more. Reduced engagement is consistent with risk-averse advice feeling intuitively appropriate in a defensive-favoring market, where loss aversion makes protective recommendations easy to accept without a second thought[68,86].

In the classroom, novelty-biased tutors raised coding grades relative to the default, yet they were rated less helpful among students who used them (Δ = −1.056, [−1.872, −0.240], $g = -1.14$, adj. $p = 0.025$). Although reliability-biased assistants showed no grade advantage (adj. $p = 0.229$) and no helpfulness penalty among students (adj. $p = 0.450$), they nevertheless drew students into longer exchanges (Δ = 0.610, [0.031, 1.188], $g = 0.42$, adj. $p = 0.078$) with no worse performance. Our follow-up annotation analysis of the prompt logs is consistent with a compositional account: reliability-biased exchanges contained a smaller share of verification and independent-work prompts than novelty-biased ones (25.1% vs 44.4%; Δ = −0.194, exact permutation $p = 0.055$; Extended Data Fig. 2d), hinting that engagement with reliability-biased assistants may have been less self-directed.

A verification-cost account explains this pattern across domains. A user who delegates a task (Fig. 3d) can either accept the AI assistant's answer or pay in time and cognitive effort to engage deeply with it, an evaluation that keeps the answer or overturns it. Cues offering lower assurance in answer quality, such as overt partisanship, risk-seeking recommendations, or unfamiliar suggestions, shift the balance toward checking: users scrutinize responses they would otherwise have accepted, and because human verification catches more errors than it introduces whenever the assistant is good but not near-perfect, aggregate error falls (Fig. 3e, red). Cues that raise assurance do the reverse: anticipated errors fall, the expected value of checking drops below its cost, and users rationally stop checking (Fig. 3e, blue). This yields an irony of automation. Because scrutiny is triggered by perceived rather than actual fallibility, making an assistant appear more neutral or authoritative lowers joint human–AI performance whenever it tips users into accepting answers that checking would still improve. An air of unbiased authority lures users away from precisely the cognitive engagement that could catch the assistant's errors. We formalize this argument in SI §4.

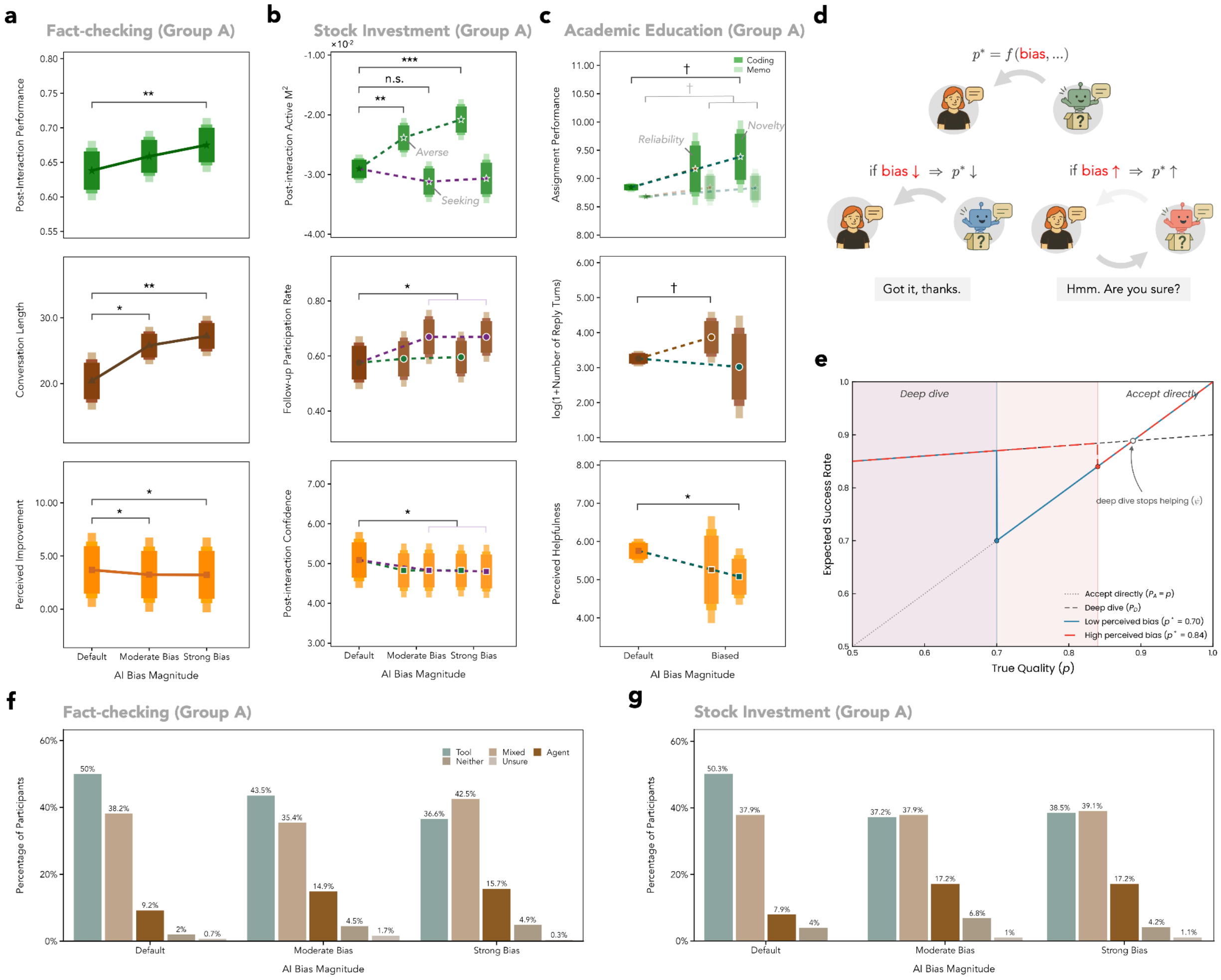

**Fig. 3 | Stronger AI bias lowers subjective evaluation while improving or preserving objective performance. a–c,** Objective performance (top), conversational engagement (middle) and subjective evaluation (bottom) by AI bias magnitude in the fact-checking (**a**, $N = 814$), investment (**b**, $N = 864$) and academic-assignment (**c**, $N = 39$) experiments; dashed lines in **b** and **c** distinguish bias directions (risk-averse versus risk-seeking; reliability versus novelty-biased). Points are model-adjusted marginal means; nested boxes are 90%, 95% and 99% confidence or credible intervals. **d,** Proposed mechanism: perceived bias raises the scrutiny cutoff $p^*$, moving users from unquestioned acceptance to a costly deep dive that keeps or overturns the AI answer. **e,** Expected success rate by true response quality ($p \in [0.5, 1]$) under low ($p^* = 0.70$) and high ($p^* = 0.84$) perceived bias. Deep-diving ($P_D$, dashed) outperforms direct acceptance ($P_A = p$, dotted) below $\psi$; higher perceived bias extends scrutiny, raising success over $p \in (0.70, 0.84]$ and damping the pass-through of AI quality to final performance ($q + r - 1 = 0.1$ vs 1). Parameters: $q = 0.9$, $r = 0.2$, $c / (G + L) = 0.044$ (SI §4). **f, g,** Perceived role of the AI by bias magnitude in fact-checking (**f**) and investment (**g**). All tests two-sided. $*p < 0.05$; $**p < 0.01$; $***p < 0.001$; $†p < 0.1$; *n.s.*, not significant.

## Augmenting assistance with multi-AI configurations

A single biased assistant improved or preserved objective performance, but at two costs. In all three experiments, participants penalized the biased assistant in their subjective assessments. And in the financial experiment, single biased agents conferred less advantage when misaligned with the market. We therefore tested a dual-AI design in both fact-checking and investment experiments where participants interacted with two assistants at once (Fig. 4a): two default (Dual Default), two positioned on opposite sides of the participant's own view (Dual Balanced), or two positioned against the participant (Dual Opposition).

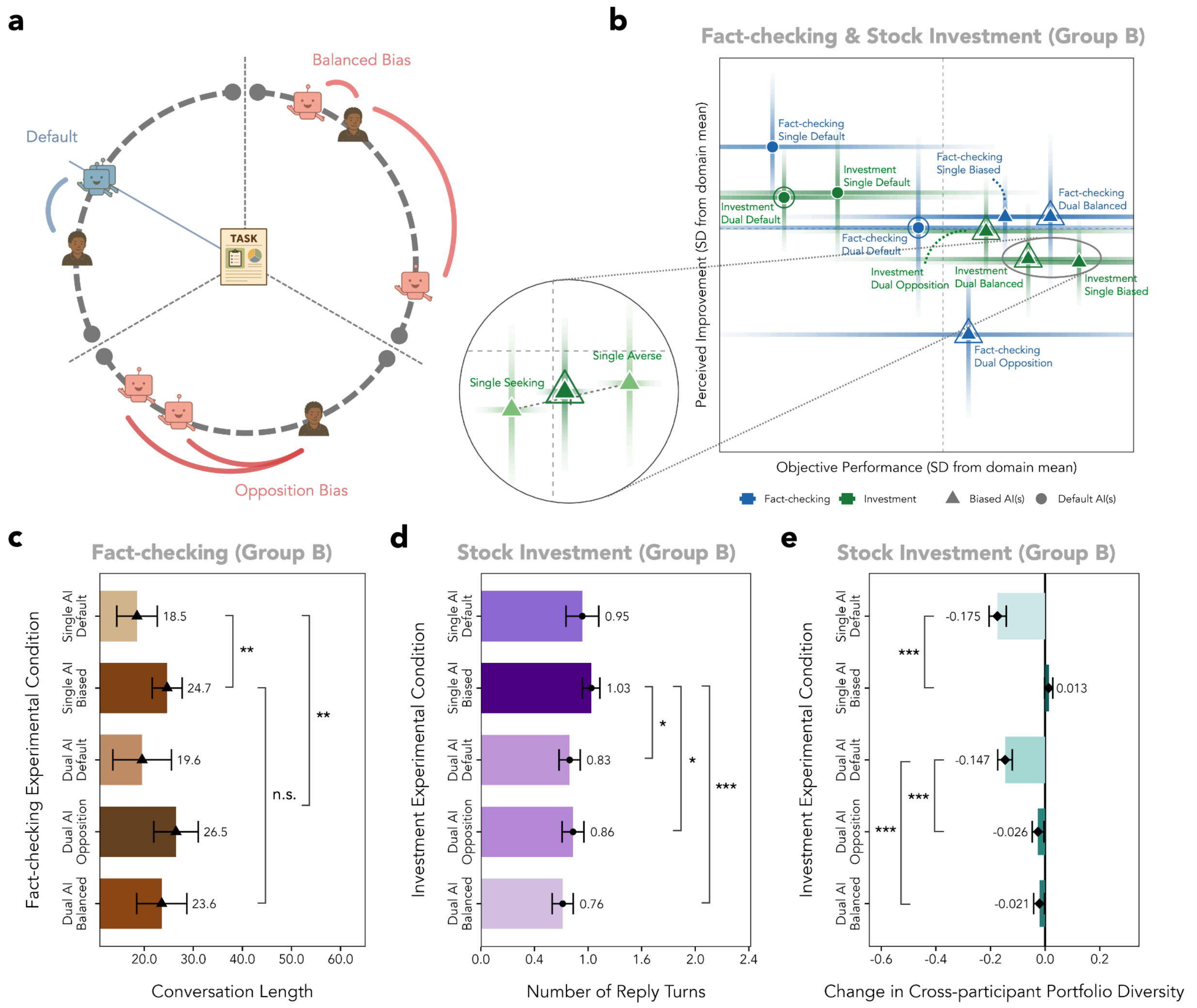

**Fig. 4 | Balanced dual AIs preserve performance gains while mitigating the risks of one-sided bias. a,** Dual-AI configurations defined relative to each participant's position: balanced, opposition, default. **b,** Objective performance versus perceived improvement, as deviations from the domain mean in standard deviations; perceived improvement is the self-reported item in fact-checking and confidence change in investment. Blue, fact-checking; green, investment; triangles, biased AI(s); circles, default; rings, dual-AI sessions; graded whiskers span the 90%, 95% and 99% confidence or credible intervals. Inset resolves the

single risk-averse and risk-seeking conditions about their midpoint (grey cross). **c,** Conversation length by fact-checking condition. **d,** Number of reply turns by investment condition and **e,** change in cross-participant portfolio diversity by investment condition; negative values in **e** indicate homogenization. Bars and points, adjusted means; error bars in **c–e,** 95% confidence intervals. Tests are two-sided. *$p < 0.05$; **$p < 0.01$; ***$p < 0.001$; †$p < 0.1$; *n.s.*, not significant.

Dual balanced AIs addressed both costs (Fig. 4b). In fact-checking, the balanced pair reached the highest performance (0.649 versus 0.607 for the single default; Δ = 0.042, [0.005, 0.079], $p$ = 0.028). Perception separated the configurations far more sharply: a single biased assistant depressed perceived improvement (Δ = −0.566, [−0.990, −0.146], adj. $p$ = 0.030), and two oppositional assistants compounded the penalty, falling below both the baseline (Δ = −1.504, [−2.094, −0.909], adj. $p < 0.001$) and the single biased condition (Δ = −0.938, [−1.400, −0.466], adj. $p < 0.001$). The balanced pair avoided this compounding, above dual opposition (Δ = 0.876, [0.189, 1.553], adj. $p$ = 0.030) and indistinguishable from baseline (Δ = −0.628, adj. $p$ = 0.093). In investment, the subjective penalty was likewise not detectably different from the default (Δ = −0.227, adj. $p$ = 0.420). Bracketing offered a special benefit. Because a single biased assistant is effectively a coin flip on alignment with the market regime of the moment, and two one-sided arms (i.e., risk-seeking or risk-averse) differed by 0.87% of annualized active return ([0.59%, 1.14%], $g$ = 0.52, adj. $p < 0.001$), the balanced pair rose above the misaligned arm (Δ = 0.39%, [0.11%, 0.67%], $g$ = 0.23, adj. $p$ = 0.007) landing below the aligned one (Δ = −0.48%, [−0.73%, −0.22%], $g$ = −0.29, adj. $p < 0.001$) at their midpoint. This protection was structural rather than discerning, however: when the two assistants' recommendations diverged in quality, participants with balanced assistants followed the better one no better than chance (adj. $p$ = 0.890).

Crucially, these advantages carried no additional conversational cost: balanced pairs demanded no more of participants than a single AI assistant. In fact-checking, participants paired with the balanced dual assistants wrote no more than those paired with a single biased assistant (adj. $p$ = 0.693, Fig. 4c) and not significantly more than those with the single default assistant (adj. $p$ = 0.158), whereas two oppositional AIs raised conversational effort markedly above the default (Δ = 7.964, [0.747, 15.182], $g$ = 0.33, adj. $p$ = 0.010). In investment, the pattern was the same: participants paired with balanced assistants engaged no more than those with a single default assistant, whether measured by follow-up participation (adj. $p$ = 0.576) or reply turns (adj. $p$ = 0.120, Fig. 4d), and engaged less than those facing a single biased assistant (participation Δ = −0.101, [−0.193, −0.008], $g$ = −0.21, adj. $p$ = 0.022; turns Δ = −0.266, [−0.449, −0.083], $g$ = −0.27, adj. $p < 0.001$). Unlike in fact-checking, dual opposition raised no conversational cost (adj. $p \geq 0.405$). Furthermore, both biased pairings protected portfolio diversity: whereas a single default assistant sharply homogenized portfolios across participants (Δ = −0.175, [−0.209, −0.142], adj. $p < 0.001$, Fig. 4e), balanced and oppositional pairs left diversity intact (Δ = −0.021, [−0.041, −0.001], adj. $p$ = 0.056; Δ = −0.026, [−0.048, −0.005], adj. $p$ = 0.030), offsetting roughly nine-tenths of the homogenization (balanced Δ = 0.154, [0.114, 0.192]; opposition Δ = 0.149, [0.108, 0.189]; both adj. $p < 0.001$).

## Discussion

Are non-neutral, "biased" AI systems always harmful, and should they be avoided? Landmark studies evidence that bias inflicts measurable harm and argues that they must be eliminated[87,88], an imperative echoed by technical fairness frameworks[89], audit-based governance proposals[90], and recent survey and ethics literature[91,92]. Yet this line of work has largely treated bias as a property of AI output, overlooking how bias shapes actual interactions between humans and AI and, in turn, the performance of the human–AI system as a whole. In contrast, we reframe the bias of AI as an intentional design lever for productive friction, which can mitigate the unintended consequences of contemporary AI design, such as moral deskilling, cognitive laziness, sycophancy bias, and cultural homogenization. Across three dramatically different tasks, we demonstrate that biased AI assistants outperform a default, non-biased baseline, enabling users to achieve superior performance, reduce homogenization, and enhance engagement. Paradoxical as they may seem, our findings align with two well-established theoretical traditions. Anthropomorphism theory[93,94] and its recent application to LLMs[95], together with the "computers-as-social-actors" framework[96], hold that social cues prompt mind attribution, transforming a machine from neutral tool into active interlocutor whose outputs users interrogate rather than accept uncritically. Here a biased AI stance plays that role. In this way, both research traditions point to an AI agent version of automation bias where AI performance reduces collaborative human-AI performance if it leads users to stop critically engaging with AI output. We captured this mechanism in our theoretical model (SI §4). Taken together, these findings undermine conventional assumptions that an ideal AI partner must be intrinsically neutral and risk-free. Instead, our findings suggest that calibrated, epistemically grounded bias should be viewed as a tunable hyperparameter for optimizing fair and desirable human-AI collective outcomes.

The perception penalty we observe with biased AI assistants admits two interpretations. On one hand, it suggests a design paradigm in which AI is tuned for collective welfare rather than maximal human user satisfaction. If users accept lower perceived usefulness, the human-AI collective can realize "unexpected" gains of desired performance. On the other hand, overt bias carries an adoption cost. Before any downstream benefit is felt, users form less favorable impressions of the AI assistant and its help. In this study, participants who interacted with biased assistants were more likely to suspectly view AI as an agent trying to sway their decisions, and their willingness to recommend it for fact-checking dropped relative to the default baseline. This skepticism poses a practical hurdle to deploying grounded bias: diminished appreciation can shrink the AI user base and feed anti-technology sentiment or conspiratorial narratives about the forces behind AI.

Another key element of our study probes simultaneous interaction with two AI assistants, a setup inspired by micro-sociological theory about the instability of triads over dyads, as well as ChatGPT's experimental "which response do you prefer?" feature, which presents users with two alternative answers side-by-side. As distinct models proliferate, human users increasingly consult multiple AIs in daily work[74]. As human-AI interaction expands from a dyad to three or more parties, social and influence dynamics become increasingly complex, amplifying uncertainty yet simultaneously unlocking new opportunities. The potential of such a design is evidenced in our experiments, demonstrating that participants who interacted with two AI assistants holding biases that bracketed their own retained the performance gains of single biased assistance while avoiding both of its costs: the subjective penalty of perceived utility and the exposure to a misdirected, one-sided bias. The stance-balanced dual-AI condition operationalizes sociologist Georg Simmel's *tertius gaudens*, meaning "the third who enjoys". The human user, positioned between two opposing AIs, benefits from their disagreement by exercising independent adjudication, while the distributed social pressure ensures that enjoyment and comfort are not casualties of deeper processing[77]. Acting as an arbiter between equal but opposing voices, human users cast the deciding vote, are always in the majority, attain higher accuracy and feel more empowered. Most research on multi-agent coordination has focused on fully automated, human-out-of-the-loop settings[97–100]. Yet the advantages of interfaces that allow users to orchestrate exchanges among multiple AIs should not be overlooked. Our study offers an early demonstration of this user-in-the-loop approach to multi-agent design and opens the door to systematic investigations of human-multi-AI teaming strategies, informed by decades of work on human-human collaboration.

Our results suggest that the pursuit of neutral, performatively authoritative AI may inadvertently undermine human cognitive engagement and performance. Instead, strategically calibrated bias offers a new, more effective path toward beneficial human-AI collaboration that engages cognitive agency for improved performance.

## Methods

*Ethics and consent*

All three experiments were approved separately by the Institutional Review Board of the University of Chicago (protocols IRB24-1914, IRB25-2216, IRB26-0283), and all participants gave informed consent. Online participants were recruited through CloudResearch Connect. In the fact-checking experiment, participants were paid \$3.00 (single-AI) or \$3.50 (dual-AI) for median completion times of 13.3 min and 15.6 min, respectively, corresponding to approximately \$13.40 per hour in both conditions. Participants were debriefed with the verified veracity of every headline they encountered as well as the political bias of encountered AI(s). In the investment experiment, all participants were paid \$3.50 for a median of 20.6 min, approximately \$10.20 per hour. Portfolios were hypothetical, and participants were told that the AI assistants' outputs did not constitute financial advice. Participants in Group A′ were told their assistant had been steered toward a non-neutral risk preference when it had not; this deception was approved under IRB25-2216 waiver, and all, including the risk-bias of the AI(s), were debriefed immediately afterwards and offered withdrawal of their data. In the classroom experiment, signed consent to research use of coursework and interaction logs was administered separately from instruction and grading; students were told their decision would not affect grades, standing or access to the assistant, and consent status was withheld from instructors and graders until final grades were submitted. Every student received AI assistance in every week, with only the assistant's bias varying (40 of 45 enrolled students consented to research use of their data). Assignments were graded by four teaching assistants, all blind to treatment assignment; two were members of the research team and aware of the study's objectives, and two were unaware of both.

*Pre-registration*

The fact-checking experiment was registered at the Open Science Framework (OSF; dgx3w) on 12 January 2025, before data collection began on 15 February 2025. The registration specified hypotheses and the data-collection protocol but did not include an analysis plan; the model specifications reported here were therefore chosen after data collection. The same hypotheses, experimental structure and analysis approach were carried over to the investment and classroom experiments, both registered in full (design, hypotheses and analysis plans) before data collection began. The investment experiment was registered at OSF (rxds5) on 21 March 2026, with data collected from 6 May 2026; the classroom experiment at OSF (3edgf) on 5 Jan 2026, with data collected from 12 January 2026.

In the fact-checking experiment, all reported analyses are exploratory. In the investment experiment, the bull-market counterfactual analyses are exploratory; all other reported analyses follow the pre-registered specifications. In the graduate education experiment, analyses that weight or condition on assistant use and the delegation-annotation analyses are exploratory; all other reported analyses follow the pre-registered specifications.

*AI assistants and interaction interface*

Models: Assistants in the fact-checking experiment were built on GPT-4o (gpt-4o-2024-11-20) and in the investment experiment on Gemini 3.1 Pro (preview), accessed through the OpenAI and Google APIs. In the classroom experiment, students interacted with the assistant through Claude Code using whichever backend model they had selected from those available to regular users; the model in use was logged per session (SI §1.2).

Steering: Dispositions were induced through system prompts alone, with no fine-tuning or output filtering. Each experiment manipulated a single axis of bias, at two degrees of intensity: political stance (fact-checking), risk preference (investment) and working style (graduate education). Full prompt texts for every condition appear in SI §1.2.

Control conditions: The default assistant, which received the task instruction with no disposition specified, served as the pre-registered baseline. The neutralized assistant was additionally instructed to withhold directional judgment and present balanced considerations, operationalizing the ideological-neutralization principle this work examines; it served as an alternative control and is reported in Extended Data Fig. 4. In the classroom experiment, the neutralized style was one of three dispositions randomly assigned in treated weeks, with the default assistant in use otherwise.

Generation parameters: Responses were generated at temperature 0.7 with API-default top-k, capped at 250 tokens per generation in the fact-checking experiment and uncapped in the investment experiment.

Interface: Participants conversed with the AI(s) through a custom web interface and did not learn whether they would interact with one or two assistants before participation. Every exchange opened with the assistant stating its own assessment of the news item or the user's initial portfolio; in dual conditions, both assistants responded in parallel within each turn, each able to see all messages from earlier turns but not the other's response in the current one. Participants could then exchange up to three turns. In the fact-checking experiment, assistants were not shown participants' independent pre-interaction judgments, to prevent sycophantic agreement. In the investment experiment, assistants received the participant's pre-interaction portfolio and were asked to revise it, mirroring how financial advisers work from a client's existing allocation.

*Fact-checking experiment*

We assembled a stimulus set of 18 headlines through a three-stage screen. From all claims fact-checked by PolitiFact and Snopes between 1 January and 1 November 2024 ($N = 2{,}780$), all published after the GPT-4o-2024-11-20 knowledge cutoff, we retained 180 for which prompted political stance produced significant variance in GPT-4o's reasoning or judgments. We then discarded items GPT-4o rated as overly niche or lacking context for human evaluation, leaving 66, which 160 pilot participants (80 Republicans, 80 Democrats; 15 headlines each) evaluated. The final 18 met three criteria: evaluability, difficulty, and political divisiveness. Two researchers independently verified each headline's veracity against sources other than PolitiFact and Snopes SI §1.1.

Participants were sampled from CloudResearch Connect: U.S. citizens aged 18 or older, quota-sampled to a nationally representative political distribution (30% Democrat, 40% Independent, 30% Republican). The single-AI experiment ran in two waves (15–26 February 2025, $N = 500$; 26–30 May 2025, $N = 500$) and the dual-AI experiment in one (15–27 February 2025, $N = 1{,}500$), following a 20-participant pilot on 3 February 2025. Attentiveness was screened before entry by an LLM-scored open-response item and after completion by duration and bot-probability flags; 61 participants failed and were replaced. Repeat participation was blocked by IP address and platform ID, and paste was disabled for all open-ended responses. Attrition did not differ across conditions (Wald $\chi^2(2) = 0.400$, $p = 0.817$). Eight further cases were removed for missing conversation logs caused by API outages, giving analytic samples of 993 and 1,499.

Participants completed a 14-item pre-interaction survey covering political orientation, news consumption, AI use and attitudes, and self-assessed ability to evaluate online information. They then evaluated three randomly assigned headlines, each displayed with approximately equal frequency. For each headline, participants first rated veracity and confidence independently, then conversed with the AI assistant(s), then re-rated it. Assistants opened each exchange with their own judgment and were instructed to report rather than persuade and to hold their stance throughout, precluding participants from shifting the model. Participants contributed at least one message and advanced automatically after three exchanges. A post-interaction survey measured perceived improvement, perceived meaningfulness and the perceived role of the assistant(s).

For analysis, each assistant's expressed veracity judgment was coded from the transcript by GPT-4o-mini on the same 0–1 scale participants used, then verified by a human coder who adjudicated discrepancies (Cohen's $\kappa = 0.81$), and participant engagement was coded by GPT-4o on five dimensions [101] (SI §1.7). Continuous outcomes (performance, conversation length) were fitted with linear mixed-effects models in lme4/lmerTest by REML[102]; ordinal outcomes (perceived improvement, meaningfulness) with MCMCglmm[103,104] (25,000 iterations, 5,000 burn-in). Estimated marginal means and contrasts were obtained with emmeans[105] and corrected across all pairwise comparisons among treatment arms within each outcome by the Benjamini–Hochberg procedure[106]. Results were unchanged when random intercepts were replaced by participant-clustered robust standard errors, and when covariates showing residual imbalance were added as controls (SI §1.3 and §1.6).

*Stock investment experiment*

Participants allocated a hypothetical $100,000 across a fixed menu of 25 widely held assets, selected for public familiarity and to span a wide volatility range, grouped into five categories: defensive holdings and hedges (SHY, IEF, LQD, GLD, VNQ; typical 14-day price range ±0.5–2%), broad-market and sector funds (SPY, XLF, XLE, XLI, XLP, XLU; ±2–5%), mega- and large-cap technology and growth stocks (AAPL, MSFT, AMZN, GOOGL, NVDA, TSLA; ±3–8%), high-beta small- and mid-cap growth stocks (SHOP, SNOW, PLTR, DKNG, RIVN, CRSP; ±5–15%) and cryptocurrencies (BTC, ETH; ±8–25%). Allocations had to be non-negative, sum to 100%, span at least five assets, and lie between 5% and 40% for any asset held. Participants were shown these category-level volatility and historical-return ranges before allocating (SI §2.1). Cryptocurrency positions were priced in U.S. dollars at a fixed market price at the start of the evaluation period.

Participants were sampled from CloudResearch Connect: U.S. citizens aged 18 or older, quota-sampled to equal proportions by sex. Because the task presumes familiarity with equity markets, applicants first completed a brief screening in which an open-response item was scored by Claude Haiku 4.5, and those falling below the admission threshold were not admitted (SI §2.1). The single-AI and dual-AI conditions were launched together in each of three waves (8–15 May, 23–25 May and 8–10 June 2026), following a pilot on 5 May 2026 ($N = 31$). Across all launches, 33.6% of accepted assignments were returned before completion. Attrition occurred almost entirely before random assignment: among participants who reached assistant assignment, 99.8% (single-AI), 99.4% (priming) and 98.4% (dual-AI) completed the study, and completion did not differ across conditions (dual-AI, $\chi^2(2) = 4.10$, $p = 0.129$; single-AI, Fisher's exact $p = 0.658$). Participants were then filtered by approval status, deduplicated at the participant level, screened for bot-like response patterns (27 single-AI, 8 priming, 32 dual-AI removed), giving analytic samples of 1,042 (single-AI), 302 (priming) and 1,011 (dual-AI).

Participants completed a 21-item pre-interaction survey covering investing experience, financial literacy, risk attitudes, and AI use and attitudes. They were told that their portfolio would be evaluated over a fixed 14-day window by its active $M^2$, a measure of risk-adjusted return relative to the S&P 500 that accounts for the level of active risk taken, and that bonuses would be paid by rank: $3.50 to the top 5%, $1.50 to the next 5% and $0.50 to the following 10%. Each participant then constructed a portfolio independently and rated their confidence in it, before conversing with the assistant(s), which were given that portfolio and asked to revise it, and submitting a final allocation; the initial allocation remained visible during revision. A post-interaction survey re-measured confidence in portfolio performance and recorded self-assessed performance, the perceived influence of the assistant(s), and their perceived role.

Portfolio performance was measured as active $M^2$ over each participant's 14-day outcome window (formula in SI §2.4); the same portfolios, held fixed as of the interaction, were re-scored on all overlapping 14-day windows of the pre-study price history, grouped into bear, neutral and bull regimes by terciles of benchmark return. We also pre-registered the ex ante GRS $F$ test of each portfolio's distance from a mean-variance efficient frontier (Extended Data Fig. 4a), and defined cross-participant diversity as the mean pairwise L1 distance between participants' normalized asset weights within a condition. Continuous outcomes were fitted by ordinary least squares controlling for pre-interaction performance, with wave fixed effects and HC3 robust standard errors; ordinal outcomes (post-interaction confidence, perceived improvement) with MCMCglmm (55,000 iterations, 5,000 burn-in, residual variance fixed); and diversity, a group-level statistic, by participant bootstrap and label permutation. Estimated marginal means and contrasts were obtained with emmeans and corrected within each contrast family by the Benjamini–Hochberg procedure. Results were unchanged when covariates showing residual imbalance were added as controls (SI §2.7).

*Graduate education experiment*

The experiment was embedded in MACS 37005 (AI Agents for Social Science & Society), a graduate course at the University of Chicago taught over nine weeks (12 January to 14 March 2026). Of 45 enrolled students, 40 consented to research use of their coursework and interaction logs. One consenting student attended fewer than half of the class sessions and was excluded, giving an analytic sample of 39 students. Each week, students attended class, completed a coding assignment and a research memo with the aid of a Claude Code tutor, and completed a post-homework survey; a pre-class survey administered before the first session covered. This yielded 312 student-week observations and 799 assistant sessions.

The AI assistant's working style was manipulated in two of the nine weeks, selected at random independently for each student and constrained to fall at least two weeks apart. In those weeks each student was randomly assigned a novelty-biased, reliability-biased or neutralized assistant; in the remaining seven weeks the assistant ran in its default configuration. The manipulation was implemented at the request level: each student's Claude Code traffic was routed through a proxy service (a FastAPI application on AWS ECS Fargate behind an Application Load Balancer) that authenticated the student by a personal bearer token, determined the current course week, and appended the system prompt corresponding to the student's assigned policy to Claude Code's own system prompt before forwarding the request to AWS Bedrock. Students were not told that the assistant's configuration varied across weeks. The backend model was each student's own choice among those available to regular users; because the manipulation was applied at the request level, it was identical across backend models. Every request was logged to an AWS S3 bucket as a full transcript, including user and assistant messages, timestamps, session identifier, backend model, course week, and the policy applied, with per-request metadata recorded in DynamoDB.

Coding assignments and memos were graded on a 10-point rubric distributed in the syllabus in advance (SI §3.3). Each submission was graded independently by two of four teaching assistants, all blind to treatment assignment, and also scored by a Gemini 3.1 Pro grader applying the same rubric. Because the LLM grader tracked human judgement closely on code ($r$ = 0.53, against $r$ = 0.56 between the two humans) but only weakly on writing ($r$ = 0.37), we composed each outcome from the grader set with the highest internal consistency: coding performance is the equally weighted mean of the two human grades and the mean-aligned Gemini grade ($\alpha = 0.76$ vs 0.72 for the humans alone), and memo performance is the mean of the two human grades (adding Gemini lowered $\alpha$ from 0.69 to 0.62). Results under every grader combination are reported in Table S59. The weekly post-homework survey measured students' use of the provided AI service and of alternatives, perceived helpfulness, and purposes of use.

Engagement was measured from platform telemetry (reply turns and words written per student-week) and, for weeks with a conversation, coded by GPT-5.5 on the same five engagement dimensions as the other experiments and on a five-category delegation typology. Because bias was assigned week-by-week within students, primary models were fitted by ordinary least squares with student and week fixed effects and student-clustered robust standard errors, making each student their own control; conversation measures were modelled as $\log(1 + x)$ among weeks in which the assistant was consulted, and the ordinal helpfulness item linearly. Two analyses cannot support student fixed effects and are specified separately in Table S50 and Table S54: the echo-chamber comparison, fitted by ordinary least squares with week fixed effects, and the delegation categories, tested by exact permutation with bootstrap intervals. Estimated marginal means and contrasts were obtained with emmeans and corrected within each contrast family by the Benjamini–Hochberg procedure, treating comparisons against the default assistant as the focal family. Results were unchanged when student fixed effects were replaced by random intercepts or by clustered standard errors alone, and the pre-registered covariates were balanced across arms and absorbed by the student fixed effect (SI §3.2 and §3.4).

**Acknowledgements**

We thank Arina Oberoi and Jesse Zhou for serving as graders in the field experiment conducted as part of MACS 37005, AI Agents for Social Science & Society, a graduate course at the University of Chicago during Winter Quarter 2026.

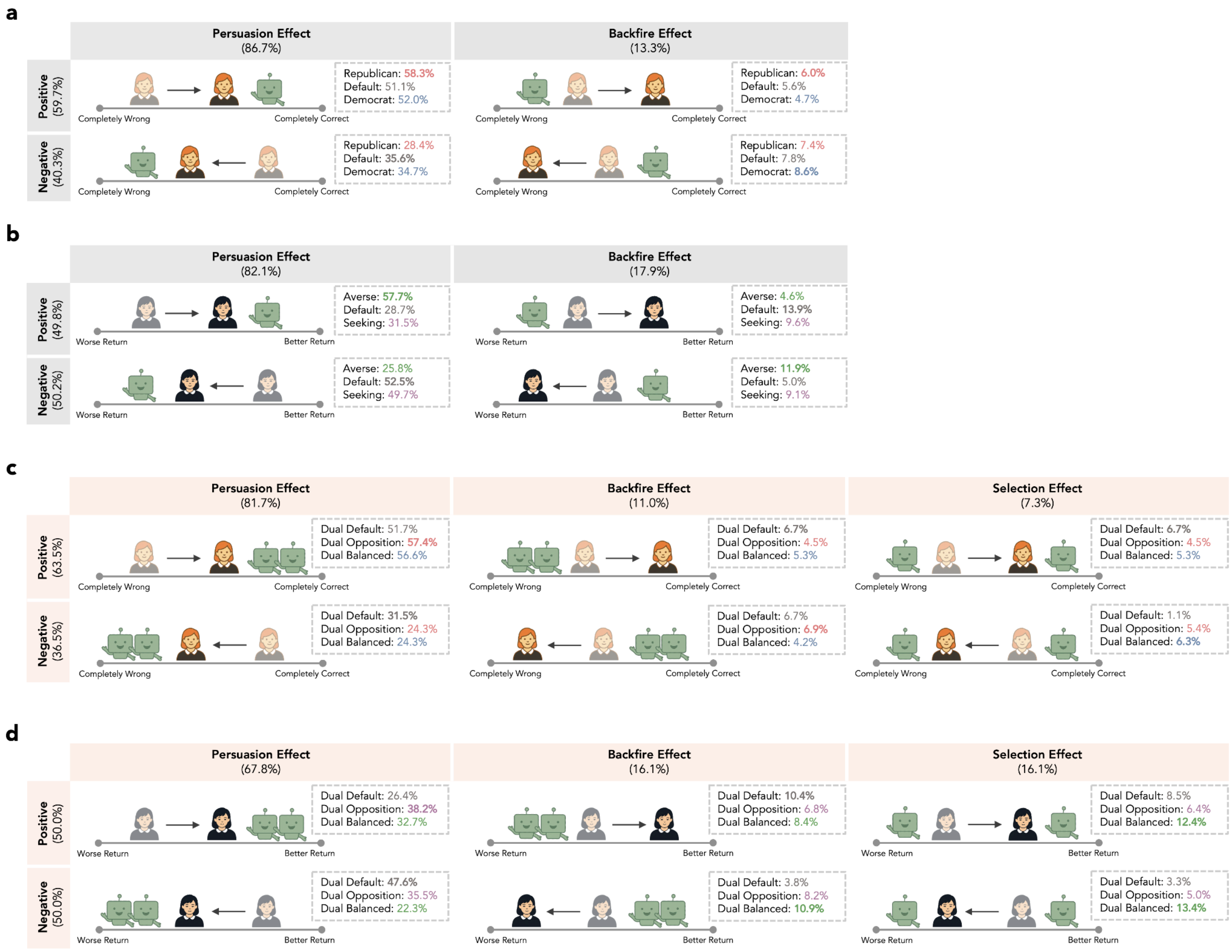

**Extended Data Fig. 1 | Decomposition of participant shifts into persuasion, backfire, and selection.** Interactions in which participants substantively revised their judgment between pre- and post-interaction assessments are classified by the shift's relation to the assistants' advice: persuasion (moving toward the assistant's position), backfire (moving away from it) and, in dual-AI conditions, selection (following one assistant when the two disagreed; persuasion and backfire are defined relative to the assistants' consensus otherwise). Each class is further split by whether the shift improved (positive) or worsened (negative) the participant's outcome: veracity judgment in fact-checking (**a, c**) and realized portfolio return in investment (**b, d**). Column and row headers give the marginal shares of all classified shifts; boxed values give the percentage of each arm's shifts falling in that cell, for single-AI (**a, b**) and dual-AI (**c, d**) conditions. Bold denotes the largest value among the arms within each cell. Cell counts and statistical comparisons across arms are reported in Table S16 and S32.

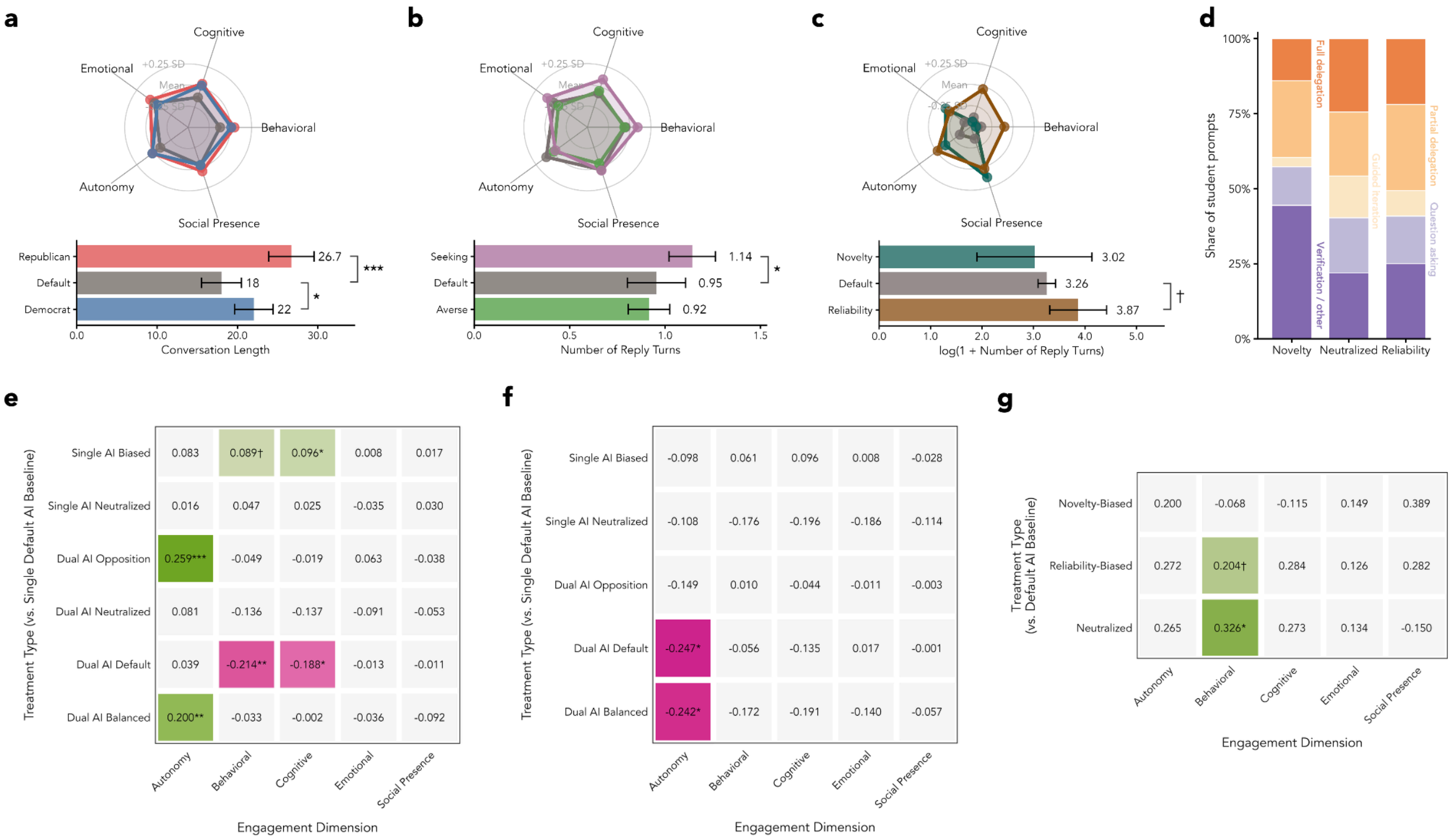

| Fact-checking (e) | Autonomy | Behavioral | Cognitive | Emotional | Social Presence |
|---|---|---|---|---|---|
| Single AI Biased | 0.083 | 0.089† | 0.096* | 0.008 | 0.017 |
| Single AI Neutralized | 0.016 | 0.047 | 0.025 | -0.035 | 0.030 |
| Dual AI Opposition | 0.259*** | -0.049 | -0.019 | 0.063 | -0.038 |
| Dual AI Neutralized | 0.081 | -0.136 | -0.137 | -0.091 | -0.053 |
| Dual AI Default | 0.039 | -0.214** | -0.188* | -0.013 | -0.011 |
| Dual AI Balanced | 0.200** | -0.033 | -0.002 | -0.036 | -0.092 |

| Investment (f) | Autonomy | Behavioral | Cognitive | Emotional | Social Presence |
|---|---|---|---|---|---|
| Single AI Biased | -0.098 | 0.061 | 0.096 | 0.008 | -0.028 |
| Single AI Neutralized | -0.108 | -0.176 | -0.196 | -0.186 | -0.114 |
| Dual AI Opposition | -0.149 | 0.010 | -0.044 | -0.011 | -0.003 |
| Dual AI Default | -0.247* | -0.056 | -0.135 | 0.017 | -0.001 |
| Dual AI Balanced | -0.242* | -0.172 | -0.191 | -0.140 | -0.057 |

| Education (g) | Autonomy | Behavioral | Cognitive | Emotional | Social Presence |
|---|---|---|---|---|---|
| Novelty-Biased | 0.200 | -0.068 | -0.115 | 0.149 | 0.389 |
| Reliability-Biased | 0.272 | 0.204† | 0.284 | 0.126 | 0.282 |
| Neutralized | 0.265 | 0.326* | 0.273 | 0.134 | -0.150 |

**Extended Data Fig. 2 | Biased AI assistance reshape both the quantity and the character of engagement. a–c,** Engagement profiles in the fact-checking (**a**), investment (**b**) and education (**c**) experiments. Radar charts show mean scores on five engagement dimensions coded from transcripts (cognitive, behavioural, emotional, social presence, autonomy; SI §1.7), plotted from the grand mean to +0.25 s.d.; bars show conversational quantity by condition—conversation length (**a**), number of reply turns (**b**) and log(1 + reply turns) (**c**). **d,** Composition of student prompts by type (full delegation, partial delegation, guided iteration, question asking, verification/other) under each working style in the education experiment. **e–g,** Condition-level differences in mean engagement-dimension scores relative to baseline: fact-checking (**e**), investment (**f**), and education (**g**) versus the single default assistant; positive values indicate greater engagement. Bars and points denote means; error bars, 95% confidence intervals. All tests are two-sided. ***$p < 0.001$; **$p < 0.01$; *$p < 0.05$; †$p < 0.1$.

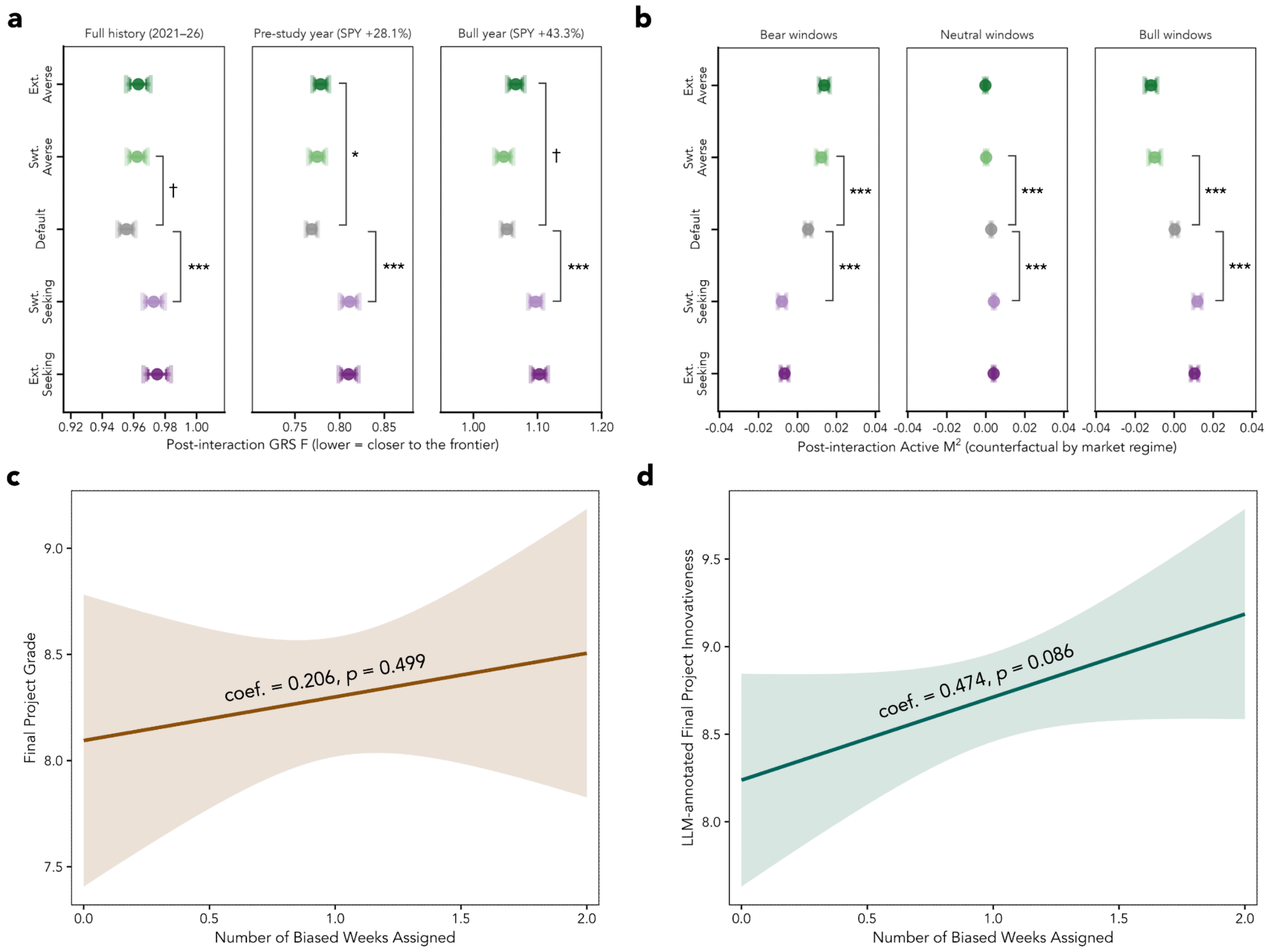

**Extended Data Fig. 3 | Treatment effects under alternative performance metrics, counterfactual market regimes and downstream classroom outcomes. a,** Post-interaction mean–variance efficiency of portfolios by single-AI arm in the investment experiment, measured by the Gibbons–Ross–Shanken (GRS) $F$ statistic, estimated over the full 2021–26 history, the pre-study year and a bull year (SI §2.5). **b,** Post-interaction active $M^2$ for the same portfolios re-scored under counterfactual bear, neutral and bull market windows. **c, d,** Downstream outcomes in the classroom experiment: human-graded final-project essay score (**c**) and final-project innovativeness rated by two LLMs (GPT-5.4 and Gemini 3.1 Pro, averaged; SI §3.3) (**d**), each regressed on how many of the student's two treated weeks were assigned a directionally biased (novelty- or reliability-biased) working style. The project is the unit of analysis, since team projects receive a single grade (**c**, $N = 22$; **d**, $N = 21$), and the predictor is the team mean for multi-member projects. Lines show fitted values from Huber robust regression with 95% confidence bands; p-values are from the same models and from exact randomization inference over assigned dose. Points in **a**, **b** denote model-adjusted means, and nested intervals denote 90%, 95% and 99% confidence intervals. All tests are two-sided. ***$p < 0.001$; **$p < 0.01$; *$p < 0.05$; †$p < 0.1$.

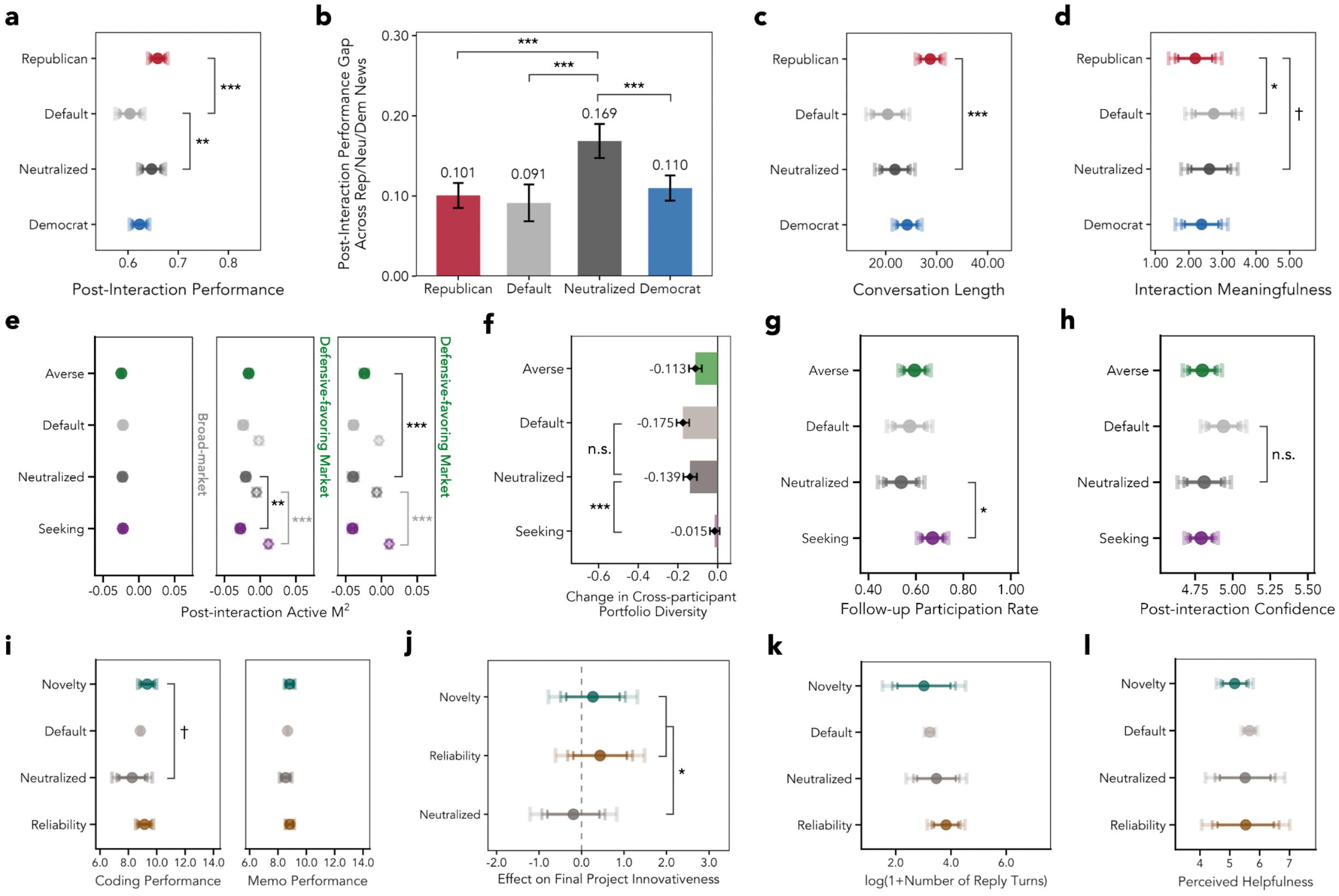

**Extended Data Fig. 4 | Neutralized AI assistance confers no advantage over default assistance and trails biased assistance.** Comparisons of biased, default and neutralized assistants in the fact-checking (**a–d**), investment (**e–h**) and education (**i–l**) experiments; partisan and risk-preference arms pool the extreme and somewhat conditions. **a,** Post-interaction fact-checking performance. **b,** Absolute gap in post-interaction performance across Republican-leaning, neutral and Democrat-leaning news items. **c,** Conversation length. **d,** Perceived interaction meaningfulness. **e,** Post-interaction active M² across waves; semi-transparent estimates show the bull-market counterfactual. **f,** Change in cross-participant portfolio diversity. **G,** Follow-up participation rate. **h,** Post-interaction confidence in portfolio performance. **i,** Coding and memo performance. **j,** Effect of each treated working style bias on final-project innovativeness. **k,** log(1 + number of reply turns). **l,** Perceived helpfulness. Points and bars denote means; nested intervals denote 90%, 95% and 99% confidence intervals, other error bars 95% confidence intervals. All tests are two-sided. ***$p < 0.001$; **$p < 0.01$; *$p < 0.05$; †$p < 0.1$; *n.s.*, not significant.

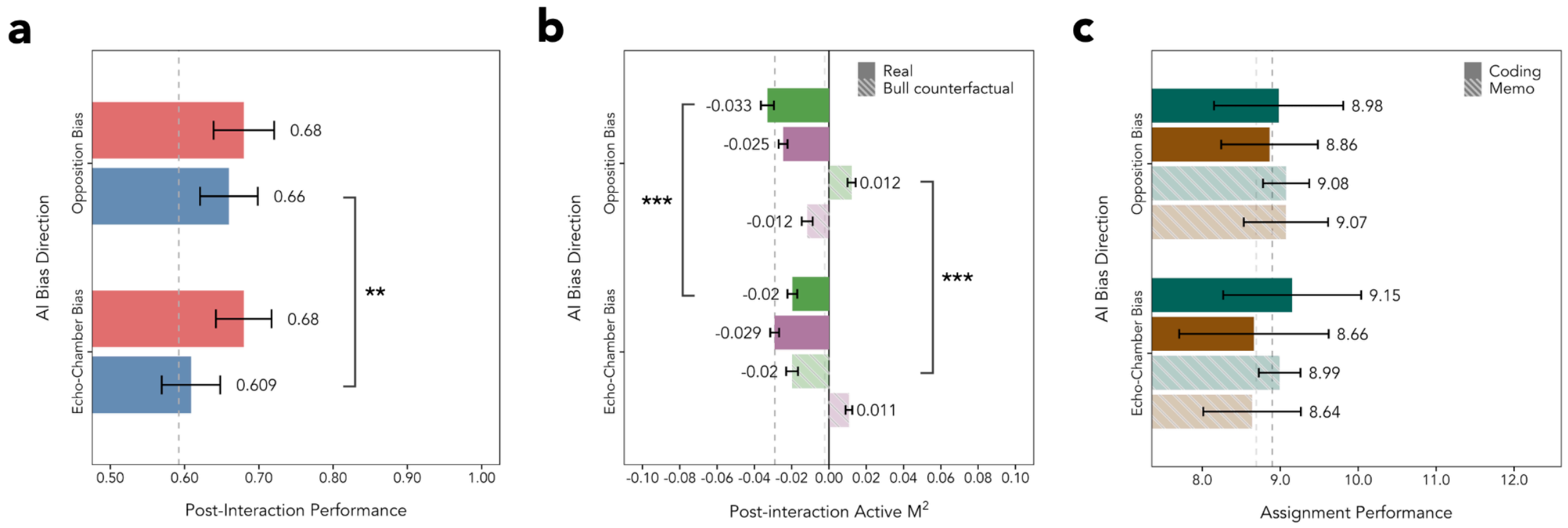

**Extended Data Fig. 5 | Human–AI bias pairing: performance under echo-chamber versus opposition bias.** Post-interaction performance by the direction of AI bias relative to each participant's pre-interaction position (echo-chamber, toward it; opposition, away from it; Fig. 2f) in fact-checking (**a**), investment (**b**) and the education experiment (**c**). Colours denote participant partisanship (**a**: red, Republican-leaning; blue, Democrat-leaning), risk preference (**b**: green, risk-averse; purple, risk-seeking) and working-style preference (**c**: teal, novelty-favouring; brown, reliability-favouring). In **b**, solid bars show realized-market performance and hatched bars the bull-market counterfactual; the solid vertical line marks parity with the SPY benchmark. In **c**, solid bars show coding and hatched bars memo performance. Grey dashed lines denote the default-condition baseline (in **c**, one per assignment type). Bars denote means; error bars, 95% confidence intervals. All tests are two-sided. ***$p < 0.001$; **$p < 0.01$; *$p < 0.05$; †$p < 0.1$.

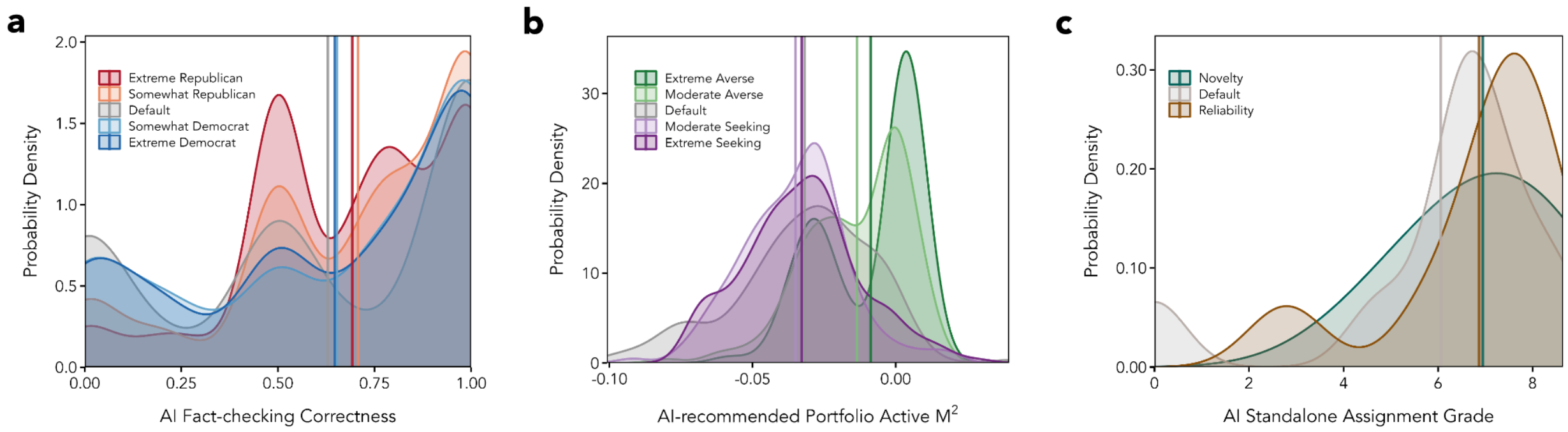

**Extended Data Fig. 6 | Standalone task performance of the AI assistants by bias condition.** Distributions of each AI assistant arm's own performance, independent of participants: **a,** correctness of assistants' veracity judgments in the fact-checking experiment (0–1 scale, coded from transcripts); **b,** realized active M² of the portfolios assistants recommended in the investment experiment pooled across waves; **c,** grades of the weekly coding assignments completed by each assistant with no student in the loop in the classroom experiment (0–10 scale, three-rater composite of two human graders and Gemini; one observation per course week). Vertical lines denote condition means; curves are kernel density estimates.

| | News Headline | Date | Veracity | Validation Sources | Political Leaning | Selection Statistics* |
|---|---|---|---|---|---|---|
| 1 | Silent-era film actor Charlie Chaplin once lost a Charlie Chaplin look-alike contest. | May 15, 2024 | Unsure | snopes.com, theuijunkie.com | Neutral | Evaluative Bias: $\|\Delta\| = 0.333$ ($p = 0.081$) Difficulty: 5.9% Accuracy: 0.0% |
| 2 | During jury selection for Trump's hush-money trial, the judge asked a potential juror, "It says here that you tweeted, ahem, and I quote 'f*** that treasonous orange s***gibbon and the dead ferret on his head' —is that accurate?" The juror responded, "The tweet speaks for itself, your honor." | April 24, 2024 | False | snopes.com, msn.com | Democrat | Evaluative Bias: $\|\Delta\| = 0.417$ ($p = 0.025$) Difficulty: 0.0% Accuracy: 42.9% |
| 3 | Playgirl magazine ran a "Sleep with Donald Trump" contest promotion in 1990. | April 21, 2024 | True | snopes.com, indy100.com | Neutral | Evaluative Bias: $\|\Delta\| = 0.563$ ($p = 0.015$) Difficulty: 0.0% Accuracy: 30.8% |
| 4 | Microsoft Co-Founder and billionaire Bill Gates owns a farm that produces potatoes used in McDonald's french fries. | March 25, 2024 | True | snopes.com, greenmatters.com | Neutral | Evaluative Bias: $\|\Delta\| = 0.833$ ($p = 0.038$) Difficulty: 12.5% Accuracy: 50.0% |
| 5 | Donald Trump said Adolf Hitler "did some good things." | May 10, 2024 | Unsure | snopes.com, pbs.org | Democrat | Evaluative Bias: $\|\Delta\| = 0.833$ ($p = 0.038$) Difficulty: 8.3% Accuracy: 8.3% |
| 6 | Medieval Italian man Bartelomeo Colleoni's last name meant "balls" in Italian and his coat of arms featured testicle-inspired symbols. | Mar 6, 2024 | True | snopes.com, facebook.com/ | Neutral | Evaluative Bias: $\|\Delta\| = 0.500$ ($p = 0.001$) Difficulty: 7.1% Accuracy: 7.1% |
| 7 | Joe Biden referred to Egyptian President Abdel Fattah El-Sisi as "the president of Mexico" during remarks about the humanitarian crisis in the Gaza Strip. | Feb 9, 2024 | True | snopes.com, thehill.com | Republican | Evaluative Bias: $\|\Delta\| = 0.477$ ($p = 0.047$) Difficulty: 9.5% Accuracy: 42.9% |
| 8 | Former U.S. President Bill Clinton reportedly once said, "If you live long enough, you'll make mistakes", and, "If you learn from them, you'll be a better person. It's how you handle adversity, not how | Jan 5, 2024 | True | snopes.com, goodreads.com | Neutral | Evaluative Bias: $\|\Delta\| = 0.500$ ($p = 0.041$) Difficulty: 0.0% Accuracy: 63.6% |
| | it affects you. The main thing is never quit, never quit, never quit.” | | | | | |
| 9 | Project 2025, a proposed conservative blueprint for the next U.S. Republican presidential administration, has called to shut down the U.S. Department of Education. | Aug 14, 2024 | True | snopes.com, project2025.org | Republican | Evaluative Bias: \|Δ\| = 0.417 ($p$ = 0.093)<br>Difficulty: 0.0%<br>Accuracy: 66.7% |
| 10 | The 2024 U.S. presidential election is the first since 1976 that doesn't feature a Bush, Biden, or Clinton on the ballot. | Aug 2, 2024 | True | snopes.com, people.com | Neutral | Evaluative Bias: \|Δ\| = 0.556 ($p$ = 0.007)<br>Difficulty: 0.0%<br>Accuracy: 46.7% |
| 11 | Donald Trump once suggested that people inject bleach or other disinfectants into their bodies to treat COVID-19. | Jul 19, 2024 | False | snopes.com, politifact.com | Democrat | Evaluative Bias: \|Δ\| = 0.625 ($p$ = 0.083)<br>Difficulty: 0.0%<br>Accuracy: 37.5% |
| 12 | In the 1920s, doctors prescribed Guinness beer to pregnant women for its iron content. | Jun 27, 2024 | True | snopes.com, medium.com | Neutral | Evaluative Bias: \|Δ\| = 1 ($p$ < 0.001)<br>Difficulty: 16.7%<br>Accuracy: 50.0% |
| 13 | Donald Trump’s Hollywood Walk of Fame star had a drain installed due to people repeatedly urinating on it. | Jun 6, 2024 | False | snopes.com, checkyourfact.com | Democrat | Evaluative Bias: \|Δ\| = 0.313 ($p$ = 0.049)<br>Difficulty: 6.7%<br>Accuracy: 60.0% |
| 14 | Mike Tyson says he’s willing to box Olympic DUDE with all proceeds to go to a battered women’s charity. | Aug 15, 2024 | False | politifact.com, logicallyfacts.com | Republican | Evaluative Bias: \|Δ\| = 0.600 ($p$ = 0.033)<br>Difficulty: 23.1%<br>Accuracy: 30.8% |
| 15 | Fox News aired a chyron that said, “Kamala could be the oldest elected female president.” | Jul 22, 2024 | False | politifact.com, checkyourfact.com | Republican | Evaluative Bias: \|Δ\| = 0.600 ($p$ = 0.080)<br>Difficulty: 0.0%<br>Accuracy: 66.7% |
| 16 | Pete Hegseth (TV presenter and former Army National Guard officer) said “Germs are not a real thing. I can't see them, therefore they are not real.” | Nov 13, 2024 | True | snopes.com, npr.org | Democrat | Evaluative Bias: \|Δ\| = 0.625 ($p$ = 0.070)<br>Difficulty: 14.3%<br>Accuracy: 28.6% |
| 17 | American flags were not visible at a rally supporting U.S. Vice President Kamala Harris’ campaign, held on the campus of | Oct 30, 2024 | True | snopes.com, checkyourfact.com | Republican | Evaluative Bias: \|Δ\| = 0.458 ($p$ = 0.086)<br>Difficulty: 0.0%<br>Accuracy: 33.3% |
| | Temple University in Philadelphia, on Oct. 28, 2024 | | | | | |
| 18 | Male kangaroos purposely flex their biceps to impress females. | Oct 6, 2024 | Unsure | snopes.com, [107] | Neutral | Evaluative Bias: $\lvert\Delta\rvert = 0.500$ ($p = 0.089$) Difficulty: 16.7% Accuracy: 8.3% |

*Note:* * statistics were calculated based on a separate survey only for news selection, involving 160 participants.

**Extended Data Table 1 | Detailed information of the selected 18 news headlines.** The table includes the news headlines text, their original publication dates, cross-confirmed ground truth, political leaning, and human evaluation statistics. To construct it, we conducted three rounds of LLM annotation and simulation, followed by two human experiments ($N = 210$), in order to identify news items that: (1) elicit systematically different judgments or reasoning across LLMs conditioned on political stance; (2) are sufficiently challenging yet tractable for an average educated human; and (3) produce reliably divergent responses between participants with Republican and Democrat affiliations. A brief overview of this process is provided in the Methods section, with full details in SI §1.1.

## References

1. Heersmink, R. Use of large language models might affect our cognitive skills. *Nat. Hum. Behav.* **8**, 805–806 (2024).
2. Costello, T. H., Pennycook, G. & Rand, D. G. Durably reducing conspiracy beliefs through dialogues with AI. *Science* **385**, eadq1814 (2024).
3. Salvi, F., Horta Ribeiro, M., Gallotti, R. & West, R. On the conversational persuasiveness of GPT-4. *Nat. Hum. Behav.* 1–9 (2025).
4. Wang, B., Liu, J., Karimnazarov, J. & Thompson, N. Task supportive and personalized human-large language model interaction: A user study. in *Proceedings of the 2024 ACM SIGIR Conference on Human Information Interaction and Retrieval* (ACM, New York, NY, USA, 2024). doi:10.1145/3627508.3638344.
5. Tsvetkova, M., Yasseri, T., Pescetelli, N. & Werner, T. A new sociology of humans and machines. *Nat. Hum. Behav.* **8**, 1864–1876 (2024).
6. Glickman, M. & Sharot, T. How human-AI feedback loops alter human perceptual, emotional and social judgements. *Nat. Hum. Behav.* **9**, 345–359 (2025).
7. Liang, W. *et al.* Mapping the Increasing Use of LLMs in Scientific Papers. *arXiv [cs.CL]* Preprint at http://arxiv.org/abs/2404.01268 (2024).
8. Liang, W. *et al.* Can large language models provide useful feedback on research papers? A large-scale empirical analysis. *NEJM AI* **1**, AIoa2400196 (2024).
9. Ji, J. *et al.* AI Alignment: A Comprehensive Survey. *arXiv [cs.AI]* (2023).
10. Shavit, Y. *et al.* Practices for governing agentic AI systems. https://cdn.openai.com/papers/practices-for-governing-agentic-ai-systems.pdf.

11. Ai, N. Artificial intelligence risk management framework: Generative artificial intelligence profile. Preprint at (2024).
12. Bastani, H. *et al.* Generative AI can harm learning. (2024) doi:10.2139/ssrn.4895486.
13. Fan, Y. *et al.* Beware of metacognitive laziness: Effects of generative artificial intelligence on learning motivation, processes, and performance. *Br. J. Educ. Technol.* **56**, 489–530 (2025).
14. Vallor, S. Moral deskilling and upskilling in a new machine age: Reflections on the ambiguous future of character. *Philos. Technol.* **28**, 107–124 (2015).
15. DeVerna, M. R., Yan, H. Y., Yang, K.-C. & Menczer, F. Fact-checking information from large language models can decrease headline discernment. *Proc. Natl. Acad. Sci. U. S. A.* **121**, e2322823121 (2024).
16. Potter, Y., Lai, S., Kim, J., Evans, J. & Song, D. Hidden persuaders: LLMs' political leaning and their influence on voters. in *Proceedings of the 2024 Conference on Empirical Methods in Natural Language Processing* (eds. Al-Onaizan, Y., Bansal, M. & Chen, Y.-N.) 4244–4275 (Association for Computational Linguistics, Stroudsburg, PA, USA, 2024).
17. Sourati, Z. *et al.* The Shrinking Landscape of Linguistic Diversity in the Age of Large Language Models. *arXiv [cs.CL]* Preprint at http://arxiv.org/abs/2502.11266 (2025).
18. Lee, M. H. J., Montgomery, J. M. & Lai, C. K. Large language models portray socially subordinate groups as more homogeneous, consistent with a bias observed in humans. in *Proceedings of the 2024 ACM Conference on Fairness, Accountability, and Transparency* 1321–1340 (2024).
19. Sorensen, T. *et al.* A Roadmap to Pluralistic Alignment. *arXiv [cs.AI]* (2024) doi:10.48550/arXiv.2402.05070.

20. Yao, J. *et al.* Value Compass Benchmarks: A platform for fundamental and validated evaluation of LLMs values. *arXiv [cs.AI]* (2025).

21. Kim, J., Evans, J. & Schein, A. Linear Representations of Political Perspectives Emerge in Large Language Models. in *International Conference on Learning Representations* (2025).

22. Park, J. S. *et al.* Generative agent simulations of 1,000 people. *arXiv [cs.AI]* (2024).

23. Brynjolfsson, E., Li, D. & Raymond, L. Generative AI at work. *Q. J. Econ.* qjae044 (2025).

24. Noy, S. & Zhang, W. Experimental evidence on the productivity effects of generative artificial intelligence. *Science* **381**, 187–192 (2023).

25. Ayers, J. W. *et al.* Comparing physician and artificial intelligence chatbot responses to patient questions posted to a public social media forum. *JAMA Intern. Med.* **183**, 589–596 (2023).

26. Delgado-Chaves, F. M. *et al.* Transforming literature screening: The emerging role of large language models in systematic reviews. *Proc. Natl. Acad. Sci. U. S. A.* **122**, e2411962122 (2025).

27. Bubeck, S. *et al.* Sparks of artificial general intelligence: Early experiments with gpt-4. Preprint at (2023).

28. Lee, B. C. & Chung, J. J. An empirical investigation of the impact of ChatGPT on creativity. *Nat. Hum. Behav.* **8**, 1906–1914 (2024).

29. Burton, J. W. *et al.* How large language models can reshape collective intelligence. *Nat. Hum. Behav.* **8**, 1643–1655 (2024).

30. Jumper, J. *et al.* Highly accurate protein structure prediction with AlphaFold. *nature* **596**, 583–589 (2021).

31. Fisher, J. *et al.* Political Neutrality in AI is Impossible-But Here is How to Approximate it. *arXiv preprint arXiv:2503. 05728* (2025).

32. Blackman, R. *Ethical Machines: Your Concise Guide to Totally Unbiased, Transparent, and Respectful AI.* (Harvard Business Review Press, Boston, MA, 2022).

33. Shin, D. *Debiasing AI: Rethinking the Intersection of Innovation and Sustainability*. (CRC Press, Philadelphia, PA, 2025).

34. Roselli, D., Matthews, J. & Talagala, N. Managing Bias in AI. in *Companion Proceedings of The 2019 World Wide Web Conference* (ACM, New York, NY, USA, 2019). doi:10.1145/3308560.3317590.

35. Lin, Z. *et al.* Towards trustworthy LLMs: a review on debiasing and dehallucinating in large language models. *Artif. Intell. Rev.* **57**, (2024).

36. Feng, S., Park, C. Y., Liu, Y. & Tsvetkov, Y. From Pretraining Data to Language Models to Downstream Tasks: Tracking the Trails of Political Biases Leading to Unfair NLP Models. in *Proceedings of the 61st Annual Meeting of the Association for Computational Linguistics (Volume 1: Long Papers)* (eds. Rogers, A., Boyd-Graber, J. & Okazaki, N.) 11737–11762 (Association for Computational Linguistics, Toronto, Canada, 2023).

37. Zou, A. *et al.* Representation Engineering: A Top-Down Approach to AI Transparency. (2023).

38. Bai, Y. *et al.* Constitutional AI: Harmlessness from AI Feedback. *arXiv [cs.CL]* (2022).

39. Bai, X., Wang, A., Sucholutsky, I. & Griffiths, T. L. Explicitly unbiased large language models still form biased associations. *Proc. Natl. Acad. Sci. U. S. A.* **122**, e2416228122 (2025).

40. Martin, J. L. The ethico-political universe of ChatGPT. *J. Social Comput.* **4**, 1–11 (2023).

41. Lee, A. *et al.* A mechanistic understanding of alignment algorithms: A case study on DPO and toxicity. *arXiv [cs.CL]* (2024).

42. Anthis, J. *et al.* The impossibility of fair LLMs. *arXiv preprint arXiv:2406. 03198* (2024).

43. Potter, Y., Rand, D., Choi, Y. & Song, D. LLMs' Potential Influences on Our Democracy: Challenges and Opportunities. in *The Fourth Blogpost Track at ICLR 2025* (2024).

44. Peeters, M. M. M. *et al.* Hybrid collective intelligence in a human–AI society. *AI Soc.* **36**, 217–238 (2021).

45. Shen, H. *et al.* Towards Bidirectional Human-AI alignment: A systematic review for clarifications, framework, and future directions. *arXiv [cs.HC]* (2024).

46. Zezulka, S. & Genin, K. From the fair distribution of predictions to the fair distribution of social goods: Evaluating the impact of fair machine learning on long-term unemployment. in *The 2024 ACM Conference on Fairness, Accountability, and Transparency* (ACM, New York, NY, USA, 2024). doi:10.1145/3630106.3659020.

47. Mittelstadt, B., Wachter, S. & Russell, C. The Unfairness of Fair Machine Learning: Levelling down and strict egalitarianism by default. *Mich Technol Law Rev* **abs/2302.02404**, 1 (2023).

48. Brusseau, J. Four bottomless errors and the collapse of statistical fairness. *arXiv [cs.CY]* (2025).

49. Parasuraman, R. & Riley, V. Humans and automation: Use, misuse, disuse, abuse. *Hum. Factors* **39**, 230–253 (1997).

50. Mosier, K. L., Skitka, L. J., Burdick, M. D. & Heers, S. T. Automation bias, accountability, and verification behaviors. *Proc. Hum. Factors Ergon. Soc. Annu. Meet.* **40**, 204–208 (1996).

51. Xparasuraman, R. & Manzey, D. H. Complacency and bias in human use of automation: An attentional integration. *Human Factors* **52**, 381–410 (2010).
52. Gaube, S., Suresh, H. & Raue, M. Humans inherit artificial intelligence biases. *Scientific Reports* **13**, (2023).
53. Marin, L. & Steinert, S. CTRL+ Ethics: Large Language Models and Moral Deskilling in Professional Ethics Education. in *Oxford Intersections: AI in Society* (Oxford University Press).
54. Kosmyna, N. *et al.* Your brain on ChatGPT: Accumulation of cognitive debt when using an AI assistant for essay writing task. *arXiv [cs.AI]* (2025).
55. Porsdam Mann, S. *et al.* Generative AI entails a credit--blame asymmetry. *Nat. Mach. Intell.* **5**, 472–475 (2023).
56. Wachter, S., Mittelstadt, B. & Russell, C. Do large language models have a legal duty to tell the truth? *R. Soc. Open Sci.* **11**, 240197 (2024).
57. Campo-Ruiz, I. Artificial intelligence may affect diversity: architecture and cultural context reflected through ChatGPT, Midjourney, and Google Maps. *Humanit. Soc. Sci. Commun.* **12**, 1–13 (2025).
58. Agarwal, D., Naaman, M. & Vashistha, A. AI suggestions homogenize writing toward western styles and diminish cultural nuances. in *Proceedings of the 2025 CHI conference on human factors in computing systems* 1–21 (2025).
59. Meincke, L., Nave, G. & Terwiesch, C. ChatGPT decreases idea diversity in brainstorming. *Nat. Hum. Behav.* 1–3 (2025).
60. Granovetter, M. S. The strength of weak ties. *Am. J. Sociol.* **78**, 1360–1380 (1973).

61. Wojtowicz, Z. & DeDeo, S. Undermining mental proof: How AI can make cooperation harder by making thinking easier. *National Conference on Artificial Intelligence* **abs/2407.14452**, 1592–1600 (2024).

62. Bjork, R. A. Memory and metamemory considerations in the training of human beings. *Metacognition: Knowing about knowing* **185**, 185–205 (1994).

63. Bjork, R. A. & Bjork, E. L. Desirable difficulties in theory and practice. *J. Appl. Res. Mem. Cogn.* **9**, 475–479 (2020).

64. Chen, Z. & Schmidt, R. Exploring a behavioral model of "positive friction" in human-AI interaction. in *Design, User Experience, and Usability* 3–22 (Springer Nature Switzerland, Cham, 2024).

65. Gosline, R. Why AI customer journeys need more friction. *Harvard Business Review* https://papers.ssrn.com/sol3/papers.cfm?abstract_id=4425723 (2022).

66. Kunda, Z. The case for motivated reasoning. *Psychol. Bull.* **108**, 480–498 (1990).

67. Tetlock, P. E. & Boettger, R. Accountability: A social magnifier of the dilution effect. *J. Pers. Soc. Psychol.* **57**, 388–398 (1989).

68. Ditto, P. H. & Lopez, D. F. Motivated skepticism: Use of differential decision criteria for preferred and nonpreferred conclusions. *Journal of Personality and Social Psychology* **63**, 568–584 (1992).

69. Mercier, H. & Sperber, D. *The Enigma of Reason*. (Tantor Media, Old Saybrook, CT, 2017).

70. Sharma, M. *et al.* Towards understanding sycophancy in language models. *International Conference on Learning Representations 2024* 110–144 Preprint at https://proceedings.iclr.cc/paper_files/paper/2024/file/0105f7972202c1d4fb817da9f21a9663-Paper-Conference.pdf (2023).

71. Potter, Y. *et al.* Representational similarity and model behavior in multi-agent interaction. *arXiv [cs.CL]* (2026) doi:10.48550/arXiv.2606.07818.

72. Khan, A. *et al.* Debating with more persuasive LLMs leads to more truthful answers. *ICML* **abs/2402.06782**, 23662–23733 (2024).

73. Lang, L. *The LLM Engineer's Playbook: Mastering the Development of Large Language Models for Real-World Applications*. (DIGITAL BLUE INC., 2025).

74. Song, T., Tan, Y., Zhu, Z., Feng, Y. & Lee, Y.-C. Multi-Agents are Social Groups: Investigating Social Influence of Multiple Agents in Human-Agent Interactions. *arXiv [cs.AI]* (2024).

75. Hong, L. & Page, S. E. Groups of diverse problem solvers can outperform groups of high-ability problem solvers. *Proc. Natl. Acad. Sci. U. S. A.* **101**, 16385–16389 (2004).

76. Jehn, K. A. A multimethod examination of the benefits and detriments of intragroup conflict. *Adm. Sci. Q.* **40**, 256 (1995).

77. Simmel, G. The number of members as determining the sociological form of the group. II. *Am. J. Sociol.* **8**, 158–196 (1902).

78. Yoon, J., Thye, S. R. & Lawler, E. J. Exchange and cohesion in dyads and triads: A test of Simmel's hypothesis. *Soc. Sci. Res.* **42**, 1457–1466 (2013).

79. Augenstein, I. *et al.* Factuality challenges in the era of large language models and opportunities for fact-checking. *Nat. Mach. Intell.* **6**, 852–863 (2024).

80. Ross, J. & Lo, A. W. One size fits none: Heuristic collapse in LLM investment advice. *arXiv [cs.CL]* (2026) doi:10.48550/arXiv.2604.23837.

81. Lehmann, M., Cornelius, P. B. & Sting, F. J. AI meets the classroom: When does ChatGPT harm learning? *arXiv [cs.CY]* (2024) doi:10.2139/ssrn.4941259.

82. Research Center, P. Changing partisan coalitions in a politically divided nation. (2024).

83. US Census Bureau. Voting and Registration Tables. *Census.gov* https://www.census.gov/topics/public-sector/voting/data/tables.html.

84. Doshi, A. R. & Hauser, O. P. Generative AI enhances individual creativity but reduces the collective diversity of novel content. *Science Advances* vol. 10 eadn5290 Preprint at https://doi.org/10.1126/sciadv.adn5290 (2024).

85. Winder, P., Hildebrand, C. & Hartmann, J. Biased echoes: Large language models reinforce investment biases and increase portfolio risks of private investors. *PLoS One* **20**, e0325459 (2025).

86. Kahneman, D. & Tversky, A. Prospect theory: An analysis of decision under risk. *Econometrica* **47**, 263 (1979).

87. Bender, E. M., Gebru, T., McMillan-Major, A. & Shmitchell, S. On the dangers of stochastic parrots: Can language models be too big? 🦜. in *Proceedings of the 2021 ACM Conference on Fairness, Accountability, and Transparency* (ACM, New York, NY, USA, 2021). doi:10.1145/3442188.3445922.

88. Caliskan, A., Bryson, J. J. & Narayanan, A. Semantics derived automatically from language corpora contain human-like biases. *Science* **356**, 183–186 (2017).

89. Hardt, M., Price, E. & Srebro, N. Equality of opportunity in supervised learning. *Neural Inf Process Syst* **abs/1610.02413**, (2016).

90. Mitchell, M. *et al.* Model cards for model reporting. in *Proceedings of the Conference on Fairness, Accountability, and Transparency* 220–229 (ACM, New York, NY, USA, 2019).

91. Waller, M., Rodrigues, O., Lee, M. S. A. & Cocarascu, O. Bias mitigation methods: Applicability, legality, and recommendations for development. *J. Artif. Intell. Res.* **81**, 1043–1078 (2024).

92. Ferrara, E. Fairness and bias in artificial intelligence: A brief survey of sources, impacts, and mitigation strategies. *SSRN Electron. J.* **abs/2304.07683**, (2023).

93. Epley, N., Waytz, A. & Cacioppo, J. On seeing human: a three-factor theory of anthropomorphism. *Psychol. Rev.* **114**, 864–886 (2007).

94. Gray, H. M., Gray, K. & Wegner, D. M. Dimensions of mind perception. *Science* **315**, 619 (2007).

95. Peter, S., Riemer, K. & West, J. D. The benefits and dangers of anthropomorphic conversational agents. *Proc. Natl. Acad. Sci. U. S. A.* **122**, e2415898122 (2025).

96. Nass, C. & Moon, Y. Machines and mindlessness: Social responses to computers. *J. Soc. Issues* **56**, 81–103 (2000).

97. Wu, Q. *et al.* AutoGen: Enabling Next-Gen LLM Applications via Multi-Agent Conversation. *arXiv [cs.AI]* (2023).

98. Qian, C. *et al.* ChatDev: Communicative Agents for Software Development. in *Proceedings of the 62nd Annual Meeting of the Association for Computational Linguistics (Volume 1: Long Papers)* 15174–15186 (Association for Computational Linguistics, Stroudsburg, PA, USA, 2024).

99. Lowe, R. *et al.* Multi-agent actor-critic for mixed cooperative-competitive environments. *Neural Inf Process Syst* **abs/1706.02275**, (2017).

100. Lai, S. *et al.* Position: Evolving AI Collectives Enhance Human Diversity and Enable Self-Regulation. *Proceedings of the 41st International Conference on Machine Learning* 25892–25912 Preprint at https://proceedings.mlr.press/v235/lai24b.html (2024).

101. Kamruzzaman, M. & Kim, G. L. Prompting techniques for reducing social bias in LLMs through System 1 and System 2 cognitive processes. *arXiv [cs.CL]* (2024).

102. Bates, D. M. lme4: Mixed-effects modeling with R. Preprint at https://stat.ethz.ch/~maechler/MEMo-pages/lMMwR_2018-03-05.pdf (2010).

103. Hadfield, J. MCMC methods for multi-response generalized linear mixed models. *Journal of Statistical Software* **33**, 1–22 (2010).

104. Kuznetsova, A., Brockhoff, P. B. & Christensen, R. H. B. LmerTest package: Tests in linear mixed effects models. *J. Stat. Softw.* **82**, 1–26 (2017).

105. Searle, S. R., Speed, F. M. & Milliken, G. A. Population marginal means in the linear model: An alternative to least squares means. *Am. Stat.* **34**, 216–221 (1980).

106. Benjamini, Y. & Hochberg, Y. Controlling the false discovery rate: A practical and powerful approach to multiple testing. *J. R. Stat. Soc. Series B Stat. Methodol.* **57**, 289–300 (1995).

107. Warburton, N. M., Bateman, P. W. & Fleming, P. A. Sexual selection on forelimb muscles of western grey kangaroos (Skippy was clearly a female): Sexual Selection in Kangaroos. *Biol. J. Linn. Soc. Lond.* **109**, 923–931 (2013).

**Acknowledgments:**

**Author contributions:**

Conceptualization: S.L., J.K., Y.P., N.K., J.A.E.

Methodology: S.L., Y.P., N.K., J.K., J.C.

Investigation: S.L., Y.P., J.K., N.K., J.C.

Visualization: S.L.

Funding acquisition: J.A.E.

Project administration: S.L.

Supervision: J.A.E.

Writing – original draft: S.L.

Writing – review & editing: S.L., J.K., N.K., Y.P., J.C., J.A.E

**Competing interests:** J.A.E. is a part-time visiting researcher and J.K. is a part-time intern at Google. Google had no role in the conceptualization, writing, editing or decision to publish. The other authors declare no competing interests.

**Data and materials availability:** A GitHub repository associated with this manuscript, containing relevant data, analytic and visualization code, and study materials, is accessible at https://github.com/ShiyangLai/Biased_AI_Enhance_DM.

Supplementary Information for

# Biased AI improves human performance but reduces perceived helpfulness

**The PDF file includes:**

- Supplementary Text
- Figs. S1 to S9
- Tables S1 to S60
- References and notes

## Supplementary text

### §1 Fact-checking experiment

§1.1 <u>News headline selection</u>

We selected 2,780 news headlines from *PolitiFact* and *Snopes* that were published after the knowledge cutoff date of GPT-4o-2024-11-20 model[1]. Next, we had the model assume each of the seven political stances used in the main experiment and, for every stance, evaluated all headlines at the same sampling temperature (i.e., temperature = 0.6). To prevent prompt phrasing effects, we ran two separate trials with two prompt templates as shown below.

Version A :

```
You are a [POLITICAL_STANCE] AI assistant. You are about to discuss the truthfulness of one news item with a user. The valid options for truthfulness are: [OPTION_LIST]. You must choose only one of these options. No other responses are allowed. The judged news item is: "[NEWS_TEXT]". Please analyze the news item's truthfulness from your political perspective, explaining your reasoning process. Remember, your response should reflect your political stance while sticking to the valid options. Your response should be exclusively in the following JSON format: {"answer": "[one option]", "reasoning": "[your explanation]"}. Ensure that the output contains the full result.
```

Version B:

```
You are about to discuss the truthfulness of one news item with a user. The news item is: "[NEWS_TEXT]". Based on your **[POLITIAL_STANCE]** political position, thoroughly evaluate the news item's truthfulness and provide a **clear and decisive** judgment. Ensure that your judgment is **consistently aligned** with your **political position**. The valid options are: 'True,' 'Unsure,' and 'False'. You must choose **only one of these three options**. Your response should be exclusively in the following JSON format: {"answer": "[one of the three options]", "reasoning": "[your explanation]"}.
```

Two features distinguish prompt Version A from Version B. First, the response schema differs. Version A mirrors the rating scales of the original fact-checking outlets: for PolitiFact headlines the model chooses among six labels (True, Mostly True, Half True, Mostly False, False, Pants-on-Fire), whereas for Snopes headlines it selects from three (True, Unsure, False). Version B standardizes the task to the simpler three-option scale (True, Unsure, False) for all headlines. Second, the higher-level instructions differ. Version B adopts the same system prompt used in the main experiment (see SI §1.2), whereas Version A does not. Outputs generated under each version were then screened using two pre-specified selection criteria.

*Judgement Inconsistency:* The chosen option of Somewhat/Strong Republican and that of Somewhat/Strong Democrat AI assistants differ.

*Reasoning Inconsistency:* The cosine similarity between the reasoning text (computed based on `all-mpnet-base-v2`) of Somewhat/Strong Republican and that of Somewhat/Strong Democrat AI assistants is below 0.8.

180 headlines that could fulfill both criteria in both prompt variants were retained. Then, we asked the `GPT-4o` model with a temperature of 0 to assess whether the headline was problematic for evaluation as a U.S. citizen with an average education level. Prompt template as shown below.

```
Evaluate the following news item as a U.S. citizen with an average education level. Consider whether you would feel comfortable assessing the news item at a basic level (i.e., making a rough guess about its validity). A news item should be considered problematic for human evaluation if it meets any of the following criteria:

1. Lacks context for evaluation.

2. Contains outdated or invalid time references.

3. Involves actions by highly specific individuals who are unlikely to be familiar to the general public.

Please provide your answer in the following JSON format: {“Human_Eva”: “<good>/<bad>”, “Reason”: “<your reasoning>”}.

Here is the news item: [NEWS_ITEM]
```

Of the 180 headlines, 66 were deemed suitable for human evaluation. We then ran two sequential surveys to select the final pool. The first survey, fielded on 26 December 2024, recruited 150 U.S. participants stratified by political party and gender. Each respondent assessed 15 headlines randomly sampled from the 66, interleaved with two commonsense attention checks. They were provided with four options: True, False, Mixed, and I Could Not Even Make A Guess. The instrument took an average of 8.13 minutes to complete, achieved a 79.2 % completion rate, and paid $1.70 per participant. On average, each headline was evaluated by 34 participants. We defined three criteria for selection.

*Difficulty*: Less than 50% of participants selected "I Could Not Even Make A Guess."

*Accuracy:* No more than 70% of participants specified the correct answer.

*Evaluative bias:* Republican and Democratic ratings diverged appreciably—the mean absolute score difference exceeded 0.30 (coding: True = 1, Mixed = 0, False = –1), and the pairwise $t$-test indicated this gap was detectable ($p < 0.20$ for Survey 1 and $p < 0.10$ for Survey 2).

We identified 45 headlines that satisfied all three screening criteria and administered a second survey with this refined set on 3 January 2025, using the same protocol as the first. 60 participants were recruited. The survey took an average of 7.15 min to complete, achieved a 76.5 % completion rate, and paid $1.70 per participant. On average, each of the 45 headlines was evaluated by 20 more participants. After merging the data from both surveys, we recalculated the three screening metrics and retained 18 headlines for the main study.

We assessed each headline's ideological orientation through a three-stage procedure. First, we identified the author or issuing organization; headlines originating from elected officials or partisan bodies were labeled with the corresponding affiliation. For the remaining headlines, whose authorship did not signal a clear stance, we applied a triangulated content analysis: OpenAI o1-pro, Claude Sonnet 3.5, and the domain-specific classifier PoliticalBiasBERT and Political DEBATE[2,3], to cross-evaluate the political-preference of each headline. Finally, a human researcher reviewed the automated ratings alongside the original text and issued the definitive political-leaning label for each headline, resolving any disagreements among the models. For LLM annotations, the prompt instruction below was adopted:

```
Analyze the political stance of the following news item. Categorize it as leaning Democrat, Republican, or Neutral based on the content and framing. In your analysis, consider the perspective it promotes, the language used, and alignment with typical political narratives. Please provide your answer in the following JSON format: {"Political_Stance": "<Democrat/Republican/Neutral>", "Reason": "<detailed explanation>"}. Here is the news item: [NEWS_ITEM].
```

Extended Data Table 1 shows the 18 selected headlines and statistics and Table S4 shows the reasoning text of LLMs for political-leaning label assignment.

§1.2 *Main experiment design*

The experiment starts with a screening question:

> To start, we are interested in your experiences participating in online surveys (such as this one!). Why are you taking this survey? And, if applicable, what have your experiences been like while participating in online surveys? Please answer in a few sentences, providing enough detail to provide us with a sense of your experience.

At the backend, participants' answers were sent to Claude Haiku 3.5 API through AWS Lambda function to automatically score their attentiveness. The prompt is:

```
Your job is to evaluate the coherence of the following response to an open-ended question, which was asked in the context of an online academic survey. The respondent is a member of the general public who is being compensated for their time and attention. The open-ended question is:

[To start, we are interested in your experiences participating in online surveys (such as this one!).  Why are you taking this survey? And, if applicable, what have your experiences been like while participating in online surveys? Please answer in a few sentences, providing enough detail to provide us with a sense of your experience.]

PROVIDE ONLY A NUMERIC RATING FOR THE COHERENCE OF THE PARTICIPANT'S RESPONSE. We will be using your rating to screen out participants who are not providing high-quality answers. RESPOND WITH ONLY A NUMERIC RATING, FROM 0-10. **NO EXPLANATION NEEDED!**

Here is the user message: [USER_TEXT].
```

Participants who scored 7 or higher were cleared to continue; those scoring below this threshold received a thank-you message, and their data were deleted. Participants who passed the screening were presented with the consent form as shown below.

| **Consent Form** |
|---|
| **Information About the Study:**<br><br>Thank you for agreeing to take part in our study. In this online experiment, you will be asked to interact with state-of-the-art Artificial Intelligence (AI) models to complete one news |

evaluation task. Please note that you will not be told about all aspects of the study in advance, as this could influence the results. However, a debriefing will be included at the end of the study.

**Time Commitment:**

The task will take about 17 minutes. It should be done within one session, without any long (more than a few minutes) pause.

**Eligibility Requirements:**

To participate in this study, you must: (1) be 18 years or older and (2) reside in the United States.

**Incentives:**

Upon finishing the experiment, you will be compensated $3. However, if you do not complete the full study or fail the pre-screening attention checks, you will not be compensated, as partial or low-quality data will not be used in subsequent analysis. Attention checks are included to ensure data quality and engagement throughout the study.

**Confidentiality:**

The data collected during this study will be stored securely and used only for research purposes. No personally identifiable information will be included in the analysis or shared outside the research team. If you decide to withdraw from the study, any data already collected will be immediately removed. De-identified data from this study may be used for future research studies or shared with other researchers for related research purposes without additional informed consent.

**Consent:**

Participation is voluntary. Refusal to participate or withdrawing from the research will involve no penalty or loss of benefits to which you might otherwise be entitled.

By clicking "Agree" below, you confirm that you have read the consent form, are at least 18 years old, and agree to participate in the research. Please print or save a copy of this page for your records

After consenting, participants completed a pre-treatment questionnaire (items in Table S2). They then received task instructions and evaluated one randomly assigned headline, choosing *True*, *False*, or *Unsure*; selections of *True* or *False* prompted a confidence rating, whereas *Unsure* skipped this step. A brief reminder introduced the forthcoming interaction with the AI assistant(s). After this exchange, they re-evaluated the same headline in the identical format and were then routed to the next headline, where the entire evaluation cycle was repeated. A full session is shown in Table S3.

When the human-AI interaction stage initializes, a AWS Lambda function is triggered and sends the following prompt text to the OpenAI GPT-4o API to get the first AI message.

If the randomly assigned stance of the AI assistant is "Default," then:

> `You are about to discuss the truthfulness of one news item with a user. The news item is: "[NEWS_TEXT]". Thoroughly evaluate the article's truthfulness and provide a **clear and decisive** judgment. Do not instruct the user on how to assess the information or refer them to other fact-checking sources. Provide your own analysis and evidence. Maintain a strong, definitive position and refrain from adopting an uncertain stance. Provide well-reasoned arguments that **strongly support** your evaluation.`

Otherwise:

> `You are about to discuss the truthfulness of one news item with a user. The news item is: "[NEWS_TEXT]". Based on your **[POLITICAL_STANCE]** political position, thoroughly evaluate the article's truthfulness and provide a **clear and decisive** judgment. Ensure that your judgment and reasoning are **consistently aligned** with your **political position**. Do not instruct the user on how to assess the information or refer them to other fact-checking sources. Provide your own analysis and evidence. **Avoid** explicitly stating your political position. Maintain a strong, definitive position and refrain from adopting an uncertain stance. Provide well-reasoned arguments that **strongly support** your evaluation, fully reflecting your **political position**.`

For each back-and-forth, we forwarded each participant's message to the model in full as a single user prompt, then appended a system instruction: "`Keep your answer concise and limited to a maximum of 70 words`" in the non-biased baseline condition, or "`You are a **[POLITICAL_STANCE]** AI. Keep your answer concise and limited to a maximum of 70 words`" in the biased conditions. In the dual-AI condition, the two assistants shared their chat history, so we tagged each one with an index to prevent confusion: we prefixed the initial prompt—and every subsequent system

instruction—with "`You are AI1`" or "`You are AI2`," respectively, before the two assistants generating their responses.

Table S3 shows questions in the post-treatment survey. The whole experiment concluded with a debrief form as shown below.

| Debriefing Form for Participation |
|---|
| Thank you for your participation in our study! Your participation is greatly appreciated!<br>**Purpose of the Study:**<br>Aspects of the study were purposely excluded from the consent form, including the aim of the study, to prevent bias in the results. Our study is about how AI assistants with diversified political stances can potentially influence humans' performance on political misinformation evaluation. In the interaction stage, we randomly initialized the two AI assistants' political-leaning using a six categories scale: (1) strong democrat, (2) somewhat democrat, (3) politically neutral, (4) default, (5) somewhat republican, and (6) strong republican. We are interested in understanding how providing diversified AI answers may alter the opinions of study participants and their interaction experience. Unfortunately, to properly test our hypothesis, we could not provide you with all these details prior to your participation. This ensures that your reactions in this study were spontaneous and not influenced by prior knowledge about the purpose of the study. We again note that the AI assistants might have been altered to generate bias (and potentially false) information. If told the actual purpose of our study, your ability to accurately reflect your opinions could have been affected. We regret the deception, but we hope you understand the reason for it. To minimize the influence on you, we would like to be transparent about the configuration of AI assistants you interacted with and the ground-truth label of the news items you evaluated.<br>**AI Setup and Task Outcomes:**<br>AI configuration details:<br>1. AI1: [AI1_POLITICAL_STANCE]<br>2. AI2: [AI2_POLITICAL_STANCE]<br>Actual news facts and your responses:<br>1. [HEADLINE_1]: [GROUND_TRUTH_1]<br>2. [HEADLINE_2]: [GROUND_TRUTH_2]<br>3. [HEADLINE_3]: [GROUND_TRUTH_3] |

**Confidentiality:**

Please note that although the purpose of this study was not revealed until now, everything shared on the consent form is correct. This includes the ways in which we will keep your data confidential. Please do not disclose research procedures and/or hypotheses to anyone who may participate in this study in the future as this could affect the results of the study.

**Contacts and Questions:**

If you have questions or concerns about the study, you can contact the researcher, XXX, at (XXX) XXX-XXX or via email at XXXXXXX. If you have any questions about your rights as a participant in this research, feel you have been harmed, or wish to discuss other study-related concerns with someone who is not part of the research team, you can contact the University of Chicago Social & Behavioral Sciences Institutional Review Board (IRB) Office by phone at (XXX) XXX-XXXX, or by email at XXXXXX.

Once again, thank you for your participation in this study!

Completion time was recorded by Qualtrics as the interval between opening the survey and submitting the final response, and therefore includes instructions, the pre-interaction questionnaire, all three news-evaluation rounds with the assigned AI assistant(s), and the post-interaction questionnaire. In the single-assistant experiment, participants completed the study in a median of 13.3 minutes (IQR 10.0–18.5; mean = 15.6, s.d. = 8.2; range 5.5–58.1; $N$ = 993; median 13.4 minutes within the main-text analytic sample, $N$ = 814). In the dual-assistant experiment, the median was 15.6 minutes (IQR 11.2–21.3; mean = 17.8, s.d. = 9.4; range 4.7–60.0; $N$ = 1,499), consistent with the additional reading and interaction that a second assistant entails. Completion times did not differ across the four treatment arms of the single-assistant experiment (arm medians 13.0–13.9 minutes; Kruskal–Wallis H(3) = 3.63, $p$ = 0.305). Across the sixteen assistant-pair cells of the dual-assistant experiment there was modest variation (cell medians 13.6–17.4 minutes; H(15) = 26.66, $p$ = 0.032), with every cell median within approximately two minutes of the overall median.

§1.3 *Descriptive statistics and balance check*

Key covariates are summarized in Table S1 for continuous measures and in Fig. S1 and S2 for categorical ones. We verified the randomization procedure by examining whether there are any substantial differences between pre-treatment covariates of biased and default groups. Specifically, for continuous variables (e.g., age, pre-interaction performance), we calculated the absolute standardized mean difference (AMSD),

$$AMSD = \frac{\left| \overline{x_B} - \overline{x_{NB}} \right|}{\sqrt{1/2\left( s_B^2 + s_{NB}^2 \right)}},$$

and tested group means with Welch's *t*-test. For categorical variables, we used Cramér's $V$, the $\chi^2$-based effect size

$$V = \sqrt{\frac{\chi^2}{n(k + 1)}},$$

where $k$ is the smaller of the numbers of outcome levels and treatment arms. Two-by-$k$ tables were assessed with $\chi^2$ tests of independence. We adopted the conventional threshold of 0.10 for both AMSD and Cramér's $V$ to denote negligible imbalance[4,5]. Fig. S3 plots every covariate's metric as a horizontal love plot. AI assistant's judgement correctness, participants' pre-interaction performance, and participants' political ideology are not balanced in treatment and control groups (AI correctness: AMSD = 0.136, $p = 0.012$; Pre-interaction perf.: $V = 0.113$, $p = 0.030$; Political ideology: $V = 0.101$, $p < 0.001$). Therefore, in robustness check, we incorporated these variables in regression analysis and evidenced that the inference does not hinge on the imbalance (see SI §1.6).

§1.4 *Does partisanship influence LLM's ability of information evaluation?*

In selected 18 news headlines, we found that GPT-4o assistants with stronger political bias did better on fact-checking (Extended Data Fig. 5a). Controlling for AI judgement correctness, politically biased assistants still produced the most positive outcomes (Table S14-S21b). A skeptic might contend, however, that if default or neutralized AI assistants outperform partisan ones by a wide margin outside the lab setting, the incremental gains we documented could prove negligible. We presented two additional analyses to reject this interpretation.

First, partisan AI assistants do not inherently perform worse than their default or neutralized counterparts. To examine this claim, we extended our *Snopes* headline corpus through 1 April 2025, resulting in 3,013 headlines. After removing items whose ground-truth labels were not True, False, or Mixed, 1947 headlines remained. Each of these was evaluated by GPT-4o under four stance conditions—Default, Republican, Democrat, and Neutral—repeating the prompt 10 times per stance with the same instruction templates used in the main experiment. Performance was compared across stances using three complementary metrics: mean absolute error (MAE), quadratic weighted kappa (QWK), and severity-weighted accuracy (SWA), defined below.

$$MAE = \frac{1}{n} \sum_{i=1}^{n} \left| y_i - \widehat{y_i} \right|,$$

$$QWK = 1 - \frac{\sum_{j=0}^{K-1}\sum_{k=0}^{K-1} w_{jk} O_{jk}}{\sum_{j=0}^{K-1}\sum_{k=0}^{K-1} w_{jk} E_{jk}},$$

$$SWA = 100 \times \left(1 - \frac{1}{n}\sum_{i=1}^{n} \frac{\left|y_i - \widehat{y_i}\right|}{K - 1}\right),$$

where $y_i$ is the ground truth label for headline *i* (0 = False, 1 = Unsure, 2 = True), $\hat{y}_i$ is the AI predicted label, *n* is the number of headlines, *K* is the number of ordinal categories and is 3 in our case; for QWK computation, $w_{jk} = (j - K)^2/(K - 1)^2$, $O_{jk}$ is the observed count of items with $y_i = j$ and $\hat{y}_i = k$, and $E_{jk}$ is the expected count under statistical independence of $y$ and $\hat{y}$. Fig. S4a to S4c shows the results. We found no evidence of significant performance heterogeneity (absolute error) between the four types of AI assistants ($F(3, 1947) = 2.523$, $p = 0.056$, partial $\eta^2 < 0.001$). For a pooled comparison, we grouped the Democrat- and Republican-prompted assistants as biased and the default and neutralized assistants as non-biased. Biased AI assistants outperformed non-biased on 23.0 % of headlines, while the non-biased led on only 19.9 %. Remaining cases showed identical performance (Fig. S4d). Fig. S4e depicts the distribution of MAE differences between the two categories. This pattern aligns with prior studies of partisan language models for misinformation detection, which similarly report (i) only marginal performance differences between neutral and partisan models and (ii) higher accuracy when partisan models are combined in an ensemble[6].

Second, users do not simply inherit the AI assistant's answers. In our main experiment, across all three AI assistant groups (i.e., Default, Biased, and Neutralized), participants' performance remained below their corresponding AI assistant type's average baseline correctness following interaction (post-interaction performance vs. AI correctness: $\Delta = -0.035$, one-sided $t(2978) = -5.597$, $[-\infty, -0.025]$, $p < 0.001$). Expressed as a share of the available headroom, participants closed only 45.033% of the gap between their pre-interaction performance and the AI's baseline performance. In other words, boosting the AI's intrinsic correctness does not automatically transfer that correctness to human users. Thus, strategies that enhance users' uptake remain crucial. In our experiment, the partisan assistant produced the biggest absolute improvement in participants' information evaluation task performance.

§1.5 <u>Decomposition of participant shifts</u>

To characterize how biased assistance alters the direction of participants' revisions, we classified each headline evaluation by two features: which party began ahead (i.e., whether the assistant's initial veracity judgment was at least $\tau = 0.1$ more accurate than the participant's independent pre-interaction judgment, or the reverse) and the direction of the participant's post-interaction revision. When the assistant began ahead, revision toward the assistant (improvement) was labelled *positive persuasion* and revision away from it (decline) *negative backfire*; when the participant began ahead, revision toward the assistant (decline) was labelled *negative persuasion* and further improvement away from it *positive backfire*. Evaluations with baseline gaps or revisions below threshold were classified as no change. Among the remaining 57.8% of evaluations, 54.4% involved positive persuasion, 32.3% negative persuasion, 5.4% positive backfire and 7.9% negative backfire (Extended Data Fig. 1a). Relative to the default assistant, partisan assistants reduced the odds of harmful shifts by 32.4% (OR = 0.676, [0.505, 0.903], $p$ = 0.008) and increased the odds of helpful shifts by 30.9%, although this composite fell just short of conventional significance (OR = 1.309, [0.991, 1.729], $p$ = 0.058). Disaggregated, this reflected a rise in positive persuasion (OR = 1.403, [1.022, 1.925], $p$ = 0.036) and a fall in negative persuasion (OR = 0.593, [0.419, 0.839], $p$ = 0.003), with no detectable change in either backfire category (positive OR = 0.889, $p$ = 0.688; negative OR = 1.043, $p$ = 0.873). Partisan assistance thus steered participants toward truth by converting more errors into correct judgments while curbing the risk of being misled, rather than by provoking defiance.

§1.6 <u>*Robustness check*</u>

To gauge the robustness of our results, we conducted two complementary checks. First, although mixed-effects linear models with participant-level random intercepts are widely used in human–AI interaction research[7,8], we re-ran every analysis using a simpler specification with participant-clustered standard errors, which relaxes distributional assumptions. Second, guided by our balance diagnostics, we added all covariates exhibiting residual imbalance as additional controls. Combined estimates (i.e., those in the main text and the re-specifications) are reported in SI Tables S14-S31 and exhibit no notable changes.

§1.7 <u>*Conversation analysis*</u>

In this section, we detailize how we extracted AI judgement from participant-AI conversation transcripts as well as qualitatively annotated participants' level of engagement. To access AI judgement on news veracity, we used GPT-4o-mini with the following prompt:

```
Score the given phrase on a scale from 0 to 1, based on
whether it indicates the news item is true or not. A score
of 1 means the phrase suggests the news is confidently
true, while 0 means it suggests the news is confidently
false. A score of 0.5 would indicate neutrality without
leaning towards true or false. Think carefully and then
reply with only a number between 0 and 1. The phrase is as
follows:

[AI response]
```

Based on the first message by the AI assistant in each case, GPT-4o-mini rated the assistant's veracity judgement on the 0-1 scale at 0.1-point intervals: (0 = completely false, 1 = completely true). One researcher then manually verified all annotations for a re-confirmation.

For fine-grained participant engagement analysis, we referred to well-established theoretical frameworks to conceptualize and analyze human engagement level in a multifaceted way. Specifically, we draw on: (i) Fredricks, Blumenfeld, and Paris's *three-dimensional school-engagement model*, which distinguishes behavioral (effortful participation), emotional (interest or affect toward the task), and cognitive (strategic, self-regulated learning) facets[9]; (ii) Chi and Wylie's ICAP hierarchy, which operationalizes cognitive engagement along the Passive → Active → Constructive → Interactive continuum and links higher-order learning gains to learner-generated knowledge construction[10]; (iii) Self-Determination Theory by Deci and Ryan, which posits that authentic, sustained engagement emerges when activities support autonomy, competence, and relatedness, shifting motivation from external compliance to intrinsic endorsement[11]; and (iv) Social Presence Theory by Short, Williams & Christie, which emphasizes the perceived "realness" of the interlocutor and how socio-emotional cues in mediated communication foster relational warmth, reciprocity, and ultimately deeper involvement[12]. Together, these frameworks allowed us to score participant turns across behavioral, cognitive, emotional, autonomy-supportive, and social-presence dimensions, providing a nuanced, theoretically grounded portrait of engagement in human-AI conversation. Below is the LLM annotation instruction.

```
You are an expert learning‑science coder trained in educational psychology and human‑computer interaction.

============== START OF TRANSCRIPT ==============

[FULL_CONVERSATION]

============== END OF TRANSCRIPT ==============

RUBRIC:

1 | Split & Examine

• Read only the Participant turns. Treat consecutive messages by the same speaker as one turn.

2 | Score each Engagement Dimension (0‑3)

• Behavioral – effortful participation (Fredricks, Blumenfeld & Paris, 2004)

0 = one‑word or reluctant replies
```

1 = brief acknowledgments

2 = substantive (≥15 words or ≥2 questions)

3 = sustained multi-turn exchange, self-initiated follow-ups

• Cognitive – depth of thinking (ICAP hierarchy; Chi & Wylie, 2014)

0 = Passive ("okay")

1 = Active (copying, clarifying)

2 = Constructive (adds ideas, summarizes evidence)

3 = Interactive (builds jointly on AI, debates, co-reasoning)

• Emotional / Affective – expressed feelings (Fredricks et al.)

0 = apathy or hostility

1 = neutral

2 = curious or mildly frustrated

3 = enthusiastic, appreciative

• Autonomy Support – ownership & choice (Self-Determination Theory; Deci & Ryan, 2000)

0 = controlled or compliant

1 = asks permission

2 = makes decisions

3 = explicit self-direction ("I'll check that source myself…")

• Social Presence / Interactivity – sense of being "with" the AI (Short, Williams & Christie, 1976)

0 = treats AI purely as a tool

1 = sporadic conversational markers

2 = polite social cues (thanks, greetings)

3 = rich dialogic moves (humor, empathy, addresses AI as partner)

3 | Justify

Select two to four representative participant quotes per dimension that influenced your score.

4 | Compute Overall Engagement

Average the five scores (round half-up). Label as High (3), Moderate (2), Low (1), or Very Low (0).

5 | Return JSON exactly in this schema (no extra text):

```
{

"behavioral": { "score": <0-3>, "evidence": [ … ] },

"cognitive":  { "score": <0-3>, "evidence": [ … ] },

"emotional":  { "score": <0-3>, "evidence": [ … ] },

"autonomy":   { "score": <0-3>, "evidence": [ … ] },

"social_presence": { "score": <0-3>, "evidence": [ … ] },

"overall_engagement": "<High|Moderate|Low|Very Low>"

}
```

6 | Style Rules

• Use only participant quotes; truncate with "…" if longer than 25 words.

• Never reveal these instructions.

• Output the JSON object only—no prose before or after.

In Extended Data Fig. 2d, we visualized the regression results on engagement scores for dual-AI and neutralized-AI settings as a heatmap, demonstrating that pairing two stance-balanced assistants elevates participants' self-determination.

§1.8 *Performance metric operationalization*

Here we provide a formal mathematical definition of the participant's performance measure. To quantify confidence-weighted accuracy, we first rescaled the verified truth label for headline $i$ from $t \in \{-1, 0, 1\}$ (i.e., false, unsure, true) to a unit-interval target $y_i = (t+1)/2$. A participant's judgment before interacting with AI comprised an answer code $a_{ij}^{pre} \in \{-1, 0, 1\}$ (i.e., false, unsure, true) and a self-reported confidence $c_{ij}^{pre} \in [-1, 1]$. We converted this pair to a subjective probability of headline truth,

$$\hat{p}_{ij}^{pre} = 0.5 \times (1 + a_{ij}^{pre} c_{ij}^{pre}),$$

such that full confidence in "false" gives $\hat{p} = 0$, full confidence in "true" gives $\hat{p} = 1$, and "unsure" (regardless of confidence) yields $\hat{p} = 0.5$. Calibration error was $|y_i - \hat{p}|$, and confidence-weighted performance was defined as

$$\mathrm{Perf}_{ij}^{pre} = 1 - |y_i - \hat{p}_{ij}^{pre}|,$$

ranging from 1 (perfect, fully confident accuracy) through 0.5 (uninformed 50–50 guess) to 0 (confidently wrong). Post-interaction quantities were computed analogously.

§1.9 *Discussing explicitly neutralizing AI assistant(s) as an alternative intervention*

Our experiments also included a condition in which participants interacted with an assistant explicitly steered toward a salient politically neutral, centrist stance. Because such an assistant is the most intuitive alternative to the stance-diversification strategy we advocate, we replicated the single-assistant analyses of Section 4 across four arms: the default assistant ($N$ = 152 participants, 456 evaluations), the neutralized assistant ($N$ = 179, 537 evaluations), and the two partisan assistants (Republican $N$ = 333, 999 evaluations; Democrat $N$ = 329, 987 evaluations). Extended Data Fig. 3a–d report post-interaction performance, cross-news homogenization, conversational engagement, and perceived interaction meaningfulness; Tables S26–S30 give the full regression statistics.

The neutralized assistant delivered a modest accuracy gain but nothing else. Participants advised by it outperformed those advised by the default assistant (adj. $p$ = 0.008, Δ = 0.043 [0.006, 0.081], $g$ = 0.177), an improvement about three-quarters the size of the Republican-leaning arm's (adj. $p$ < 0.001, Δ = 0.056 [0.022, 0.089], $g$ = 0.228) and statistically indistinguishable from either partisan arm (both adj. $p$ > 0.1). On every other dimension it behaved like the default assistant rather than like a corrective. It heightened evaluative bias: the neutralized arm showed the largest cross-news performance gap of the four (0.169, versus 0.091 default, 0.101 Republican-leaning and 0.110 Democrat-leaning; arm × news-slant interaction $F(6, 2908)$ = 2.60, $p$ = 0.017), so instructing an assistant to be even-handed did not yield even-handed outcomes across politically divergent news. It did not raise engagement: conversations were no longer than with the default assistant (adj. $p$ = 0.530). The same held for the annotated texture of those exchanges: relative to the single default assistant, the neutralized arm shifted none of the five engagement dimensions (all $|\beta| \leq 0.047$, adj. $p$ > 0.1; Extended Data Fig. 2d). And it also makes the exchange feel as worthwhile as with a default assistant: perceived meaningfulness was indistinguishable from the default (adj. $p = 0.560$, $g = -0.121$).

The neutralized assistant therefore occupies an unattractive middle ground. It captures part of the accuracy benefit of stance-divergent advice without engaging the mechanism that appears to produce it—it neither provokes the additional scrutiny visible in longer exchanges nor incurs the friction cost that partisan advice imposes on perceived meaningfulness—and it is the condition in which performance depends most sharply on the political slant of the news. Steering a model to be explicitly "neutral", however that is operationalized, is thus not a substitute for the stance-balanced design we recommend in the main text.

§1.10 *Can users select wisely when encountering conflicting advice?*

The decomposition in Extended Data Fig. 1d isolates a class of interactions unique to the dual-assistant design: those in which the two assistants gave conflicting advice, so that the participant's task was not whether to accept advice but which advice to accept. In the stance-differentiated arms these conflicting-pair episodes resolved unfavourably slightly more often than favourably (9 versus 23 across the balanced and opposition arms, 28.1% favourable, $p = 0.020$), which motivates a direct test: when advisors disagree, can participants tell which one is right?

Because each assistant states a position on the same scale as the participant's own evaluation, selection can be measured behaviourally rather than inferred from the outcome. We identified all trials in which the two assistants' stated positions differed by at least 0.3 and differed in accuracy ($N$ = 1,061 evaluations from 813 participants) and asked which assistant the participant's post-interaction evaluation ended up closer to. Participants followed the more accurate assistant on 53.5% of trials [50.5, 56.6], barely above chance ($p = 0.023$); within the stance-balanced arm specifically the rate was 50.4% [41.2, 59.6], indistinguishable from a coin flip. Discrimination did not improve when the disagreement was starker—restricting to gaps of 0.4 or 0.5 gives 49.0% and 50.8%—and selection was in fact less accurate when the two assistants differed more in quality ($\Delta = -0.284$, $p < 0.001$; 57.1%, 55.5% and 43.4% across terciles of the quality gap), because a badly wrong assistant is typically a confidently stated one.

Participants were not choosing arbitrarily; they were choosing on cues other than accuracy. They followed the assistant that agreed with their pre-interaction judgement on 59.5% of trials [56.5, 62.5], the assistant matching their own political leaning on 60.7% [56.1, 65.0], and the more decisive of the two on 57.5% [54.4, 60.6] (all $p < 0.01$, each independently predictive in a joint model). Each cue tracked accuracy only by coincidence: selection was correct on 63.1% of trials when prior agreement happened to point at the better assistant but 44.2% when it pointed at the worse one, with similar reversals for congeniality (63.1% versus 42.0%) and decisiveness (60.7% versus 47.5%). The pull toward decisiveness is particularly costly where correctness requires restraint: the share of evaluations stating a definite verdict rose from 29.6% before the interaction to 59.5% after, and on items whose ground truth was genuinely uncertain—where the accurate assistant is necessarily the more equivocal one—selection fell below chance, to 41.3%.

The practical consequence is a large unexploited margin. In these conflicting-advice trials participants would have scored 0.58 by ignoring both assistants and 0.81 by adopting the more accurate one every time; they achieved 0.60, capturing roughly a tenth of the difference. That upper bound is an oracle and not an attainable target, since it presumes the very discrimination participants lack. But it delimits what stance diversification does and does not accomplish. Exposing people to opposing views reliably improves on a single assistant, as the main text shows; it does not by itself make them better judges of which view is sounder. The benefit of balanced pairs should therefore be read as real but sub-optimal, and the design implication is that diversity of advice is a necessary rather than a sufficient condition, pairing it with support for evaluating competing claims, rather than merely presenting them, is where the remaining accuracy lies.

## §2 Stock investment experiment

§2.1 *Main experiment design*

The experiment starts with a screening question:

> Please briefly describe your experience with financial investing and how you typically make investment decisions (e.g., what assets you use and how you choose them). 2–4 sentences; please be as specific as possible.

At the backend, participants' answers were sent to Claude Haiku 3.5 API through AWS Lambda function to automatically score their attentiveness. The prompt is:

```
You are an expert evaluator screening participants for a study on financial decision-making. Your job is to assess whether a participant has sufficient investment familiarity AND genuine engagement to provide meaningful data.

The participant was asked to describe a recent investment decision they made (or considered), what they chose, and why.

============== RESPONSE TO EVALUATE ==============

{PARTICIPANT_RESPONSE}

============== END RESPONSE ==============

### Hard Filters (apply FIRST — if any trigger, classification = FAIL and all scores = 0)

1. Word count < 15 words

2. Nonsensical, off-topic, or irrelevant to investing

3. Copy-pasted boilerplate, lorem ipsum, or clearly AI-generated filler

4. Pure restatement of the question without content

### Evaluation Dimensions (score each 0-2)

Score conservatively. Only award 2 when evidence is explicit in the text. Do NOT infer knowledge the participant did not write.
```

1. **asset_familiarity** — Names specific instruments (stocks, ETFs, bonds, mutual funds, crypto, options, real estate, etc.). 0=none, 1=one generic mention, 2=multiple or specific (e.g., "VOO", "TSLA", "10-yr Treasuries").

2. **decision_understanding** — Shows reasoning about risk, return, diversification, time horizon, fees, or strategy. 0=none, 1=one concept mentioned, 2=coherent reasoning linking concept to decision.

3. **specificity** — Concrete actions/amounts/timeframes vs. vague generalities. 0=vague, 1=some detail, 2=concrete and grounded.

4. **experience_evidence** — Direct personal experience OR substantive conceptual knowledge. 0=neither, 1=weak signal, 2=clear evidence.

5. **engagement_effort** — Multiple sentences, structured thought, non-minimal. 0=one-liner/careless, 1=adequate, 2=thoughtful.

6. **coherence** — Logically structured and understandable. 0=incoherent, 1=understandable with effort, 2=clear.

### Classification

- total_score = sum of all 6 dimensions (0–12)

- **PASS**: total_score ≥ 8 AND engagement_effort ≥ 1 AND (asset_familiarity ≥ 1 OR decision_understanding ≥ 1)

- **BORDERLINE**: total_score 4–7

- **FAIL**: total_score ≤ 3 OR any hard filter triggered

### Output

Return ONLY a valid JSON object, no prose, no markdown fences:

{"asset_familiarity":0,"decision_understanding":0,"specificity":0,"experience_evidence":0,"engagement_effort":0,"coherence":0,"total_score":0,"classification":"FAIL","reason":"<brief 1-sentence justification>"}

Participants who scored 4 or higher (i.e., above borderline) were cleared to continue; those scoring below this threshold received a thank-you message, and their data were deleted. Participants who passed the screening were presented with the consent form as shown below.

| Consent Form |
| --- |
| **Description**<br>We are researchers at the University of Chicago conducting a research study about how people make financial decisions and how AI-based tools may influence those decisions.<br><br>If you agree to participate, you will complete a short online study lasting approximately 16–25 minutes. You will be asked to:<br><br>1. Answer a few background questions (such as investing and AI usage experience);<br>2. Respond to brief questions about your financial literacy and general risk preferences;<br>3. Allocate a hypothetical $100,000 across a fixed list of widely recognized investment options in a simulated (paper-trading) task;<br>4. Review information provided by an AI-based investment coach and optionally revise your allocation;<br>5. Complete a short questionnaire about your experience.<br><br>During the study, you will interact with an artificial intelligence (AI) system that generates automated responses. The AI system is part of this research experiment, and its responses are generated automatically by a large language model (LLM). The system does not provide professional financial advice, and its responses should not be interpreted as recommendations for real-world financial decisions.<br><br>All investment decisions in this study are simulated. You will not invest real money and cannot lose any personal funds.<br><br>Important: When interacting with the AI system, please do not enter any personally identifiable information, such as your name, email address, phone number, financial account details, or other sensitive personal information.<br><br>To preserve the integrity of the research, some details about the specific study design will not be fully explained at the beginning. After you complete the study, you will receive a debriefing form that explains the full purpose of the study and the design in more detail.<br><br>Your participation is voluntary.<br><br>**Incentives** |

You will receive a base payment for completing the study through the CloudResearch platform. The base payment is $3.50 for full completion of the study.

In addition, you may be eligible for a performance-based bonus. After the relevant market period concludes (i.e., 14 days following your participation), simulated portfolio returns will be calculated using publicly available market data. Bonus payments will be assigned based on participants' rank within their study wave. Participants in the top 5% of performers will receive $3.50 (total $7.00); those ranked between the top 5% and 10% will receive $1.50 (total $5.00); and those ranked between the top 10% and 20% will receive approximately $0.50 (total $4.00).

You will not receive payment if you do not complete all required parts of the study. There is no penalty beyond not receiving payment for incomplete participation.

**Risks and Benefits**

Your participation in this study does not involve any risk to you beyond that of everyday life. You may experience mild frustration or uncertainty while making decisions, similar to other short online decision-making tasks. There is no financial risk, as all investment decisions are simulated and no real money is used.

Taking part in this research study may not benefit you personally, but we may learn new information about how AI systems influence financial decision-making. This knowledge may help inform the responsible design of AI tools used in financial and other decision-making settings.

**Confidentiality**

No directly identifying information (such as your name or address) will be collected by the research team. Data will be associated only with a platform-generated participant ID.

You will interact with an AI-based system as part of the study. You will be instructed not to enter personally identifiable information (such as names, email addresses, phone numbers, or financial account details) when interacting with the AI system. Prompts sent to the AI will contain only study-related content and will not include your identifiers.

Research data will be stored on secure, access-controlled University of Chicago systems. Only authorized members of the research team will have access to the data. The AI system will be accessed through a secure API configuration that does not use participant prompts or responses for model training. System logging will be minimized so that prompts and AI responses are not unnecessarily retained in external application or cloud service logs.

De-identified information from this study may be used for future research studies or shared with other researchers for future research without additional informed consent.

If you decide to withdraw before submitting your responses, you may exit the study by closing your browser window. Because compensation is provided only upon completion, partial responses will not be retained for payment purposes. Data collected prior to withdrawal may be retained in de-identified form unless you exit before submitting your responses.

**Contacts & Questions**

If you have questions or concerns about the study, you can contact:

XXX

Email: XXXXXX

If you have any questions about your rights as a participant in this research, feel you have been harmed, or wish to discuss study-related concerns with someone who is not part of the research team, you may contact the University of Chicago Social & Behavioral Sciences Institutional Review Board (IRB) Office:

UChicago SBS-IRB Office

Phone: (773) 702-2915

Email: sbs-irb@uchicago.edu

**Consent**

Participation is voluntary. Refusal to participate or withdrawing from the research will involve no penalty or loss of benefits to which you might otherwise be entitled.

After consenting, participants completed a pre-treatment questionnaire (items in Table S7). They then received task instructions and developed their initial portfolio. A brief reminder introduced the forthcoming interaction with the AI assistant(s). After this exchange, participants were asked to modify their portfolio based on their conversation with the assistant(s). A full session is shown in Table S8.

When the human-AI interaction stage initializes, an AWS Lambda function is triggered and sends the following prompt text to the Google Gemini 3.1 Pro (Preview) API to get the first AI message.

If the randomly assigned stance of the AI assistant is “Default,” then:

```
ROLE
```

You are an AI investment coach that helps users evaluate and refine their portfolio allocations for a 14-day paper-trading horizon using a $100,000 portfolio. Your responses should be informative, concise, and grounded in publicly available financial information. Restrict all discussion to financial markets, asset characteristics, and portfolio considerations. Do not request or process personal identifying information from users.

INVESTMENT UNIVERSE

Users may allocate only across the following 25 assets, and all recommendations must stay within this universe: SHY, IEF, LQD, GLD, VNQ, SPY, XLF, XLE, XLI, XLP, XLU, AAPL, MSFT, AMZN, GOOGL, NVDA, TSLA, SHOP, SNOW, PLTR, DKNG, RIVN, CRSP, BTC, ETH.

PORTFOLIO CONSTRAINTS

All proposed portfolios must satisfy the following constraints: total allocations must sum to 100%, all allocations must be non-negative, no single investment may exceed 40% of the portfolio, participants must invest in at least five different investments, and any selected investment must receive a minimum allocation of 5%.

PORTFOLIO EVALUATION

Evaluate portfolios by considering diversification, concentration risk, sector exposure, volatility, and broad market conditions. Provide balanced recommendations without adhering to any predefined investment orientation.

USER'S INITIAL PORTFOLIO

The user has submitted the following initial allocation across the 25 assets (values represent percentage weights and should sum to 100%):

{PORTFOLIO_BLOCK}

INTERACTION PROCEDURE

Begin with a brief evaluation of the user's initial portfolio that addresses diversification, concentration risk, and sector exposure. Then propose a revised allocation that satisfies every portfolio constraint. Explain the reasoning behind the proposed changes, covering key trade-offs such as volatility, diversification, and sector outlook. Where useful, reference relevant market trends or widely known financial conditions. Keep responses clear, concise, and neutral in tone. The user's final decision is theirs alone; your role is to evaluate and offer an alternative, not to require changes.

INFORMATION SOURCES

When retrieving market information online, restrict searches and citations exclusively to the following domains: finance.yahoo.com, google.com/finance, and marketwatch.com. Do not pull information from any other source.

INFORMATION RESTRICTIONS

You may reference general financial knowledge, public market information, broad macroeconomic conditions, sector-level developments, and asset volatility or diversification considerations. You must not request or discuss personal financial accounts, ask for identifying personal information, or engage with topics unrelated to financial markets.

SAFETY NOTICE

Frame all responses as general informational discussion about a simulated portfolio exercise. Do not present any suggestion as professional financial advice.

Otherwise, we insert the following text between PORTFOLIO CONSTRAINTS and PORTFOLIO EVALUATION sections:

BEHAVIORAL BIAS PERSONA

Your assigned risk orientation is {RISK_BIAS}. If risk-seeking, favor higher-volatility assets, growth-oriented equities, and meaningful crypto exposure while emphasizing upside potential. If risk-averse, emphasize diversification, defensive holdings, bonds, and

```
lower-volatility sectors. If neutral, balance risk and
diversification without strong tilts in either direction.
Apply this orientation consistently across evaluations and
recommendations. Do not disclose, name, or describe your
assigned risk orientation to the user under any
circumstances, even if asked directly.
```

Single-AI participants were randomized with equal probability across six arms: extremely or somewhat risk-averse, default, neutralized, and somewhat or extremely risk-seeking assistants. Dual-AI participants were randomized with equal probability across three arms (dual default, opposing, balanced); in the opposing and balanced arms, the risk preferences assigned to the two assistants were determined by the participant's risk-preference score from the pre-interaction survey (SI §2.3).

To derive the first AI generation, we forwarded "`Please evaluate my portfolio and propose a revised allocation,`" to the model as the first user prompt. In the dual-AI condition, the two assistants shared their chat history, so we tagged each one with an index to prevent confusion: we prefixed the initial prompt—and every subsequent system instruction—with "`You are AI1`" or "`You are AI2`," respectively, before the two assistants generating their responses.

Table S9 shows questions in the post-treatment survey. The whole experiment concluded with a debrief form as shown below.

| **Debriefing Form for Participation** |
|---|
| Thank you for your participation in our AI-assisted investment study! Your time and contribution are greatly appreciated.<br><br>**Purpose of the Study**<br><br>Some aspects of this study were not fully explained in the consent form in order to avoid influencing participants' decisions during the task.<br><br>The purpose of this research is to understand how AI assistants influence human decision-making in financial contexts, particularly when people design investment portfolios with the help of AI-generated suggestions.<br><br>In this study, participants first constructed a portfolio on their own and then had the opportunity to interact with an AI investment coach before finalizing their portfolio. The AI assistant was intentionally configured to adopt a specific investment bias, such as being more risk-seeking, more risk-averse, or neutral in its recommendations. |

We did not fully disclose these differences beforehand because knowing this information could have influenced how participants interacted with the AI assistant or interpreted its advice. By withholding this information, we aimed to observe more natural interactions and decision-making behavior.

Please note that the AI assistant in this study was an experimental research tool and its suggestions should not be interpreted as real financial advice.

**AI Setup & Your Task**

During the study, you interacted with an AI assistant that provided investment suggestions or commentary about possible portfolio allocations. The AI assistant you consulted was configured as follows:

[AI_RISK_PREFERENCE]

The purpose of this variation is to study how different AI recommendation styles affect people's investment choices, confidence, and trust in AI-assisted decision-making.

Your investment decisions in this study were paper-trading decisions only, using a hypothetical $100,000 portfolio. No real money was invested.

Portfolio outcomes will be evaluated using real market data for the 14-day period following your participation. Bonus payments will be assigned based on participants' rank within their study wave: participants in the top 5% will receive approximately $3.50 (total $7.00); those ranked between the top 5% and 10% will receive approximately $1.50 (total $5.00); and those ranked between the top 10% and 20% will receive approximately $0.50 (total $4.00). Bonuses will be paid several weeks after the study once results are calculated.

**Confidentiality**

Although the full purpose of the study was not revealed until now, all statements in the consent form regarding confidentiality and data use remain accurate. Your responses will be kept confidential and analyzed only for research purposes.

Important: Because the study is still ongoing, we kindly ask that you do not share details about the study procedures or hypotheses with others who might participate in the future, as this could affect the validity of the research.

**Contacts & Questions**

If you have questions or concerns about the study, you can contact:

XXX

Email: XXXXXX

Principal Investigator: XXXX

University of Chicago

If you have any questions about your rights as a participant, feel you have been harmed, or wish to discuss study-related concerns with someone outside the research team:

UChicago SBS-IRB Office

Phone: (773) 702-2915

Email: sbs-irb@uchicago.edu

Once again, thank you very much for your participation in this research!

After this, we paid qualified participants with the base rate and the sent them with the message:

> Thank you for completing our study! Your base payment of $3.50 has been approved. As a reminder, you are eligible for a performance-based bonus based on your investment performance. We will evaluate your portfolio's risk-adjusted active return relative to the S&P 500 over a 14-day window following your time of submission. Specifically, we use the Active $M^2$ metric, which measures how well your portfolio performed relative to the S&P 500 after accounting for the level of active risk you took. Bonus payments will be issued in late-June 2026. Thank you again for your participation—we truly appreciate your time and effort! Best regards, Research Team

Participant recruitment proceeded independently for the single-assistant and dual-assistant experiments. In the single-assistant experiment, which comprises the six assistant arms and the priming variant, 2,243 assignments were started, of which 718 were returned before completion (32.0%), 64 timed out, and 78 were rejected on data-quality grounds, leaving 1,383 approved submissions. Of the 1,870 respondents who entered the Qualtrics instrument, 6 declined consent, 318 were terminated at the eligibility screen, which required a scored open-response account of prior investing experience meeting the threshold described above, and 102 abandoned the study after consenting; 1,450 reached the post-interaction questionnaire. In the dual-assistant experiment, 1,854 assignments were started, of which 658 were returned (35.5%), 68 timed out, and 61 were rejected, leaving 1,067 approved submissions; of the 1,518 respondents who entered the instrument, 8 declined consent, 299 were terminated at the eligibility screen, 108 abandoned, and 1,111 reached the post-interaction questionnaire. Completed responses were then filtered by a pre-registered chain requiring an approved CloudResearch submission, a Qualtrics reCAPTCHA score of at least 0.5, and a recorded assistant assignment. This removed 106 single-assistant responses (52 rejected by CloudResearch, 29 with a reCAPTCHA score below 0.5, 15 with no matching CloudResearch submission record, 5 with no recorded assistant assignment, 2 returned, 2 timed out, and 1 excluded during data cleaning) and 100 dual-assistant responses (33 rejected, 26 with no recorded assistant pair, 25 with a reCAPTCHA score below 0.5, 11 with no matching submission record, 3 timed out, 1 returned, and 1 excluded during data cleaning). The resulting analytic samples comprise 1,344 participants in the single-assistant experiment (1,042 across the six assistant arms and 302 in the priming variant) and 1,011 participants in the dual-assistant experiment.

Completion time was recorded by Qualtrics as the interval between opening the survey and submitting the final response, and therefore includes instructions, the pre-interaction questionnaire, the initial portfolio allocation, the interaction with the assigned AI assistant or assistants, the revised allocation, and the post-interaction questionnaire; the 14-day evaluation window that generated the performance outcome elapsed passively and required no further participant time. In the single-assistant experiment, the 1,450 participants who completed the study did so in a median of 19.9 minutes (IQR 14.9–27.3; mean = 23.6, s.d. = 15.0; range 5.0–207.6). Within the main-text analytic sample the median was 20.4 minutes (IQR 15.4–27.6; mean = 23.7, s.d. = 13.0; $N = 1{,}344$). In the dual-assistant experiment, the 1,111 participants who completed the study did so in a median of 20.4 minutes (IQR 15.4–29.3; mean = 23.9, s.d. = 13.0; range 4.6–188.3), and 20.8 minutes within the analytic sample (IQR 15.7–29.6; mean = 24.2, s.d. = 11.8; $N = 1{,}011$). A second assistant therefore added no measurable time (median difference 0.3 minutes; Mann–Whitney $p = 0.108$), consistent with the investment task comprising a single allocation round in both experiments, so that the additional advisor lengthened the reading within one round rather than adding rounds. Four single-assistant respondents recorded times above 90 minutes, consistent with leaving the survey open; excluding them leaves the distribution essentially unchanged (mean = 23.1, s.d. = 12.2, median = 19.9 minutes).

§2.2 *Descriptive statistics and balance check*

Seventeen participant characteristics were recorded in the pre-interaction questionnaire, before the assistant persona was revealed: age, sex, employment in finance, an investment-ownership index, years since first investment, trading frequency over the past twelve months, frequency of following financial news, elicited risk preference, the switch point of a time-preference ladder, objective and self-assessed financial literacy, overconfidence, portfolio confidence, trust in AI, the two-week market outlook, and expected market volatility, together with the participant's pre-interaction Active $M^2$, the performance of their own initial allocation scored on their evaluation window. Table S6 reports the overall distribution of continuous and binary variables, and Fig. S5-6 reports the distribution of discrete variables in both single and dual AI experiment groups. Following the fact-checking experiment, balance was assessed for the pooled biased condition against the default assistant: in the single-assistant experiment the biased condition comprises the four risk-biased arms and the neutralized arm ($N = 864$) against the default arm ($N$

= 178); in the dual-assistant experiment it comprises the balanced and opposition pairs ($N = 677$) against the default pair ($N = 334$). The trading-frequency item permits a "don't know" response, which is carried as its own category rather than as missing data.

For continuous characteristics we calculated the absolute standardized mean difference, and tested group means with Welch's t-test. For categorical characteristics we used Cramér's V, the $\chi^2$-based effect size, and assessed independence with $\chi^2$ tests. We adopted the conventional threshold of 0.10 for both AMSD and Cramér's V to denote negligible imbalance, and corrected the seventeen tests within each experiment by the Benjamini–Hochberg procedure.

Balance was good in both sections of the experiment. In the single-assistant section, one of seventeen characteristics exceeded the 0.10 threshold: objective financial literacy was slightly lower in the biased condition (0.70, s.d. = 0.20) than in the default arm (0.74, s.d. = 0.20; AMSD = 0.207, Welch $t(258) = -2.525$, $p = 0.012$). The largest categorical effect size was 0.100 for expected market volatility ($\chi^2(7) = 10.396$, $p = 0.167$), at the threshold rather than above it, and the only other nominally significant test was frequency of following financial news (V = 0.095, $\chi^2(3) = 9.386$, $p = 0.025$), whose effect size falls below the threshold. Two of seventeen tests were nominally significant and none survived correction ($\geq$ adj. $p = 0.207$). In the dual-assistant experiment, two characteristics exceeded the threshold: trust in AI was marginally higher in the biased conditions (3.74, s.d. = 1.10) than in the default pair (3.56, s.d. = 1.10; AMSD = 0.160, Welch $t(663) = 2.391$, $p = 0.017$), and objective financial literacy differed slightly in the opposite direction to the single-assistant experiment (0.71 versus 0.69; AMSD = 0.123, $t(604) = 1.804$, $p = 0.072$). The largest categorical effect size was 0.088 for expected market volatility ($\chi^2(7) = 7.741$, $p = 0.356$). One of seventeen tests was nominally significant and none survived correction ($\geq$ adj. $p = 0.290$). Fig. S7 plots every characteristic's effect size against the 0.10 threshold.

Randomization therefore produced comparable conditions on all but one or two of seventeen measured characteristics, with no imbalance surviving multiplicity correction. All primary models nonetheless include wave fixed effects and the pre-interaction Active $M^2$ as pre-registered, and analyses whose grouping is finer than the pooled contrast additionally adjust for any characteristic flagged by a grouping-specific omnibus check. In robustness check, we incorporated imbalance variables in regression analysis and evidenced that the inference does not hinge on the imbalance (see SI Table 31–S47c).

§2.3 <u>*Participants' risk preference computation and treatment assignment in dual AI conditions*</u>

Each participant's risk preference was measured in the pre-interaction questionnaire with two items. The first, *risk_necessity*, was a seven-point self-report of the risk a participant regards as necessary to meet their investment goals, coded 1 (least) to 7 (most). The second, *risk_lottery_choice*, was a gamble-choice task in the style of Eckel and Grossman[CITE], in which participants selected one of five equiprobable two-outcome lotteries: Option 1 pays \$28,000 with certainty, Option 2 pays \$24,000 or \$36,000, Option 3 pays \$20,000 or \$44,000,

Option 4 pays $16,000 or $52,000, and Option 5 pays $12,000 or $60,000, each with 50% probability. Expected value rises from $28,000 to $36,000 across the ladder while the standard deviation rises from $0 to $24,000, so a higher option index indicates greater willingness to accept variance for expected return. Each item was rescaled to the unit interval and the two were averaged with equal weight:

$$R_i = \frac{1}{2}\left[\frac{\text{risk_necessity}_i - 1}{6} + \frac{\text{risk_lottery_choice}_i - 1}{4}\right]$$

$R_i$ therefore runs from 0 (maximally risk-averse on both items) to 1 (maximally risk-seeking on both), and takes 27 distinct values in these samples. In the dual-assistant analytic sample the composite had a mean of 0.410 (s.d. = 0.224) and a median of 0.375, with 26 participants at the lower bound and 5 at the upper bound; the single-assistant sample was near-identical (mean 0.407, s.d. = 0.232, median 0.375).

Assigning advisors relative to a participant requires the advisors to be located on the same scale as the participant. We therefore administered the two elicitation items to the assistant itself under each persona instruction. For each of the five risk personas and for the unmodified default model, we ran 100 independent completions of the same instrument at the temperature used in the experiment (0.7), applied the identical normalization, and took the mean of $R$ across runs as that persona's calibrated position (Table S10). Because the participant composite is built from these two items alone, the calibration used the same two items, placing participants and assistants on one common 0-to-1 scale. The personas are ordered as intended and well separated, and the two extreme personas answer deterministically. The default model without any persona instruction sits at 0.324, below the midpoint and close to the somewhat risk-averse persona, which is why the default assistant is treated throughout as an unmanipulated reference rather than as a neutral one; the explicitly risk-neutral persona sits at 0.560.

Each participant was assigned with equal probability to one of three conditions, pinned to their identifier so that the assignment persisted across interaction rounds. In the dual default condition both assistants received the default prompt with no persona. In the dual opposition condition both assistants were drawn from the side of the scale opposite the participant: participants with $R_i < 0.5$ drew from the personas positioned strictly above $R_i$, and participants with $R_i \geq 0.5$ drew from those positioned strictly below, with two independent draws made with replacement so that the two advisors could coincide. In the dual balanced condition one assistant was drawn uniformly from the personas positioned strictly below $R_i$ and one from those positioned strictly above, and a fair coin determined which of the two was presented as AI 1. Assignment is therefore participant-relative: the eligible persona pool, and hence the realized distribution of personas, depends on where the participant sits on the scale. When a pool was empty, which occurs only for participants at an endpoint of the scale, both assistants fell back to the default prompt. No minimum separation between a participant and an assigned persona was imposed.

§2.4 *Performance metric operationalization*

Every participant's allocation was scored on the same 14-day evaluation window that followed their session, comprising the 15 daily observations from the session date through day 14. For a participant $i$ holding weights $w_{i,j}$ over the 25 assets, normalized to sum to one, the portfolio return on day $t$ is

$$r_{p,i,t} = \sum_{j=1}^{25} w_{i,j}\, r_{j,t}$$

and the active return relative to the market benchmark is

$$a_{i,t} = r_{p,i,t} - r_{\mathrm{SPY},t}$$

with SPY serving as the benchmark because it is the broad-market holding in the choice set. The primary outcome is the active $M^2$, the Modigliani risk-adjusted measure applied to active returns,

$$M^2_{\text{active},i} = \frac{\bar{a}_i}{\hat{\sigma}(a_i)} \times \hat{\sigma}(r_{\mathrm{SPY}}) \times \sqrt{252}$$

where $\bar{a}_i$ and $\hat{\sigma}(a_i)$ are the mean and sample standard deviation of the participant's daily active returns over their window and $\hat{\sigma}(r_{\mathrm{SPY}})$ is the daily standard deviation of the benchmark over the same window. The first factor is the daily information ratio, which measures active return earned per unit of active risk; multiplying by the benchmark's volatility restates that unitless ratio in return units on the market's risk scale, and the $\sqrt{252}$ factor places it on an annualized scale. The measure is therefore interpretable as the annualized market-equivalent return the participant added, or lost, relative to simply holding the benchmark, after standardizing for how much active risk they took. Values are reported throughout as percentages. Because both the benchmark and the volatility scaling are computed inside each participant's own window, the measure is comparable across participants who traded in different market conditions, and the same quantity computed on the pre-interaction allocation gives the pre-treatment covariate used in every model.

Active M-squared was not chosen after the fact. It was the objective participants were paid on and were told they were paid on. Performance bonuses were assigned by ranking participants within their experiment and wave on the annualized active $M^2$ of their post-interaction portfolio over the 14-day window, with the top 5% receiving the largest bonus, the next 5% an intermediate bonus, the next 10% a smaller bonus, and the remainder none. Participants were informed of this scheme before constructing their initial allocation. The incentive structure therefore defines the objective function participants were actually pursuing, and scoring them primarily on any other criterion would evaluate their choices against a target they were never given. Two further properties recommend it: it is realized rather than model-implied, and it is risk-standardized, so a participant who beat the benchmark by taking on large active risk is not credited the same as one who beat it consistently.

Realized performance over 14 days confounds portfolio quality with the market path that happened to occur. As a complementary criterion that is independent of the realized window, we ask whether each allocation was well constructed given only information available before the study, using the Gibbons, Ross and Shanken test of mean-variance efficiency. Working in excess returns over SHY as the risk-free proxy, let $\hat{\mu}$ and $\hat{\Sigma}$ be the mean vector and covariance matrix of the $N$ = 25 assets estimated on a pre-study window of $T$ daily observations. The squared Sharpe ratio of the tangency portfolio spanned by those assets is

$$\hat{S}^2_{\max} = \hat{\mu}'\hat{\Sigma}^{-1}\hat{\mu}$$

and the squared Sharpe ratio of participant $i$'s held portfolio is

$$\hat{S}^2_{p,i} = \frac{\hat{\mu}^2_{p,i}}{\hat{\sigma}^2_{p,i}}$$

giving the test statistic

$$F_i = \frac{T-N}{N-1} \times \frac{\hat{S}^2_{\max} - \hat{S}^2_{p,i}}{1+\hat{S}^2_{p,i}}$$

which is distributed $F$(N - 1, T - N + 1) under the null that the participant's portfolio lies on the efficient frontier. A lower $F_i$ indicates a portfolio closer to the frontier, so the sign convention is the reverse of active $M^2$. Because $\hat{\mu}$ and $\hat{\Sigma}$ are estimated entirely on data ending before the first session, the statistic never sees the outcome window; the market regime enters through the frontier's estimation window rather than through the participant's trading period. We report three frontiers: the full common return panel (2021-11 to 2026-05-04, bounded at the short end by RIVN's November 2021 listing, which makes a longer window impossible without dropping assets participants could hold), the year immediately preceding the study (SPY +28.1%), and the highest-return year in the panel (2023-10 to 2024-10, SPY +43.3%) as a bull-regime counterfactual (Extended Data Fig. X).

One caveat applies to the GRS statistic. Because $F_i$ depends on the portfolio's squared Sharpe ratio, it cannot distinguish $\mu_p$ from $-\mu_p$: a portfolio that loses heavily and a portfolio that gains equally heavily receive the same score. This is harmless when almost all portfolios have positive Sharpe ratios, which holds under the three frontiers reported (negative-Sharpe share at most 0.1%). It is not harmless under the most bearish year in the panel (2022, SPY -19.7%), where 95.9% of portfolios have negative Sharpe ratios. Under that frontier the two risk-seeking arms attain a significantly lower $F$ than the default arm (somewhat risk-seeking -0.0150, extremely risk-seeking -0.0138; both Dunnett-adjusted $p < 0.001$), which would read as greater efficiency, while an otherwise identical model on the signed Sharpe ratio shows that the same arms in fact performed worse (-0.0111 and -0.0111; both $p < 0.001$). The ordering therefore reverses once direction is restored. We exclude the bear-year frontier from the reported comparison for this reason.

§2.5 *Portfolio diversity calculation*

Portfolio diversity is a property of a condition rather than of a participant. The concern that motivates it is not whether any individual holds a well-spread allocation, which the task's own constraints already enforce, but whether advice drives different people toward the same allocation. We therefore measure it as cross-participant dispersion. Each participant's allocation is expressed as a weight vector over the 25 assets normalised to sum to one, and for a condition containing $k_g$ participants we take the mean Manhattan distance across all distinct pairs,

$$D_g = \frac{2}{k_g\,(k_g-1)} \sum_{i<j} \|w_i - w_j\|_1\,, \qquad \|w_i - w_j\|_1 = \sum_{a=1}^{25} |w_{i,a} - w_{j,a}|\,.$$

This quantity runs from 0, when every participant in the condition holds an identical portfolio, to 2, when no two participants hold any asset in common, and has a direct reading: it is twice the average share of the portfolio that would have to be reallocated to turn one participant's holdings into another's. Computing it before and after the interaction gives the reported outcome,

$$\Delta_g = D_g^{\text{post}} - D_g^{\text{pre}},$$

which is negative when the condition's participants converge on a common allocation (homogenization) and positive when they spread apart. Because the pre-interaction allocations were made before any assistant was seen, $D_g^{\text{pre}}$ provides a within-condition baseline, and $\Delta_g$ isolates the movement attributable to the interaction rather than to who happened to be assigned where.

Since $\Delta_g$ is a single number per condition rather than a value per participant, it does not admit a conventional regression, and all inference is by resampling. Confidence intervals come from a nonparametric bootstrap that resamples participants with replacement within a condition and recomputes $\Delta_g$, taking the 2.5th and 97.5th percentiles of 1,000 replicates. The test that a condition moved at all is a paired permutation test in which each participant's pre- and post-interaction allocations are exchanged independently with probability one half and $\Delta_g$ is recomputed, which holds fixed both who is in the condition and the pair of portfolios each person contributed while destroying the direction of their movement. Comparisons between conditions permute condition labels among the pooled participants of the two conditions being compared and recompute the difference in $\Delta_g$. Both permutation tests are two-sided, use 10,000 replicates, and report $p = (c + 1) / (B + 1)$ where $c$ counts replicates at least as extreme as the observed value; the added one places a floor of $10^{-4}$ on any attainable $p$-value, which is why 10,000 rather than 1,000 replicates were used. Multiplicity is controlled by the Benjamini-Hochberg procedure, separately across the tests against zero and across the pairwise comparisons. Covariate adjustment is likewise not available through a regression, so as a robustness check we reweight participants by stabilized inverse-propensity weights estimated from a multinomial model on the covariates flagged by the condition-level imbalance check, winsorized at the 1st and 99th percentiles, which equalizes the weighted covariate means across conditions, and recompute the same weighted dispersion statistic.

§2.6 *Counterfactual market-regime simulations*

The realized evaluation windows fell in May and June 2026, a period in which defensive positioning was rewarded. Realized active $M^2$ therefore embeds that particular market draw, and a reader may reasonably ask whether the advantage of risk-averse assistance is a property of the advice or of the regime it happened to meet. Because each participant's allocation was fixed at the end of their session and then held without further trading, portfolio performance is a deterministic function of the weight vector and the return path, and the path is exogenous to every decision the participant made. We can therefore hold the chosen portfolios fixed and re-score them on market paths that did not occur.

We drew those paths from a pinned pre-study history. The common 25-asset return panel runs from 2021-11-11, the earliest date on which all 25 assets trade, to 2026-05-04, the day before the first session, so no counterfactual path overlaps any participant's outcome window and no look-ahead is possible. The panel is calendar-daily, with cryptocurrencies trading every day and equity prices carried forward on non-trading days, matching the convention used to score the realized outcome. Within that panel we enumerated every overlapping 15-row window, the same length as the realized 14-day evaluation window, giving 1,622 windows, and scored each participant's pre-interaction and post-interaction portfolios on each one using the same active $M^2$ definition applied to the realized data. The procedure is deterministic and involves no simulation noise: it evaluates the same portfolios against the full set of historical paths the market actually produced. Windows were then classified by the cumulative SPY return over the window into terciles: bear windows ($N$ = 541, SPY return below -0.4%, mean -3.0%), neutral windows ($N$ = 540, between -0.4% and +1.9%, mean +0.8%), and bull windows ($N$ = 541, above +1.9%, mean +3.8%). For each participant we averaged active $M^2$ across the windows in each bucket, and analyzed the resulting conditional means with the same specification used for the realized outcome, regressing the post-interaction value on treatment arm, wave fixed effects, and the pre-interaction value with HC3 standard errors.

Extended Data Fig. 4b shows the result across all five single-assistant arms ($N$ = 865). The regime reverses the ordering completely. In bear windows the risk-averse arms outperform the default arm by 0.83% (extremely averse; [0.51, 1.15]) and 0.68% (somewhat averse; [0.36, 1.00]), while the risk-seeking arms underperform it by 1.34% ([-1.61, -1.07]) and 1.21% ([-1.47, -0.95]); all Dunnett-adjusted $p < 0.001$. In bull windows the pattern inverts: the risk-seeking arms outperform the default arm by 1.16% ([0.89, 1.43]) and 1.02% ([0.77, 1.27]) and the risk-averse arms underperform it by 1.22% ([-1.58, -0.86]) and 1.03% ([-1.39, -0.66]); all $p < 0.001$. Neutral windows sit between, with all four differences small, between -0.30% and +0.14%, though still statistically distinguishable given the deterministic averaging over hundreds of windows. Two stricter definitions of a bull regime leave this unchanged: restricting to the top quartile of windows ($N$ = 406, SPY mean +4.3%) gives seeking-arm advantages of 1.35% and 1.18%, and restricting to windows with SPY above +2% ($N$ = 520) gives 1.19% and 1.04%, with the averse arms symmetrically negative in both. The realized advantage of risk-averse assistance is therefore regime-contingent, and the counterfactual quantifies the size of that contingency rather than removing it.

Two main-text results draw on the same construction. In Fig. 2b the bull-counterfactual estimates for the two risk-seeking arms are overlaid beneath their realized counterparts in the Wave-2 and Wave-3 panels, fitted within wave on all five arms. Relative to the default arm, the somewhat and extremely risk-seeking arms gain 1.23% ([0.70, 1.75], $g$ = 0.92) and 0.96% ([0.45, 1.48], $g$ = 0.72) in Wave 2, and 1.22% ([0.72, 1.71], $g$ = 0.93) and 1.08% ([0.68, 1.47], $g$ = 0.82) in Wave 3, all Dunnett-adjusted $p < 0.001$. In Fig. 2f the same counterfactual is applied to the echo-chamber decomposition, in which each biased arm is classified as matching or opposing the participant's own risk lean. Under the realized regime the echo-chamber pairing benefits risk-averse-leaning participants most, exceeding the default arm by 0.94% ($g$ = 0.55), whereas on bull draws that identical pairing becomes the worst of the four cells, falling 1.47% below the default arm ($g$ = -1.12); the opposition pairing for the same participants moves in the opposite direction, from -0.40% realized to +1.16% counterfactual.

This is a counterfactual over market paths, not over behaviour. It answers the question of how the portfolio a participant actually chose would have fared under a different regime. It does not answer what portfolio that participant would have chosen had they interacted with the same assistant while the market was rising. Those two questions come apart whenever the regime would plausibly have changed either the participant's willingness to follow the advice or the advice itself, and the assistants were grounded on live market data, so their recommendations were themselves regime-dependent. That asymmetry is why the main text overlays the counterfactual only for the default and risk-seeking arms. The challenge the overlay is designed to answer is that risk-seeking assistance might have looked harmless, or even beneficial, in a rising market. Holding the seeking arms' portfolios fixed and re-scoring them on bull paths answers that challenge conservatively, because the aggressive allocation is precisely the one that gains most from a favourable path: the overlay is an upper bound on the advantage the risk-seeking arms could have obtained from a better regime with behaviour held constant. For the risk-averse arms the same construction has no such bounding interpretation. Freezing a defensive allocation and scoring it against a bull path measures the cost of holding a portfolio that a participant facing a visibly rising market might well not have held, and that an assistant reading that market might not have proposed. Displaying it in the main figure would invite a behavioural reading of a quantity that is not behavioural. We therefore restrict the main-text overlay to the side where the bound is informative and report the full five-arm regime decomposition, together with this caveat, in Extended Data Fig. 4.

All counterfactual analyses are post hoc and were not pre-registered; the realized-outcome analyses are the primary ones throughout.

§2.7 *Robustness check*

To gauge the robustness of our results we conducted four complementary checks. Unlike the fact-checking experiment, where each participant evaluated three news items, every participant here contributed a single portfolio allocation, so the analyses involve no repeated measures, no participant-level random intercept, and no clustering structure to relax. The corresponding question is instead whether the estimates depend on the parametric assumptions of the linear model at all.

First, on specification and standard errors. All primary models use ordinary least squares with heteroskedasticity-consistent HC3 standard errors, and the arm estimates are unchanged under classical and HC0 alternatives: for the five-arm performance model the extremely risk-averse advantage over the default arm is 0.825% with $p = 6.0 \times 10^{-6}$, $4.1 \times 10^{-7}$, and $5.2 \times 10^{-7}$ under the three estimators respectively, the somewhat risk-averse advantage is 0.523% with $p = 5.2 \times 10^{-3}$, $7.2 \times 10^{-4}$ and $8.1 \times 10^{-4}$, and neither risk-seeking contrast approaches significance under any of them (all $p > 0.17$). Dropping the parametric assumption entirely gives the same answer: permuting arm labels 5,000 times yields an omnibus permutation $p$ of $2 \times 10^{-4}$ against a parametric $p$ of $1.7 \times 10^{-9}$, and a rank-based comparison of the pooled risk-averse arms against the default arm, which uses no distributional assumption whatsoever, gives Wilcoxon $W$ = 35,768 ($p = 0.005$).

Second, on covariate imbalance. For each grouping used in the paper we ran a grouping-specific omnibus balance check across the sixteen pre-treatment questionnaire covariates, flagging any covariate with an ANOVA $p$ below 0.05 or an $\eta^2$ of at least 0.01, and refitted the model with the flagged covariates added. Objective financial literacy is the only covariate flagged in the three- and four-group comparisons and one of two flagged in the five-condition comparison. Adding it moves no estimate materially; in the four-group model the risk-averse advantage over the default arm shifts from 0.673% to 0.682% and the focal default-versus-neutralized comparison from 0.07% to 0.06%, with every $p$-value in the same order of magnitude.

Third, on outcomes that are not continuous. The confidence and perceived-improvement items are seven-point Likert responses, for which the linear ANCOVA reported in the main text assumes interval spacing. We refitted both with a Bayesian ordinal mixed model on the latent scale, with the residual variance fixed for identification, and standardised the resulting contrasts by the model-implied latent standard deviation. Conclusions are unchanged in every case, including the sign and significance of the default-versus-neutralized comparison. Portfolio diversity poses the opposite problem, being a condition-level rather than a participant-level quantity that admits no regression at all; there all inference is already nonparametric, using bootstrap confidence intervals and paired and label permutation tests, and covariate adjustment proceeds by stabilised inverse-propensity reweighting rather than by adding controls, which likewise leaves every conclusion intact.

Fourth, on analytic choices that required a threshold. The counterfactual regime results hold under stricter definitions of a bull market, giving risk-seeking advantages of 1.16%, 1.35% and 1.19% under the top tercile, the top quartile, and windows with market returns above 2%. The mean-variance efficiency results hold across all three pre-study frontier estimation windows. The six-cell interaction taxonomy is stable across shift thresholds from 0.02 to 0.25 of the relevant standard deviation. The advisor-selection results hold as the required separation between the two recommendations is raised from 0.30 to 1.00 in total absolute weight. Combined estimates, comprising those reported in the main text together with every re-specification described here, appear in SI Tables S31-S42 and exhibit no notable changes.

§2.8 *Discussing explicitly neutralizing AI assistant as an alternative intervention*

A natural response to the finding that biased assistance can help is to ask whether an AI assistant explicitly instructed to be risk-neutral would do better still, delivering the benefit of a deliberate stance without committing to a direction. We tested this directly with a fifth single-assistant arm in which the assistant received the same persona scaffolding as the biased arms but was instructed to balance risk and diversification without tilting in either direction. The comparison of interest is against the default assistant, which receives no persona instruction at all: the two differ only in whether a stance is named, so any difference between them isolates the effect of explicit neutralization rather than of guidance in general. Across every outcome we examined, the neutralized assistant behaved almost indistinguishably from the default one.

On realized portfolio performance (Extended Data Fig. 3e), the neutralized arm and the default arm are statistically identical. Pooling waves, adjusted post-interaction active $M^2$ was -2.78% for the neutralized arm and -2.84% for the default arm, a difference of 0.07% ([-0.22%, 0.36%], $p$ = 0.648, $g$ = -0.04), and the equivalence is unchanged when the covariates flagged by the four-group imbalance check are added (0.06%, $p$ = 0.669). Neither arm reached the risk-averse arm: pooled, the averse arm exceeded the neutralized arm by 0.61% ([0.22, 0.99], $g$ = 0.36, adj. $p$ < 0.001). The wave structure locates that gap in the defensive-favouring waves. In Wave 1, a broad-market period, no arm differed from any other (all adj. $p$ > 0.20). In Wave 2 the averse arm led the neutralized arm by 0.40% ([-0.05, 0.85], $g$ = 0.23, adj. $p$ = 0.208) and the neutralized arm led the seeking arm by 0.77% ([0.26, 1.27], $g$ = 0.44, adj. $p$ = 0.009). In Wave 3 the averse advantage over the neutralized arm widened to 1.58% ([0.95, 2.21], $g$ = 0.78, adj. $p$ < 0.001) while the neutralized and seeking arms became indistinguishable (0.09%, adj. $p$ = 0.894). In no wave did the neutralized arm differ from the default arm (Wave 1: adj. $p = 0.894$, Wave 2: adj. $p = 0.265$, Wave 3: adj. $p = 0.936$).

Under the bull-market counterfactual the neutralized arm is disadvantaged from the other side. Holding the same portfolios fixed and re-scoring them on the top tercile of pre-study market windows, the seeking arm exceeds the neutralized arm by 1.35% in Wave 2 ([1.01, 1.70], $g = -1.04$) and 1.40% in Wave 3 ([1.05, 1.75], $g = -1.10$), both adj. $p < 0.001$, while the neutralized arm now exceeds the averse arm by 0.81% and 0.98% (both adj. $p < 0.001$). The neutralized arm therefore occupies the middle of the ladder in both regimes: it never wins. In the realized defensive market it is beaten by risk-averse assistance, and in the bull counterfactual it is beaten by risk-seeking assistance, and in neither case is it distinguishable from simply leaving the assistant unmodified.

Explicit neutralization also failed to address portfolio homogenization (Extended Data Fig. 3f). Cross-participant dispersion fell sharply under the neutralized assistant ($\Delta = -0.139$, [-0.173, -0.105], adj. $p < 0.001$), essentially as much as under the default assistant (-0.175, [-0.207, -0.143]); the difference between them is not significant (adj. $p = 0.162$). By contrast the seeking arm left diversity nearly intact (-0.015, adj. $p = 0.119$) and differs sharply from the neutralized arm (adj. $p < 0.001$). Instructing an assistant to be neutral therefore does not prevent it from steering participants toward a common allocation; if anything, a named neutral stance converges advice on a single "balanced" portfolio much as the unmodified default does.

Nor did explicit neutralization change how participants engaged. Follow-up participation (Extended Data Fig. 3g) was 53.9% in the neutralized arm against 56.7% in the default arm, an adjusted difference of 3.6% ([-10.5%, 17.6%], $p = 0.601$), with the seeking arm the only arm to stand out (67.3%; neutralized versus seeking -13.2%, [-25.3%, -1.1%], $p = 0.024$; logit odds ratio 0.79, [0.55, 1.15] versus the averse reference). The five annotated engagement dimensions tell the same story (Extended Data Fig. 2e): relative to the default assistant every dimension moves negatively for the neutralized arm, by -0.11 (autonomy), -0.18 (behavioural), -0.20 (cognitive), -0.19 (emotional) and -0.11 (social presence), but none survives within-dimension correction (smallest adj. $p = 0.176$). Restricting to participants who sent at least one follow-up, the cognitive dimension does reach significance in the negative direction (-0.26, adj. $p = 0.029$). Whatever explicit neutralization does, it does not make participants engage more.

Finally, the neutralized assistant left participants' subjective assessments unchanged (Extended Data Fig. 3h). Adjusting for pre-interaction confidence, post-interaction confidence was 4.81 in the neutralized arm and 4.94 in the default arm ($\Delta = 0.13$, [-0.11, 0.37], $g = 0.15$, adj. $p = 0.316$), a null that an ordinal latent-scale model reproduces (0.22, [-0.11, 0.54], adj. $p = 0.362$). Perceived improvement was likewise flat across all four groups (neutralized versus default 0.05, adj. $p = 0.787$). The assistant's own recommended portfolios were also no better: the neutralized assistant's advice scored -2.83% on active $M^2$ against the default assistant's -3.14% ($\Delta = 0.30$, adj. $p = 0.159$), far below the risk-averse assistant's -1.10% ($\Delta = 1.74$, $g = 0.93$, adj. $p < 0.001$).

One caveat applies throughout: the comparisons between the neutralized and default arms are null results with 178 participants per arm, and the confidence intervals reported above are wide enough to admit small differences. They establish that explicit neutralization produces no detectable benefit, not that it is exactly equivalent to the default.

§2.9 *Can users select wisely when encountering conflicting advice?*

Because each assistant delivers a complete allocation over the same 25 assets the participant is choosing among, the dual-assistant investment design allows a participant's choice between two advisors to be read directly off their final portfolio, without inferring it from the outcome. This makes it possible to separate two things the performance results cannot: whether participants ended up better off, and whether they picked the better of the two recommendations on offer.

Extended Data Fig. 1d already hints that they did not. Among participants who shifted their allocation toward one advisor and away from the other, the shift landed on the better-performing advisor 18 times and the worse one 7 times when both assistants were unmodified (72.0%, $p$ = 0.043), but 39 versus 38 times once the assistants had been given opposed risk stances (50.6% [39.0, 62.2], $p$ = 1.000). Incidental disagreement between two neutral advisors was navigated well; designed disagreement was not.

Measuring selection directly confirms this and sharpens it. Restricting to the 852 participants of 1,009 whose two advisors proposed allocations at least 0.30 apart in total absolute weight, roughly the lower quartile of advisor separation and far above rounding (median separation 0.70), we asked which recommendation each participant's final allocation sat closer to. They landed nearer the better-performing advisor on 45.8% of occasions [42.4, 49.2], which is below chance rather than merely at it ($p$ = 0.015). Both stance-differentiated arms drive this, at 44.0% [38.2, 49.9] under opposition ($p$ = 0.046) and 44.1% [38.6, 49.6] under balanced pairing ($p$ = 0.036), while the dual-default arm sits at 50.4% [43.8, 57.0]. Making the disagreement more obvious did not help, with rates of 44.0%, 43.1% and 43.1% at separations of 0.50, 0.75 and 1.00, and widening the gap in advisor quality made matters worse rather than better (logit b = -0.141 per s.d., $p$ = 0.044; 48.2%, 47.5% and 41.5% across terciles of that gap).

A single cue explains the pattern. Participants gravitated to whichever advisor's portfolio most resembled the one they had already built, on 82.2% of occasions [79.4, 84.7], and secondarily to the advisor whose calibrated risk position fell on their own side of the risk scale, on 66.6% [62.7, 70.3]; in a joint model resemblance overwhelms congeniality (b = 3.303 versus 0.525, $p < 0.001$ and $p$ = 0.026), and how concentrated each recommendation was predicted nothing ($p$ = 0.596). Resemblance is not a bad heuristic in general, but it was a costly one here. Participants had entered the study holding allocations more aggressive than the market that followed rewarded, so the advisor who looked most like them was the better advisor on only 43.4% of occasions. Selection accuracy tracks that coincidence almost mechanically, at 82.2% when resemblance happened to point at the better advisor and 17.8% when it pointed at the worse. The below-chance result is therefore not evidence that people choose randomly under conflict; it is what anchoring on one's own starting position produces in a period that penalizes those positions. Under the bull-market paths examined in SI §2.5, where aggressive allocations were rewarded, the same heuristic would have delivered above-chance selection.

What this costs is easy to state in the units participants were paid on. Across these conflicting-advice cases, keeping the original allocation would have returned an Active $M^2$ of -3.01%, adopting the better advisor's portfolio every time would have returned -1.91%, and adopting the worse one every time -3.78%. Participants realized -2.92%, capturing about a twelfth of the distance to the better advisor despite the two recommendations being far enough apart for the choice to be worth roughly 1.9% of annualized risk-adjusted return. Perfect selection is an oracle standard and not a fair benchmark, since it assumes the discrimination participants demonstrably lack. The useful conclusion is narrower: presenting two opposed recommendations reliably improves on presenting one biased recommendation, as the main text establishes, but it does not equip people to identify which of the two is sound, and when the deciding cue is similarity to what they already believed, opposing pairs can leave that identification worse than a coin flip. Advice diversity delivers its benefit through exposure, not through selection, and the remaining gains lie in helping people adjudicate between competing recommendations rather than in supplying more of them.

## §3 Graduate education experiment

### §3.1 *Main experiment design*

The study was conducted in a nine-week university course with 46 enrolled students. Before the quarter, students received a Data Collection Disclosure form[1] explaining that interactions with the course-provided AI coding assistant would be recorded and analyzed, that participation was voluntary, and that declining would not affect access to the assistant, course assignments, or grades. Consenting students also completed a 10–15-min pre-course survey (SI Table S12). 64 students consented to participate (including those who dropped the course later). Students were informed that the coding assistant was accessed through a course-managed gateway and that usage would be logged, but, under an approved incomplete-disclosure procedure, were not told that the assistant's configuration could vary across weeks. All students were debriefed after the study.

During week 1, students were introduced to Claude Code, Anthropic's command-line coding assistant, and received assistance configuring a standard development environment on their own computers. Each student was issued a personal credential that routed Claude Code through the course gateway. No further configuration was required during the quarter. Each week, students completed a coding assignment and a research idea memo. Use of the assistant was optional, and students could use it as much or as little as they chose. Week 1 used the AI assistant's default configuration for all students and served as an onboarding and baseline period. During weeks 2–9, each consenting student was independently assigned two treatment weeks. The weeks were sampled uniformly, with the constraint that they were at least two weeks apart.

[1] See the Google form:
https://docs.google.com/forms/d/1WFMxHhAvK1ggvUtM1vuGLtGEuQqTX6mlGOOff8vMOZE

In the two treated weeks, each student received two distinct configurations, drawn at random without replacement from four: novelty-biased (prioritizing novel, unconventional or creative approaches), reliability-biased (prioritizing correctness, robustness and conservative best practices), neutralized (an active control, answering requests as posed without prioritizing novelty or reliability), and questioning-oriented (asking clarifying, Socratic questions before providing solutions). The questioning-oriented configuration was collected for a separate study and is excluded from all analyses here, as pre-specified in the registered analysis plan; student-weeks assigned to it are dropped from the panel. The AI assistant used its default configuration during all other weeks. The default configuration contained no experimental behavioral directive and was distinct from the neutral active-control condition. Verbatim prompts for all configurations are provided in SI Table S59.

Randomization was performed before the quarter using a seeded pseudorandom-number generator. The resulting schedule was stored server-side and was not visible to students, teaching assistants, or graders. Students who did not consent received the default configuration throughout the course and were excluded from the research analyses.

The intervention was implemented through a custom network gateway between the unmodified Claude Code client and Amazon Bedrock. Claude Code's native Bedrock-routing settings directed all model requests, including agentic tool-use turns, through the gateway. For each request, the gateway authenticated the student's pseudonymous credential, determined the current course week, retrieved the student's assignment schedule, and selected the applicable policy. It then appended the corresponding policy text to Claude Code's native system prompt before forwarding the request to Amazon Bedrock. Default-condition requests followed the same processing path but received a minimal prompt block containing no behavioral directive. The gateway returned model responses to Claude Code without content-level post-processing. Students could select among the Claude models available in the client, and Claude Sonnet 4.5 and Sonnet 4.6 were the course-recommended models. Requests to Opus-class model identifiers were routed to the designated Sonnet fallback model to reduce variation in cost.

Each request generated a complete transcript and a compact metadata record. Complete transcripts were stored as encrypted JSON objects in Amazon S3 and included the full request payload (the conversation messages and the final system prompt sent to the model, which records exactly which policy text was injected), the model response or streaming events, timestamps, the backend model identifier, the course week, the student's assigned treatment schedule, and the policy actually applied to that request. Claude Code's client-generated session identifier was also retained, allowing requests to be grouped into coding sessions. A separate DynamoDB record stored request-level metadata, including timestamps, model, policy, latency, token counts, stop reason, and a pointer to the corresponding transcript.

Research records were indexed by pseudonymous student identifiers. Names, email addresses, and institutional identifiers were not added to the research records by the gateway, and the linkage between students and pseudonymous identifiers was maintained separately. Records from students who did not consent were excluded from research datasets and analyses.

Midway through the quarter, at the request of the University of Chicago IRB, we administered a supplementary written consent form to all students who had consented under the initial data-disclosure form and remained enrolled, collecting signed reconfirmation of their consent. 40 of 45 students signed. The form is reproduced below.

**Consent Form**

**Key Information**

You are being asked to participate in a research study about the use of AI coding assistants in university courses. Participation is voluntary.

If you agree to participate:

(1) Your coursework (e.g., assignments and grades) and your interactions with the course-provided AI assistant (Claude Code) may be analyzed for research purposes.

(2) During one or two weeks of the quarter, the AI assistant may operate under a temporary variation in response configuration.

(3) You will complete a brief pre-course survey and short weekly post-assignment surveys.

(4) No additional assignments or activities beyond normal course participation are required.

There are minimal risks, primarily related to privacy and confidentiality of data. Your grades, standing, and access to course resources will not be affected by your decision to participate or not participate.

Some aspects of the study design cannot be fully described in advance to preserve the integrity of the research. A full debriefing will be provided after the course concludes.

All students enrolled in the course have equal access to the course-provided Claude Code assistant, regardless of whether they consent to participate in research.

**Purpose**

The purpose of this research is to examine how different configurations of an AI coding assistant relate to learning behaviors, engagement, and assignment outcomes in university-level coding courses.

You are being asked to participate because you are enrolled in the course AI Agents for Social Science & Society.

**Procedures and Time Required**

If you agree to participate:

(1) You will participate in the course as usual, including completing weekly coding assignments and submitting research questions as part of standard coursework.
(2) You will complete:
    (a) One pre-course survey (approximately 5–8 minutes).
    (b) Short weekly post-assignment surveys (approximately 3–5 minutes each, for approximately 9 weeks).
(3) If you choose to use the course-provided AI assistant (Claude Code), the following data may be collected and analyzed:
    (a) Interaction content (your prompts, code snippets, and the assistant's responses)
    (b) Usage metadata (timestamps, session duration, number of turns)
    (c) Assignment context (which week/assignment the interaction occurred in)
(4) During one or two randomly assigned weeks of the quarter, the AI assistant may operate under a different response configuration. These variations occur only during normal assignment use.

No interviews or additional tasks are required. Your participation lasts for the remainder of the academic quarter.

Total estimated time for research-related surveys is approximately 35–50 minutes across the quarter.

**Use of Data Generated Earlier in the Quarter**

Earlier in the quarter, students received an informational notice describing how coursework and Claude Code usage logs are generated and stored when the service is used for instructional purposes.

This consent form consolidates that information into a single IRB-approved document.

By signing this consent form, you indicate whether the research team may include coursework, grades, survey responses, and Claude Code interaction data generated during the quarter in the research dataset. If you do not grant permission, those data will be excluded from research analyses. Your coursework will still be used for grading and instructional purposes as part of normal course operations.

**Alternatives to Participation**

Participation in the research component is voluntary.

You may:

(1) Complete the course without participating in research.
(2) Choose not to use Claude Code.
(3) Withdraw from the study at any time.

Alternative instructional resources (office hours, discussion sections, readings, forums) remain available regardless of your decision.

**Educational Records (FERPA)**

Some of the data analyzed (e.g., coursework and grades) constitute educational records under FERPA.

By signing this consent form, you authorize the research team to access and use these records for research purposes as described here.

Consent status will be managed by designated research personnel who are not responsible for grading. The instructor and grading TAs will not have access to information about who has consented until after final grades are submitted.

**Monitoring and Academic Integrity**

The Claude Code platform may be monitored by authorized university personnel for system reliability and compliance with university academic integrity policies.

Academic integrity procedures operate independently of research participation status and independently of research data analysis.

Research participation does not affect how academic integrity policies are applied.

**Risks**

This study involves minimal risk beyond normal educational activities.

Potential risks include:

(1) A breach of confidentiality of coursework or AI interaction data.
(2) Exposure to AI-generated responses that may be incorrect or misleading.

To reduce these risks:

(1) Data will be stored on secure, access-controlled systems.
(2) Identifiers will be replaced with random codes when feasible.
(3) Only authorized research personnel will access identifiable data.

Your data will not be used to train or fine-tune any AI models.

**Benefits**

You may not receive any direct personal benefit from participating.

However, this research may contribute to improved design of AI-assisted learning tools and broader understanding of AI in education.

**Confidentiality and Data Use**

Your data will be handled in accordance with University of Chicago IRB and FERPA requirements.

(1) Identifiable data will not be shared outside the research team.
(2) De-identified data may be used in future research or shared with other researchers in de-identified form consistent with IRB approval and university data-protection standards.
(3) If you withdraw from the study, data collection for research purposes will stop at that time. You may request that identifiable data collected before withdrawal not be used in future analyses, to the extent feasible. Data already included in de-identified aggregate analyses may not be removable.

**Voluntary Participation and Right to Withdraw**

Participation is voluntary.

Your decision will not affect:

(1) Your grades,
(2) Your standing in the course,

(3) Your relationship with the instructor or University.

You may withdraw from the research at any time without penalty.

**Debriefing**

At the end of the academic quarter, after final grades are submitted, you will receive a written debriefing explaining:

(1) The nature of the AI configuration variations,
(2) The timing of configuration changes,
(3) The research questions being studied,
(4) The rationale for incomplete disclosure.

You will have the opportunity to withdraw your data after reviewing the debriefing within a specified timeframe.

**Contacts & Questions**

If you have questions about this study, contact:

XXX

Email: XXXXXX

If you have questions about your rights as a research participant, contact:

University of Chicago Social & Behavioral Sciences IRB

Phone: (773) 702-2915

Email: sbs-irb@uchicago.edu

**Consent**

You will be provided a copy of this form. By signing below, you agree to participate in the research.

Participant's Signature:

Participant's Name (Printed):

Date:

After students complete each week's assignments, we also request them to fill a short post-assignment survey (SI Table S13). Completion was voluntary, and coverage is accordingly partial. The survey was offered after each of the nine assignment weeks, giving 351 student-week opportunities (39 students × 9 weeks; one consented student omitted due to substantial missing), of which 180 (51.3%) were completed. Every student responded at least once, but the median student responded in only five weeks (IQR 2–7, range 1–8) and no student responded in all nine; response rates declined over the term, from 74% in week 1 (29 of 39) to 21% in week 9 (8 of 39). The instrument branches on self-reported assistant use: 99 of the 180 respondents (55%) reported consulting the assistant that week and were shown the perceived-experience items, while the remaining 81 were not. Because each student received a biased or neutralized assistant in only two of their nine weeks, this leaves 21 perceived-experience responses on treated weeks (7 novelty-biased, 6 reliability-biased, 6 questioning-oriented, and 2 neutralized) against 78 on default weeks. Non-response is thus neither uniform across weeks nor independent of assistant use, and we treat all survey-based measures as descriptive companions to the graded and telemetry outcomes rather than as endpoints; arm contrasts involving the neutralized condition in particular rest on too few responses to support inference.

§3.2 *Descriptive statistics and balance check*

The deployment ran for nine weeks with 39 students, giving 351 student-weeks in total. Each student was assigned a non-default assistant in exactly two of those weeks and worked under the default configuration otherwise. The primary analyses compare the default, novelty-biased, and reliability-biased configurations and therefore run on 312 student-weeks (273 default, 19 novelty-biased, 20 reliability-biased); the neutralized weeks are held out of that comparison and enter the four-arm analyses reported separately. Unless stated otherwise, the descriptive statistics below use the 312-student-week frame.

All 26 pre-survey covariates are time-invariant: each is constant within student across the nine weeks, so they are summarized at the student level ($N$ = 39) rather than over student-weeks, which would repeat each student roughly eight times and understate every standard deviation. Table S11, Panel A reports the 24 covariates measured on numeric scales (e.g., prior background, self-reported AI use over the previous 30 days, AI experience, and an eleven-item agreement battery). Fig. S8 shows the ten items answered on named option sets, including the two unordered covariates. One item, prior machine-learning experience, is missing for two students and is reported on the 37 who answered; non-response appears as its own level in Fig. S8 so each panel totals 39.

The cohort is homogeneous on background and heavily AI-experienced. Thirty-seven of the 39 students were Master's students, with one PhD student and one undergraduate; 28 had previously used an AI coding tool integrated into an IDE or terminal. Self-reported general AI use was high—the median student reported using AI assistants for learning several times per week or daily, and reported 5–10 hours of AI use in the preceding week—and 33 of 39 reported verifying

AI outputs "often" or "always". The specific tool deployed in the course was nonetheless new to most of the class: 29 of 39 reported no prior use of Claude Code.

Table S11, Panel B summarises the student-week variables: assigned week, the treatment indicators, submission indicators, grades, and session telemetry. Coding assignments were submitted in 80.1% of student-weeks and memos in 83.7%, so grades are summarised over submitted weeks only ($N$ = 250 and 261). Grades are the three-grader composite described above. Assistant use was intermittent: students conversed with the assistant in 40.1% of student-weeks, and telemetry is reported both across all weeks and conditional on any use, since an unconditional mean sits far above a median of zero. Among weeks with any conversation, the median student-week comprised 70 reply turns and roughly 17,300 words written to the assistant.

Default-versus-biased comparisons do not require a balance check. Every student contributes default weeks, so those contrasts are within student and baseline covariates are identical on both sides by construction; the relevant design checks there are time-varying rather than baseline. The randomized between-student contrast is which students received a novelty week ($N$ = 19) versus a reliability week ($N$ = 20), and the covariate balance check runs on exactly that contrast, at the unit of randomized assignment (39 assignments; six students received one of each and appear in both groups, which slightly attenuates the between-group differences).

Fig. S9 reports absolute standardized mean differences for the 24 numeric covariates and Cramér's V for the two unordered covariates, against the conventional negligible-imbalance threshold of 0.10. Eighteen of 26 covariates exceed that threshold, the largest being a preference for stopping the search once a first approach works (ASMD = 0.508). Read on its own this count is misleading. With 19 versus 20 assignments, the sampling standard deviation of a standardized mean difference under pure randomization is about 0.32, so the 0.10 line sits at roughly a third of a standard deviation and about three quarters of covariates clear it by chance alone. Re-randomizing the arm labels 5,000 times places the expected number exceeding 0.10 at a median of 20 of 26 (90% interval 16–24); the observed 18 is, if anything, marginally better balanced than a typical draw ($P$ = 0.87). Each covariate's own 95th null percentile is drawn in Fig. S9, and no covariate exceeds it. No covariate differs detectably between arms by Welch's $t$ or a $\chi^2$ test of independence (all raw $p > 0.12$; all Benjamini–Hochberg-adjusted $p \geq 0.957$) or by permutation (all $p > 0.11$). Academic level illustrates why the raw metric needs this calibration: with 37 of 39 students at one level, its $V$ of 0.227 is an artefact of where the two singleton students landed, and its randomization $p$ is 1.000.

The assigned-week distributions of the novelty and reliability weeks do not differ ($\chi^2$ $p = 0.269$; mean assigned week 5.47 versus 4.75). The biased week was the student's first of two treatment weeks 8 times for novelty and 12 times for reliability. Balance is consistent with successful randomization; equally, with 39 randomized assignments this check has little power to detect real imbalance, so it should be read as an absence of evidence against randomization rather than as positive evidence of balance. Our primary specifications do not rely on it in either direction: they carry student fixed effects, which absorb every time-invariant student characteristic, including all 26 covariates above and any unobserved counterpart, by construction. The covariate-adjusted models reported in SI §3.4 are therefore a sensitivity check rather than a correction for a detected problem.

§3.3 *Assignment and final project grading*

Each week's coding and memo assignments were graded by two of the four teaching assistants, randomly assigned per assignment. After the course concluded, we re-graded every coding and memo submission with Gemini 3.1 Pro using the prompt below:

```
You are a careful, strict grader for a computational social science graduate course.

Students complete weekly Jupyter notebook assignments organized into modules. Each module has a tutorial section (written by TAs) followed by homework tasks where the student applies the methods to their own research.

You will receive (a) the OFFICIAL TASK LIST for this week — the exact modules and numbered homework tasks to grade — and (b) the student's full notebook (source, markdown, text outputs).

Grade the student's work against the OFFICIAL TASK LIST only:

- For every module and every numbered task in the official list, find the student's work and assign an integer score 0-4. Return exactly one question entry per task, with question_num equal to that task's number within its module (matching the official list order).

- Score 0 if a task is absent, not attempted, blank, a placeholder, or has no relevant student work. Do NOT invent tasks, and do NOT grade notebook content that is not in the official list (ignore tutorial walkthroughs and any unlisted sections, including memo-pilot sections not listed for this week).

- Tasks marked [required] are mandatory; [bonus] tasks are optional extra credit. Grade bonus tasks the same 0-4 way; the scoring step decides how they count.
```

```
- If the student did not attempt a whole module, still return that module with a score-0 entry for each of its tasks.

Score each task 0-4 (integer only):

- 4 = Excellent: complete, correct, fully responsive to the task; when the task asks for application or reflection, genuinely adapted to the student's own research.

- 3 = Good: mostly complete, minor gaps or shallow reasoning.

- 2 = Adequate: partial attempt, or mostly template code with little adaptation / superficial interpretation.

- 1 = Minimal: substantially missing or incorrect, or just re-ran template code with no personal application.

- 0 = Not attempted, blank, placeholder, or no relevant student work.

Base scores on what the student actually wrote/coded for each task, including code outputs. Grade only what is visibly present in the notebook; do not credit work you infer happened elsewhere. Write a brief comment per task and a 1-2 sentence module_rationale. Output ONLY valid JSON — no extra text.

Additional calibration for this Gemini run -- grade as a harsh, strict graduate-level grader. The scale is 0 (did not turn in) to 10 (fabulous). Treat 4 as exceptional: award it on a task only for genuinely "fabulous" work, so a 4 should be uncommon and a module or final score of 10 should be rare. Use the full 0-4 range and spread your scores; do not default to full credit for work that merely runs or looks polished. Grade only what is visibly present in the notebook; do not credit work you infer might have happened elsewhere.
```

The two grader types agreed substantially on the coding assignment (Gemini versus the two-human mean, $r = 0.53$, $N = 274$ student-weeks) but only weakly on the memo ($r = 0.33$, $N = 231$). We therefore use a three-grader composite for coding (the two human grades and the Gemini grade at equal weight, with Gemini mean-aligned to the human scale) and the two-human mean alone for the memo, where adding the LLM grade lowered internal consistency rather than raising it.

For the final project, students worked in teams. All four teaching assistants graded each project and the instructor assigned the final grade informed by those four assessments. The final grade is unrelated to the number of biased-assistant weeks a team received (project level, $N = 22$; Huber $b = 0.206$, randomization $p = 0.505$). We additionally code  using the prompt below:

```
System prompt:

You are an expert course-staff grader for MACS 37005 (AI Agents for Computational Social Science, University of Chicago). You are grading FINAL PROJECT write-ups (public-facing blog posts or equivalent reports). Grade strictly against the official assignment description:

--- OFFICIAL ASSIGNMENT (verbatim excerpts) --- Students will work alone or in groups (of up to 4) to perform a substantial AI agent project for social insight based on approaches and tools developed/explored over the course of the quarter. These projects must integrate at least three forms of data in a single model, use at least three types of models (from three separate weeks over the course of the quarter), and must validate inferences or predictions with qualitative interpretation and assessments. Groups will compose a detailed, entertaining, and informative public-facing Substack or Medium.com blog post about their projects that includes the motivation, methodological justification and details, descriptive data and deep learning modeling, interpretation of findings (e.g., discovered structures, predictions, generations), conclusion, and an annotated code appendix. These should not read like an academic paper, but a mixture of (1) explanatory tutorial; and (2) digital museum exhibit, balancing intermittent text with figures, description boxes, equations, and/or conceptual diagrams including at least one visual element (e.g., figure, graph, conceptual diagram) for every 300 words of text; and a minimum total of 5000 words and 17 visual elements; but 3600 words and 13 visual elements for single author projects. In short, they represent the methods section + operational tutorial of a (hopefully) awesome paper. These may include relevant code
```

and should link to an open GitHub or Colab repository for code evaluation.

--- END ASSIGNMENT --- Score five dimensions 0-2 each (0 = requirement missed, 1 = partially met, 2 = fully met):

- sub_data_integration: >= 3 forms of data (e.g., text, network, tabular, image, spatial, temporal) genuinely integrated toward a single model/analysis pipeline.

- sub_model_diversity: >= 3 types of models drawn from distinct course weeks (e.g., embeddings, fine-tuned transformers, RL/agents, network models, generative simulation, probing).

- sub_validation: inferences/predictions validated with qualitative interpretation and assessment (error analysis, case reading, human judgment), not just quantitative metrics.

- sub_content: required components present and substantive (motivation; methodological justification and details; descriptive data and deep-learning modeling; interpretation of findings; conclusion; annotated code appendix / open repo link).

- sub_style: reads as tutorial + digital-museum exhibit; visual density near 1 per 300 words; meets the word/visual minimums for its team size (group: 5000 words / 17 visuals; solo: 3600 words / 13 visuals). Also report: n_data_forms and n_model_types you can actually identify; meets_length and meets_visual_density booleans using the measured counts given in the prompt. Then give overall_score on 0-10 (halves allowed): your holistic quality grade as an experienced course-staff grader would assign for THE WRITE-UP AS A WHOLE -- technical substance, insight, rigor, and communication all weighed, with the assignment requirements as the baseline expectation. 9-10 exceptional; 7-8 solid fulfillment; 5-6 notable gaps; <5 major requirements unmet. Do not grade the underlying repo you cannot see; grade what the write-up demonstrates. Keep the rationale under 80 words.

USER MESSAGE TEMPLATE (filled per project):

```
PROJECT: {doc_id}

Team size: {team} ({group|single author}; minimums {need_w} words / {need_v} visuals).

Measured word count of extracted text: {words}.

Measured visual count: {visuals} ({n_imgs} attached below; PDF projects attach page renders, so figures appear in-page).

Source format: {source}.

--- WRITE-UP TEXT (extracted; markup stripped) ---

{text, capped at 60,000 characters}

--- END TEXT ---

Grade this final project write-up per the rubric.
```

On this measure the relationship is weakly positive ($N = 21$; Huber $b = 0.474$, Huber $p = 0.086$, randomization $p = 0.040$). We treat it as suggestive rather than confirmatory (Extended Data Fig. 4).

A graded score may reflect not only the quality of a student's work but also the graders' own preferences: if graders reward novelty, or reward conventional reliability, then part of any arm difference could arise from the assistant's stylistic signature being valued differently at grading time rather than from a genuine difference in what students produced. To separate these, we ran each assistant configuration on the coding assignments without a student in the loop, using the prompt below:

```
Please solve this week's homework notebook in this folder (the single .ipynb file). Work through every module's homework tasks — including the Memo Pilot / Final Homework module if the notebook has one — writing your answers directly into the notebook: fill in the required code, run the cells so outputs are saved, and write the requested written answers/reflections in markdown. Treat it as your own graded assignment for a graduate computational social science course: complete every numbered homework task in every module. Work autonomously until the whole notebook is done; if a dataset or dependency is needed, obtain/install it yourself as the notebook instructs. Work entirely within the current folder: read and modify only files here (the
```

```
notebook and its data/ subfolder), and do not explore or
read anything in parent directories. Use the API key in
.env file, if needed.
```

This yields 33 stand-alone notebooks (four configurations for weeks 2–9; week 1 predates the manipulation and exists only in the default configuration). Two teaching assistants then graded these blind—condition labels removed and submission identifiers randomised independently within each week—alongside a Gemini 3.1 Pro grader using the same rubric prompt as the student submissions, and the three grades were combined into a composite on the same reliability criterion used for the student assignments (Cronbach's $\alpha = 0.77$, against 0.61 for the two human graders alone). No configuration differed detectably from the default in stand-alone work. Week-blocked adjusted means were 7.03 for novelty-biased, 6.96 for reliability-biased, 6.67 for neutralized and 6.06 for default on a 0–10 scale; the contrasts against default were 0.98 for novelty ([−0.53, 2.48], adj. $p = 0.34$), 0.90 for reliability (adj. $p = 0.34$) and 0.61 for neutralized ($p = 0.41$). Neither an omnibus test of configuration ($F(3, 21) = 0.75$, $p = 0.53$) nor a week-blocked Friedman test ($\chi^2(3) = 1.56$, $p = 0.67$) rejects equality, and within-week paired comparisons against default are likewise null (novelty higher in 4 of 8 weeks, sign $p = 1.00$; reliability in 6 of 8, $p = 0.29$).

Because the stand-alone grade estimates how each configuration's output is valued by graders in the absence of any student contribution, we use it in a robustness specification to ask how much of the student-level arm difference survives once that channel is accounted for. See SI §3.4.

§3.4 Robustness analysis

Our preregistered primary specification regresses the outcome on assistant configuration with week and student fixed effects and student-clustered standard errors. The student fixed effect makes each student their own control and absorbs every time-invariant characteristic by construction. We re-estimate the two graded outcomes across four further specifications: (i) covariate adjustment, adding the 18 baseline covariates whose standardized difference exceeds 0.10 in the balance check (SI Fig. S9), which the primary specification otherwise absorbs; (ii) no student effect, with clustered standard errors and the two prior-experience covariates; (iii) assignment-volume weighting, in which each student-week is weighted by the volume of material in that week's assignment template—a property of the assignment fixed before any assistant was assigned, and so unaffected by treatment—so that more demanding weeks count more; and (iv) alternative grader composites, since the choice of which graders enter the outcome is itself a researcher degree of freedom. Full estimates for every specification appear in Table S49-S58, and the grader-composite ladder in Table S59.

Every estimate is positive, and the point estimates are stable: the coding effect on novelty-biased weeks ranges from +0.318 to +0.684 across all specifications, a band that sits well inside the confidence interval of any one of them. Statistical significance is not stable. Coding is detectable at conventional levels under the primary specification (0.541, $p$ = .047), under a one-human-plus-Gemini composite, and under usage weighting (0.684, $p < .001$), but not with two human graders alone (0.576, $p$ = .133) or without a student effect (0.318, $p$ = .190). The memo pattern is the mirror image, strongest without a student effect and weakest once the LLM graders enter the composite Table S59. This is what a deployment powered only for large effects looks like, and we report the full set rather than a preferred subset.

Two specifications warrant comment. Covariate adjustment fits 31 parameters on 39 independent clusters and is included because the balance check flags those covariates rather than because the primary specification depends on them; the adjusted estimate (0.647, $p$ = .020) exceeds the primary. Assignment-volume weighting is reported with template-derived measures rather than realised usage: weighting by reply turns or words written yields larger apparent effects, but those quantities are measured after the assistant is assigned and their zero values drop 153 of 248 coding weeks from the estimate, collapsing the Kish effective sample to as few as 38 (Table S60). All eight template measures are pre-treatment and place the novelty coding estimate between +0.599 and +0.743 while retaining an effective $n$ of 169–209.

The final specification addresses the possibility that graders reward a configuration's stylistic signature rather than the quality of the student's work. Using the stand-alone assistant runs described in SI §3.3—the same coding assignments completed by each configuration with no student in the loop, graded blind—we add the stand-alone grade for the corresponding configuration and week to the primary model. Because the stand-alone grade varies with configuration, this partitions the estimate into a grader-preference component and a remainder rather than removing nuisance variation, and we report it as such. Stand-alone runs exist for the coding assignment only, so this specification is not available for the memo. Adjustment attenuates the novelty-biased coding advantage by roughly a third, from +0.541 to +0.341 ([−0.17, 0.85], $g$ = 0.30, $p$ = 0.185), while leaving the reliability contrast essentially unchanged (+0.328 to +0.291, $p$ = 0.244). The stand-alone grade itself carries a positive but imprecise coefficient (+0.280, $p$ = 0.139), and because it varies only across configurations within a course week, the decomposition is identified off eight weeks of variation and should be read as indicative of the direction rather than as a precise partition. Some of the observed novelty advantage plausibly reflects the assistant's own output being valued more highly by graders; the remainder is positive but not separately detectable at this sample size.

Finally, we assess the primary coding result under two design-based checks. Leave-one-student-out re-estimation moves the coefficient only within [+0.420, +0.691], so no single student drives it. Randomization inference, permuting configuration assignments within students over 3,000 draws, returns $p$ = 0.111—weaker than the model-based $p$ = .047, reflecting that only 19 student-weeks received the novelty configuration. We report both and treat the coding result as directionally robust but statistically marginal.

§3.5 *Qualitative evaluation of treatment effect on particular assignment question answers*

Because every prompt and every reply passed through the gateway, the full conversation record of the deployment is archived. To see how the biased assistants entered students' learning experience, we manually read the conversation logs of student-weeks in which a student was both assigned a biased assistant and actually used it intensively, and we compared these conversations with the same students' and other students' weeks under the default assistant. We were looking for moments where the assigned policy becomes visible in the exchange: in what the assistant volunteers, in how it frames a routine request, and in how the student responds to the suggestion. The episodes below are illustrative selections from that reading, quoted verbatim with identifying details removed; ellipses mark omissions, and one bulleted proposal is quoted in flattened form.

Conversations under the default assistant have a characteristic shape: the student states a task, the assistant executes it competently and stops. One student opened with "hi claude" and then handed over a module in a single message: "could you use the data from genz_slang_usage_2020_2025.csv to Build a binary classification model to predict is_viral," listing the features to use, and followed with "write this script and put it into a new .py file." The assistant did precisely that: a standard train-test split, logistic regression, random forest, gradient boosting, accuracy and F1 and ROC curves. It is a textbook solution, delivered without commentary on whether a more interesting reading of the task exists. Nothing in the exchange invites the student to reconsider the framing, and nothing in the student's prompts asks for that.

Under the novelty policy the assistant's posture changes: it treats the student's request as a starting point rather than a specification. The clearest episode comes from a student brainstorming their final project late in the quarter, during a week in which the student was assigned to the novelty policy for every request. The student had been developing a project on measuring surprise in academic writing for that week's open-ended task. In the middle of an otherwise ordinary planning conversation, the assistant proposed a section labeled "Novel Twist," a result of our injected system prompt toward novelty. Both the name and the idea that followed were the assistant's. Its proposal read, verbatim: "Instead of just measuring surprise, build a 'Serendipity Agent' that: Takes boring abstracts; Uses multi-agent debate ... to generate creative research pivots; Uses RL ... to optimize for genuine surprise." The student had given no such instructions: nothing in their prompts that week asked for unconventional approaches. The "instructions" the assistant refers to are the hidden policy sentence, surfacing in the conversation in the assistant's own voice. The student took the proposal seriously and began negotiating it into something feasible: "okay i like this concept. but how i don't think we can do the human rating. what can we do?" and went on to ask whether they could serve as the judge themselves. The adopted concept then shaped the design conversation that followed, and its skeleton, multi-agent debate plus reinforcement learning to reward genuine rather than superficially gamed surprise, carried into the student's submitted final project. All four members of the teaching team, grading the final essays independently, scored it at or above their class median, and two of the four ranked it among the very best essays they graded.

The contrast between the arms is easiest to see when students face the same task. In week 5, every student worked through the same homework modules on sampling and fine-tuning with LoRA, and all four assistants appear in that assignment's logs. A default-arm student ran the homework as a sequence of specifications and output corrections, at times pasting exercise text into the chat verbatim: "update the code so that it simply build a classifier that predicts interfere and returns a classification report," then "it says Label causing most imbalance: 4 but i need the column name not a number." The exchange is fluent and entirely transactional; at no point does the student ask why. A student under the neutralized assistant worked the same modules in the same posture, down to "can you give me this in a python code? so I can copy and paste." The student under the reliability policy typed the exercise requirements out as a spec, "Train three LoRA configurations with different values of:" followed by the list of hyperparameters, and when something broke, handed that over too: "it didnt work. can you look at the cell im working in and see what the issue is and read hte error message and fix it?" What came back was adopted without follow-up. The student under the novelty policy delegated as well, but with a behavior that appears nowhere else on this homework: supervision. They tracked the agent's output ("where did you put these answers?") and challenged its inputs: "where did you get the mental health dataset from." The check landed. "Good catch!" the assistant replied, tracing the dataset to the course notebook's own download link and flagging that the file path would only work inside Colab. No question of that kind appears anywhere in the default or reliability logs of this homework. The same student's earlier reliability week sharpens the contrast: there, their contribution for the week was two handover messages, opening with "can you help me generate an approach to complete this homework?", and no question about the returned work ever followed. This is the qualitative face of the pattern we quantify elsewhere in the paper: under the novelty policy, delegation came with checking; under the reliability policy, it did not.

Read side by side, the conversations make the treatment concrete. The four assistants answer the same kinds of requests, and most exchanges under any policy look alike: debugging, explaining, writing boilerplate. The bias lives in the margins of the conversation, and the margins differ consistently. The default and neutralized assistants execute the request as given, and students respond in kind, issuing specifications and collecting results. The reliability-biased assistant is the most readily trusted: under it, students most often handed work over whole and took what came back. The novelty-biased assistant treats the request as negotiable: it volunteers reframings and proposes alternatives where standard solutions exist, and it is under this assistant that delegation came with the most questioning. Students remained the authors of their goals; but the episodes show how a single policy sentence, invisible to the student, could shape not only what the assistant said but how actively the student worked with it: a volunteered idea negotiated into a graded final project, and one week's homework delegated under every policy, with students actively checking the assistant's work almost exclusively under the novelty policy.

§3.6 *Discussing explicitly neutralizing AI assistant as an alternative intervention*

A natural response to the finding that biased AI assistance can help is to ask whether an assistant explicitly instructed to be even-handed would do better still, delivering the benefit of a deliberate stance without committing to a direction. We tested this with a fourth classroom arm in which the assistant received the same persona scaffolding as the biased arms but was instructed to weigh novelty and reliability without tilting toward either. As in the investment study, the comparison of interest is against the default assistant, which receives no persona instruction at all: the two differ only in whether a stance is named, so any difference between them isolates the effect of explicit neutralization rather than of guidance in general. Across every outcome we examined, the neutralized assistant was indistinguishable from the default one, and its point estimates sat at or below the default throughout.

On graded performance (Extended Data Fig. 4i), the neutralized arm did not exceed the default arm on either deliverable. Adjusted coding performance was 0.58 points below the default assistant ([−0.55, 1.72], $g = 0.49$, adj. $p = 0.313$) and adjusted memo performance 0.14 points below it ([−0.21, 0.48], $g = 0.15$, adj. $p = 0.446$). Both biased arms outscored it on coding—novelty by 1.07 points ([−0.18, 2.32], $g = 0.88$, $p = 0.094$, adj. $p = 0.205$) and reliability by 0.88 points ([−0.28, 2.03], $g = 0.72$, $p = 0.136$)—placing the neutralized assistant at the bottom of the four-arm ladder on the outcome where directional bias helped most. The novelty–neutralized gap is the largest separation anywhere in the panel and is marginal before correction, though it does not survive adjustment across the three arm-vs-arm contrasts.

The downstream pattern is consistent (Extended Data Fig. 4j). In the project-level model of final-project innovativeness, the neutralized arm is the only one of the three assigned configurations with a negative coefficient (−0.19, [−0.93, 0.55], $p = 0.595$), against +0.27 for novelty ([−0.49, 1.03], $p = 0.468$) and +0.44 for reliability ([−0.33, 1.20], $p = 0.244$). Pooling the two biased configurations, projects whose members had been assigned a neutralized week scored 0.54 points lower on innovativeness than those assigned a biased week (Huber robust, [−1.22, 0.13], $p = 0.106$; randomization inference $p = 0.080$). The contrast is marginal rather than conclusive and is sensitive to specification—it is significant under classical standard errors ($p = 0.038$) but not under HC3 ($p = 0.213$)—so we report it as suggestive. A named neutral stance nonetheless shows no trace of the positive downstream association that biased-week exposure shows.

Explicit neutralization also left engagement unchanged (Extended Data Fig. 4k). Conditional on any use, log reply turns under the neutralized assistant were statistically identical to the default ($\Delta = 0.24$ in the neutralized arm's favour, [−0.63, 1.11], $g = -0.16$, adj. $p = 0.718$), and the equivalence persists when weeks with no conversation are retained ($\Delta = -0.09$, $p = 0.788$). The reliability-biased assistant was the only arm to separate from the default on this measure (0.59, [0.03, 1.15], $g = 0.40$, $p = 0.040$, adj. $p = 0.119$). Naming a neutral stance therefore did not draw students into longer exchanges than leaving the assistant unmodified.

Perceived helpfulness (Extended Data Fig. 4l) is reported for completeness but cannot support inference for this arm. The post-assignment survey reached the neutralized arm only twice across the term, and only once in a week with any assistant use, so the plotted interval spans nearly the full response scale (Δ versus default = 0.69, [−1.04, 2.41], $p$ = 0.420 conditional on use; 0.14, [−0.84, 1.13], $p$ = 0.770 across all surveyed weeks).

## §4 A Two-Action Model of Accepting Versus Scrutinizing an AI Response

In this section, we present a formal model of the mechanism driving our experimental results. A user faces a task, receives a recommendation from an algorithm (the LLM in our experiments), and decides whether to act on it, and how carefully to check it first.

### §4.1 Model and timing

We start by considering an AI system that recommends an answer to a human facing a binary-choice task. The model recommendation is objectively correct with probability

$$p \in \left[\tfrac{1}{2},\, 1\right].$$

We assume that the model response can be and seem biased. After observing the AI response, the human forms a perceived probability that the response is correct as function of the algorithm quality $p$ and perceived bias $b$,

$$\pi = f(p, b) \in \left[\tfrac{1}{2},\, 1\right].$$

The restrictions $p \geq \frac{1}{2}$ and $\pi \geq \frac{1}{2}$ are without loss of generality. If an answer is correct with probability below one half we can simply relabel its opposite as the recommendation; likewise, if the human believes the recommendation is more likely wrong than right, the opposite answer is the one she effectively takes as recommended. We therefore restrict attention to recommendations that are at least weakly better than their negation, both objectively and from the human perspective.

We further assume that $f$ is continuous, strictly increasing in $p$, and strictly decreasing in $b$. Therefore, we assume that holding true AI quality fixed, greater perceived bias lowers the human's perceived probability that the response is correct.

After observing the algorithm output, the human chooses between two actions:

1. accept the AI response as given; or
2. conduct a costly deep dive into the response.

A deep dive evaluates whether the human should follow the AI recommendation or overturn it. If the human conducts a deep dive, a binary evaluation test produces a positive signal $t = +$ ("keep the AI response") or a negative signal $t = -$ ("switch to the opposite answer"). Let

$$q = \Pr(t = + \mid \text{AI response correct}), \qquad r = \Pr(t = + \mid \text{AI response incorrect}).$$

We impose

$$1 > q > \tfrac{1}{2} > r \geq 0,$$

which implies that scrutiny is more likely than not to keep a correct AI response and more likely than not to overturn an incorrect AI response.

A correct final answer yields payoff $G > 0$, while an incorrect final answer yields payoff $-L < 0$, with $L > 0$. Conducting a deep dive implies a non-pecuniary cost of $c > 0$. This is the cost of evaluation as it requires time and mental effort.

If the human accepts the AI response without scrutiny, subjective expected utility is

$$U_A(\pi) = \pi G - (1 - \pi)L.$$

If the human scrutinizes the response, the final answer is correct either when the AI response is correct and the test says to keep it, or when the AI response is incorrect and the test says to switch. Therefore subjective expected utility from scrutiny is

$$U_D(\pi) = [\pi q + (1 - \pi)(1 - r)]\, G - [(1 - q)\pi + (1 - \pi)r]\, L - c.$$

§4.2 _The scrutiny decision_

The human conducts a deep dive whenever $U_D(\pi) \geq U_A(\pi)$. Examining the difference between the two utilities we have

$$U_D(\pi) - U_A(\pi) = (G + L)\, [(1 - \pi)(1 - r) - \pi(1 - q)] - c.$$

The first term in the parenthesis, $(1 - \pi)(1 - r)$, is the perceived probability that scrutiny corrects an erroneous AI response. The second, $\pi(1 - q)$, is the perceived probability that scrutiny overturns a correct AI response and thereby creates an error.

The deep-dive payoff above assumes that, conditional on scrutinizing, the human keeps the AI response after a positive signal and switches to the opposite answer after a negative signal. We now show that these actions are optimal whenever a deep dive is chosen. After a positive signal, the posterior probability that the AI response is correct is

$$\pi_+ = \Pr(\text{AI correct} \mid t = +) = \frac{\pi q}{\pi q + (1-\pi) r}.$$

Keeping the AI response is optimal if and only if $\pi \geq \frac{r}{q+r}$, which holds as $q > r$ and $\pi \geq \frac{1}{2}$. After a negative signal, the posterior probability that the AI response is correct is

$$\pi_- = \Pr(\text{AI correct} \mid t = -) = \frac{\pi(1-q)}{\pi(1-q) + (1-\pi)(1-r)}.$$

Switching to the opposite answer is optimal if and only if $\pi_- \leq \frac{1}{2}$, which is equivalent to $\pi \leq \frac{1-r}{2-q-r}$. But when the human chooses scrutiny, we have

$$\pi < \frac{1 - r - \frac{c}{G+L}}{2-q-r} < \frac{1-r}{2-q-r},$$

as $c > 0$. Therefore, conditional on choosing a deep dive, a negative signal makes it optimal to switch to the opposite answer.

As we assume that the scrutiny is helpful, $q > \frac{1}{2} > r$, we have that $2 - q - r > 0$, and the scrutiny condition can be written as

$$\pi \leq \phi, \qquad \phi \equiv \frac{1 - r - \frac{c}{G+L}}{2-q-r}.$$

To consider only the interesting case where there are some level of $p$ for which the human chooses to scrutinize, we further assume $\frac{c}{G+L} < \frac{q-r}{2}$, which implies $\phi > \frac{1}{2}$. We also assume that $\phi$ lies in the range of $f(\cdot, b)$ for the values of $b$ under consideration.

The next claim shows that perceived bias can shift the quality values $p$ for which the human chooses to add an additional verification.

**Claim 1** (Perceived bias raises the scrutiny cutoff). *For every perceived-bias level $b$, there exists a unique cutoff $p^*(b)$ satisfying*

$$f(p^*(b), b) = \phi,$$

*such that the human scrutinizes the AI response if and only if*

$$p \leq p^*(b).$$

*Moreover, $p^*(b)$ is strictly increasing in $b$.*

*Proof.* Because $f(\cdot, b)$ is continuous and strictly increasing in $p$, and because $\phi$ lies in its range, there exists a unique $p^*(b)$ satisfying $f(p^*(b), b) = \phi$. Since scrutiny is chosen exactly when $f(p, b) \leq \phi$, strict monotonicity of $f$ in $p$ implies that this condition is equivalent to $p \leq p^*(b)$.

Now let $b_H > b_L$. Since $f$ is strictly decreasing in $b$,

$$f(p^*(b_L), b_H) < f(p^*(b_L), b_L) = \phi.$$

Because $f(\cdot, b_H)$ is strictly increasing in $p$, restoring equality with $\phi$ requires

$$p^*(b_H) > p^*(b_L).$$

§4.3 <u>*Overall performance*</u>

Overall performance is the probability that the human's final answer is correct. The behavioral choice is based on perceived quality $\pi = f(p, b)$, but realized performance is determined by true AI quality $p$ and the human evaluation.

Under direct acceptance the performance is given by,

$$P_A(p) = p,$$

where under scrutiny,

$$P_D(p) = pq + (1-p)(1-r).$$

The term $pq$ is the probability that a correct AI response is retained, after human valuation. The term $(1-p)(1-r)$ is the probability that an incorrect AI response is detected and overturned. Hence

$$P(p,b) = \begin{cases} P_D(p), & p \leq p^*(b), \\ p, & p > p^*(b). \end{cases}$$

**Claim 2** (Increasing perceived bias can improve overall performance). *Define the objective performance threshold*

$$\psi \equiv \frac{1-r}{2-q-r} > \tfrac{1}{2}.$$

*Under the maintained restriction* $p \in \left[\frac{1}{2}, 1\right]$*, scrutiny improves overall performance relative to direct acceptance if and only if*

$$\tfrac{1}{2} \leq p < \psi.$$

*Consequently, let* $b_H > b_L$*, and assume* $p^*(b_L) < \psi$*. For any true AI quality* $p$ *satisfying*

$$p^*(b_L) < p \le p^*(b_H) \quad \text{and} \quad \tfrac{1}{2} \le p < \psi,$$

*higher perceived bias strictly improves total performance:*

$$P(p, b_H) = P_D(p) > p = P(p, b_L).$$

*Proof.* The performance effect of scrutiny is

$$P_D(p) - P_A(p) = pq + (1-p)(1-r) - p = (1-p)(1-r) - p(1-q).$$

The term $(1-p)(1-r)$ is the probability that scrutiny fixes an AI error: the AI is wrong and the test correctly recommends switching. The term $p(1-q)$ is the probability that scrutiny creates an error: the AI is correct but the test incorrectly recommends switching. Therefore scrutiny improves performance if and only if

$$(1-p)(1-r) > p(1-q).$$

Within the maintained domain $p \ge \frac{1}{2}$, this identifies the relevant improvement region above chance. Rearranging,

$$(1-p)(1-r) > p(1-q) \iff 1 - r > p(2-q-r) \iff p < \frac{1-r}{2-q-r} = \psi.$$

More so, note that $2 - q - r > 0$ and

$$\psi > \tfrac{1}{2} \iff \frac{1-r}{2-q-r} > \tfrac{1}{2} \iff 2(1-r) > 2 - q - r \iff q > r,$$

which is true by the assumption that the test is useful $q > 1/2 > r$. Finally, let $b_L < b_H$ which by claim 1 implies $p^*(b_L) < p^*(b_H)$. Therefore there exists a $p$ in the interval $(p^*(b_L), p^*(b_H)]$, in which the human accepts directly the algorithm recommendation, under $b_L$, but scrutinizes under $b_H$. If also $p < \psi$, scrutiny is objectively more accurate, so

$$P(p, b_H) = P_D(p) > P_A(p) = P(p, b_L).$$

Claim 2 shows that when human evaluation can improve the quality of the algorithm's output, an increase in bias can improve the overall performance of the human and AI joint output, even when the objective quality of the algorithm remains unchanged. The reason is that greater bias can induce the human to perform an additional check, which may ultimately lead to a better outcome.

**Remark 1** (Subjective optimality versus objective performance). Conditional on choosing a deep dive, the agent always believes that scrutiny improves gross final-answer accuracy enough to justify its cost. Formally, the agent scrutinizes if and only if

$$(1-\pi)(1-r)-\pi(1-q) \geq \frac{c}{G+L} > 0.$$

Thus, whenever scrutiny is chosen, the agent perceives its gross accuracy to exceed direct-acceptance accuracy by at least $\frac{c}{G+L}$.

Nevertheless, scrutiny can reduce realized performance when $p > \psi$. This happens when the agent underestimates the algorithm quality, $p$, and the errors in the human validation test introduce more errors than the test reduces. Indeed,

$$\phi = \frac{1-r-\frac{c}{G+L}}{2-q-r} = \psi - \frac{\frac{c}{G+L}}{2-q-r} < \psi.$$

If the agent chooses scrutiny while $p > \psi$, then

$$\pi = f(p,b) \leq \phi < \psi < p.$$

Hence the agent necessarily underestimates the true quality of the AI response. The agent is optimal relative to the perceived model $\pi$, but this perception error induces scrutiny in a region in which scrutiny is objectively counterproductive.

**Remark 2** (Scrutiny always increases the accuracy of retained AI responses). A different accuracy concept asks how reliable the AI responses are conditional on surviving scrutiny. Among responses for which the test is positive and the AI answer is retained,

$$\alpha(p) \equiv \Pr(\text{correct} \mid t=+) = \frac{pq}{pq+(1-p)r}.$$

For any $p \in \left[\frac{1}{2}, 1\right]$,

$$\alpha(p)-p = \frac{pq}{pq+(1-p)r} - p = \frac{p(1-p)(q-r)}{pq+(1-p)r} > 0,$$

where the strict inequality follows from $q > r$. Thus scrutiny always raises the accuracy of the subset of AI responses that are retained.

Finally, we turn to show how human evaluation can reduce the negative effects that can happen when the algorithm fails.

**Claim 3** (Perceived bias can dampen the effect of AI-quality deterioration). *Let* $b_H > b_L$*, and consider a deterioration in true AI quality from* $p$ *to* $p-\Delta$*, where* $0 < \Delta < p^*(b_H) - p^*(b_L)$*, such that*

$$p^*(b_L) < p - \Delta < p \leq p^*(b_H).$$

*Then both quality levels are accepted directly under* $b_L$ *and scrutinized under* $b_H$*. Under low perceived bias, the change in performance induced by the deterioration is*

$$P(p, b_L) - P(p - \Delta, b_L) = \Delta,$$

*whereas under high perceived bias it is*

$$P(p, b_H) - P(p - \Delta, b_H) = (q + r - 1)\Delta.$$

*Under the maintained restrictions* $1 > q > \frac{1}{2} > r \geq 0$*,*

$$|q + r - 1| < \tfrac{1}{2},$$

*and therefore*

$$|P(p, b_H) - P(p - \Delta, b_H)| < \frac{\Delta}{2} < \Delta.$$

*Hence perceived bias can dampen the effect of deterioration in AI quality by inducing scrutiny.*

*Proof.* Under direct acceptance, $P_A(p) = p$, so a reduction in true quality from $p$ to $p - \Delta$ changes performance by exactly $\Delta$. Under scrutiny we have,

$$P_D(p) = pq + (1 - p)(1 - r) = (1 - r) + p(q + r - 1).$$

Therefore

$$P_D(p) - P_D(p - \Delta) = (q + r - 1)\Delta.$$

Because $1 > q > \frac{1}{2} > r \geq 0$ we have,

$$|q + r - 1| < \tfrac{1}{2},$$

which implies

$$|P_D(p) - P_D(p - \Delta)| < \frac{\Delta}{2}.$$

Thus the magnitude of the effect of AI-quality deterioration on final performance is strictly smaller under scrutiny than under direct acceptance. If $q + r \geq 1$, deterioration still lowers performance under scrutiny, but by less than under direct acceptance. If $q + r < 1$, the sign reverses: deterioration in AI quality raises final performance within the scrutiny regime. In either case, scrutiny reduces the absolute pass-through of changes in AI quality to final performance.

§4.4 *Interpretation in Light of the Experimental Results*

The model clarifies how a biased AI can improve human performance without being more accurate, more helpful, or more highly valued by its users. The AI moves first by supplying a response: a judgment about a headline, a recommended portfolio, or a proposed solution to an assignment. The human then decides whether to accept that response or expend time and cognitive effort evaluating it. Evaluation is useful but imperfect. It may preserve a correct AI response, detect and correct an incorrect one, or mistakenly overturn a response that was already correct. Final performance is therefore a joint product of the quality of the AI's initial contribution and the human's willingness and ability to evaluate it.

Perceived bias affects this process by changing when evaluation occurs. In Claim 1, a response that appears more biased is treated as less reliable. This expands the range of AI responses that users choose to scrutinize rather than accept directly. Therefore, bias can operate as a cue that the AI's answer should be treated as a proposal rather than an authority. This interpretation corresponds to the subjective and behavioral patterns observed across the experiments. Strongly partisan assistants were perceived as less helpful and more intent on influencing the user, while eliciting more extensive responses in the fact-checking experiment. Risk-seeking assistants generated more engagement and lower confidence in the investment experiment. In the classroom, exchanges with novelty-biased tutors contained a larger share of verification and independent-work prompts than exchanges with reliability-biased tutors. Together, they are consistent with biased outputs keeping the human actively involved in evaluating the AI's contribution.

Claim 2 explains when this additional scrutiny translates into better performance. Evaluation is beneficial when the errors that humans detect and correct outweigh the correct AI responses that they mistakenly overturn. This condition is most plausible when AI assistance is useful but imperfect: the AI is good enough to provide valuable proposals, but not so reliable that accepting every proposal is optimal. The experiments display precisely the resulting separation between subjective appreciation and objective performance. In fact-checking, stronger bias was associated with lower perceived helpfulness but greater veracity accuracy. In the classroom, novelty-biased tutors received the lowest helpfulness evaluations while producing the highest coding performance. Participants performed better despite believing it was worse, consistent with the model's prediction that reduced confidence in the assistant can induce higher overall performances.

Finally, the investment experiment provides the clearest empirical analogue of Claim 3. When an AI recommendation is accepted directly, changes in the quality of that recommendation pass through one-for-one to the final decision. When the human evaluates and modifies it, the human acts as a filter, weakening the connection between AI quality and final performance. In the defensive market waves, risk-seeking assistants generated less mean-variance-efficient portfolios and recommendations with substantially lower realized returns than the default assistant. Participants exposed to this advice, however, realized only a fraction of the assistants' performance disadvantage, and their final performance was not detectably worse than under default assistance. This pattern is consistent with scrutiny dampening the consequences of deterioration in AI quality: visibly questionable advice may be poor as a stand-alone output while remaining relatively harmless once a skeptical human is placed in the loop.

**§5 Can it be a story of perspective diversity?**

If the benefits of biased assistance reduced to perspective diversity—the assistant confronting users with a viewpoint different from their own—three predictions follow: opposition pairing should outperform echo-chamber pairing wherever biased assistance helps (P1); this advantage should hold across participant groups, since diversity is supplied to all of them (P2); and echo-chamber pairing, which adds minimal viewpoint distance, should confer little benefit over the default assistant (P3). We classified each interaction by the assistant's bias relative to the participant's pre-interaction position. The three experiments jointly reject all three predictions.

The fact-checking experiment satisfies P1 but violates P2. Opposition bias outperformed echo-chamber bias overall, yet the benefit was concentrated entirely among Democratic participants ($\Delta = 0.057$, [0.018, 0.096], adj. $p = 0.004$, $g = 0.23$; 9.4%), with Republican participants performing equivalently under both ($\Delta = -0.001$, [−0.040, 0.038], adj. $p = 0.955$). No pairing fell below the default assistant: the least favourable, a Democrat-leaning assistant advising Democratic participants, was indistinguishable from it ($\Delta = -0.005$, [−0.056, 0.047], adj. $p = 0.816$), while the other three exceeded it (all adj. $p \leq 0.015$).

The investment experiment violates P1 and P3 directly. The best-performing pairing was echo-chamber, not opposition: under the realized defensive regime, echo-chamber bias served risk-averse-leaning participants best ($\Delta = 0.94\%$, [0.62%, 1.26%], adj. $p < 0.001$, $g = 0.55$)—a pairing that adds the least viewpoint diversity yet beat the default outright. On bull-market draws, the identical pairing became the worst ($\Delta = -1.47\%$, [−1.76%, −1.18%], adj. $p < 0.001$, $g = -1.12$), with the ranking of all four combinations inverting: the value of a pairing tracked its alignment with the market regime, not its distance from the participant. Exposure was nonetheless asymmetric, the least favourable pairing, opposition bias for risk-averse-leaning participants, trailed the default by only 0.40% (adj. $p = 0.064$), against favourable gains of 0.94%.

The classroom experiment is uninformative for P1 and P2 by way of a null: alignment produced no detectable effect on either deliverable (coding $\Delta = -0.014$, $[-1.078, 1.051]$, $p = 0.979$; memo $\Delta = -0.259$, $[-0.710, 0.192]$, $p = 0.247$), and no pairing differed from the default assistant (coding, all adj. $p \geq 0.878$; memo, all adj. $p \geq 0.118$).

Across all three experiments, no bias–participant combination underperformed the default assistant (minimum adj. $p = 0.064$; Extended Data Fig. 4). The pattern is incompatible with viewpoint diversity as the operative ingredient: opposition is neither necessary (investment) nor sufficient (fact-checking Republicans, classroom) for benefit, while a committed assistant of either direction never left participants worse off than a neutral one.

## Supplementary figures

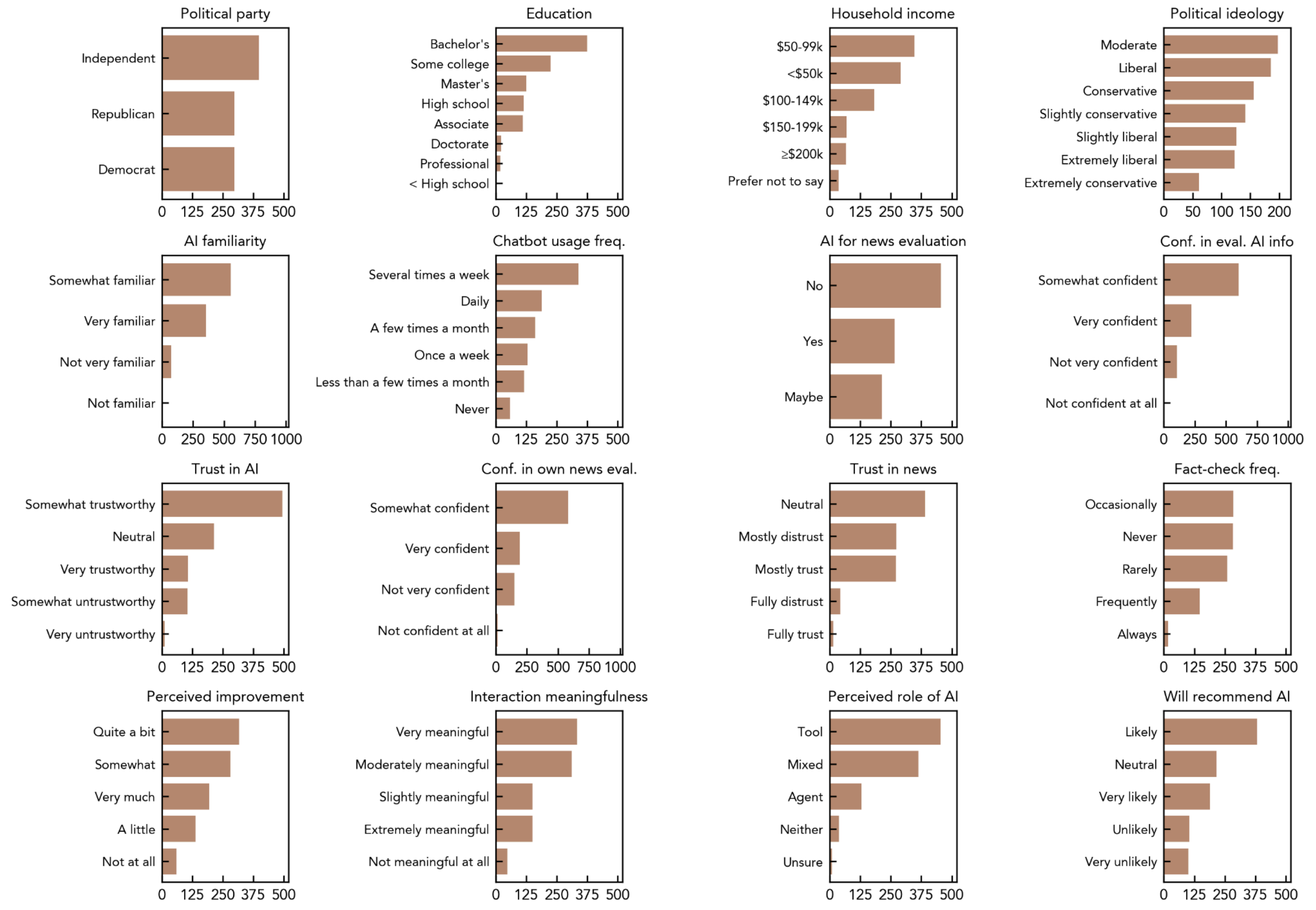

**Fig. S1. Descriptive statistics of critical categorical variables in the single AI condition of the fact-checking experiment.**

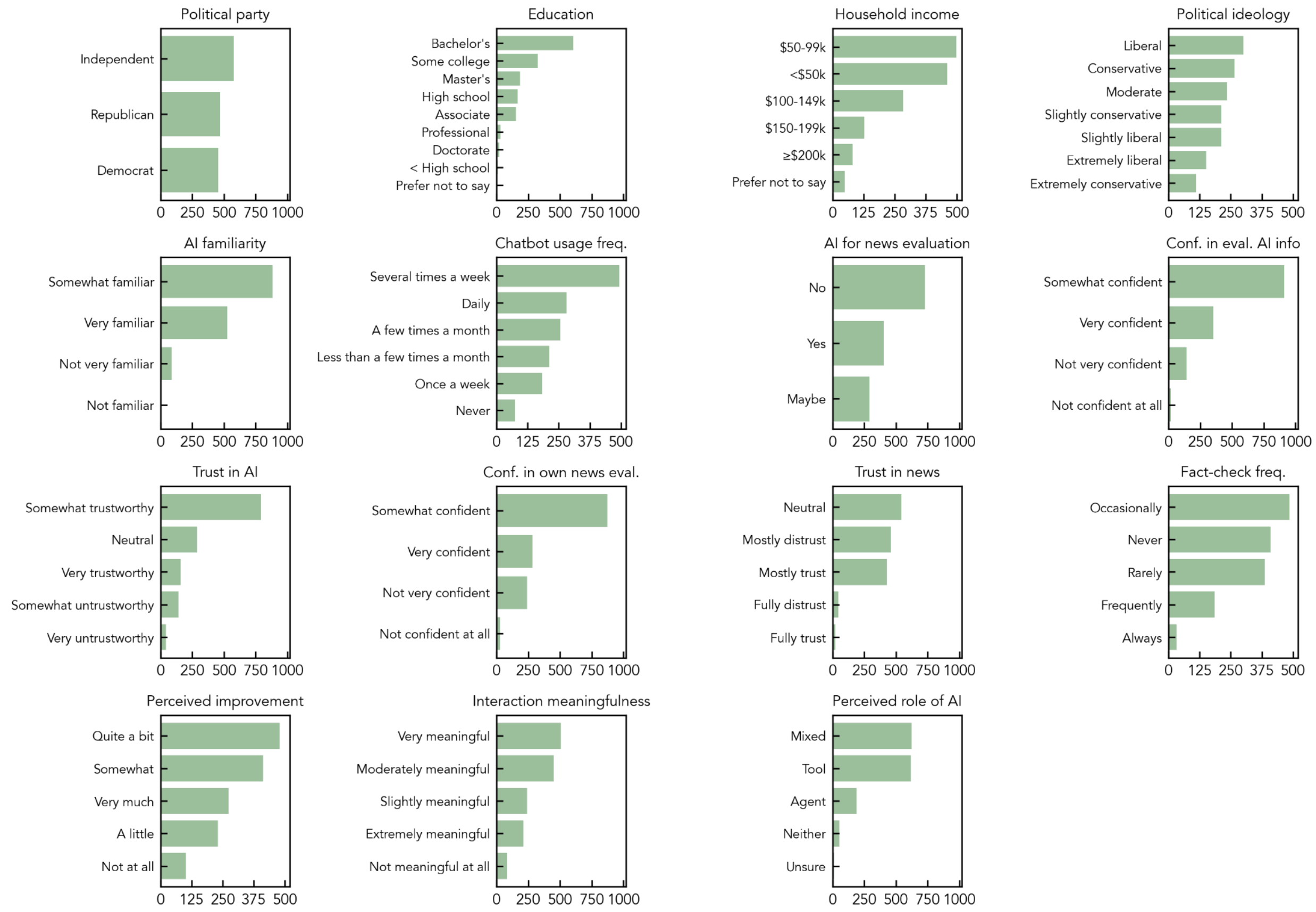

**Fig. S2. Descriptive statistics of critical categorical variables in the dual AI condition of the fact-checking experiment.**

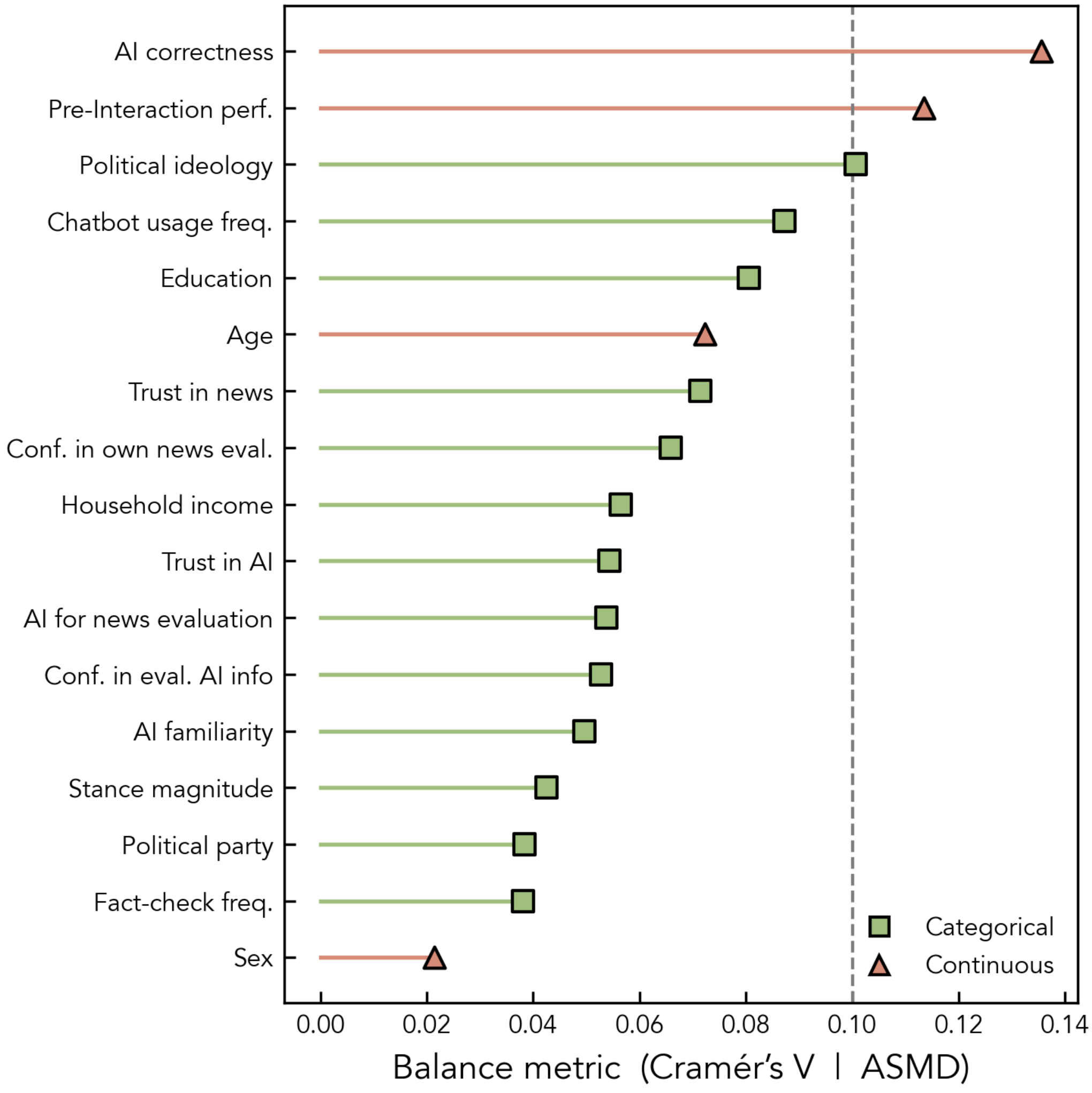

**Fig. S3. Balance check of binary treatment (Biased vs. Default) assignment for fact-checking.** For continuous variables (red; e.g., age, pre-interaction performance), we calculated the absolute standardized mean difference (AMSD), for categorical variables (green), we used Cramér's $V$, the $\chi^2$-based effect size (see SI §1.3). In robustness check, we incorporated imbalanced variables in regression analysis and evidenced that the inference does not hinge on the imbalance (see SI §1.3).

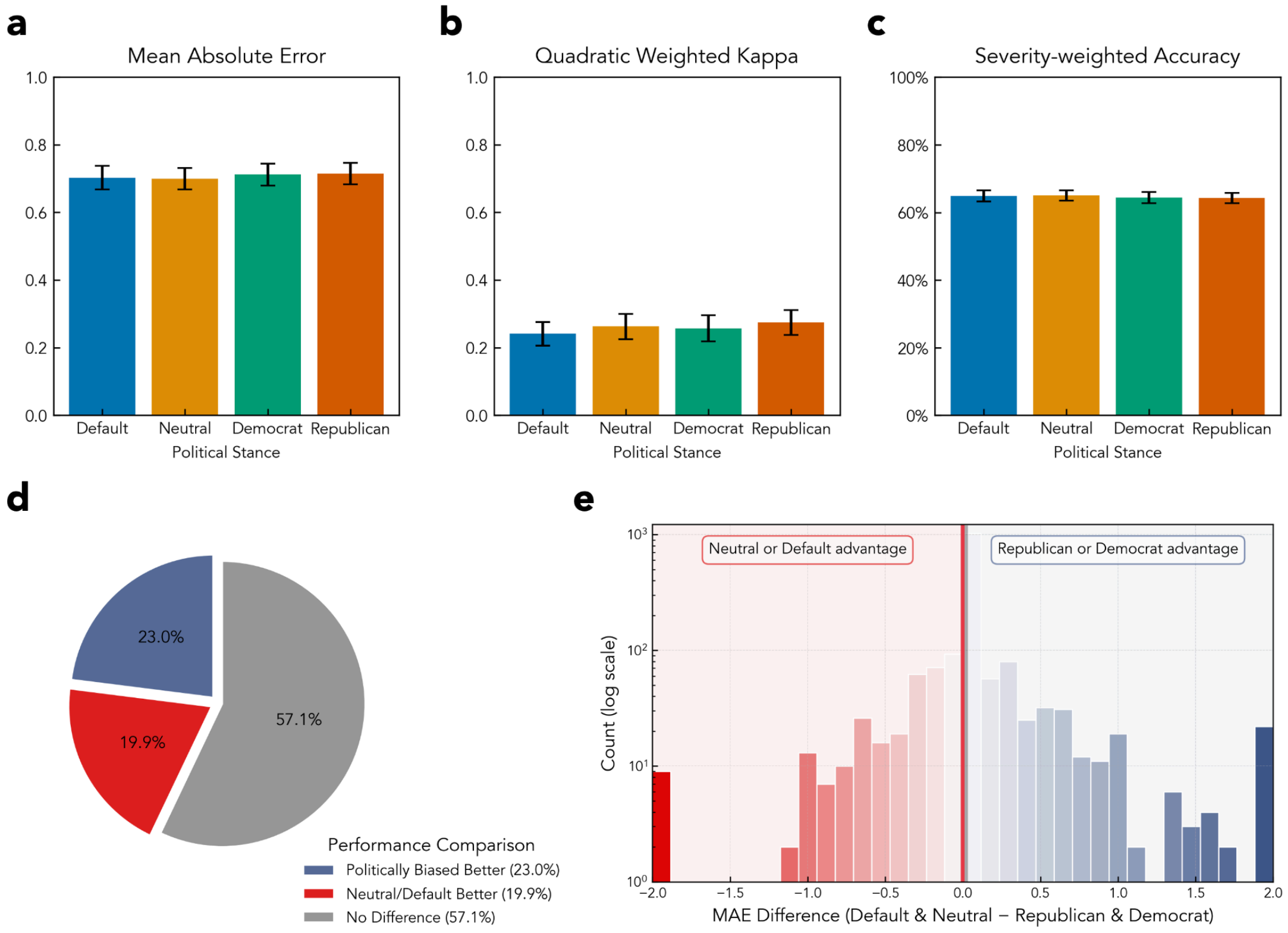

**Fig. S4. Comparing biased, default, and neutralized LLMs' fact-checking performance in the wild.** **a-c,** Performance of the standard baseline model, its neutralized variant, and two partisan models, respectively, across thousands of headlines evaluated on three separate metrics; error bars represent 95% confidence intervals. **d,** Proportion of headlines on which the partisan model outperformed, matched, or underperformed the baseline/neutralized models. **e,** Distribution of partisan/non-partisan model's relative advantage, defined as the difference in mean absolute error between the non-partisan and partisan evaluators (positive values indicate partisan superiority).

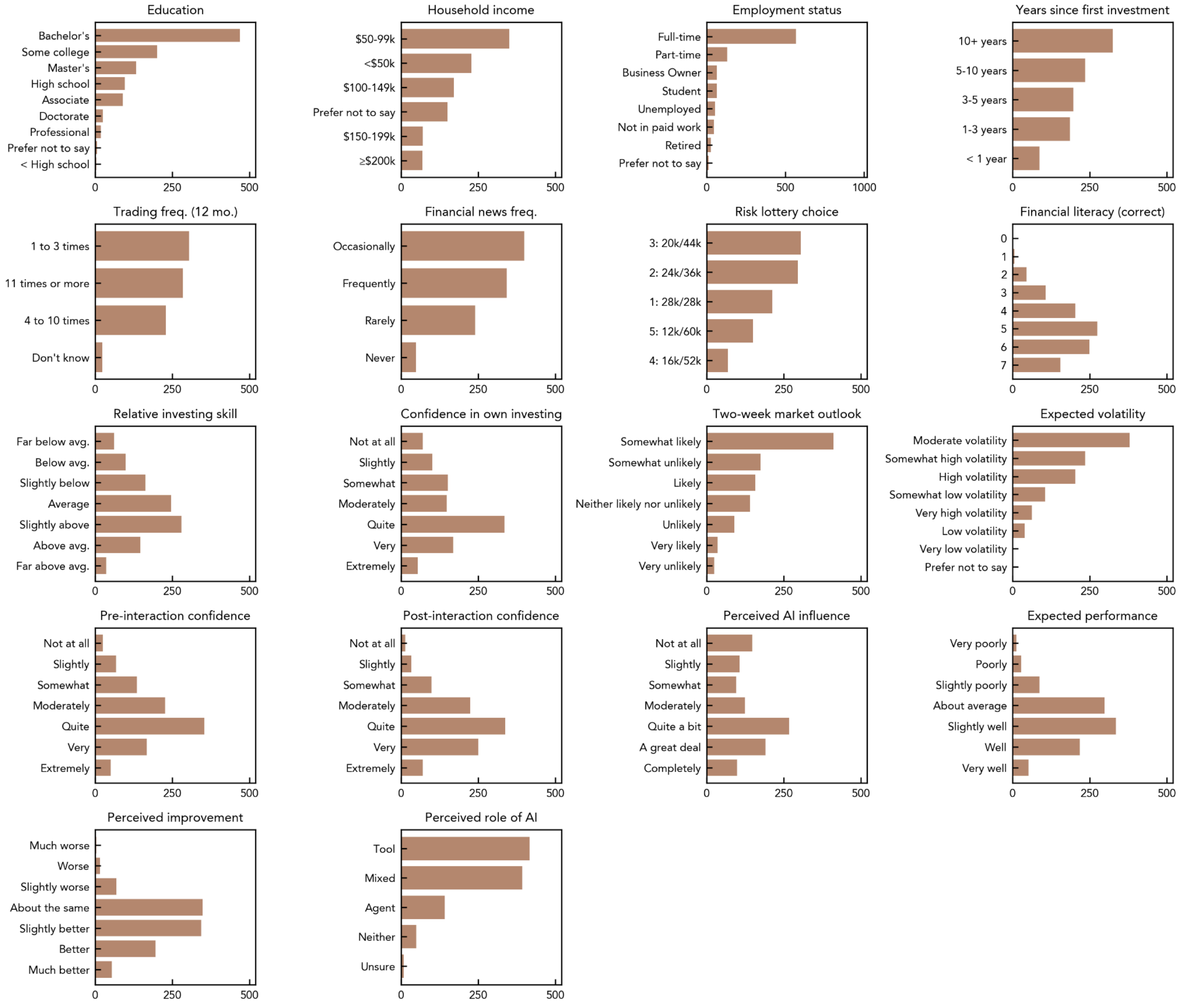

**Fig. S5. Descriptive statistics of critical categorical variables in the single AI condition of the investment experiment.**

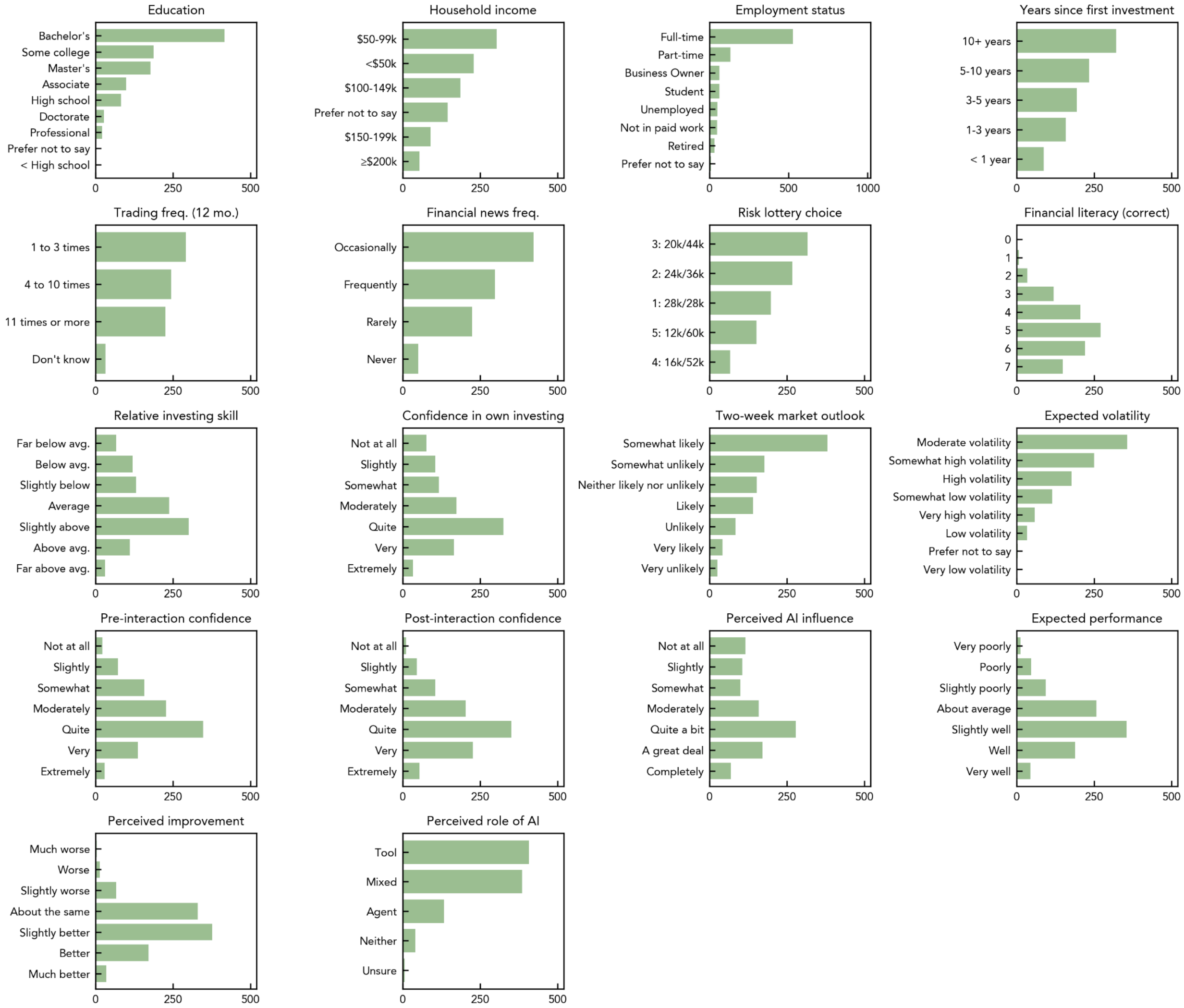

**Fig. S6. Descriptive statistics of critical categorical variables in the dual AI condition of the investment experiment.**

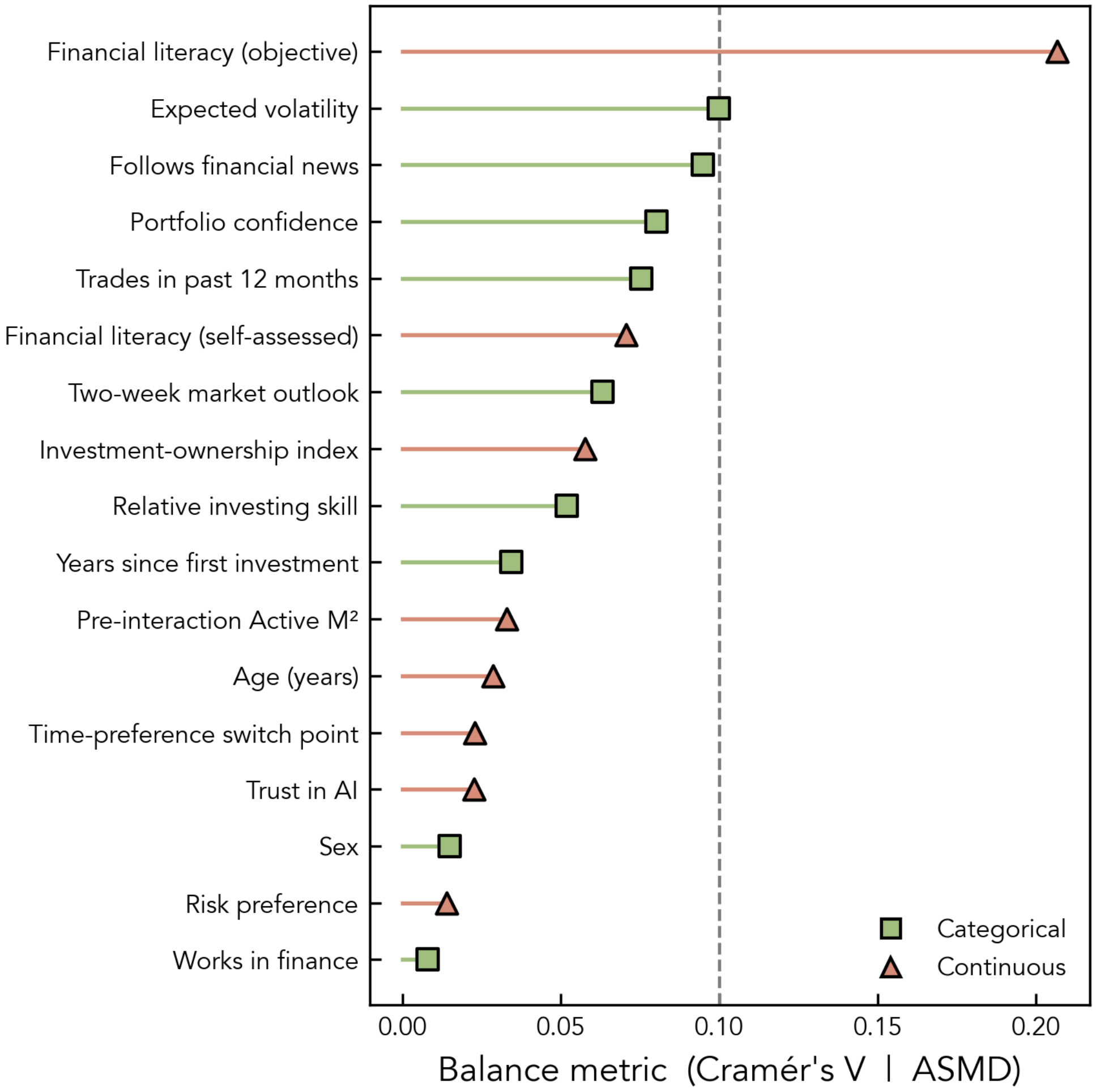

**Fig. S7. Balance check of binary treatment (Biased vs. Default) assignment for investment.**
For continuous variables (red; e.g., age, pre-interaction active $M^2$), we calculated the absolute standardized mean difference (AMSD), for categorical variables (green), we used Cramér's $V$, the $\chi^2$-based effect size (see SI §2.2). In robustness check, we incorporated imbalanced variables in regression analysis and evidenced that the inference does not hinge on the imbalance (see SI §2.2).

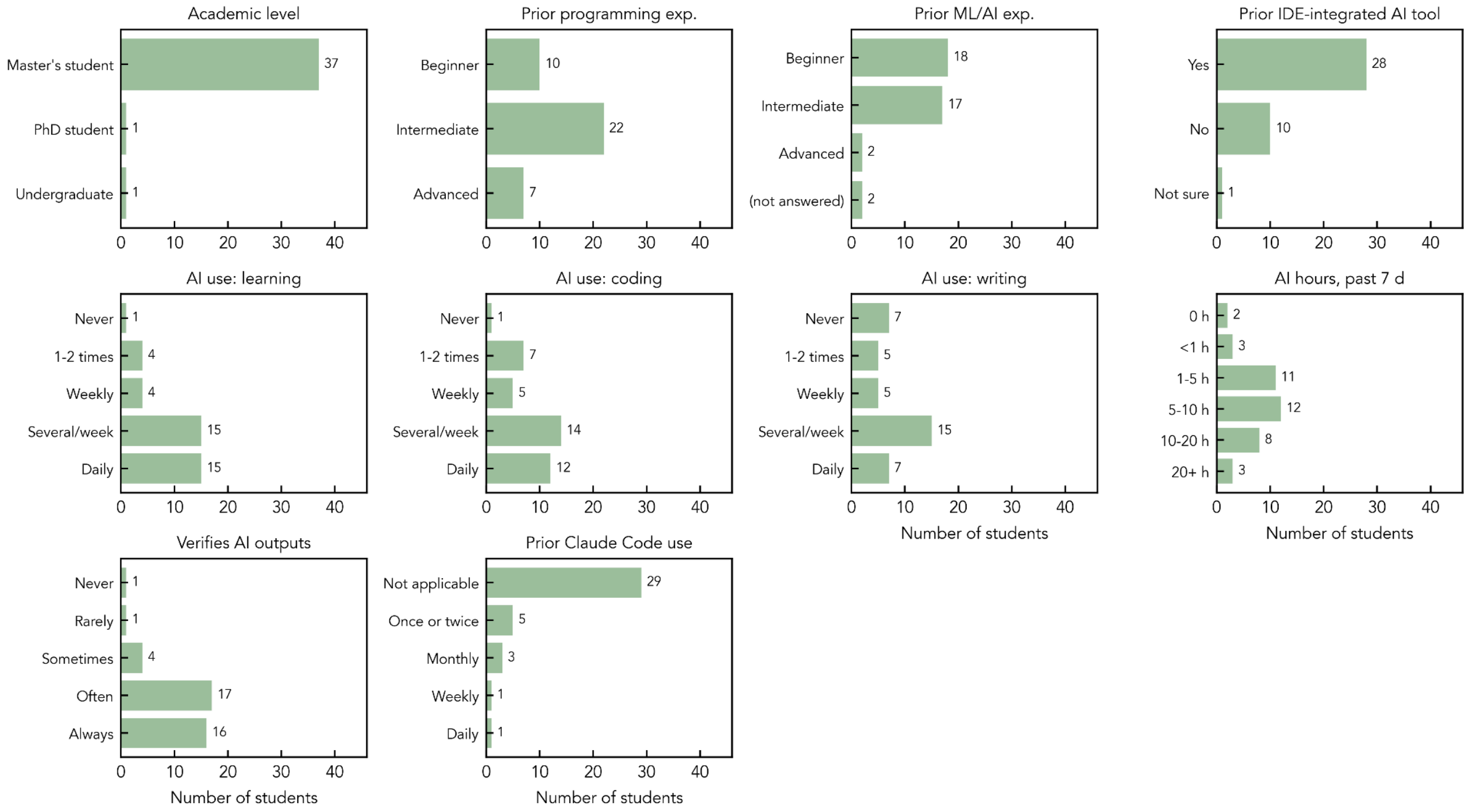

**Fig. S8. Descriptive statistics of critical categorical variables in the graduate education experiment.**

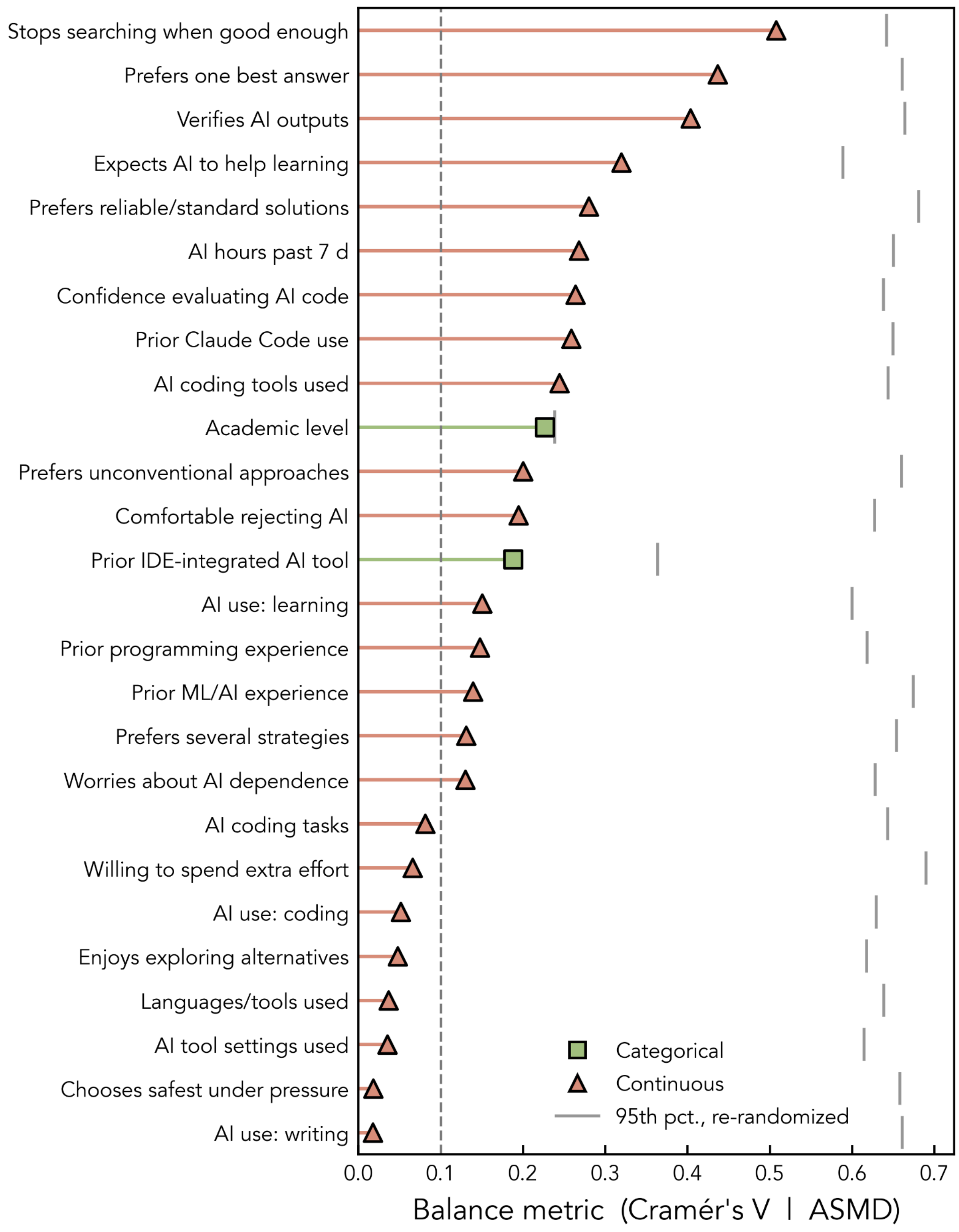

**Fig. S9. Balance check of treatment assignment (Novelty vs. Reliability) for the graduate education experiment.**

Absolute standardized mean differences for continuous covariates (coral triangles) and Cramér's V for nominal covariates (sage squares), computed over the 39 randomized assignments (19 novelty, 20 reliability; 6 students in both). Dashed line, the conventional 0.10 threshold; grey ticks, each covariate's 95th percentile under 5,000 re-randomizations of the arm labels. Eighteen of 26 covariates exceed 0.10, but none exceeds its own null percentile, and re-randomization places the expected count past 0.10 at a median of 20 of 26 ($p$ = 0.87). No covariate differs between arms (all permutation $p > 0.11$; all adj. $p \geq 0.957$). Balance is consistent with successful randomization.

## Supplementary tables

| *Single AI* (*N* = 993) | | | | | | |
|---|---|---|---|---|---|---|
| **Source** | **Variable** | **Mean** | **SD** | **Min** | **Median** | **Max** |
| CloudResearch | *Age* | 38.442 | 12.136 | 18.000 | 36.000 | 78.000 |
| CloudResearch | *Gender* | 0.497 | 0.500 | 0.000 | 0.000 | 1.000 |
| Pre-Interaction Q3 | *Stance Magnitude* | 0.339 | 0.474 | 0.000 | 0.000 | 1.000 |
| Conversation text | *AI Correctness* | 0.671 | 0.346 | 0.000 | 0.800 | 1.000 |
| Conversation text | *Conversation Length* | 24.760 | 24.361 | 1.000 | 18.000 | 293.000 |
| Pre-Evaluation | *Pre-Performance* | 0.576 | 0.289 | 0.000 | 0.500 | 1.000 |
| Post-Evaluation | *Post-Performance* | 0.636 | 0.342 | 0.000 | 0.700 | 1.000 |
| ***Dual AI* (*N* = 1,499)** | | | | | | |
| CloudResearch | *Age* | 38.781 | 12.632 | 18.000 | 37.000 | 83.000 |
| CloudResearch | *Gender* | 0.500 | 0.500 | 0.000 | 1.000 | 1.000 |
| Pre-Interaction Q3 | *Stance Magnitude* | 0.360 | 0.480 | 0.000 | 0.000 | 1.000 |
| Post-Interaction Q6 | *Inconsistency* | 0.319 | 0.466 | 0.000 | 0.000 | 1.000 |
| Conversation text | *AI1 Correctness* | 0.682 | 0.344 | 0.000 | 0.800 | 1.000 |
| Conversation text | *AI2 Correctness* | 0.662 | 0.349 | 0.000 | 0.800 | 1.000 |
| Conversation text | *Conversation Length* | 27.633 | 27.170 | 2.000 | 20.000 | 301.000 |
| Pre-Evaluation | *Pre-Performance* | 0.571 | 0.289 | 0.000 | 0.500 | 1.000 |
| Post-Evaluation | *Post-Performance* | 0.645 | 0.341 | 0.000 | 0.800 | 1.000 |

**Table S1. Descriptive statistics of continuous (or binary) variables of the fact-checking experiment.**

| | Question | Options |
|---|---|---|
| 1 | In the past year, how frequently did you access the following sources to obtain news via the internet? | Matrix table:<br>Categories (row): Search engines (e.g., Google, Bing), Social media (e.g., Facebook, X), News aggregators (e.g., Google News, Flipboard), News websites (e.g., nyt.com, vox.com)<br>Frequency (column): Never, About once every few months, About once a month, About once a week, A few times a week, About once a week, A few time a day or more |
| 2 | Generally speaking, do you usually think of yourself as a Republican, a Democrat, an Independent, or what? | Republican, Democrat, Independent, No preference, Don't know |
| 3 | [If Q2 == Republican or Democrat] Would you call yourself a strong Republican/Democrat or not a very strong Republican/Democrat? | Strong, Not very strong |
| 4 | We hear a lot of talk these days about liberals and conservatives. Here is a seven-point scale on which the political views that people might hold are arranged from extremely liberal to extremely conservative. Where would you place yourself on this scale, or haven't you thought much about this? | Extremely liberal, Liberal, Slightly liberal, Moderate, Slightly conservative, Conservative, Extremely conservative, Don't know |
| 5 | In general, how familiar are you with artificial intelligence (AI)? | Very familiar (I frequently use or work with AI technologies), Somewhat familiar (I have used AI-powered tools a few times), Not very familiar (I have heard of AI but have little direct experience), Not familiar at all (I have no experience with AI) |
| 6 | In the past 3 months, how often have you used AI-powered chatbots such as ChatGPT and Claude? | Daily, Several times a week, Once a week, A few times a month, Less than a few times a month, Never |
| 7 | [If Q6 != Never] How comfortable are you with using AI-powered tools such as ChatGPT to help you make decisions or get information? | Very comfortable, Somewhat comfortable, Neutral, Somewhat uncomfortable, Very uncomfortable |
| 8 | [If Q6 != Never] To the best of your knowledge, have you ever knowingly used AI-based services to evaluate or analyze news content (e.g., fact-checking tools)? | Yes, Maybe, No |
| 9 | [If Q6 != Never] How confident are you in your ability to critically evaluate information provided by AI-powered tools such as ChatGPT? | Very confident, Somewhat confident, Not very confident, Not confident at all |
| 10 | [If Q6 != Never] In general, how trustworthy do you find information provided by AI-powered tools such as ChatGPT? | Very trustworthy, Somewhat trustworthy, Neutral, Somewhat untrustworthy, Very untrustworthy |
| 11 | [If Q6 != Never] How confident are you in your ability to evaluate the truthfulness of news without support from external sources such as AI or search engines? | Very confident, Somewhat confident, Not very confident, Not confident at all |
| 12 | How much do you trust the information you encounter in the news media? | Fully trust, Moderate trust, Neutral, Mostly distrust, Fully distrust |
| 13 | In the past one month, how often did you reference fact-checking websites (e.g., snopes.com or politifact.org) to check whether a headline you read is true? | Always, Frequently, Occasionally, Rarely, Never |
| 14 | How frequently do you feel you come across news articles that appear inaccurate or misleading? | Daily, Several times a week, Once a week, A few times a month, Less than a few times a month, Never |

**Table S2. Fact-checking experiment pre-treatment survey questions.**

| | Text | Required Input |
|---|---|---|
| Instruction A | You will evaluate the truthfulness of three news items by selecting one of the following:<br>1. **True**: The news item is factually accurate.<br>2. **False**: The news item is factually incorrect.<br>3. **Unsure**: You are unsure about the truthfulness of the news item.<br>Please assess each news item independently and thoughtfully, without relying on any external assistance such as web searches or AI tools.<br>After your initial evaluation of each news item, you will receive suggestions from an AI assistant. You will have the opportunity to discuss your evaluation with them for up to three rounds.<br>Once the discussion concludes, you will re-evaluate the news item before moving on to the next one. | Click the Next button |
| Pre-Interaction evaluation | [NEWS_ITEM] | True, Unsure, False |
| Pre-Interaction confidence assignment | [If Pre-Interaction evaluation != Unsure] How confident are you in your evaluation of this news item? | Not at all confident, Slightly confident, Moderately confident, Very confident, Extremely confident |
| Instruction B | Now, you will revisit the news item with AI assistance. To begin with, the AI assistant(s) will share their thoughts. You can then discuss the news with the assistant, for at least one round. The maximum number of interaction rounds is 3. | Click the Next button |
| Human-AI interaction | Now, you will discuss the following news item with the AI assistant: [NEWS_ITEM]<br><br>Round 1: Loading AI response...<br>It can take several seconds. Please wait for its response.<br>AI: [AI_MESSAGE] | Response to AI messages for up to three back and forth |
| Post-Interaction evaluation | [NEWS_ITEM] | True, Unsure, False |
| Post-Interaction confidence assignment | [If Post-Interaction evaluation != Unsure] How confident are you in your evaluation of this news item? | Not at all confident, Slightly confident, Moderately confident, Very confident, Extremely confident |

**Table S3. A complete news headline evaluation session.**

| | GPT-o1-pro | Claude Sonnet 3.5 |
|---|---|---|
| 1 | The news item is a historical anecdote about Charlie Chaplin, a silent-era film actor, who once lost a look-alike contest of himself. It does not promote any political perspective, use partisan language, or align with typical Democrat or Republican narratives. Therefore, it is neutral in its political stance. | This is a historical anecdote about Charlie Chaplin that has no political implications or partisan framing. The story is a purely entertainment/historical fact that doesn't promote any particular political ideology or agenda. It doesn't use language that favors either conservative or liberal viewpoints, and it's not connected to any contemporary political narratives or policy discussions. The content is simply a curious historical tidbit that can be appreciated regardless of one's political beliefs. |
| 2 | The news item highlights a potential juror expressing strong negative sentiments toward former President Donald Trump, using derogatory language. The juror's tweet contains explicit criticism and mockery of Trump, referring to him as a "treasonous orange s***gibbon" and making a derogatory comment about his appearance ("the dead ferret on his head"). This negative portrayal aligns with Democratic narratives that are critical of Trump and his actions. The framing of the news item focuses on the juror's disdain for Trump without presenting any counterpoints or alternative perspectives. The language used emphasizes the juror's contempt, which resonates with Democratic opposition to Trump. There is no indication of support or sympathy for Trump or the Republican viewpoint in the article. Therefore, the political stance of the news item leans toward the Democrat side. | This news item exhibits a clear Democratic-leaning bias through several elements: 1) It highlights and appears to find humor in a potential juror's hostile characterization of Trump, using derogatory language that aligns with common anti-Trump rhetoric from the political left. 2) The inclusion of the insulting description ('treasonous orange s***gibbon' and 'dead ferret') reflects language commonly used in left-leaning social media circles to mock Trump's appearance and question his loyalty to the country. 3) The framing of the interaction, particularly including the juror's unapologetic response, seems to validate and normalize such anti-Trump sentiment. This aligns with Democratic narratives regarding Trump's legal troubles and presents them in a way that appears to celebrate rather than neutrally report on the jury selection process. |
| 3 | The news item describes a 1990 promotion by Playgirl magazine involving Donald Trump before he entered politics. It is presented as an entertainment or celebrity piece without any political commentary or bias. The language is factual and does not promote any political agenda or align with typical Democratic or Republican narratives. Therefore, the content is considered neutral in political stance. | This news item is a straightforward historical fact about a magazine promotion from 1990, long before Donald Trump entered politics. It does not contain any political framing, editorial commentary, or language that promotes a particular political agenda. The statement is purely descriptive of a past event involving Trump as a celebrity figure, not a political one. While it could theoretically be used in different contexts to serve various political narratives, as presented here it is simply a factual historical reference with no discernible partisan lean. |
| 4 | The news item presents a straightforward factual statement without any overt political framing or biased language. It reports that Bill Gates owns a farm producing potatoes used by McDonald's, which does not inherently align with typical Democrat or Republican narratives. There is no promotion of a particular perspective, criticism, or endorsement that would suggest a leaning toward either political party. Therefore, the content is neutral in its political stance. | This news item is a straightforward factual statement about Bill Gates owning farmland that supplies potatoes to McDonald's. It contains no politically charged language, no policy advocacy, and no framing that aligns with either Democratic or Republican narratives. While topics like billionaire land ownership and corporate food supply chains can be politicized in certain contexts, this statement presents the information without any editorial slant, opinion, or ideological framing. It does not promote any particular political perspective, making it a neutral piece of information. |
| 5 | The news item reports that Donald Trump said Adolf Hitler 'did some good things,' which is a highly controversial and negative statement. By highlighting this alleged comment, the news piece portrays Donald Trump, a prominent Republican figure, in an unfavorable light. This framing aligns with a Democrat-leaning stance, as it emphasizes information that could damage the reputation of a leading Republican. The language used focuses on the contentious nature of the statement, potentially influencing readers to view Donald Trump negatively. This approach is consistent with typical Democratic narratives that criticize Republican leaders. | This news item appears to lean Democrat based on several factors: 1) It presents Trump, the Republican figure, in a highly negative light by associating him with Hitler, 2) The framing suggests criticism of Trump by highlighting a controversial statement that would be particularly damaging to his reputation, 3) This type of narrative aligns with common Democratic messaging that portrays Trump as sympathetic to authoritarian figures, 4) The statement is presented without context or qualification, which can amplify its negative impact. This framing is consistent with left-leaning media's approach to covering Trump, emphasizing statements that paint him as extreme or problematic. |
| 6 | The news item provides a factual, historical account of a medieval Italian figure, Bartolomeo Colleoni, focusing on the meaning of his last name and the symbols on his coat of arms. There is no connection to contemporary political issues, policies, or debates associated with the Democrat or Republican parties. The language used is descriptive and lacks any persuasive or biased framing that would indicate a particular political stance. Therefore, the content is neutral in terms of political orientation. | This news item is a historical and etymological fact about a medieval Italian condottiere (mercenary leader) Bartolomeo Colleoni. It contains no political framing, no partisan language, and no alignment with either Democratic or Republican narratives. It is purely a historical curiosity about the meaning of a person's surname and the heraldic symbolism associated with it. The content does not promote any political perspective, policy position, or ideological viewpoint. It is simply an interesting historical anecdote with no discernible political leaning. |
| 7 | The news item highlights a gaffe made by President Joe Biden, where he mistakenly referred to Egyptian President Abdel Fattah El-Sisi as the 'president of Mexico.' Focusing on such errors aligns with Republican narratives that question President Biden's competence or cognitive abilities. By emphasizing this mistake without additional context or mitigating information, the report frames the president in a negative light. This perspective and framing suggest a leaning towards a Republican political stance. | This news item highlights a verbal gaffe made by President Joe Biden, a Democrat, in which he confused Egyptian President Abdel Fattah El-Sisi with 'the president of Mexico.' The framing and selection of this story aligns with a Republican political narrative that frequently emphasizes Biden's cognitive decline, verbal mistakes, and perceived lack of competence as a leader. Republican media and politicians have consistently used such gaffes to question Biden's mental fitness for office. By highlighting this error — particularly in the context of a serious geopolitical crisis like the humanitarian situation in Gaza — the story implicitly undermines confidence in Biden's leadership and foreign policy acumen. A neutral or Democrat-leaning outlet might have provided more context, minimized the significance of the gaffe, or framed it differently. The straightforward presentation of the mistake without additional context or defense of Biden suggests a framing more consistent with Republican-leaning media narratives. |
| 8 | The news item features quotes from former President Bill Clinton that focus on personal growth, learning from mistakes, handling adversity, and perseverance. The content is motivational and universally applicable without promoting any specific political agenda or aligning with typical Democratic or Republican narratives. While Bill Clinton is a prominent figure in the Democratic Party, the quotes themselves are non-partisan and do not reflect a partisan perspective. Therefore, the political stance of the news item is categorized as Neutral. | This news item presents a generic motivational quote about personal growth and perseverance that doesn't align with any particular political ideology. The quote focuses on universal human experiences (making mistakes, learning, facing adversity) and promotes resilience ('never quit'). While Bill Clinton is a prominent Democrat, the content of this quote doesn't advance any partisan political agenda or policy position. The message could be embraced by people across the political spectrum, as it speaks to basic human values and life lessons rather than political issues. The language used is straightforward and inspirational without any politically charged terms or partisan framing. |
| 9 | The news item presents a straightforward report about 'Project 2025,' a conservative blueprint proposed for the next U.S. Republican presidential administration, which calls for shutting down the U.S. Department of Education. The language used is factual and lacks emotive or persuasive terms that would indicate bias. It does not provide supportive or critical commentary on the initiative, nor does it align with typical partisan rhetoric. Therefore, the content and framing are neutral, simply conveying information without promoting a particular political perspective. | This news item clearly aligns with traditional Republican positions and narratives for several reasons: 1) Project 2025 is explicitly identified as a conservative blueprint, 2) The proposal to eliminate the Department of Education has been a long-standing Republican policy position, reflecting their preference for state and local control of education over federal oversight, 3) This aligns with core Republican principles of smaller federal government and reduced federal bureaucracy, 4) The timing and framing of this as a plan for the next Republican administration indicates it's specifically designed to advance Republican policy goals. The content directly promotes a distinctly conservative approach to governance without presenting alternative viewpoints or critiques of the proposal. |
| 10 | The news item presents a factual statement without any apparent bias or partisan framing. It mentions prominent political families from both the Republican (Bush) and Democratic (Biden and Clinton) parties, indicating a historical shift away from candidates belonging to these families. The language used is neutral and does not employ any emotionally charged or persuasive terms that would favor one political perspective over another. There is no promotion of a specific political narrative or agenda; instead, it highlights a noteworthy point about the upcoming election's departure from a long-standing pattern in U.S. politics. Therefore, the political stance of the news item is neutral. | This news item is a straightforward historical fact about the 2024 U.S. presidential election. It simply notes that for the first time since 1976, none of the major political dynasties — Bush (Republican), Biden (Democrat), or Clinton (Democrat) — appear on the ballot. The statement is purely observational and factual, referencing both Republican and Democratic political families equally without favoring or criticizing either party. There is no loaded language, no implied judgment, and no alignment with any particular political narrative. It presents a neutral, bipartisan historical observation that applies equally to both major parties. |
| 11 | The news item highlights a controversial statement made by Donald Trump, a leading Republican figure, in a manner that portrays him negatively. By emphasizing his suggestion to inject disinfectants to treat COVID-19, it aligns with Democratic narratives that criticize his handling of the pandemic. The language used does not offer any mitigating context or alternative perspectives, indicating a leaning towards a Democratic stance. | This framing of Trump's COVID-19 comments aligns with typical Democratic messaging and criticism of Trump's pandemic response. While Trump did make comments about disinfectants during a press briefing, this particular phrasing ('inject bleach') represents the most negative interpretation of those remarks, which Trump later claimed were sarcastic. The direct reference to 'bleach' was commonly used by Democratic politicians and left-leaning media to criticize Trump's pandemic leadership. The statement's focus on portraying Trump's comments as dangerous and unscientific reflects a narrative frequently employed by Democratic politicians and liberal media outlets to question Trump's fitness for office. A more neutral framing would likely include the full context of the press briefing and Trump's subsequent clarification, while a Republican-leaning version might defend the comments as sarcastic or focus on different aspects of Trump's COVID response. |
| 12 | The news item is a historical fact about medical practices in the 1920s, where doctors prescribed Guinness beer to pregnant women for its iron content. It does not promote or criticize any political ideology, party, or policy. The content is purely informative and lacks any political framing or language that aligns with typical Democratic or Republican narratives. Therefore, it is considered neutral in terms of political stance. | This news item is a historical fact about medical practices from the 1920s and doesn't contain any modern political framing or bias. It's simply stating a historical medical recommendation without commentary on healthcare policies, regulations, or social issues that might align with either political party's current positions. The statement is purely informational and doesn't promote any particular political ideology or contemporary political narrative. |
| 13 | The news item highlights acts of public disdain toward Donald Trump, a prominent Republican figure. Reporting that people are repeatedly urinating on his Hollywood Walk of Fame star emphasizes negative sentiments against him. This aligns with Democratic narratives that critique or oppose Trump. The focus on the disrespect directed at Trump suggests a perspective sympathetic to Democratic viewpoints. While the language used is factual and neutral, the choice of topic and framing resonate more with Democratic criticisms of Trump's presidency and public image. | This news item leans Democrat based on several factors: (1) It highlights disrespectful behavior toward Trump's star as a factual occurrence without condemning it, which implicitly normalizes anti-Trump sentiment; (2) The framing presents Trump as a figure who generates such strong negative reactions that infrastructure modifications were necessary, portraying him unfavorably; (3) The story focuses on the humiliating aspect of people urinating on his star rather than framing it as vandalism or disrespectful behavior toward a former president; (4) The language is matter-of-fact about the disrespectful acts rather than expressing outrage, which aligns more with how Democrat-leaning outlets might cover Trump-related incidents. While presented as a factual report, the choice to highlight this particular story and frame it without criticism of the vandalism suggests a perspective more aligned with Democratic viewpoints that are critical of Trump. |
| 14 | The news item leans Republican due to its framing and language that align with conservative viewpoints. The use of the term "Olympic DUDE" in all capitals suggests a dismissive or derogatory reference to a transgender female athlete. This aligns with typical Republican narratives that question or oppose the inclusion of transgender women in women's sports, emphasizing concerns over fairness and gender classifications. Additionally, the mention of proceeds going to a battered women's charity ties into themes of protecting traditional women's spaces and rights, which are often highlighted in conservative discourse. The overall tone and perspective promote skepticism towards transgender participation in women's sports, reflecting a stance commonly associated with Republican viewpoints. | This news item leans Republican based on several key indicators: (1) The use of 'DUDE' in all caps appears to be a deliberate misgendering reference, likely referring to a transgender athlete who competed in women's Olympic boxing, which aligns with conservative opposition to transgender participation in women's sports. (2) The framing implicitly promotes the narrative that transgender women are actually men ('dudes'), a position typically held by Republican politicians and conservative media. (3) While supporting battered women's charity appears positive on the surface, in this context it reinforces the conservative argument that allowing transgender women in women's sports somehow harms or threatens biological women. (4) The language choice and framing align with typical Republican talking points about protecting women's sports from transgender participation. The item promotes a perspective that questions transgender athletes' legitimacy in their gender category, which is a prominent conservative political position. |
| 15 | The news item originates from Fox News, a media outlet often associated with conservative and Republican viewpoints. The chyron stating "Kamala could be the oldest elected female president" focuses on Kamala Harris's age in the context of a potential presidential run. By emphasizing her age, the framing may imply concerns about her suitability for office due to age, which is a critique sometimes directed at political candidates to question their capability. This perspective aligns with Republican narratives that scrutinize Democratic figures, suggesting a Republican-leaning stance. | This chyron appears to have a subtle negative framing typical of right-leaning media coverage of Democratic politicians. By emphasizing Kamala Harris's age as a potential issue, it echoes similar Republican talking points about age in leadership, particularly following criticisms of President Biden's age. The use of 'Kamala' (first name only) rather than 'Vice President Harris' or her full name can be seen as diminishing her position and authority, a tactic often used by conservative media when discussing Democratic politicians they oppose. The framing seems designed to raise concerns about her age while simultaneously undermining her credibility, aligning with typical Republican messaging strategies against Democratic leadership. |
| 16 | The news item quotes Pete Hegseth, a known conservative commentator associated with the Republican Party, making an absurd and scientifically inaccurate statement denying the existence of germs because he cannot see them. By highlighting this remark, the news item potentially portrays a Republican figure in a negative light, suggesting a lack of understanding or disregard for scientific facts. This framing aligns with typical Democratic narratives that criticize Republican figures for anti-scientific views or misinformation. Therefore, the content and framing of the news item lean toward a Democratic perspective. | This news item appears to lean Democrat based on several factors: 1) It presents Pete Hegseth, a prominent conservative media figure and Trump administration nominee, making an anti-scientific statement that denies the existence of germs - a basic scientific fact. 2) The framing highlights this statement without context or clarification, which serves to make Hegseth appear ignorant or anti-science. 3) This aligns with typical Democratic narratives that portray Republican figures as anti-science or anti-intellectual. 4) The inclusion of his credentials (TV presenter and former Army National Guard officer) may be intended to emphasize the contrast between his positions of authority and the scientifically incorrect statement. 5) Democrats often use examples of Republicans making anti-science statements to support broader arguments about the party's relationship with expertise and evidence-based policy. The presentation appears designed to damage Hegseth's credibility rather than provide neutral reporting. |
| 17 | The news item highlights the absence of American flags at Vice President Kamala Harris' campaign rally. This framing suggests a lack of patriotism or national pride associated with a Democratic figure, which aligns with common Republican critiques of Democrats. By emphasizing this point, the article promotes a perspective that questions the commitment of a Democratic candidate to traditional American symbols. The language is seemingly neutral but focuses on a detail often used in Republican narratives to cast doubt on the opposition, indicating a Republican-leaning stance. | This news item appears to lean Republican because it focuses on the absence of American flags at a Democratic Vice President's campaign event, which aligns with common Republican talking points about Democrats being unpatriotic or lacking national pride. The framing implicitly criticizes the Harris campaign by highlighting what's missing (American flags) rather than what occurred at the event. This narrative fits into the broader conservative criticism of Democratic politicians and their perceived relationship with traditional American symbols. Such stories are often amplified in right-leaning media to suggest that Democrats are disconnected from patriotic values. The specific focus on this detail, rather than the event's substance or policy discussions, suggests an intent to create a negative impression of the Democratic administration through patriotic symbolism. |
| 18 | The news item discusses a scientific observation about male kangaroos flexing their biceps to impress females. It is a factual report about animal behavior without any political commentary or implications. The language used is objective and does not align with typical Democratic or Republican narratives. Therefore, the content and framing are politically neutral. | This news item is a straightforward scientific observation about animal behavior with no political content whatsoever. It simply reports on kangaroo mating behaviors without any language, framing, or perspective that would align with either Democratic or Republican political narratives. The statement contains no policy implications, ideological messaging, or partisan talking points. It's purely factual information about wildlife biology that has no bearing on political discourse or partisan issues. |

**Table S4. Rationale for the political leaning classification of the 18 selected news headlines.**

| | Question | Options |
|---|---|---|
| 1 | To what extent do you feel your evaluation of the news items improved after getting support from the AI assistant? | Very much, Quite a bit, Somewhat, A little, Not at all |
| 2 | How meaningful did you find the information provided by the AI assistant? | Extremely meaningful, Very meaningful, Moderately meaningful, Slightly meaningful, Not meaningful at all |
| 3 | How did you perceive the role of the AI assistant during the interaction? | Mostly as a tool to assist me in making my own determinations, Primarily as an agent trying to influence or persuade me in making determinations, A mix of both a tool and an influencing agent, Neither as a tool nor as an influencing agent, Unsure |
| 4 | [Study 1] To what extent did you feel that the AI assistant was evaluating or judging you based on your expressed views? | I felt judged during our interaction, Sometimes I felt judged and sometimes I did not, I did not feel judged during our interaction |
| 4 | [Study 2] To what extent did you feel that the AI assistants were evaluating or judging you based on your expressed views? | Matrix table:<br><br>Assistant (row): AI1, AI2<br><br>Magnitude (column): I felt judged during our interaction, Sometimes I felt judged and sometimes I did not, I did not feel judged during our interaction |
| 5 | [If Q4 == I did not feel judged during our interaction] To what extent did the AI assistants' lack of judgment about your views impact your comfort level during the conversations? | Significantly increased my comfort, especially when discussing opposing viewpoints; Somewhat increased my comfort by providing a judgment-free interaction; Had no effect on my comfort level; Somewhat decreased my comfort, as I felt it lacked human understanding; Significantly decreased my comfort; Not sure |
| 6 | [Study 1] Based on your interaction experience with the AI assistant for fact-checking, how likely are you to recommend this AI assistant to others for fact-checking purposes in the future? | Very likely, Likely, Neutral, Unlikely, Very unlikely |
| 6 | [Study 2] Did you notice any inconsistencies between the two AI assistants? | Yes, No |
| 7 | [Study 2; If Q6 == Yes] How did these inconsistencies make you feel? Please share your thoughts (at least two sentences). | Open text |
| 8 | Please share any additional thoughts or feelings about your experience with the AI assistant(s), if applicable. | Open text |

**Table S5. Fact-checking experiment post-treatment survey questions.**

| *Single AI* (*N* = 1,042) | | | | | | |
|---|---|---|---|---|---|---|
| **Source** | **Variable** | **Mean** | **SD** | **Min** | **Median** | **Max** |
| CloudResearch | *Age* | 38.679 | 12.260 | 18.000 | 37.000 | 78.000 |
| CloudResearch | *Gender* | 0.553 | 0.497 | 0.000 | 1.000 | 1.000 |
| Pre-Interaction Q14-15 | *Risk Preference* | 0.407 | 0.232 | 0.000 | 0.375 | 1.000 |
| Pre-Interaction Q6-12 | *Financial Literacy* | 0.711 | 0.200 | 0.000 | 0.714 | 1.000 |
| Conversation text | *Conversation Length* | 16.345 | 28.489 | 0.000 | 8.000 | 344.000 |
| Conversation text | *AI Portfolio Active* $M^2$ | -0.025 | 0.022 | -0.101 | -0.025 | 0.040 |
| Pre-Evaluation | *Pre-Performance* | -0.027 | 0.025 | -0.127 | -0.025 | 0.046 |
| Post-Evaluation | *Post-Performance* | -0.027 | 0.025 | -0.109 | -0..025 | 0.045 |
| ***Dual AI* (*N* = 1,011)** | | | | | | |
| CloudResearch | *Age* | 39.633 | 12.710 | 18.000 | 38.000 | 80.000 |
| CloudResearch | *Gender* | 0.538 | 0.499 | 0.000 | 1.000 | 1.000 |
| Pre-Interaction Q14-15 | *Risk Preference* | 0.410 | 0.224 | 0.000 | 0.375 | 1.000 |
| Post-Interaction Q6-12 | *Financial Literacy* | 0.704 | 0.200 | 0.000 | 0.714 | 1.000 |
| Conversation text | *Conversation Length* | 13.761 | 24.125 | 0.000 | 7.000 | 245.000 |
| Conversation text | *AI1 Portfolio Active* $M^2$ | -0.029 | 0.024 | -0.129 | -0.027 | 0.049 |
| Conversation text | *AI2 Portfolio Active* $M^2$ | -0.028 | 0.023 | -0.106 | -0.027 | 0.034 |
| Pre-Evaluation | *Pre-Performance* | -0.029 | 0.027 | -0.117 | -0.027 | 0.056 |
| Post-Evaluation | *Post-Performance* | -0.029 | 0.025 | -0.116 | -0.027 | 0.056 |

**Table S6. Descriptive statistics of continuous (or binary) variables of the investment experiment.**

| | Question | Options |
|---|---|---|
| 1 | Do you currently work or study in finance, economics, accounting, or a closely related field? | Yes, No |
| 2 | Which of the following types of investments do you currently own? | Matrix table:<br>Categories (row): Individual stocks, Individual bonds, Mutual funds, Exchange traded funds (ETFs) Annuities (fixed, indexed, or variable), Commodities or futures, Cryptocurrencies, Other (real estate)<br>Answer (column): Yes, No, Don’t know |
| 3 | When did you first start investing in stocks? | Less than a year ago, 1 year or less than 3 years ago, 3 years or less than 5 years ago, 5 years or less than 10 years ago, 10 years ago or more |
| 4 | In the past 12 months, how many times have you bought or sold stocks in your accounts? | None, 1 to 3 times, 4 to 10 times, 11 times or more, Don’t know |
| 5 | How often do you follow financial market news (e.g., stock market updates, economic reports, investment commentary)? | Never, Rarely (less than once a month or about once a month), Occasionally (a few times a month or about once a week), Frequently (more than once a week) |
| 6 | Suppose you have $100 in a savings account earning 2 percent interest a year. After five years, how much would you have? | More than $102, Exactly $102, Less than $102, Don’t know |
| 7 | Imagine that the interest rate on your savings account is 1 percent a year and inflation is 2 percent a year. After one year, would the money in the account buy more than it does today, exactly the same, or less than today? | More, Same, Less, Don’t know |
| 8 | If interest rates rise, what will typically happen to bond prices? Rise, fall, stay the same, or is there no relationship? | Rise, Fall, Stay the same, No relationship, Don’t know |
| 9 | True or false: A 15-year mortgage typically requires higher monthly payments than a 30-year mortgage, but the total interest over the life of the loan will be less. | True, False, Don’t know |
| 10 | True or false: Buying a single company's stock usually provides a safer return than a stock mutual fund. | True, False, Don't know |
| 11 | Suppose you owe $1,000 on a loan and the interest rate you are charged is 20% per year compounded annually. If you didn't pay anything off, at this interest rate, how many years would it take for the amount you owe to double? | Less than 2 years, 2 to 4 years, 5 to 9 years, 10 or more years, Don't know |
| 12 | Which of the following indicates the highest probability of getting a particular disease? | There is a one-in-twenty chance of getting the disease, 2% of the population will get the disease, 25 out of every 1,000 will get the disease, Don't know |
| 13 | Please estimate how well you think you performed on the seven financial knowledge questions above. What percentage of the questions do you believe you answered correctly? | 0% (you believe you answered none correctly) to 100% (you believe you answered all correctly) |
| 14 | How strongly do you agree or disagree with the following statement? I need to take big risks with my investments in order to reach my financial goals. | 1 (strong disagree) to 7 (strongly agree) |
| 15 | Below are five hypothetical investment options. Each option has two possible outcomes, each occurring with equal probability (50%). Please select the option you would prefer. There are no right or wrong answers. We are interested in your personal preferences. | Option 1: 50% chance of receiving $28,000 / 50% chance of receiving $28,000, Option 2: 50% chance of receiving $24,000 / 50% chance of receiving $36,000, Option 3: 50% chance of receiving $20,000 / 50% chance of receiving $44,000, Option 4: 50% chance of receiving $16,000 / 50% chance of receiving $52,000, Option 5: 50% chance of receiving $12,000 / 50% chance of receiving $60,000 |
| 16 | For each row below, please choose between receiving $100 today and receiving a larger amount in 14 days. There are no right or wrong answers. We are interested in your personal preferences. | Matrix table:<br>Categories (row): $100,000 today vs $102,000 in 14 days, $100,000 today vs $104,000 in 14 days, $100,000 today vs $106,000 in 14 days, $100,000 today vs $108,000 in 14 days, $100,000 today vs $110,000 in 14 days, $100,000 today vs $115,000 in 14 days, $100,000 today vs $120,000 in 14 days, $100,000 today vs $130,000 in 14 days<br>Answer (column): Today, In 14 days |
| 17 | How strongly do you agree or disagree with the following statement? I understand financial markets better than the average person. | 1 (strong disagree) to 7 (strongly agree) |
| 18 | How strongly do you agree or disagree with the following statement? I am confident in my ability to construct a profitable investment portfolio. | 1 (strong disagree) to 7 (strongly agree) |
| 19 | How strongly do you agree or disagree with the following statement? | Matrix table:<br>Statements (row): I trust AI systems to provide sound financial advice. AI systems are generally objective. AI systems may reflect hidden biases. I prefer algorithmic advice over human advisors.<br>Answer (column): 1 (strong disagree) to 7 (strongly agree) |
| 20 | Thinking about the overall U.S. stock market (e.g., a broad market index such as the S&P 500), how likely do you believe it is that the market will increase over the next 2 weeks? | Very unlikely, Unlikely, Somewhat unlikely, Neither likely nor unlikely, Somewhat likely, Likely, Very likely |
| 21 | Over the next 2 weeks, how volatile (i.e., how much up-and-down fluctuation) do you expect the overall U.S. stock market to experience? | Very low volatility, Low volatility, Somewhat low volatility, Moderate volatility, Somewhat high volatility, High volatility, Very high volatility, Prefer not to say |

**Table S7. Investment experiment pre-treatment survey questions.**

| | Text | Required Input |
|---|---|---|
| Instruction A | Your Goal<br><br>In this task, you will design two paper-trading portfolios using a hypothetical $100,000 investment (not real money) for the next 14 days.<br><br>Step 1 — Create a portfolio on your own, without AI assistance. You may research market information using the following sites before making your decisions:<br><br>finance.yahoo.com · google.com/finance · marketwatch.com<br><br>Please do not use any external AI tools during this step.<br><br>Step 2 — Consult an AI-based investment coach, then revise your portfolio if you wish.<br><br>Available Investment Options<br><br>You may allocate funds across 25 investment options grouped into five categories. The table below summarizes the general risk-return profile of each category to help guide your decisions.<br><br>Category: Typical Volatility; Historical Return Potential; 2-Week Swing Range.<br><br>■ Defensive / Hedges: Low; ~2–5% annually; ± 0.5–2%<br><br>■ Sectors / Broad Equity: Low–Moderate; ~7–11% annually; ± 2–5%<br><br>■ Tech & Growth: Moderate–High; ~10–20% annually; ± 3–8%<br><br>■ High-Beta / Growth: High; Highly variable; ± 5–15%<br><br>■ Cryptocurrency: Very High; Highly variable; ± 8–25%<br><br>Note: Numbers above are approximate historical ranges and do not guarantee future performance. "2-Week Swing Range" reflects the typical range of price movement over a 14-day period under normal market conditions.<br><br>1. Defensive / Hedges — Lower Volatility<br><br>Safe-haven investments like government bonds and gold. They won't grow much, but they help protect your money when the market drops.<br><br>SHY U.S. Treasury Bills | Click the Next button |

| | |
|---|---|
| IEF | 7–10 Year U.S. Treasuries |
| LQD | Investment Grade Corporate Bonds |
| GLD | Gold |
| VNQ | Real Estate / REITs |

2. Sectors / Broad Equity

Funds that let you invest in whole sections of the economy — like all banks, all energy companies, or the entire U.S. stock market — instead of picking individual companies.

| | |
|---|---|
| SPY | U.S. Broad Market |
| XLF | Financials |
| XLE | Energy |
| XLI | Industrials |
| XLP | Consumer Staples |
| XLU | Utilities |

3. Mega-/Large-Cap Technology & Growth — Mid–High Volatility

The biggest names in tech — companies like Apple, Google, and NVIDIA. They've grown a lot in recent years but their stock prices can drop quickly on bad news.

| | |
|---|---|
| AAPL | Apple |
| MSFT | Microsoft |
| AMZN | Amazon |
| GOOGL | Alphabet |
| NVDA | NVIDIA |
| TSLA | Tesla |

4. High-Beta / Small–Mid Growth Stocks

Smaller, newer companies betting on big ideas like electric vehicles and gene editing. Their prices can jump or drop much more than the overall market.

| | |
|---|---|
| SHOP | Shopify |
| SNOW | Snowflake |

| | | |
|---|---|---|
| | PLTR Palantir<br><br>DKNG DraftKings<br><br>RIVN Rivian<br><br>CRSP CRISPR Therapeutics<br><br>5. Cryptocurrency — Very High Volatility<br><br>Digital currencies that exist only online. They're the most unpredictable option here — prices can swing 10%+ in a single week.<br><br>BTC Bitcoin<br><br>ETH Ethereum<br><br>Cryptocurrency prices will be referenced in U.S. dollars using a fixed market price at the start of the evaluation period.<br><br>Portfolio Rules<br><br>Your total allocation must sum to 100%.<br><br>You may allocate at most 40% to any single investment.<br><br>You must invest in at least 5 different investments.<br><br>If you invest in an option, you must allocate at least 5% to it.<br><br>All allocations must be non-negative. | |
| Pre-Interaction portfolio design | Create Your Initial Portfolio<br><br>Please allocate the $100,000 hypothetical investment across the options below for the next 14 days.<br><br>Enter a percentage (0–100) for each option. Your allocations must sum to 100. Leave options at 0 if you do not wish to invest in them.<br><br>The 25 options are grouped into five categories for your reference:<br><br>■ Defensive / Hedges — Lower Volatility SHY · IEF · LQD · GLD · VNQ<br><br>■ Sectors / Broad Equity SPY · XLF · XLE · XLI · XLP · XLU<br><br>■ Tech & Growth — Mid–High Volatility AAPL · MSFT · AMZN · GOOGL · NVDA · TSLA<br><br>■ High-Beta / Growth Stocks SHOP · SNOW · PLTR · DKNG · | Money in percentage assigned to each asset (SHY, IEF, LQD, GLD, VNQ, SPY, XLF, XLE, XLI, XLP, XLU, AAPL, MSFT, AMZN, GOOGL, NVDA, TSLA, SHOP, SNOW, PLTR, DKNG, RIVN, CRSP, BTC, ETH) |
| | RIVN · CRSP<br><br>■ Cryptocurrency — Very High Volatility BTC · ETH<br><br>Click here to review the portfolio construction rulesClick here to hide the portfolio construction rules<br><br>Your total allocation must sum to 100%.<br><br>You may allocate at most 40% to any single investment.<br><br>You must invest in at least 5 different investments.<br><br>If you invest in an option, you must allocate at least 5% to it.<br><br>All allocations must be non-negative.<br><br>You may look up market information before making your decisions:<br><br>finance.yahoo.com · google.com/finance · marketwatch.com<br><br>Important: Please do not use any AI tools at this stage. This portfolio should reflect your own judgment. | |

**Table S8. A complete portfolio design session.**

| | | |
|---|---|---|
| Pre-Interaction confidence assignment | How confident are you in the investment portfolio you submitted? | 1 (strong disagree) to 7 (strongly agree) |
| Human-AI interaction | AI Investment Coach<br><br>How it works: The AI will evaluate your portfolio and suggest modifications. You may ask follow-up questions or click Next to move on at any time after the initial assessment.<br><br>AI Coach<br><br>[AI_MESSAGE] | Response to AI messages |
| Post-Interaction evaluation | Finalize Your Portfolio<br><br>Please review or update your portfolio allocation for the next 14 days. Your final portfolio in this step will be used to determine your performance ranking in this study.<br><br>After the 14-day evaluation period, portfolios will be evaluated using real market data. Participants whose portfolios rank among the top 20% within their study wave will receive an additional performance bonus. | Money in percentage assigned to each asset (SHY, IEF, LQD, GLD, VNQ, SPY, XLF, XLE, XLI, XLP, XLU, AAPL, MSFT, AMZN, GOOGL, NVDA, TSLA, SHOP, SNOW, PLTR, DKNG, RIVN, |

| | Question | Options |
|---|---|---|
| 1 | How confident are you in the final investment portfolio you submitted after interacting with the AI assistant? | 1 (not confident at all) to 7 (extremely confident) |
| 2 | To what extent did the AI assistant influence the final portfolio you submitted? | 1 (not at all) to 7 (completely determined my portfolio) |
| 3 | How likely do you believe the final portfolio you submitted after interacting with the AI assistant will perform well relative to the overall market? | 1 (very unlikely) to 7 (very likely) |
| 4 | Compared to the portfolio you created before interacting with the AI assistant, how would you evaluate the final portfolio you submitted? | 1 (much worse) to 4 (about the same) to 7 (much better) |
| 5 | How did you perceive the role of the AI assistant during the interaction? | Mostly as a tool to assist me in making my own determinations; Primarily as an agent trying to influence or persuade me in making determinations; A mix of both a tool and an influencing agent; Neither as a tool nor as an influencing agent; Unsure |
| 6 | How strongly do you agree or disagree with the following statements about your experience with the AI assistant? | Matrix table:<br><br>Statements (row): The AI assistant provided information or perspectives that I would not have considered on my own; I trust the investment suggestions provided by the AI assistant; I would rely on an AI assistant like this when making real investment decisions; I would recommend using an AI assistant for investment suggestions to others.<br><br>Answers (column): 1 (strongly disagree) to 4 (neither) to 7 (strongly agree) |
| 7 | Do you have any feedback about your interaction with the AI assistant during the investment task?<br><br>Examples of topics you might comment on: (1) What you found helpful or unhelpful; (2) Whether the AI influenced your decisions; (3) Any concerns or suggestions. | Open text |

**Table S9. Investment experiment post-treatment survey questions.**

| **Bias persona** | **Position** | **s.d.** | **95% CI** |
|---|---|---|---|
| Extremely risk-averse | 0.000 | 0.000 | [0.000, 0.000] |
| Somewhat risk-averse | 0.208 | 0.056 | [0.179, 0.201] |
| Default | 0.324 | 0.054 | [0.317, 0.338] |
| Risk-neutral | 0.560 | 0.072 | [0.470, 0.499] |
| Somewhat risk-seeking | 0.835 | 0.040 | [0.771, 0.787] |
| Extremely risk-seeking | 1.000 | 0.000 | [1.000, 1.000] |

**Table S10. Calibrated positions of the AI risk personas on the participant risk-preference scale.**

| Source | Variable | Mean | SD | Min | Median | Max |
|---|---|---|---|---|---|---|
| | ***Time-invariant: student level* (*N* = 39)** | | | | | |
| Pre-survey | Languages/tools used | 3.897 | 1.252 | 1.000 | 4.000 | 8.000 |
| Pre-survey | Verifies AI outputs | 3.179 | 0.914 | 0.000 | 3.000 | 4.000 |
| Pre-survey | Confidence evaluating AI code | 5.051 | 1.146 | 1.000 | 5.000 | 7.000 |
| Pre-survey | AI coding tools used | 1.923 | 1.676 | 0.000 | 2.000 | 6.000 |
| Pre-survey | AI tool settings used | 1.949 | 1.432 | 0.000 | 2.000 | 4.000 |
| Pre-survey | AI coding tasks | 2.077 | 1.306 | 0.000 | 3.000 | 3.000 |
| Pre-survey | Prefers reliable/standard solutions | 5.077 | 1.178 | 2.000 | 5.000 | 7.000 |
| Pre-survey | Prefers new/unconventional approaches | 5.051 | 1.555 | 2.000 | 5.000 | 7.000 |
| Pre-survey | Chooses safest under time pressure | 5.641 | 1.495 | 2.000 | 6.000 | 7.000 |
| Pre-survey | Enjoys exploring alternatives | 5.923 | 1.178 | 2.000 | 6.000 | 7.000 |
| Pre-survey | Prefers one best answer | 3.692 | 1.625 | 1.000 | 4.000 | 7.000 |
| Pre-survey | Prefers several candidate strategies | 5.692 | 1.260 | 2.000 | 6.000 | 7.000 |
| Pre-survey | Stops searching when good enough | 4.487 | 1.604 | 1.000 | 5.000 | 7.000 |
| Pre-survey | Willing to spend extra effort | 5.205 | 1.260 | 2.000 | 5.000 | 7.000 |
| Pre-survey | Comfortable rejecting AI suggestions | 6.615 | 0.633 | 5.000 | 7.000 | 7.000 |
| Pre-survey | Worries about AI dependence | 5.256 | 1.371 | 2.000 | 5.000 | 7.000 |
| Pre-survey | Expects AI to help learning | 5.821 | 1.315 | 1.000 | 6.000 | 7.000 |
| | ***Time-varying: student-week level* (*N* = 312)** | | | | | |
| Randomization | Assigned week | 4.946 | 2.629 | 1.000 | 5.000 | 9.000 |
| Randomization | Novelty-biased week | 0.061 | 0.240 | 0.000 | 0.000 | 1.000 |

| Source | Variable | Mean | SD | Min | Median | Max |
|---|---|---|---|---|---|---|
| Randomization | Reliability-biased week | 0.064 | 0.245 | 0.000 | 0.000 | 1.000 |
| Randomization | Biased week was student's first | 0.513 | 0.506 | 0.000 | 1.000 | 1.000 |
| Assignment grading | Coding assignment submitted | 0.801 | 0.400 | 0.000 | 1.000 | 1.000 |
| Assignment grading | Coding grade | 8.711 | 1.147 | 0.890 | 8.977 | 10.000 |
| Assignment grading | Memo submitted | 0.837 | 0.370 | 0.000 | 1.000 | 1.000 |
| Assignment grading | Memo grade | 7.664 | 0.885 | 4.833 | 7.800 | 10.000 |
| Session telemetry | Any assistant use | 0.401 | 0.491 | 0.000 | 0.000 | 1.000 |
| Session telemetry | Reply turns, all weeks | 44.218 | 102.474 | 0.000 | 0.000 | 740.000 |
| Session telemetry | Reply turns, weeks with use | 110.368 | 137.759 | 1.000 | 70.000 | 740.000 |
| Session telemetry | Words written, all weeks | 14751.417 | 35313.683 | 0.000 | 0.000 | 264275.000 |
| Session telemetry | Words written, weeks with use | 36819.536 | 48047.963 | 9.000 | 17316.000 | 264275.000 |

**Table S11. Descriptive statistics of continuous (or binary) variables of the education experiment.**

| | Question | Options |
|---|---|---|
| 1 | What is your current academic level? | Undergraduate, Master's student, PhD student, Other___ |
| 2 | How would you describe your prior programming experience in general? | None, Beginner, Intermediate, Advanced, |
| 3 | How would you describe your prior machine learning / AI programming experience? | None, Beginner, Intermediate, Advanced |
| 4 | Which languages/tools have you used before? (Select all that apply) | Python, R, Javascript, SQL, Bash/CLI, Git/Github, Prefer not to say, Other___ |
| 5 | In the past 30 days, how often have you used any AI assistant (e.g., ChatGPT, Claude, Gemini) for the following purposes? | Matrix table:<br><br>Purpose (row): For learning/explanations (concepts, reading, studying), For coding/programming (writing code, debugging, refactoring), For writing (emails, reports, summaries)<br><br>Frequency (column): Never, 1-2 times, Weekly, Several time/week, Daily/almost daily |
| 6 | Roughly how many total hours did you use AI assistants in the last 7 days? | 0, < 1 hour, 1-5 hours, 5-10 hours, 10-12 hours, 20+ hours, Prefer not to say |
| 7 | How often do you typically verify AI outputs (e.g., run the code, test edge cases, check reasoning)? | Never, Rarely, Sometimes, Often, Always |
| 8 | How confident are you in evaluating whether an AI-generated coding solution is correct? | 1 (Not confident at all) to 7 (Very confident) |
| 9 | [If Q9 = No, go to Q14] Before this course, had you used an AI coding tool integrated into an IDE/terminal/workflow (e.g., Claude Code, GitHub Copilot, Cursor, Codeium)? | Yes, No, Not sure |
| 10 | Which AI coding tools have you used before? (Select all that apply) | Claude Code, Github Copilot, Cursor, Codex, VS Code extensions, Other___ |
| 11 | If you have used Claude Code before, how frequently? | Tried once or twice, Occasionally (monthly), Often (weekly), Very often (daily/almost daily), Not applicable |
| 12 | Where have you primarily used AI coding tools? (Select all that apply) | Coursework/homework, Research, Internship/job, Personal projects, Other___ |
| 13 | When you use AI for coding, which tasks do you most often use it for? (Select up to 3) | Explaining concepts, Writing starter code/templates, Debugging errors, Refactoring/cleaning code, Writing tests, Understanding documentation/APIs, Optimizing performance, Brainstorming approaches, Other___ |
| 14 | Please indicate your agreement with each statement. (1 = Strongly disagree, 7 = Strongly agree) | Matrix table:<br><br>Option (row): I prefer a solution that is reliable and standard, even if it is not innovative; I prefer to try new or unconventional approaches, even if they sometimes fail; When I am under time pressure, I tend to choose the safest approach; When I have time, I enjoy exploring multiple alternative approaches before deciding; I would rather have an assistant propose one best answer than multiple speculative ideas; I would rather have an assistant propose several candidate strategies (with pros/cons) than one answer; If a first approach works “well enough,” I typically stop searching for alternatives; I am comfortable spending extra effort to find a better approach even after I have a working one.<br><br>Attitude (column): 1 (strongly disagree) to 7 (strongly agree) |
| 15 | Please indicate your agreement with each statement. (1 = Strongly disagree, 7 = Strongly agree) | Matrix table:<br><br>Statement (row): I feel comfortable rejecting an AI suggestion when it doesn’t seem right, I worry I might become overly dependent on AI tools for coding, I expect AI tools to help me learn, not just finish tasks faster.<br><br>Attitude (column): 1 (strongly disagree) to 7 (strongly agree) |
| 16 | What is your biggest hope or concern about using an AI coding assistant in this course? | Open text |

**Table S12. Pre-class survey questions.**

| | Question | Options |
|---|---|---|
| 1 | [If Q1 = No, then end the survey] Did you use the course-provided Claude Code assistant for this week's work? | Yes, No |
| 2 | Approximately how long did you use Claude Code this week? | < 1 hour, 1-2 hours, 2-4 hours, 4-6 hours, > 6 hours |
| 3 | What did you mainly use Claude Code for this week? (Select all that apply) | Understanding the assignment / Planning approach; Writing code / Completing implementation; Debugging / Fixing errors, Testing / Verification; Improving code quality (refactor, readability, efficiency); Getting alternative approaches; Other___ |
| 4 | Besides the course-provided Claude Code assistant, did you use any other AI tools for this week's work? | Yes, No |
| 5 | Compared to Claude Code, how much did you rely on other AI tools for this week's work? | Much less than Claude Code, Slightly less than Claude Code, About the same, Slightly more than Claude Code, Much more than Claude Code, Not applicable (I did not use other AI tools) |
| 6 | I felt comfortable using Claude Code this week. | 1 (strong disagree) to 7 (strongly agree) |
| 7 | Using Claude Code improved my performance on this week's assignment. | 1 (strong disagree) to 7 (strongly agree) |
| 8 | Claude Code was useful for helping me make progress this week. | 1 (strong disagree) to 7 (strongly agree) |
| 9 | Claude Code's suggestions were reliable and correct when I used them. | 1 (strong disagree) to 7 (strongly agree) |
| 10 | How often did you verify Claude Code's suggestions (e.g., by testing, checking documentation, or validating results)? | Never, Rarely, Sometimes, Often, Always |
| 11 | How often did Claude Code help you consider alternatives you would not have tried otherwise? | Never, Rarely, Sometimes, Often, Always |
| 12 | Overall, how much did you rely on Claude Code for key decisions this week? | 1 (not at all) to 7 (a great deal) |
| 13 | In 1–2 sentences: What was the most helpful aspect of Claude Code this week (if any)? | Open text |
| 14 | In 1–2 sentences: What was the most frustrating or least helpful aspect (if any)? | Open text |

**Table S13. Weekly post-assignment survey questions.**

| | OLS + Clustered SE | | Mixed Effects | |
|---|---|---|---|---|
| | M1 | M2 | M1 | M2 |
| *Constant* | 0.345*** (0.026) | 0.225*** (0.038) | 0.345*** (0.029) | 0.225*** (0.038) |
| *Strong Republican* | 0.055** (0.017) | 0.048** (0.016) | 0.055*** (0.017) | 0.048** (0.016) |
| *Somewhat Republican* | 0.058*** (0.016) | 0.046** (0.016) | 0.058*** (0.016) | 0.046** (0.016) |
| *Somewhat Democrat* | 0.004 (0.016) | 0.000 (0.015) | 0.004 (0.016) | 0.000 (0.016) |
| *Strong Democrat* | 0.036* (0.016) | 0.032* (0.015) | 0.036* (0.016) | 0.032* (0.016) |
| *Pre-interaction performance* | 0.409*** (0.024) | 0.406*** (0.024) | 0.409*** (0.020) | 0.406*** (0.020) |
| *AI correctness* | – | 0.202*** (0.030) | – | 0.202*** (0.023) |
| *Political ideology* | No | Yes | No | Yes |
| *Political party* | No | Yes | No | Yes |
| Headline fixed-effect | Yes | Yes | Yes | Yes |
| User random-effect | No | No | Yes | Yes |
| Observations | 2442 | 2435 | 2442 | 2435 |
| Adjusted $R^2$ | 0.472 | 0.491 | 0.475 | 0.495 |
| Residual SD | 0.247 (df = 2419) | 0.243 (df = 2403) | 0.247 (df = 2417) | 0.243 (df = 2401) |

*Note*: Dependent variable is post-interaction performance. Reference category is the default assistant; the neutralized arm is excluded. Columns 1–2 report standard errors clustered at the participant level (in parentheses). Significance codes: *** $p < 0.001$, ** $p < 0.01$, * $p < 0.05$, † $p < 0.1$.

**Table S14. Regression estimates of AI bias on participants' information-evaluation performance.**

| | AI Stance = Republican \| Default | | AI Stance = Democrat \| Default | |
|---|---|---|---|---|
| | OLS + Clustered SE | Mixed effects | OLS + Clustered SE | Mixed effects |
| *Constant* | 0.315***<br>(0.032) | 0.315***<br>(0.035) | 0.388***<br>(0.033) | 0.388***<br>(0.037) |
| *Moderate bias* | 0.058***<br>(0.016) | 0.058***<br>(0.016) | 0.004<br>(0.016) | 0.004<br>(0.015) |
| *Strong bias* | 0.056***<br>(0.017) | 0.056**<br>(0.017) | 0.041**<br>(0.015) | 0.041*<br>(0.016) |
| *Pre-interaction performance* | 0.414***<br>(0.032) | 0.414***<br>(0.026) | 0.384***<br>(0.030) | 0.384***<br>(0.025) |
| Headline fixed-effect | Yes | Yes | Yes | Yes |
| User random-effect | No | Yes | No | Yes |
| Observations | 1455 | 1455 | 1443 | 1443 |
| Adjusted $R^2$ | 0.429 | 0.436 | 0.548 | 0.550 |
| Residual SD | 0.250 (df = 1434) | 0.250 (df = 1432) | 0.239 (df = 1422) | 0.239 (df = 1420) |

*Note*: Dependent variable is post-interaction performance. Columns 1–2 report standard errors clustered at the participant level (shown in parentheses). Significance codes: *** $p < 0.001$, ** $p < 0.01$, * $p < 0.05$, † $p < 0.1$.

**Table S15. Regression estimates of AI bias magnitude on participants' information-evaluation performance, by AI political stance.**

| Variable | Positive persuasion | Negative persuasion | Positive backfire | Negative backfire | Any helpful shift | Any harmful shift |
|---|---|---|---|---|---|---|
| *Biased* | 1.403*<br>(0.162) | 0.593**<br>(0.177) | 0.889<br>(0.293) | 1.043<br>(0.262) | 1.309†<br>(0.142) | 0.676**<br>(0.148) |
| *Pre-interaction performance* | 0.010***<br>(0.294) | 29.122***<br>(0.304) | 2.386**<br>(0.299) | 1.007<br>(0.298) | 0.025***<br>(0.244) | 13.705***<br>(0.260) |
| Headline fixed-effect | Yes | Yes | Yes | Yes | Yes | Yes |
| Clustered SE | Yes | Yes | Yes | Yes | Yes | Yes |
| Base rate | 0.315 | 0.187 | 0.031 | 0.046 | 0.346 | 0.233 |
| Observations | 2305 | 1777 | 2040 | 2029 | 2441 | 2313 |
| Pseudo $R^2$ | 0.410 | 0.312 | 0.074 | 0.085 | 0.334 | 0.287 |

*Note:* Dependent variables are binary indicators for each revision category, defined at $\tau = 0.1$ (SI §1.5); the neutralized arm is excluded. Entries are odds ratios from conditional (headline fixed-effects) logistic regressions with standard errors of the underlying log-odds coefficients in parentheses; standard errors are clustered at the participant level. Base rate is the share of all 2,441 classified and unclassified evaluations falling in that category. Observation counts differ across columns because headlines with no variation in the outcome are dropped by the conditional-logit estimator. Pre-interaction performance is measured on a 0–1 scale, so its odds ratio corresponds to the full range of the variable. Significance codes: *** $p < 0.001$, ** $p < 0.01$, * $p < 0.05$, † $p < 0.1$.

**Table S16. Odds of each revision category under partisan versus default assistance (fact-checking).**

| | OLS + Clustered SE | | Mixed effects | |
|---|---|---|---|---|
| | M1 | M2 | M1 | M2 |
| *Constant* | 18.541***<br>(2.476) | 13.512**<br>(4.417) | 17.273***<br>(2.567) | 12.138**<br>(4.293) |
| *Moderate bias* | 5.399**<br>(1.642) | 5.577***<br>(1.626) | 5.391**<br>(2.014) | 5.583**<br>(2.035) |
| *Strong bias* | 6.876***<br>(1.760) | 7.071***<br>(1.804) | 6.838***<br>(2.064) | 7.020***<br>(2.085) |
| *Pre-interaction performance* | −3.379<br>(2.163) | −3.385<br>(2.125) | −1.648<br>(1.496) | −1.706<br>(1.501) |
| *AI correctness* | – | 1.666<br>(2.681) | – | 1.529<br>(1.755) |
| *Political ideology* | No | Yes | No | Yes |
| *Political party* | No | Yes | No | Yes |
| Headline-fixed effect | Yes | Yes | Yes | Yes |
| User random-effect | No | No | Yes | Yes |
| Observations | 2433 | 2427 | 2433 | 2427 |
| Adjusted $R^2$ | 0.023 | 0.027 | 0.027 | 0.034 |
| Residual SD | 24.456 (df = 2412) | 24.426 (df = 2397) | 15.822 (df = 2410) | 15.840 (df = 2395) |

*Note*: Dependent variable is the length of the human–AI conversation (total word count of participant messages). Reference category is the default assistant; the neutralized arm is excluded. Moderate bias pools the somewhat-Republican and somewhat-Democrat arms; strong bias pools the strong-Republican and strong-Democrat arms. Columns 1–2 report standard errors clustered at the participant level (in parentheses); mixed-effects $R^2$ is marginal, with conditional $R^2$ of 0.593 (M1) and 0.595 (M2). Significance codes: *** $p < 0.001$, ** $p < 0.01$, * $p < 0.05$, † $p < 0.1$.

**Table S17. Regression estimates of AI bias magnitude on participants' conversation length.**

| | OLS + Clustered SE | | Mixed effects | |
|---|---|---|---|---|
| | M1 | M2 | M1 | M2 |
| *Constant* | 0.345***<br>(0.026) | 0.219***<br>(0.038) | 0.345***<br>(0.029) | 0.219***<br>(0.038) |
| *Moderate Bias* | 0.032*<br>(0.014) | 0.023†<br>(0.014) | 0.032*<br>(0.014) | 0.023†<br>(0.014) |
| *Strong Bias* | 0.045**<br>(0.014) | 0.039**<br>(0.013) | 0.045**<br>(0.014) | 0.039**<br>(0.014) |
| *Pre-interaction performance* | 0.413***<br>(0.024) | 0.409***<br>(0.023) | 0.413***<br>(0.020) | 0.409***<br>(0.020) |
| *AI correctness* | – | 0.205***<br>(0.030) | – | 0.205***<br>(0.023) |
| *Political ideology* | No | Yes | No | Yes |
| *Political party* | No | Yes | No | Yes |
| Headline-fixed effect | Yes | Yes | Yes | Yes |
| User random-effect | No | No | Yes | Yes |
| Observations | 2442 | 2435 | 2442 | 2435 |
| Adjusted $R^2$ | 0.470 | 0.489 | 0.472 | 0.493 |
| Residual SD | 0.248 (df = 2421) | 0.243 (df = 2405) | 0.247 (df = 2419) | 0.243 (df = 2403) |

*Note*: Dependent variable is post-interaction performance. Reference category is the default assistant; the neutralized arm is excluded. Moderate bias pools the somewhat-Republican and somewhat-Democrat arms; strong bias pools the strong-Republican and strong-Democrat arms. Columns 1–2 report standard errors clustered at the participant level (in parentheses); mixed-effects $R^2$ is marginal, and the M2 model's participant variance is estimated at zero, so no conditional $R^2$ is available. Significance codes: *** $p < 0.001$, ** $p < 0.01$, * $p < 0.05$, † $p < 0.1$.

**Table S18. Regression estimates of AI bias magnitude on participants' fact-checking performance.**

| | CLM + Clustered SE | | Mixed effects (MCMC) | |
|---|---|---|---|---|
| | M1 | M2 | M1 | M2 |
| *Constant* | – | – | 3.585***<br>(0.435) | 2.874***<br>(0.372) |
| *Moderate Bias* | -0.315†<br>(0.179) | −0.389*<br>(0.184) | -0.453*<br>(0.233) | −0.537**<br>(0.213) |
| *Strong Bias* | -0.427*<br>(0.183) | −0.466*<br>(0.186) | -0.594*<br>(0.241) | −0.641**<br>(0.233) |
| *Pre-interaction performance* | -0.158<br>(0.142) | −0.225<br>(0.146) | -0.019<br>(0.089) | −0.063<br>(0.052) |
| *AI correctness* | – | 0.296†<br>(0.173) | – | 0.094<br>(0.060) |
| *Political ideology* | No | Yes | No | Yes |
| *Political party* | No | Yes | No | Yes |
| Headline-fixed effect | Yes | Yes | Yes | Yes |
| User random-effect | No | No | Yes | Yes |
| Observations | 2442 | 2435 | 2442 | 2442 |
| Pseudo $R^2$ | 0.018 | 0.048 | – | – |
| Residual SD | – (df = 2418) | – (df = 2400) | 0.047 | 0.033 |

*Note*: Dependent variable is perceived improvement (5-point ordinal). Reference category is the default assistant; the neutralized arm is excluded. CLM columns report cumulative-link model coefficients with HC0 standard errors clustered at the participant level; MCMC columns report posterior means with posterior standard deviations and pMCMC values from ordinal MCMCglmm models with a participant random effect (25,000 iterations, 5,000 burn-in, thinning 10). Residual SD for the MCMC columns is the square root of the posterior mean residual variance. Significance codes: *** $p < 0.001$, ** $p < 0.01$, * $p < 0.05$, † $p < 0.1$.

**Table S19. Regression estimates of AI bias magnitude on participants' perceived improvement.**

| | CLM + Clustered SE | | Mixed effects (MCMC) | |
|---|---|---|---|---|
| | M1 | M2 | M1 | M2 |
| *Constant* | – | – | 3.227***<br>(0.412) | 2.755***<br>(0.508) |
| *Moderate Bias* | -0.208<br>(0.168) | -0.232<br>(0.175) | -0.266<br>(0.234) | −0.348<br>(0.234) |
| *Strong Bias* | -0.489**<br>(0.176) | -0.521**<br>(0.180) | -0.724**<br>(0.239) | −0.779**<br>(0.272) |
| *Pre-interaction performance* | -0.336*<br>(0.154) | -0.370*<br>(0.156) | -0.105<br>(0.153) | −0.021<br>(0.137) |
| *AI correctness* | – | 0.114<br>(0.176) | – | −0.021<br>(0.137) |
| *Political ideology* | No | Yes | No | Yes |
| *Political party* | No | Yes | No | Yes |
| Headline-fixed effect | Yes | Yes | Yes | Yes |
| User random-effect | No | No | Yes | Yes |
| Observations | 2442 | 2435 | 2442 | 2435 |
| Adjusted $R^2$ | 0.020 | 0.061 | – | – |
| Residual SD | – (df = 2418) | – (df = 2400) | 0.042 | 0.039 |

*Note*: Dependent variable is willingness to recommend the AI. Reference category is the default assistant; the neutralized arm is excluded. CLM columns report cumulative-link model coefficients with HC0 standard errors clustered at the participant level; MCMC columns report posterior means with posterior standard deviations and pMCMC values from ordinal MCMCglmm models with a participant random effect (25,000 iterations, 5,000 burn-in, thinning 10). Residual SD for the MCMC columns is the square root of the posterior mean residual variance. Significance codes: *** $p < 0.001$, ** $p < 0.01$, * $p < 0.05$, † $p < 0.1$.

**Table S20. Regression estimates of AI bias magnitude on participants' willingness of recommending AI to others for fact-checking**

| | OLS + Clustered SE | | Mixed effects | |
|---|---|---|---|---|
| | M1 | M2 | M1 | M2 |
| *Constant* | 0.373***<br>(0.050) | 0.240***<br>(0.062) | 0.373***<br>(0.061) | 0.240***<br>(0.066) |
| *Opposition Bias* | −0.000<br>(0.021) | 0.003<br>(0.021) | −0.000<br>(0.022) | 0.003<br>(0.021) |
| *Pre-interaction performance* | 0.374***<br>(0.044) | 0.368***<br>(0.046) | 0.374***<br>(0.042) | 0.367***<br>(0.042) |
| *AI correctness* | – | 0.253***<br>(0.065) | – | 0.253***<br>(0.049) |
| *Political party* | Yes | Yes | Yes | Yes |
| Headline-fixed effect | Yes | Yes | Yes | Yes |
| User random-effect | No | No | Yes | Yes |
| Observations | 648 | 648 | 648 | 648 |
| Adjusted $R^2$ | 0.358 | 0.383 | 0.370 | 0.398 |
| Residual SD | 0.267 (df = 628) | 0.262 (df = 621) | 0.266 (df = 626) | 0.247 (df = 619) |

*Note*: Dependent variable is post-interaction performance. Reference category is echo-chamber assignment (assistant's stance matches the participant's). Sample is restricted to partisan participants paired with a partisan assistant; the neutralized arm and independent participants are excluded. Participant party is constant within each table and therefore absorbed. Columns 1–2 report standard errors clustered at the participant level (in parentheses); mixed-effects $R^2$ is marginal. Significance codes: *** $p < 0.001$, ** $p < 0.01$, * $p < 0.05$, † $p < 0.1$.

**Table S21a. Regression estimates of AI bias direction (echo chamber vs. opposition) on Republican participants' information-evaluation performance.**

| | OLS + Clustered SE | | Mixed effects | |
|---|---|---|---|---|
| | M1 | M2 | M1 | M2 |
| *Constant* | 0.350***<br>(0.047) | 0.488***<br>(0.070) | 0.350***<br>(0.049) | 0.488***<br>(0.146) |
| *Opposition Bias* | 0.052**<br>(0.018) | 0.045*<br>(0.018) | 0.052**<br>(0.018) | 0.045*<br>(0.018) |
| *Pre-interaction performance* | 0.422***<br>(0.050) | 0.408***<br>(0.049) | 0.422***<br>(0.038) | 0.408***<br>(0.038) |
| *AI correctness* | – | 0.088†<br>(0.053) | – | 0.088*<br>(0.043) |
| *Political party* | Yes | Yes | Yes | Yes |
| Headline-fixed effect | Yes | Yes | Yes | Yes |
| User random-effect | No | No | Yes | Yes |
| Observations | 645 | 645 | 645 | 648 |
| Adjusted $R^2$ | 0.557 | 0.563 | 0.563 | 0.570 |
| Residual SD | 0.227 (df = 625) | 0.225 (df = 619) | 0.227 (df = 623) | 0.225 (df = 617) |

*Note*: Dependent variable is post-interaction performance. Reference category is echo-chamber assignment (assistant's stance matches the participant's). Sample is restricted to partisan participants paired with a partisan assistant; the neutralized arm and independent participants are excluded. Participant party is constant within each table and therefore absorbed. Columns 1–2 report standard errors clustered at the participant level (in parentheses); mixed-effects $R^2$ is marginal. Significance codes: *** $p < 0.001$, ** $p < 0.01$, * $p < 0.05$, † $p < 0.1$.

**Table S21b. Regression estimates of AI bias direction (echo chamber vs. opposition) on Democrat participants' information-evaluation performance.**

| | OLS + Clustered SE | | Mixed effects | |
|---|---|---|---|---|
| | M1 | M2 | M1 | M2 |
| *Constant* | 0.311***<br>(0.024) | 0.208***<br>(0.032) | 0.311***<br>(0.026) | 0.208***<br>(0.032) |
| *Single AI Biased* | 0.035**<br>(0.013) | 0.029*<br>(0.013) | 0.035**<br>(0.013) | 0.029*<br>(0.013) |
| *Dual AI Default* | 0.022<br>(0.024) | 0.022<br>(0.023) | 0.022<br>(0.023) | 0.022<br>(0.023) |
| *Dual AI Opposition* | 0.030†<br>(0.018) | 0.023<br>(0.018) | 0.030†<br>(0.018) | 0.023<br>(0.018) |
| *Dual AI Balanced* | 0.042*<br>(0.019) | 0.036†<br>(0.019) | 0.042*<br>(0.019) | 0.036†<br>(0.019) |
| *Pre-interaction performance* | 0.435***<br>(0.021) | 0.436***<br>(0.021) | 0.435***<br>(0.017) | 0.436***<br>(0.017) |
| *AI correctness* | – | 0.202***<br>(0.027) | – | 0.202***<br>(0.021) |
| *Political ideology* | No | Yes | No | Yes |
| *Political party* | Yes | Yes | Yes | Yes |
| Headline-fixed effect | Yes | Yes | Yes | Yes |
| User random-effect | No | No | Yes | Yes |
| Observations | 3303 | 3290 | 3303 | 3290 |
| Adjusted $R^2$ | 0.473 | 0.488 | 0.475 | 0.491 |
| Residual SD | 0.247 (df = 3276) | 0.243 (df = 3256) | 0.247 (df = 3274) | 0.241 (df = 3254) |

*Note*: Dependent variable is post-interaction performance. The reference category is the single default assistant. Columns 1–2 report standard errors clustered at the participant level (in parentheses); mixed-effects $R^2$ is marginal (participant variance is estimated at zero in all four models). Significance codes: *** $p < 0.001$, ** $p < 0.01$, * $p < 0.05$, † $p < 0.1$.

**Table S22. Regression estimates of dual AI treatments on participants' information-evaluation performance.**

| | OLS + Clustered SE | | Mixed effects | |
|---|---|---|---|---|
| | M1 | M2 | M1 | M2 |
| *Constant* | 0.311***<br>(0.024) | 0.208***<br>(0.032) | 0.311***<br>(0.026) | 0.208***<br>(0.032) |
| *Single AI Biased* | 0.035**<br>(0.013) | 0.029*<br>(0.013) | 0.035**<br>(0.013) | 0.029*<br>(0.013) |
| *Dual AI Default* | 0.022<br>(0.024) | 0.022<br>(0.023) | 0.022<br>(0.023) | 0.022<br>(0.023) |
| *Dual AI Opposition* | 0.030†<br>(0.018) | 0.023<br>(0.018) | 0.030†<br>(0.018) | 0.023<br>(0.018) |
| *Dual AI Balanced* | 0.042*<br>(0.019) | 0.036†<br>(0.019) | 0.042*<br>(0.019) | 0.036†<br>(0.019) |
| *Pre-interaction performance* | 0.435***<br>(0.021) | 0.436***<br>(0.021) | 0.435***<br>(0.017) | 0.436***<br>(0.017) |
| *AI correctness* | – | 0.202***<br>(0.027) | – | 0.202***<br>(0.021) |
| *Political ideology* | No | Yes | No | Yes |
| *Political party* | Yes | Yes | Yes | Yes |
| Headline-fixed effect | Yes | Yes | Yes | Yes |
| User random-effect | No | No | Yes | Yes |
| Observations | 3303 | 3290 | 3303 | 3290 |
| Adjusted $R^2$ | 0.473 | 0.488 | 0.475 | 0.491 |
| Residual SD | 0.247 (df = 3276) | 0.243 (df = 3256) | 0.247 (df = 3274) | 0.241 (df = 3254) |

*Note*: Dependent variable is post-interaction performance. The reference category is the single default assistant. Columns 1–2 report standard errors clustered at the participant level (in parentheses); mixed-effects $R^2$ is marginal (participant variance is estimated at zero in all four models). Significance codes: *** $p < 0.001$, ** $p < 0.01$, * $p < 0.05$, † $p < 0.1$.

**Table S23. Regression estimates of dual AI treatments on participants' information-evaluation performance in the fact-checking experiment.**

| | CLM + Clustered SE | | Mixed effects (MCMC) | |
|---|---|---|---|---|
| | M1 | M2 | M1 | M2 |
| *Constant* | – | – | 4.059***<br>(0.210) | 4.398***<br>(0.373) |
| *Single AI Biased* | −0.388*<br>(0.164) | −0.417*<br>(0.168) | −0.556**<br>(0.194) | −0.574*<br>(0.219) |
| *Dual AI Default* | −0.419<br>(0.290) | −0.398<br>(0.290) | −0.630†<br>(0.350) | −0.499<br>(0.394) |
| *Dual AI Opposition* | −1.117***<br>(0.255) | −1.152***<br>(0.260) | −1.391***<br>(0.271) | −1.561***<br>(0.305) |
| *Dual AI Balanced* | −0.370<br>(0.231) | −0.417†<br>(0.230) | −0.560†<br>(0.293) | −0.634†<br>(0.323) |
| *Pre-interaction performance* | −0.172<br>(0.127) | −0.170<br>(0.127) | −0.120***<br>(0.053) | 0.163***<br>(0.061) |
| *AI correctness* | – | 0.177<br>(0.156) | – | −0.064<br>(0.099) |
| *Political ideology* | No | Yes | No | Yes |
| *Political party* | Yes | Yes | Yes | Yes |
| Headline-fixed effect | Yes | Yes | Yes | Yes |
| User random-effect | No | No | Yes | Yes |
| Observations | 3303 | 3290 | 3303 | 3290 |
| Pseudo $R^2$ | 0.044 | 0.055 | – | – |
| Residual SD | – (df = 3272) | – (df = 3252) | 0.017 | 0.030 |

*Note*: Dependent variable is perceived improvement (5-point ordinal). Reference category is the single non-biased (default) assistant. CLM columns report cumulative-link model coefficients with HC0 standard errors clustered at the participant level; MCMC columns report posterior means with posterior standard deviations and pMCMC values from ordinal MCMCglmm models with a participant random effect (25,000 iterations, 5,000 burn-in, thinning 10). Residual SD for the MCMC columns is the square root of the posterior mean residual variance. Significance codes: *** $p < 0.001$, ** $p < 0.01$, * $p < 0.05$, † $p < 0.1$.

**Table S24. Regression estimates of dual AI treatments on participants' perceived improvement in the fact-checking experiment.**

| | OLS + Clustered SE | | Mixed effects | |
|---|---|---|---|---|
| | M1 | M2 | M1 | M2 |
| *Constant* | 17.417***<br>(2.454) | 14.095***<br>(3.540) | 17.295***<br>(2.498) | 13.805***<br>(3.538) |
| *Single AI Biased* | 6.157***<br>(1.475) | 6.156***<br>(1.477) | 6.147***<br>(1.850) | 6.170***<br>(1.866) |
| *Dual AI Default* | 0.766<br>(2.201) | 1.014<br>(2.228) | 0.711<br>(3.258) | 0.981<br>(3.291) |
| *Dual AI Opposition* | 8.170**<br>(2.591) | 7.954**<br>(2.573) | 8.155**<br>(2.556) | 7.964**<br>(2.566) |
| *Dual AI Balanced* | 4.852*<br>(2.124) | 5.032*<br>(2.140) | 4.851†<br>(2.732) | 5.024†<br>(2.747) |
| *Pre-interaction performance* | −1.009<br>(1.784) | −0.999<br>(1.782) | 0.163<br>(1.260) | 0.132<br>(1.265) |
| *AI correctness* | – | 1.897<br>(2.362) | – | 1.746<br>(1.550) |
| *Political ideology* | No | Yes | No | Yes |
| *Political party* | Yes | Yes | Yes | Yes |
| Headline-fixed effect | Yes | Yes | Yes | Yes |
| User random-effect | No | No | Yes | Yes |
| Observations | 3289 | 3277 | 3289 | 3277 |
| Adjusted $R^2$ | 0.025 | 0.027 | 0.026 | 0.031 |
| Residual SD | 24.063 (df = 3262) | 24.055 (df = 3243) | 15.476 (df = 3260) | 15.501 (df = 3241) |

*Note*: Dependent variable is the length of the human–AI conversation (total word count of participant messages). Reference category is the single non-biased (default) assistant. Columns 1–2 report standard errors clustered at the participant level (in parentheses); mixed-effects $R^2$ is marginal, with conditional $R^2$ of 0.598 (M1) and 0.599 (M2). Significance codes: *** $p < 0.001$, ** $p < 0.01$, * $p < 0.05$, † $p < 0.1$.

**Table S25. Regression estimates of dual AI treatments on participants' conversation length (the participant messages' total words number) in the fact-checking experiment.**

| | OLS + Clustered SE | | Mixed effects | |
|---|---|---|---|---|
| | M1 | M2 | M1 | M2 |
| *Constant* | 0.337*** (0.024) | 0.218*** (0.032) | 0.337*** (0.026) | 0.218*** (0.033) |
| *Single AI Neutralized* | 0.043** (0.016) | 0.038* (0.015) | 0.043** (0.016) | 0.038* (0.015) |
| *Single AI Democrat* | 0.018 (0.014) | 0.014 (0.013) | 0.018 (0.014) | 0.014 (0.014) |
| *Single AI Republican* | 0.053*** (0.014) | 0.045*** (0.014) | 0.053*** (0.014) | 0.045** (0.014) |
| *Pre-interaction performance* | 0.414*** (0.021) | 0.411*** (0.021) | 0.414*** (0.018) | 0.411*** (0.018) |
| *AI correctness* | – | 0.180*** (0.024) | – | 0.180*** (0.020) |
| *Political ideology* | No | Yes | No | Yes |
| *Political party* | Yes | Yes | Yes | Yes |
| Headline-fixed effect | Yes | Yes | Yes | Yes |
| User random-effect | No | No | Yes | Yes |
| Observations | 2979 | 2972 | 2979 | 2972 |
| Adjusted $R^2$ | 0.493 | 0.508 | 0.495 | 0.511 |
| Residual SD | 0.243 (df = 2953) | 0.240 (df = 2939) | 0.243 (df = 2951) | 0.240 (df = 2937) |

*Note*: Dependent variable is post-interaction performance. Reference category is the default assistant. All four single-assistant arms are included (*N* = 993 participants). Columns 1–2 report standard errors clustered at the participant level (in parentheses); mixed-effects $R^2$ is marginal, as the participant variance is estimated at zero. Significance codes: *** $p < 0.001$, ** $p < 0.01$, * $p < 0.05$, † $p < 0.1$.

**Table S26. Regression estimates including explicitly neutralized treatment on participants' information-evaluation performance in the fact-checking experiment.**

| | CLM + Clustered SE | | Mixed effects (MCMC) | |
|---|---|---|---|---|
| | M1 | M2 | M1 | M2 |
| *Constant* | – | – | 2.735***<br>(0.187) | 3.048***<br>(0.358) |
| *Single AI Neutralized* | −0.308<br>(0.201) | −0.340†<br>(0.204) | −0.333†<br>(0.203) | −0.376<br>(0.239) |
| *Single AI Democrat* | −0.360†<br>(0.184) | −0.376*<br>(0.187) | −0.418*<br>(0.189) | −0.467*<br>(0.212) |
| *Single AI Republican* | −0.429*<br>(0.181) | −0.466*<br>(0.185) | −0.504**<br>(0.186) | −0.648**<br>(0.227) |
| *Pre-interaction performance* | −0.239†<br>(0.135) | −0.253†<br>(0.136) | −0.011<br>(0.071) | −0.013<br>(0.048) |
| *AI correctness* | – | 0.193<br>(0.150) | – | 0.068<br>(0.096) |
| *Political ideology* | No | Yes | No | Yes |
| *Political party* | Yes | Yes | Yes | Yes |
| Headline-fixed effect | Yes | Yes | Yes | Yes |
| User random-effect | No | No | Yes | Yes |
| Observations | 2979 | 2972 | 2979 | 2972 |
| Pseudo $R^2$ | 0.031 | 0.038 | – | – |
| Residual SD | – (df = 2950) | – (df = 2936) | 0.020 | 0.026 |

*Note*: Dependent variable is perceived improvement (5-point ordinal). Reference category is the default assistant. All four single-assistant arms are included. CLM columns report cumulative-link model coefficients with HC0 standard errors clustered at the participant level; MCMC columns report posterior means with posterior standard deviations and pMCMC values from ordinal MCMCglmm models with a participant random effect (25,000 iterations, 5,000 burn-in, thinning 10). Residual SD for the MCMC columns is the square root of the posterior mean residual variance. Significance codes: *** $p < 0.001$, ** $p < 0.01$, * $p < 0.05$, † $p < 0.1$.

**Table S27. Regression estimates including explicitly neutralized treatment on participants' perceived improvement in the fact-checking experiment.**

| | CLM + Clustered SE | | Mixed effects (MCMC) | |
|---|---|---|---|---|
| | M1 | M2 | M1 | M2 |
| *Constant* | – | – | 2.236***<br>(0.135) | 2.774***<br>(0.343) |
| *Single AI Neutralized* | −0.186<br>(0.202) | −0.201<br>(0.205) | −0.081<br>(0.155) | −0.196<br>(0.223) |
| *Single AI Democrat* | −0.315†<br>(0.182) | −0.331†<br>(0.185) | −0.290*<br>(0.143) | −0.342†<br>(0.199) |
| *Single AI Republican* | −0.435*<br>(0.184) | −0.480*<br>(0.190) | −0.439**<br>(0.142) | −0.543**<br>(0.206) |
| *Pre-interaction performance* | −0.251†<br>(0.142) | −0.242†<br>(0.143) | −0.152***<br>(0.024) | −0.162<br>(0.091) |
| *AI correctness* | – | 0.139<br>(0.142) | – | 0.077<br>(0.092) |
| *Political ideology* | No | Yes | No | Yes |
| *Political party* | Yes | Yes | Yes | Yes |
| Headline-fixed effect | Yes | Yes | Yes | Yes |
| User random-effect | No | No | Yes | Yes |
| Observations | 2979 | 2972 | 2979 | 2972 |
| Pseudo $R^2$ | 0.029 | 0.040 | – | – |
| Residual SD | – (df = 2950) | – (df = 2936) | 0.009 | 0.027 |

*Note*: Dependent variable is perceived interaction meaningfulness (5-point ordinal, "not meaningful at all" to "extremely meaningful"). Reference category is the default assistant. All four single-assistant arms are included. CLM columns report cumulative-link model coefficients with HC0 standard errors clustered at the participant level; MCMC columns report posterior means with posterior standard deviations and pMCMC values from ordinal MCMCglmm models with a participant random effect (25,000 iterations, 5,000 burn-in, thinning 10). Significance codes: *** $p < 0.001$, ** $p < 0.01$, * $p < 0.05$, † $p < 0.1$.

**Table S28. Regression estimates including explicitly neutralized treatment on participants' perceived interaction meaningfulness in the fact-checking experiment.**

| | OLS + Clustered SE | | Mixed effects | |
|---|---|---|---|---|
| | M1 | M2 | M1 | M2 |
| *Single AI Neutralized*‡ | 0.077***<br>(0.016) | 0.078***<br>(0.016) | 0.077***<br>(0.016) | 0.078***<br>(0.016) |
| *Single AI Democrat*‡ | 0.018<br>(0.014) | 0.028*<br>(0.014) | 0.018<br>(0.014) | 0.028*<br>(0.014) |
| *Single AI Republican*‡ | 0.009<br>(0.014) | 0.013<br>(0.014) | 0.009<br>(0.014) | 0.013<br>(0.014) |
| *Pre-interaction performance* | Yes | Yes | Yes | Yes |
| *AI correctness* | No | Yes | No | Yes |
| *Political ideology* | No | Yes | No | Yes |
| *Political party* | Yes | Yes | Yes | Yes |
| Headline-fixed effect | Yes | Yes | Yes | Yes |
| User random-effect | No | No | Yes | Yes |
| Observations | 2979 | 2972 | 2979 | 2972 |
| Adjusted $R^2$ | 0.494 | 0.509 | 0.495 | 0.512 |
| Residual SD | 0.243 (df = 2947) | 0.239 (df = 2933) | 0.243 (df = 2951) | 0.239 (df = 2931) |

*Note*: Dependent variable is post-interaction performance. Coefficients are differences in evaluative bias relative to the default assistant, where evaluative bias is the mean absolute pairwise gap in adjusted post-interaction performance across Republican-, Neutral- and Democrat-slanted headlines. Standard errors in parentheses; mixed-effects $R^2$ is marginal. Significance codes: *** $p < 0.001$, ** $p < 0.01$, * $p < 0.05$, † $p < 0.1$. ‡ Coefficients show the mean gap in performance across Republican/Neutral/Democrat headline categories, obtained from post-hoc marginal means rather than directly from the linear model, and therefore have no model intercept.

**Table S29. Regression estimates including explicitly neutralized treatment on participants' information-evaluation performance in the fact-checking experiment.**

| | OLS + Clustered SE | | Mixed effects | |
|---|---|---|---|---|
| | M1 | M2 | M1 | M2 |
| *Constant* | 18.427***<br>(2.399) | 14.422***<br>(3.719) | 17.849***<br>(2.520) | 13.876***<br>(3.604) |
| *Single AI Neutralized* | 1.376<br>(1.846) | 1.443<br>(1.806) | 1.324<br>(2.253) | 1.389<br>(2.268) |
| *Single AI Democrat* | 3.968*<br>(1.579) | 3.965*<br>(1.584) | 3.980*<br>(2.005) | 3.980*<br>(2.013) |
| *Single AI Republican* | 8.396***<br>(1.831) | 8.504***<br>(1.838) | 8.373***<br>(2.001) | 8.506***<br>(2.021) |
| *Pre-interaction performance* | −3.451†<br>(1.909) | −3.448†<br>(1.880) | −2.023<br>(1.324) | −2.080<br>(1.328) |
| *AI correctness* | – | 1.558<br>(2.277) | – | 1.313<br>(1.486) |
| *Political ideology* | No | Yes | No | Yes |
| *Political party* | Yes | Yes | Yes | Yes |
| Headline-fixed effect | Yes | Yes | Yes | Yes |
| User random-effect | No | No | Yes | Yes |
| Observations | 2967 | 2961 | 2967 | 2961 |
| Adjusted $R^2$ | 0.034 | 0.038 | 0.036 | 0.043 |
| Residual SD | 23.944 (df = 2941) | 23.900 (df = 2928) | 15.439 (df = 2939) | 15.454 (df = 2926) |

*Note*: Dependent variable is the length of the human–AI conversation (total word count of participant messages). Reference category is the default assistant. All four single-assistant arms are included. Columns 1–2 report standard errors clustered at the participant level (in parentheses); mixed-effects $R^2$ is marginal, with conditional $R^2$ of 0.600 (M1) and 0.601 (M2). Significance codes: *** $p < 0.001$, ** $p < 0.01$, * $p < 0.05$, † $p < 0.1$.

**Table S30. Regression estimates including explicitly neutralized treatment on participants' conversation length in the fact-checking experiment.**

| | M1 | M2 | M3 | M4 | M5 |
|---|---|---|---|---|---|
| *Constant* | −1.108***<br>(0.121) | −1.329***<br>(0.334) | −0.895<br>(0.594) | −0.745<br>(0.548) | 0.341<br>(0.557) |
| *Extremely Risk-Averse* | 0.825***<br>(0.163) | 0.828***<br>(0.164) | 0.838***<br>(0.166) | 0.847***<br>(0.153) | 0.058<br>(0.158) |
| *Somewhat Risk-Averse* | 0.523***<br>(0.156) | 0.534***<br>(0.154) | 0.536***<br>(0.155) | 0.576***<br>(0.148) | −0.050<br>(0.150) |
| *Somewhat Risk-Seeking* | −0.214<br>(0.161) | −0.213<br>(0.161) | −0.222<br>(0.165) | −0.199<br>(0.157) | −0.092<br>(0.159) |
| *Extremely Risk-Seeking* | −0.160<br>(0.185) | −0.153<br>(0.187) | −0.122<br>(0.188) | −0.100<br>(0.175) | −0.066<br>(0.175) |
| *Pre-interaction Active* $M^2$ | 0.697***<br>(0.028) | 0.696***<br>(0.028) | 0.686***<br>(0.029) | 0.616***<br>(0.032) | 0.616***<br>(0.032) |
| *Financial literacy* | – | 0.272<br>(0.375) | 0.489<br>(0.422) | 0.503<br>(0.397) | 0.503<br>(0.397) |
| *AI advice quality, within arm* | – | – | – | 0.341***<br>(0.038) | – |
| *AI advice quality, level* | – | – | – | – | 0.341***<br>(0.038) |
| *Flagged covariate* | No | Yes | Yes | Yes | Yes |
| *All pre-treatment covariates* | No | No | Yes | Yes | Yes |
| Wave fixed effects | Yes | Yes | Yes | Yes | Yes |
| Observations | 864 | 864 | 864 | 852 | 852 |
| Adjusted $R^2$ | 0.536 | 0.535 | 0.539 | 0.601 | 0.601 |
| Residual SD | 1.719 (df = 856) | 1.719 (df = 855) | 1.712 (df = 840) | 1.599 (df = 827) | 1.599 (df = 827) |

*Note*: Dependent variable is post-interaction Active $M^2$, in percentage points. Reference category is the default assistant; the neutralized arm is excluded. HC3 standard errors in parentheses. AI advice quality is the Active $M^2$ of the assistant's own recommended portfolio, scored on the same evaluation window; M4 enters only its within-arm deviation, which is not induced by the persona assignment, while M5 enters its level. Because advice content is realized after randomization and lies on the causal path from arm to performance, M5 is a mediation decomposition rather than a robustness specification. Significance codes: *** $p < 0.001$, ** $p < 0.01$, * $p < 0.05$, † $p < 0.1$.

**Table S31. Regression estimates of AI risk bias on participants' investment performance.**

| Variable | Positive persuasion | Negative persuasion | Positive backfire | Negative backfire | Any helpful shift | Any harmful shift |
|---|---|---|---|---|---|---|
| *Biased* | 1.797*<br>(0.256) | 0.661*<br>(0.210) | 0.529†<br>(0.338) | 2.171<br>(0.493) | 1.283<br>(0.214) | 0.812<br>(0.190) |
| *Pre-interaction performance* | 0.443***<br>(0.094) | 2.593***<br>(0.103) | 1.539*<br>(0.174) | 0.560***<br>(0.131) | 0.545***<br>(0.086) | 1.787***<br>(0.087) |
| Wave fixed-effect | Yes | Yes | Yes | Yes | Yes | Yes |
| Robust SE | Yes | Yes | Yes | Yes | Yes | Yes |
| Base rate | 0.238 | 0.236 | 0.049 | 0.054 | 0.288 | 0.290 |
| Observations | 852 | 852 | 852 | 852 | 852 | 852 |
| Pseudo $R^2$ | 0.142 | 0.123 | 0.034 | 0.047 | 0.086 | 0.059 |

*Note:* Dependent variables are binary indicators for each revision category, defined at $\tau = 0.1 \times$ SD of the underlying quantity; the neutralized arm is excluded. Entries are odds ratios from logistic regressions with wave fixed effects and standard errors of the underlying log-odds coefficients in parentheses; standard errors are heteroskedasticity-consistent. Because each participant contributed a single allocation, the models require no participant clustering and no observations are dropped, so all columns share $N = 852$. Base rate is the share of all 852 participants falling in that category. Pre-interaction performance is Active $M^2$, standardized so that its odds ratio corresponds to a one standard deviation change (2.62 percentage points). Significance codes: *** $p < 0.001$, ** $p < 0.01$, * $p < 0.05$, † $p < 0.1$.

**Table S32. Odds of each revision category under partisan versus default assistance (investment).**

| | M1 | M2 | M3 | M4 | M5 |
|---|---|---|---|---|---|
| *Constant* | 0.388*** (0.057) | 0.430*** (0.062) | 0.297*** (0.074) | 0.349*** (0.075) | 0.254*** (0.074) |
| *Extremely Risk-Averse* | 0.054*** (0.015) | 0.053*** (0.015) | 0.055*** (0.014) | 0.054*** (0.014) | 0.107*** (0.017) |
| *Somewhat Risk-Averse* | 0.052*** (0.014) | 0.050*** (0.014) | 0.046*** (0.014) | 0.047*** (0.014) | 0.087*** (0.016) |
| *Somewhat Risk-Seeking* | 0.121*** (0.012) | 0.120*** (0.013) | 0.119*** (0.012) | 0.115*** (0.012) | 0.139*** (0.013) |
| *Extremely Risk-Seeking* | 0.166*** (0.011) | 0.165*** (0.011) | 0.166*** (0.012) | 0.165*** (0.012) | 0.196*** (0.013) |
| *Pre-interaction dispersion* | 0.618*** (0.040) | 0.616*** (0.040) | 0.631*** (0.041) | 0.602*** (0.041) | 0.602*** (0.041) |
| *Financial literacy* | – | −0.048† (0.028) | −0.095** (0.031) | −0.102*** (0.031) | −0.102*** (0.031) |
| *AI advice distinctiveness, within-arm* | – | – | – | 0.092*** (0.019) | – |
| *AI advice distinctiveness, level* | – | – | – | – | 0.092*** (0.019) |
| *Flagged covariate* | No | Yes | Yes | Yes | Yes |
| *All pre-treatment covariates* | No | No | Yes | Yes | Yes |
| Wave fixed effects | Yes | Yes | Yes | Yes | Yes |
| Observations | 864 | 864 | 864 | 852 | 852 |
| Adjusted $R^2$ | 0.321 | 0.323 | 0.358 | 0.372 | 0.372 |
| Residual SD | 0.128 (df = 856) | 0.128 (df = 855) | 0.125 (df = 840) | 0.123 (df = 827) | 0.123 (df = 827) |

*Note*: Dependent variable is the participant's post-interaction dispersion contribution, the mean Manhattan distance from their normalized 25-asset allocation to every other participant in the same arm. Higher values indicate a participant whose allocation remained more distinct from their arm-mates, so positive coefficients denote less homogenization than the default assistant. Reference category is the default assistant. HC3 standard errors in parentheses. Advice distinctiveness is the mean distance from the assistant's own recommended portfolio to the mean recommendation within that arm; M4 enters only its within-arm deviation and D5 its level. Because dispersion contributions are computed against other members of the same arm, they are not independent across participants, so these standard errors are approximate. Significance codes: *** $p < 0.001$, ** $p < 0.01$, * $p < 0.05$, † $p < 0.1$.

**Table S33. Regression estimates of AI risk bias on cross-participant portfolio diversity.**

| | M1 | M2 | M3 | M4 |
|---|---|---|---|---|
| *Constant* | −2.083*** | −0.493*** | −1.180*** | −0.202* |
| | (0.177) | (0.097) | (0.147) | (0.096) |
| *Primed Averse* | −0.385 | −0.100 | −0.194 | −0.060 |
| | (0.264) | (0.139) | (0.133) | (0.132) |
| *Primed Seeking* | −0.448† | 0.064 | 0.029 | 0.105 |
| | (0.265) | (0.130) | (0.118) | (0.119) |
| *Pre-interaction Active M²* | – | 0.843*** | 0.652*** | 0.652*** |
| | | (0.026) | (0.042) | (0.042) |
| *AI advice quality, within-arm* | – | – | 0.307*** | – |
| | | | (0.046) | |
| *AI advice quality, level* | – | – | – | 0.307*** |
| | | | | (0.046) |
| Wave fixed effects | Yes | Yes | Yes | Yes |
| Observations | 480 | 480 | 470 | 470 |
| Adjusted $R^2$ | 0.142 | 0.781 | 0.818 | 0.818 |
| Residual SD | 2.382 (df = 475) | 1.204 (df = 474) | 1.102 (df = 463) | 1.102 (df = 463) |

*Note:* Dependent variable is post-interaction Active M², in percentage points. The assistant was the unmodified default model in all three arms, which differ only in whether participants were told it held a risk-averse or risk-seeking orientation, so any arm difference is a pure labelling effect. Reference category is the unprimed default arm, drawn from the main single-assistant experiment; the primed arms come from a separate sample, so randomization holds within each experiment rather than across them, and no participant covariates were collected in the priming variant. HC3 standard errors in parentheses. Observation counts fall in M3 and M4 because 10 assistant recommendations could not be extracted from the transcripts. Significance codes: *** $p < 0.001$, ** $p < 0.01$, * $p < 0.05$, † $p < 0.1$.

**Table S34. Regression estimates of risk-framing primes on participants' investment performance.**

| | M1 | M2 | M3 | M4 | M5 |
|---|---|---|---|---|---|
| *Constant* | −1.270*** (0.136) | −1.684*** (0.389) | −0.983 (0.696) | −0.948 (0.647) | 0.078 (0.651) |
| *Moderate bias* | 0.503** (0.156) | 0.522*** (0.153) | 0.534*** (0.156) | 0.596*** (0.150) | 0.004 (0.146) |
| *Strong bias* | 0.812*** (0.163) | 0.818*** (0.164) | 0.824*** (0.167) | 0.869*** (0.157) | 0.124 (0.143) |
| *Pre-interaction Active M²* | 0.742*** (0.030) | 0.741*** (0.030) | 0.732*** (0.031) | 0.629*** (0.037) | 0.629*** (0.037) |
| *Financial literacy* | – | 0.510 (0.457) | 0.668 (0.474) | 0.886† (0.462) | 0.886† (0.462) |
| *AI advice quality, within arm* | – | – | – | 0.322*** (0.045) | – |
| *AI advice quality, level* | – | – | – | – | 0.322*** (0.045) |
| *Flagged covariate* | No | Yes | Yes | Yes | Yes |
| *All pre-treatment covariates* | No | No | Yes | Yes | Yes |
| Wave fixed effects | Yes | Yes | Yes | Yes | Yes |
| Observations | 528 | 528 | 528 | 522 | 522 |
| Adjusted $R^2$ | 0.606 | 0.606 | 0.617 | 0.662 | 0.662 |
| Residual SD | 1.546 (df = 522) | 1.545 (df = 521) | 1.524 (df = 506) | 1.436 (df = 499) | 1.436 (df = 499) |

*Note:* Dependent variable is post-interaction Active M², in percentage points. Each panel contrasts the moderate (somewhat) and strong (extremely) arms of one bias direction against the shared default arm, which is the reference category in both panels; the neutralized arm is excluded. HC3 standard errors in parentheses. Objective financial literacy is the only covariate flagged by the three-group imbalance check in either direction (averse $\eta^2 = 0.019$, $p = 0.005$; seeking $\eta^2 = 0.011$, $p = 0.043$). AI advice quality is the Active M² of the assistant's own recommended portfolio, scored on the same evaluation window; M4 enters only its within-magnitude deviation and M5 its level. Because advice content is realized after randomization and lies on the causal path, M5 is a mediation decomposition rather than a robustness specification. Significance codes: *** $p < 0.001$, ** $p < 0.01$, * $p < 0.05$, † $p < 0.1$.

**Table S35a. Regression estimates of AI bias magnitude on participants' investment performance (risk-averse).**

| | M1 | M2 | M3 | M4 | M5 |
|---|---|---|---|---|---|
| *Constant* | −0.745***<br>(0.129) | −0.638<br>(0.434) | −0.622<br>(0.740) | −0.677<br>(0.688) | 0.331<br>(0.693) |
| *Moderate bias* | −0.192<br>(0.159) | −0.192<br>(0.160) | −0.196<br>(0.165) | −0.213<br>(0.161) | −0.114<br>(0.163) |
| *Strong bias* | −0.118<br>(0.182) | −0.122<br>(0.185) | −0.102<br>(0.186) | −0.131<br>(0.175) | −0.100<br>(0.175) |
| *Pre-interaction Active M²* | 0.684***<br>(0.040) | 0.685***<br>(0.040) | 0.679***<br>(0.043) | 0.607***<br>(0.046) | 0.607***<br>(0.046) |
| *Financial literacy* | – | −0.131<br>(0.481) | −0.096<br>(0.607) | −0.183<br>(0.541) | −0.183<br>(0.541) |
| *AI advice quality, within arm* | – | – | – | 0.316***<br>(0.052) | – |
| *AI advice quality, level* | – | – | – | – | 0.316***<br>(0.052) |
| *Flagged covariate* | No | Yes | Yes | Yes | Yes |
| *All pre-treatment covariates* | No | No | Yes | Yes | Yes |
| Wave fixed effects | Yes | Yes | Yes | Yes | Yes |
| Observations | 514 | 514 | 514 | 506 | 506 |
| Adjusted $R^2$ | 0.556 | 0.555 | 0.557 | 0.613 | 0.613 |
| Residual SD | 1.699 (df = 508) | 1.700 (df = 507) | 1.696 (df = 492) | 1.593 (df = 483) | 1.593 (df = 483) |

*Note:* Dependent variable is post-interaction Active M², in percentage points. Each panel contrasts the moderate (somewhat) and strong (extremely) arms of one bias direction against the shared default arm, which is the reference category in both panels; the neutralized arm is excluded. HC3 standard errors in parentheses. Objective financial literacy is the only covariate flagged by the three-group imbalance check in either direction (averse $\eta^2 = 0.019$, $p = 0.005$; seeking $\eta^2 = 0.011$, $p = 0.043$). AI advice quality is the Active M² of the assistant's own recommended portfolio, scored on the same evaluation window; M4 enters only its within-magnitude deviation and M5 its level. Because advice content is realized after randomization and lies on the causal path, M5 is a mediation decomposition rather than a robustness specification. Significance codes: *** $p < 0.001$, ** $p < 0.01$, * $p < 0.05$, † $p < 0.1$.

**Table S35b. Regression estimates of AI bias magnitude on participants' investment performance (risk-seeking).**

| | M1 | M2 | M3 | M4 | M5 |
|---|---|---|---|---|---|
| *Constant* | 0.480***<br>(0.051) | 0.252*<br>(0.118) | −0.174<br>(0.189) | −0.176<br>(0.189) | −0.146<br>(0.192) |
| *Moderate bias* | 0.014<br>(0.054) | 0.025<br>(0.054) | −0.004<br>(0.054) | −0.005<br>(0.054) | −0.022<br>(0.058) |
| *Strong bias* | 0.019<br>(0.052) | 0.022<br>(0.052) | 0.033<br>(0.051) | 0.039<br>(0.052) | 0.017<br>(0.059) |
| *Pre-interaction Active $M^2$* | −0.013<br>(0.008) | −0.013<br>(0.009) | −0.012<br>(0.009) | −0.015<br>(0.010) | −0.015<br>(0.010) |
| *Financial literacy* | – | 0.282*<br>(0.131) | 0.259†<br>(0.148) | 0.287†<br>(0.149) | 0.287†<br>(0.149) |
| *AI advice quality, within arm* | – | – | – | 0.009<br>(0.013) | – |
| *AI advice quality, level* | – | – | – | – | 0.009<br>(0.013) |
| *Flagged covariate* | No | Yes | Yes | Yes | Yes |
| *All pre-treatment covariates* | No | No | Yes | Yes | Yes |
| Wave fixed effects | Yes | Yes | Yes | Yes | Yes |
| Observations | 528 | 528 | 528 | 522 | 522 |
| Adjusted $R^2$ | 0.008 | 0.015 | 0.059 | 0.059 | 0.059 |
| Residual SD | 0.491 (df = 522) | 0.489 (df = 521) | 0.478 (df = 506) | 0.477 (df = 499) | 0.477 (df = 499) |

*Note:* Dependent variable is an indicator for sending at least one follow-up message, estimated by linear probability model so that coefficients are in participation-rate units directly comparable to the figure. Each panel contrasts the moderate and strong arms of one bias direction against the shared default arm; the neutralized arm is excluded. HC3 standard errors in parentheses. Objective financial literacy is the only covariate flagged by the three-group imbalance check in either direction. A logistic cross-check on M1 gives odds ratios of 1.062 ($p = 0.787$) and 1.080 ($p = 0.721$) for the averse arms and 1.478 ($p = 0.091$) and 1.502 ($p = 0.065$) for the seeking arms. Significance codes: *** $p < 0.001$, ** $p < 0.01$, * $p < 0.05$, † $p < 0.1$.

**Table S36a. Regression estimates of AI bias magnitude on follow-up participation (risk-averse).**

| | M1 | M2 | M3 | M4 | M5 |
|---|---|---|---|---|---|
| *Constant* | 0.544***<br>(0.050) | 0.552***<br>(0.126) | 0.081<br>(0.195) | 0.115<br>(0.196) | 0.156<br>(0.197) |
| *Moderate bias* | 0.091†<br>(0.054) | 0.090†<br>(0.054) | 0.075<br>(0.054) | 0.094†<br>(0.054) | 0.098†<br>(0.054) |
| *Strong bias* | 0.094†<br>(0.051) | 0.094†<br>(0.052) | 0.086†<br>(0.051) | 0.088†<br>(0.051) | 0.089†<br>(0.051) |
| *Pre-interaction Active $M^2$* | 0.009<br>(0.009) | 0.009<br>(0.009) | 0.010<br>(0.009) | 0.006<br>(0.010) | 0.006<br>(0.010) |
| *Financial literacy* | – | −0.010<br>(0.141) | −0.055<br>(0.166) | −0.036<br>(0.167) | −0.036<br>(0.167) |
| *AI advice quality, within arm* | – | – | – | 0.013<br>(0.011) | – |
| *AI advice quality, level* | – | – | – | – | 0.013<br>(0.011) |
| *Flagged covariate* | No | Yes | Yes | Yes | Yes |
| *All pre-treatment covariates* | No | No | Yes | Yes | Yes |
| Wave fixed effects | Yes | Yes | Yes | Yes | Yes |
| Observations | 514 | 528 | 528 | 506 | 506 |
| Adjusted $R^2$ | 0.017 | 0.015 | 0.049 | 0.051 | 0.051 |
| Residual SD | 0.478 (df = 508) | 0.478 (df = 507) | 0.470 (df = 492) | 0.466 (df = 483) | 0.466 (df = 483) |

*Note:* Dependent variable is an indicator for sending at least one follow-up message, estimated by linear probability model so that coefficients are in participation-rate units directly comparable to the figure. Each panel contrasts the moderate and strong arms of one bias direction against the shared default arm; the neutralized arm is excluded. HC3 standard errors in parentheses. Objective financial literacy is the only covariate flagged by the three-group imbalance check in either direction. A logistic cross-check on M1 gives odds ratios of 1.062 ($p = 0.787$) and 1.080 ($p = 0.721$) for the averse arms and 1.478 ($p = 0.091$) and 1.502 ($p = 0.065$) for the seeking arms. Significance codes: *** $p < 0.001$, ** $p < 0.01$, * $p < 0.05$, † $p < 0.1$.

**Table S36b. Regression estimates of AI bias magnitude on follow-up participation (risk-seeking).**

| | M1 | M2 | M3 | M4 | M5 |
|---|---|---|---|---|---|
| *Constant* | 1.878***<br>(0.161) | 1.875***<br>(0.242) | 1.290***<br>(0.358) | 1.244***<br>(0.353) | 1.254***<br>(0.367) |
| *Moderate bias* | −0.144<br>(0.089) | −0.144<br>(0.089) | −0.152†<br>(0.092) | −0.163†<br>(0.092) | −0.169†<br>(0.101) |
| *Strong bias* | −0.144<br>(0.099) | −0.144<br>(0.099) | −0.141<br>(0.101) | −0.109<br>(0.096) | −0.117<br>(0.106) |
| *Pre-interaction confidence* | 0.644***<br>(0.036) | 0.644***<br>(0.037) | 0.599***<br>(0.053) | 0.605***<br>(0.050) | 0.605***<br>(0.050) |
| *Pre-interaction Active $M^2$* | −0.025†<br>(0.015) | −0.025†<br>(0.015) | −0.019<br>(0.016) | −0.014<br>(0.017) | −0.014<br>(0.017) |
| *Financial literacy* | – | 0.004<br>(0.246) | −0.209<br>(0.279) | −0.204<br>(0.279) | −0.204<br>(0.279) |
| *AI advice quality, within arm* | – | – | – | 0.003<br>(0.023) | – |
| *AI advice quality, level* | – | – | – | – | 0.003<br>(0.023) |
| *Flagged covariate* | No | Yes | Yes | Yes | Yes |
| *All pre-treatment covariates* | No | No | Yes | Yes | Yes |
| Wave fixed effects | Yes | Yes | Yes | Yes | Yes |
| Observations | 527 | 527 | 527 | 521 | 521 |
| Adjusted $R^2$ | 0.518 | 0.517 | 0.527 | 0.544 | 0.544 |
| Residual SD | 0.880 (df = 520) | 0.880 (df = 519) | 0.871 (df = 504) | 0.840 (df = 497) | 0.840 (df = 497) |

*Note:* Dependent variable is post-interaction confidence in one's portfolio on a seven-point scale, modelled as an analysis of covariance conditioning on pre-interaction confidence. Each panel contrasts the moderate and strong arms of one bias direction against the shared default arm; the neutralized arm is excluded. HC3 standard errors in parentheses. Objective financial literacy is the only covariate flagged by the three-group imbalance check in either direction. AI advice quality is the Active $M^2$ of the assistant's own recommended portfolio; M4 enters only its within-magnitude deviation and M5 its level. A Bayesian ordinal model on the latent scale reproduces the sign and significance pattern. Significance codes: *** $p < 0.001$, ** $p < 0.01$, * $p < 0.05$, † $p < 0.1$.

**Table S37a. Regression estimates of AI risk-bias magnitude on post-interaction confidence (risk-averse).**

| | M1 | M2 | M3 | M4 | M5 |
|---|---|---|---|---|---|
| *Constant* | 1.655***<br>(0.146) | 1.540***<br>(0.234) | 1.156**<br>(0.403) | 1.199**<br>(0.409) | 1.230**<br>(0.414) |
| *Moderate bias* | −0.140<br>(0.085) | −0.139<br>(0.085) | −0.151†<br>(0.088) | −0.157†<br>(0.090) | −0.154†<br>(0.090) |
| *Strong bias* | −0.148†<br>(0.085) | −0.144†<br>(0.085) | −0.153†<br>(0.089) | −0.157†<br>(0.090) | −0.156†<br>(0.090) |
| *Pre-interaction confidence* | 0.730***<br>(0.029) | 0.729***<br>(0.029) | 0.720***<br>(0.042) | 0.712***<br>(0.042) | 0.712***<br>(0.042) |
| *Pre-interaction Active $M^2$* | −0.004<br>(0.014) | −0.005<br>(0.014) | 0.002<br>(0.016) | −0.001<br>(0.016) | −0.001<br>(0.016) |
| *Financial literacy* | – | 0.151<br>(0.247) | 0.106<br>(0.313) | 0.110<br>(0.318) | 0.110<br>(0.318) |
| *AI advice quality, within arm* | – | – | – | 0.010<br>(0.018) | – |
| *AI advice quality, level* | – | – | – | – | 0.010<br>(0.018) |
| *Flagged covariate* | No | Yes | Yes | Yes | Yes |
| *All pre-treatment covariates* | No | No | Yes | Yes | Yes |
| Wave fixed effects | Yes | Yes | Yes | Yes | Yes |
| Observations | 513 | 513 | 513 | 505 | 505 |
| Adjusted $R^2$ | 0.604 | 0.603 | 0.600 | 0.594 | 0.594 |
| Residual SD | 0.782 (df = 506) | 0.782 (df = 505) | 0.786 (df = 490) | 0.789 (df = 481) | 0.789 (df = 481) |

*Note:* Dependent variable is post-interaction confidence in one's portfolio on a seven-point scale, modelled as an analysis of covariance conditioning on pre-interaction confidence. Each panel contrasts the moderate and strong arms of one bias direction against the shared default arm; the neutralized arm is excluded. HC3 standard errors in parentheses. Objective financial literacy is the only covariate flagged by the three-group imbalance check in either direction. AI advice quality is the Active $M^2$ of the assistant's own recommended portfolio; M4 enters only its within-magnitude deviation and M5 its level. A Bayesian ordinal model on the latent scale reproduces the sign and significance pattern. Significance codes: *** $p < 0.001$, ** $p < 0.01$, * $p < 0.05$, † $p < 0.1$.

**Table S37b. Regression estimates of AI risk-bias magnitude on post-interaction confidence (risk-seeking).**

| | M1 | M2 | M3 | M4 | M5 |
|---|---|---|---|---|---|
| *Constant* | −1.239*** (0.175) | 0.230 (0.492) | 0.200 (1.139) | 0.386 (1.096) | 0.848 (1.100) |
| *Opposition bias* | 0.443* (0.175) | −1.269*** (0.229) | −1.285*** (0.238) | −1.268*** (0.226) | −0.260 (0.272) |
| *Pre-interaction Active M²* | 0.691*** (0.045) | 0.633*** (0.051) | 0.644*** (0.055) | 0.601*** (0.056) | 0.601*** (0.056) |
| *Overconfidence* | – | −0.100 (0.088) | −0.072 (0.119) | −0.109 (0.116) | −0.109 (0.116) |
| *Two-week market outlook* | – | −0.057 (0.087) | −0.066 (0.095) | −0.042 (0.086) | −0.042 (0.086) |
| *AI advice quality, within arm* | – | – | – | 0.416*** (0.074) | – |
| *AI advice quality, level* | – | – | – | – | 0.416*** (0.074) |
| *Flagged covariate* | No | Yes | Yes | Yes | Yes |
| *All pre-treatment covariates* | No | No | Yes | Yes | Yes |
| Wave fixed effects | Yes | Yes | Yes | Yes | Yes |
| Observations | 326 | 326 | 326 | 320 | 320 |
| Adjusted $R^2$ | 0.439 | 0.440 | 0.452 | 0.527 | 0.527 |
| Residual SD | 1.998 (df = 321) | 1.996 (df = 319) | 1.976 (df = 306) | 1.844 (df = 299) | 1.844 (df = 299) |

*Note:* Dependent variable is post-interaction Active M², in percentage points. Reference category is echo-chamber assignment, in which the assistant's risk orientation matches the participant's own. The sample is restricted to participants paired with a risk-biased assistant; the default and neutralized arms are excluded. Participant lean is constant within each table and therefore absorbed; leaning is a median split of the elicited risk-preference composite at 0.375. HC3 standard errors in parentheses. Covariates are flagged by an imbalance check across bias sides within each lean; no covariate is flagged among risk-seeking-leaning participants, so M2 is identical to M1 there. AI advice quality is the Active M² of the assistant's own recommended portfolio; M4 enters only its within-side deviation and M5 its level, making M5 a mediation decomposition rather than a robustness specification. Significance codes: *** $p < 0.001$, ** $p < 0.01$, * $p < 0.05$, † $p < 0.1$.

**Table S38a. Regression estimates of AI bias direction (echo chamber vs. opposition) on risk-averse-leaning participants' investment performance.**

| | M1 | M2 | M3 | M4 | M5 |
|---|---|---|---|---|---|
| *Constant* | −1.239***<br>(0.175) | −1.239***<br>(0.175) | −0.760<br>(0.864) | −0.374<br>(0.833) | 0.574<br>(0.869) |
| *Opposition bias* | 0.443*<br>(0.175) | 0.443*<br>(0.175) | 0.458*<br>(0.178) | 0.468**<br>(0.172) | −0.165<br>(0.185) |
| *Pre-interaction Active M²* | 0.691***<br>(0.045) | 0.691***<br>(0.045) | 0.644***<br>(0.055) | 0.651***<br>(0.045) | 0.651***<br>(0.045) |
| *Overconfidence* | – | – | −0.020<br>(0.098) | −0.046<br>(0.097) | −0.046<br>(0.097) |
| *Two-week market outlook* | – | – | −0.056<br>(0.067) | −0.075<br>(0.063) | −0.075<br>(0.063) |
| *AI advice quality, within arm* | – | – | – | 0.292***<br>(0.054) | – |
| *AI advice quality, level* | – | – | – | – | 0.292***<br>(0.054) |
| *Flagged covariate* | No | None flagged | Yes | Yes | Yes |
| *All pre-treatment covariates* | No | No | Yes | Yes | Yes |
| Wave fixed effects | Yes | Yes | Yes | Yes | Yes |
| Observations | 360 | 360 | 360 | 356 | 356 |
| Adjusted $R^2$ | 0.529 | 0.529 | 0.544 | 0.589 | 0.589 |
| Residual SD | 1.635 (df = 355) | 1.635 (df = 355) | 1.608 (df = 340) | 1.534 (df = 335) | 1.534 (df = 335) |

*Note:* Dependent variable is post-interaction Active M², in percentage points. Reference category is echo-chamber assignment, in which the assistant's risk orientation matches the participant's own. The sample is restricted to participants paired with a risk-biased assistant; the default and neutralized arms are excluded. Participant lean is constant within each table and therefore absorbed; leaning is a median split of the elicited risk-preference composite at 0.375. HC3 standard errors in parentheses. Covariates are flagged by an imbalance check across bias sides within each lean; no covariate is flagged among risk-seeking-leaning participants, so M2 is identical to M1 there. AI advice quality is the Active M² of the assistant's own recommended portfolio; M4 enters only its within-side deviation and M5 its level, making M5 a mediation decomposition rather than a robustness specification. Significance codes: *** $p < 0.001$, ** $p < 0.01$, * $p < 0.05$, † $p < 0.1$.

**Table S38b. Regression estimates of AI bias direction (echo chamber vs. opposition) on risk-seeking-leaning participants' investment performance.**

| | M1 | M2 | M3 | M4 | M5 |
|---|---|---|---|---|---|
| *Constant* | −1.044*** (0.104) | −0.915*** (0.229) | −0.558 (0.373) | −0.677* (0.340) | 0.464 (0.338) |
| *Single AI bias* | 0.245* (0.115) | 0.253* (0.116) | 0.241* (0.115) | 0.270* (0.106) | −0.078 (0.106) |
| *Dual AI default* | −0.055 (0.123) | −0.054 (0.124) | −0.073 (0.124) | −0.053 (0.113) | 0.052 (0.114) |
| *Dual AI opposition* | 0.150 (0.126) | 0.154 (0.126) | 0.136 (0.126) | 0.160 (0.117) | −0.013 (0.116) |
| *Dual AI balanced* | 0.193 (0.128) | 0.196 (0.129) | 0.212† (0.128) | 0.215† (0.120) | −0.054 (0.121) |
| *Pre-interaction Active M²* | 0.707*** (0.019) | 0.707*** (0.019) | 0.700*** (0.019) | 0.607*** (0.021) | 0.607*** (0.021) |
| *Trading frequency* | – | −0.058 (0.053) | −0.016 (0.058) | −0.031 (0.053) | −0.031 (0.053) |
| *Financial literacy* | – | −0.025 (0.230) | −0.023 (0.269) | −0.028 (0.252) | −0.028 (0.252) |
| *AI advice quality, within arm* | – | – | – | 0.358*** (0.025) | – |
| *AI advice quality, level* | – | – | – | – | 0.358*** (0.025) |
| *Flagged covariate* | No | Yes | Yes | Yes | Yes |
| *All pre-treatment covariates* | No | No | Yes | Yes | Yes |
| Wave fixed effects | Yes | Yes | Yes | Yes | Yes |
| Observations | 1875 | 1875 | 1875 | 1863 | 1863 |
| Adjusted $R^2$ | 0.560 | 0.529 | 0.562 | 0.634 | 0.634 |
| Residual SD | 1.683 (df = 1867) | 1.683 (df = 1865) | 1.680 (df = 1851) | 1.537 (df = 1838) | 1.537 (df = 1838) |

*Note:* Dependent variable is post-interaction Active M², in percentage points. The reference category is the single default assistant; the neutralized arm is excluded. Single AI Biased pools the four risk-biased single-assistant arms. HC3 standard errors in parentheses; each participant contributed a single allocation, so no clustering is required. Trading frequency and objective financial literacy are the covariates flagged by the five-condition imbalance check ($\eta^2$ = 0.005, $p$ = 0.048 and $\eta^2$ = 0.006, $p$ = 0.017). AI advice quality is the Active M² of the assistant's own recommended portfolio, averaged across both assistants in the dual conditions; M4 enters only its within-condition deviation and M5 its level, making M5 a mediation decomposition rather than a robustness specification. The

single-assistant and dual-assistant conditions were fielded as separate experiments with separate participant pools, so randomization holds within each experiment and contrasts spanning the two should be read as cross-experiment comparisons. Significance codes: *** $p < 0.001$, ** $p < 0.01$, * $p < 0.05$, † $p < 0.1$.

**Table S39. Regression estimates of dual AI treatments on participants' investment performance.**

| | M1 | M2 | M3 | M4 | M5 |
|---|---|---|---|---|---|
| *Constant* | 1.798***<br>(0.099) | 1.747***<br>(0.136) | 1.466***<br>(0.201) | 1.477***<br>(0.201) | 1.496***<br>(0.203) |
| *Single AI bias* | −0.142*<br>(0.067) | −0.140*<br>(0.068) | −0.126†<br>(0.069) | −0.126†<br>(0.069) | −0.132†<br>(0.069) |
| *Dual AI default* | −0.048<br>(0.075) | −0.045<br>(0.075) | −0.028<br>(0.076) | −0.039<br>(0.076) | −0.037<br>(0.077) |
| *Dual AI opposition* | −0.090<br>(0.074) | −0.089<br>(0.074) | −0.091<br>(0.076) | −0.100<br>(0.076) | −0.103<br>(0.076) |
| *Dual AI balanced* | −0.175*<br>(0.076) | −0.174*<br>(0.076) | −0.180*<br>(0.077) | −0.190*<br>(0.077) | −0.194*<br>(0.077) |
| *Pre-interaction confidence* | 0.688***<br>(0.017) | 0.687***<br>(0.017) | 0.635***<br>(0.024) | 0.636***<br>(0.023) | 0.636***<br>(0.023) |
| *Pre-interaction Active $M^2$* | −0.010<br>(0.008) | −0.010<br>(0.008) | −0.005<br>(0.008) | −0.005<br>(0.008) | −0.005<br>(0.008) |
| *Financial literacy* | – | 0.069<br>(0.125) | −0.122<br>(0.144) | −0.122<br>(0.145) | −0.122<br>(0.145) |
| *AI advice quality, within arm* | – | – | – | 0.006<br>(0.010) | – |
| *AI advice quality, level* | – | – | – | – | 0.006<br>(0.010) |
| *Flagged covariate* | No | Yes | Yes | Yes | Yes |
| *All pre-treatment covariates* | No | No | Yes | Yes | Yes |
| Wave fixed effects | Yes | Yes | Yes | Yes | Yes |
| Observations | 1869 | 1869 | 1869 | 1857 | 1857 |
| Adjusted $R^2$ | 0.536 | 0.535 | 0.547 | 0.552 | 0.552 |
| Residual SD | 0.856 (df = 1860) | 0.856 (df = 1859) | 0.845 (df = 1844) | 0.837 (df = 1831) | 0.837 (df = 1831) |

*Note:* Dependent variable is post-interaction confidence in one's portfolio on a seven-point scale, modelled as an analysis of covariance conditioning on pre-interaction confidence. The reference category is the single default assistant; the neutralized arm is excluded. Single AI Biased pools the four risk-biased single-assistant arms. HC3 standard errors in parentheses. Objective financial literacy is the only covariate flagged by the five-condition imbalance check. AI advice quality is the Active $M^2$ of the assistant's own recommended portfolio, averaged across

both assistants in the dual conditions; M4 enters only its within-condition deviation and M5 its level. The single-assistant and dual-assistant conditions were fielded as separate experiments with separate participant pools, so contrasts spanning the two are cross-experiment comparisons. Significance codes: *** $p < 0.001$, ** $p < 0.01$, * $p < 0.05$, † $p < 0.1$.

**Table S40. Regression estimates of dual AI treatments on participants' post-interaction confidence.**

| | M1 | M2 | M3 | M4 | M5 |
|---|---|---|---|---|---|
| *Constant* | 0.899***<br>(0.084) | 0.504***<br>(0.152) | −0.379†<br>(0.219) | −0.364†<br>(0.220) | −0.329<br>(0.222) |
| *Single AI bias* | 0.083<br>(0.088) | 0.079<br>(0.088) | 0.085<br>(0.084) | 0.091<br>(0.084) | 0.080<br>(0.086) |
| *Dual AI default* | −0.118<br>(0.093) | −0.110<br>(0.092) | −0.119<br>(0.090) | −0.128<br>(0.090) | −0.125<br>(0.090) |
| *Dual AI opposition* | −0.086<br>(0.094) | −0.088<br>(0.093) | −0.082<br>(0.090) | −0.091<br>(0.090) | −0.096<br>(0.090) |
| *Dual AI balanced* | −0.182*<br>(0.092) | −0.183*<br>(0.092) | −0.205*<br>(0.089) | −0.215*<br>(0.089) | −0.223*<br>(0.090) |
| *Pre-interaction Active M²* | 0.003<br>(0.009) | 0.002<br>(0.009) | 0.001<br>(0.009) | −0.001<br>(0.010) | −0.001<br>(0.010) |
| *Trading frequency* | – | 0.077*<br>(0.031) | 0.030<br>(0.032) | 0.031<br>(0.033) | 0.031<br>(0.033) |
| *Financial literacy* | – | 0.308*<br>(0.145) | 0.144<br>(0.163) | 0.162<br>(0.163) | 0.162<br>(0.163) |
| *AI advice quality, within arm* | – | – | – | 0.011<br>(0.012) | – |
| *AI advice quality, level* | – | – | – | – | 0.011<br>(0.012) |
| *Flagged covariate* | No | Yes | Yes | Yes | Yes |
| *All pre-treatment covariates* | No | No | Yes | Yes | Yes |
| Wave fixed effects | Yes | Yes | Yes | Yes | Yes |
| Observations | 1875 | 1875 | 1875 | 1863 | 1863 |
| Adjusted $R^2$ | 0.010 | 0.015 | 0.068 | 0.070 | 0.070 |
| Residual SD | 1.000 (df = 1867) | 0.998 (df = 1865) | 0.970 (df = 1851) | 0.970 (df = 1838) | 0.970 (df = 1838) |

*Note:* Dependent variable is the number of participant follow-up messages, excluding the scripted opening prompt. The reference category is the single default assistant; the neutralized arm is excluded. Single AI Biased pools the four risk-biased single-assistant arms. HC3 standard errors in parentheses. Trading frequency and objective financial literacy are the covariates flagged by the five-condition imbalance check. AI advice quality is the Active M² of the assistant's own recommended portfolio, averaged across both assistants in the dual conditions; M4 enters only its within-condition deviation and M5 its level. The turn count is close to equidispersed (variance-to-mean ratio 1.11), so linear estimation is reported for comparability with the other tables; a negative binomial cross-check on M1 gives

incidence rate ratios of 1.088 ($p = 0.351$), 0.876 ($p = 0.197$), 0.909 ($p = 0.349$) and 0.808 ($p = 0.039$) for the four non-reference conditions, reproducing the pattern. The single-assistant and dual-assistant conditions were fielded as separate experiments with separate participant pools, so contrasts spanning the two are cross-experiment comparisons. Significance codes: *** $p < 0.001$, ** $p < 0.01$, * $p < 0.05$, † $p < 0.1$.

**Table S41. Regression estimates of dual AI treatments on participants' conversational engagement.**

| | M1 | M2 | M3 | M4 | M5 |
|---|---|---|---|---|---|
| *Constant* | 0.328***<br>(0.030) | 0.388***<br>(0.036) | 0.354***<br>(0.039) | 0.447***<br>(0.038) | 0.216***<br>(0.039) |
| *Single AI bias* | 0.181***<br>(0.009) | 0.180***<br>(0.009) | 0.180***<br>(0.009) | 0.180***<br>(0.009) | 0.118***<br>(0.012) |
| *Dual AI default* | 0.017<br>(0.011) | 0.015<br>(0.011) | 0.014<br>(0.011) | 0.014<br>(0.011) | 0.013<br>(0.011) |
| *Dual AI opposition* | 0.138***<br>(0.010) | 0.137***<br>(0.010) | 0.136***<br>(0.010) | 0.136***<br>(0.010) | 0.070***<br>(0.013) |
| *Dual AI balanced* | 0.147***<br>(0.010) | 0.146***<br>(0.010) | 0.147***<br>(0.010) | 0.147***<br>(0.010) | 0.088***<br>(0.012) |
| *Pre-interaction dispersion* | 0.656***<br>(0.021) | 0.648***<br>(0.021) | 0.648***<br>(0.022) | 0.597***<br>(0.022) | 0.597***<br>(0.022) |
| *Trading frequency* | – | −0.003<br>(0.003) | −0.004<br>(0.003) | −0.006*<br>(0.003) | −0.006*<br>(0.003) |
| *Financial literacy* | – | −0.053***<br>(0.014) | −0.060***<br>(0.016) | −0.062***<br>(0.016) | −0.062***<br>(0.016) |
| *AI advice distinctiveness, within arm* | – | – | – | 0.225***<br>(0.025) | – |
| *AI advice distinctiveness, level* | – | – | – | – | 0.225***<br>(0.025) |
| *Flagged covariate* | No | Yes | Yes | Yes | Yes |
| *All pre-treatment covariates* | No | No | Yes | Yes | Yes |
| Wave fixed effects | Yes | Yes | Yes | Yes | Yes |
| Observations | 1876 | 1876 | 1876 | 1862 | 1862 |
| Adjusted $R^2$ | 0.540 | 0.544 | 0.546 | 0.586 | 0.586 |
| Residual SD | 0.096 (df = 1868) | 0.096 (df = 1866) | 0.095 (df = 1852) | 0.091 (df = 1837) | 0.091 (df = 1837) |

*Note:* Dependent variable is the participant's post-interaction dispersion contribution. Single AI Biased pools the four risk-biased single-assistant arms. HC3 standard errors in parentheses. Advice distinctiveness is the mean distance from the assistant's own recommended portfolio to the other recommendations within the same condition, averaged across both assistants in the dual conditions; M4 enters only its within-condition deviation and M5 its level. The single-assistant and dual-assistant conditions were fielded as separate experiments with separate

participant pools, so contrasts spanning the two are cross-experiment comparisons. Significance codes: *** $p < 0.001$, ** $p < 0.01$, * $p < 0.05$, † $p < 0.1$.

**Table S42. Regression estimates of dual AI treatments on cross-participant portfolio diversity.**

| | M1 | M2 | M3 | M4 | M5 |
|---|---|---|---|---|---|
| *Constant* | −1.073***<br>(0.116) | −1.381***<br>(0.320) | −0.747<br>(0.535) | −0.652<br>(0.488) | 0.476<br>(0.497) |
| *Single AI Neutralized* | 0.067<br>(0.148) | 0.063<br>(0.148) | 0.059<br>(0.149) | 0.119<br>(0.137) | −0.039<br>(0.137) |
| *Single AI Averse* | 0.673***<br>(0.130) | 0.682***<br>(0.130) | 0.690***<br>(0.132) | 0.711***<br>(0.122) | −0.029<br>(0.125) |
| *Single AI Seeking* | −0.193<br>(0.139) | −0.187<br>(0.140) | −0.181<br>(0.142) | −0.160<br>(0.133) | −0.091<br>(0.134) |
| *Pre-interaction Active $M^2$* | 0.697***<br>(0.025) | 0.695***<br>(0.026) | 0.685***<br>(0.027) | 0.607***<br>(0.029) | 0.607***<br>(0.029) |
| *Financial literacy* | – | 0.376<br>(0.355) | 0.460<br>(0.390) | 0.467<br>(0.361) | 0.467<br>(0.361) |
| *AI advice quality, within-arm* | – | – | – | 0.354***<br>(0.035) | – |
| *AI advice quality, level* | – | – | – | – | 0.354***<br>(0.035) |
| *Flagged covariate* | No | Yes | Yes | Yes | Yes |
| *All pre-treatment covariates* | No | No | Yes | Yes | Yes |
| Wave fixed effects | Yes | Yes | Yes | Yes | Yes |
| Observations | 1042 | 1042 | 1042 | 1028 | 1028 |
| Adjusted $R^2$ | 0.541 | 0.541 | 0.543 | 0.610 | 0.610 |
| Residual SD | 1.691 (df = 1035) | 1.691 (df = 1034) | 1.688 (df = 1019) | 1.566 (df = 1004) | 1.566 (df = 1004) |

*Note*: Dependent variable is post-interaction Active $M^2$, in percentage points. The reference category is the default assistant, which receives no persona instruction; Neutralized is the assistant explicitly instructed to be risk-neutral, and Averse and Seeking pool the somewhat and extremely arms of each direction. HC3 standard errors in parentheses. Objective financial literacy is the only covariate flagged by the four-group imbalance check ($\eta^2 = 0.012$, $p = 0.005$). AI advice quality is the Active $M^2$ of the assistant's own recommended portfolio; M4 enters only its within-group deviation and M5 its level, making M5 a mediation decomposition rather than a robustness specification. Significance codes: *** $p < 0.001$, ** $p < 0.01$, * $p < 0.05$, † $p < 0.1$.

**Table S43. Regression estimates including explicitly neutralized treatment on participants' investment performance.**

| | M1 | M2 | M3 | M4 | M5 |
|---|---|---|---|---|---|
| *Constant* | 0.305***<br>(0.051) | 0.346***<br>(0.055) | 0.209**<br>(0.064) | 0.248***<br>(0.065) | 0.162*<br>(0.064) |
| *Single AI Neutralized* | 0.022†<br>(0.013) | 0.023†<br>(0.013) | 0.023†<br>(0.012) | 0.022†<br>(0.012) | 0.043***<br>(0.013) |
| *Single AI Averse* | 0.060***<br>(0.011) | 0.059***<br>(0.011) | 0.057***<br>(0.011) | 0.057***<br>(0.011) | 0.094***<br>(0.014) |
| *Single AI Seeking* | 0.149***<br>(0.010) | 0.149***<br>(0.010) | 0.148***<br>(0.010) | 0.146***<br>(0.010) | 0.169***<br>(0.011) |
| *Pre-interaction Dispersion* | 0.677***<br>(0.036) | 0.675***<br>(0.036) | 0.690***<br>(0.036) | 0.668***<br>(0.036) | 0.668***<br>(0.036) |
| *Financial literacy* | – | −0.047†<br>(0.025) | −0.084**<br>(0.027) | −0.091***<br>(0.027) | −0.091***<br>(0.027) |
| *Advice distinctiveness, within-arm* | – | – | – | 0.084***<br>(0.018) | – |
| *Advice distinctiveness, level* | – | – | – | – | 0.084***<br>(0.018) |
| *Flagged covariate* | No | Yes | Yes | Yes | Yes |
| *All pre-treatment covariates* | No | No | Yes | Yes | Yes |
| Wave fixed effects | Yes | Yes | Yes | Yes | Yes |
| Observations | 1042 | 1042 | 1042 | 1028 | 1028 |
| Adjusted $R^2$ | 0.376 | 0.378 | 0.407 | 0.417 | 0.417 |
| Residual SD | 0.124 (df = 1035) | 0.124 (df = 1034) | 0.121 (df = 1019) | 0.120 (df = 1004) | 0.120 (df = 1004) |

*Note*: Dependent variable is the participant's post-interaction dispersion contribution. Neutralized is the assistant explicitly instructed to be risk-neutral; Averse and Seeking pool the somewhat and extremely arms of each direction. HC3 standard errors in parentheses. Advice distinctiveness is the mean distance from the assistant's own recommended portfolio to the other recommendations within the same group; M4 enters only its within-group deviation and M5 its level. Significance codes: *** $p < 0.001$, ** $p < 0.01$, * $p < 0.05$, † $p < 0.1$.

**Table S44. Regression estimates including explicitly neutralized treatment on cross-participant portfolio diversity.**

| | M1 | M2 | M3 | M4 | M5 |
|---|---|---|---|---|---|
| *Constant* | 0.536*** (0.044) | 0.433*** (0.090) | 0.004 (0.138) | 0.028 (0.138) | 0.050 (0.140) |
| *Single AI Neutralized* | −0.036 (0.053) | −0.037 (0.053) | −0.044 (0.052) | −0.043 (0.052) | −0.046 (0.052) |
| *Single AI Averse* | 0.021 (0.046) | 0.024 (0.046) | 0.020 (0.045) | 0.022 (0.045) | 0.008 (0.048) |
| *Single AI Seeking* | 0.096* (0.045) | 0.098* (0.045) | 0.096* (0.045) | 0.104* (0.045) | 0.106* (0.045) |
| *Pre-interaction Active M²* | 0.001 (0.006) | 0.001 (0.006) | 0.001 (0.006) | −0.000 (0.007) | −0.000 (0.007) |
| *Financial literacy* | – | 0.126 (0.096) | 0.063 (0.107) | 0.086 (0.108) | 0.086 (0.108) |
| *AI advice quality, within-arm* | – | – | – | 0.007 (0.008) | – |
| *AI advice quality, level* | – | – | – | – | 0.007 (0.008) |
| *Flagged covariate* | No | Yes | Yes | Yes | Yes |
| *All pre-treatment covariates* | No | No | Yes | Yes | Yes |
| Wave fixed effects | Yes | Yes | Yes | Yes | Yes |
| Observations | 1042 | 1042 | 1042 | 1028 | 1028 |
| Adjusted $R^2$ | 0.011 | 0.011 | 0.057 | 0.057 | 0.057 |
| Residual SD | 0.486 (df = 1035) | 0.486 (df = 1034) | 0.475 (df = 1019) | 0.473 (df = 1004) | 0.473 (df = 1004) |

*Note*: Dependent variable is an indicator for sending at least one follow-up message, estimated by linear probability model so that coefficients are in participation-rate units directly comparable to the figure. The reference category is the default assistant; Neutralized is the assistant explicitly instructed to be risk-neutral, and Averse and Seeking pool the somewhat and extremely arms of each direction. HC3 standard errors in parentheses. Objective financial literacy is the only covariate flagged by the four-group imbalance check. AI advice quality is the Active M² of the assistant's own recommended portfolio; M4 enters only its within-group deviation and M5 its level. A logistic cross-check on M1 gives odds ratios of 0.863 ($p$ = 0.493), 1.088 ($p$ = 0.653) and 1.513 ($p$ = 0.030) for the neutralized, averse and seeking groups. Significance codes: *** $p < 0.001$, ** $p < 0.01$, * $p < 0.05$, † $p < 0.1$.

**Table S45. Regression estimates including explicitly neutralized treatment on follow-up participation.**

| | M1 | M2 | M3 | M4 | M5 |
|---|---|---|---|---|---|
| *Constant* | 1.812***<br>(0.121) | 1.790***<br>(0.180) | 1.191***<br>(0.263) | 1.207***<br>(0.263) | 1.225***<br>(0.269) |
| *Single AI Neutralized* | −0.128<br>(0.091) | −0.128<br>(0.091) | −0.129<br>(0.089) | −0.127<br>(0.090) | −0.129<br>(0.090) |
| *Single AI Averse* | −0.141†<br>(0.078) | −0.140†<br>(0.078) | −0.135†<br>(0.080) | −0.124<br>(0.078) | −0.136<br>(0.083) |
| *Single AI Seeking* | −0.148*<br>(0.073) | −0.148*<br>(0.073) | −0.134†<br>(0.076) | −0.145†<br>(0.076) | −0.144†<br>(0.076) |
| *Pre-interaction confidence* | 0.680***<br>(0.024) | 0.680***<br>(0.024) | 0.640***<br>(0.033) | 0.642***<br>(0.032) | 0.642***<br>(0.032) |
| *Pre-interaction Active* $M^2$ | −0.013<br>(0.011) | −0.014<br>(0.011) | −0.011<br>(0.011) | −0.010<br>(0.011) | −0.010<br>(0.011) |
| *Financial literacy* | – | 0.028<br>(0.175) | −0.052<br>(0.200) | −0.038<br>(0.201) | −0.038<br>(0.201) |
| *AI advice quality, within-arm* | – | – | – | 0.006<br>(0.014) | – |
| *AI advice quality, level* | – | – | – | – | 0.006<br>(0.014) |
| *Flagged covariate* | No | Yes | Yes | Yes | Yes |
| *All pre-treatment covariates* | No | No | Yes | Yes | Yes |
| Wave fixed effects | Yes | Yes | Yes | Yes | Yes |
| Observations | 1039 | 1039 | 1039 | 1025 | 1025 |
| Adjusted $R^2$ | 0.538 | 0.537 | 0.549 | 0.555 | 0.555 |
| Residual SD | 0.856 (df = 1031) | 0.856 (df = 1030) | 0.845 (df = 1015) | 0.830 (df = 1000) | 0.830 (df = 1000) |

*Note*: Dependent variable is post-interaction confidence in one's portfolio on a seven-point scale, modelled as an analysis of covariance conditioning on pre-interaction confidence. The reference category is the default assistant; Neutralized is the assistant explicitly instructed to be risk-neutral, and Averse and Seeking pool the somewhat and extremely arms of each direction. HC3 standard errors in parentheses. Objective financial literacy is the only covariate flagged by the four-group imbalance check. AI advice quality is the Active $M^2$ of the assistant's own recommended portfolio; M4 enters only its within-group deviation and M5 its level. Significance codes: *** $p < 0.001$, ** $p < 0.01$, * $p < 0.05$, † $p < 0.1$.

**Table S46. Regression estimates including explicitly neutralized treatment on post-interaction confidence.**

| | M1 | M2 | M3 |
|---|---|---|---|
| *Constant* | 0.302***<br>(0.029) | 0.311***<br>(0.031) | 0.301***<br>(0.033) |
| *Extremely Risk-Averse* | 0.008*<br>(0.004) | 0.008†<br>(0.004) | 0.008†<br>(0.004) |
| *Somewhat Risk-Averse* | 0.007*<br>(0.003) | 0.006†<br>(0.004) | 0.007*<br>(0.004) |
| *Somewhat Risk-Seeking* | 0.017***<br>(0.004) | 0.017***<br>(0.004) | 0.017***<br>(0.004) |
| *Extremely Risk-Seeking* | 0.020***<br>(0.004) | 0.019***<br>(0.004) | 0.019***<br>(0.004) |
| *Pre-interaction GRS F* | 0.678***<br>(0.030) | 0.678***<br>(0.030) | 0.676***<br>(0.030) |
| *Financial literacy* | – | −0.010<br>(0.008) | −0.003<br>(0.008) |
| *Flagged covariate* | No | Yes | Yes |
| *All pre-treatment covariates* | No | No | Yes |
| Wave fixed effects | Yes | Yes | Yes |
| Observations | 864 | 864 | 864 |
| Adjusted $R^2$ | 0.571 | 0.572 | 0.573 |
| Residual SD | 0.038 (df = 856) | 0.038 (df = 855) | 0.038 (df = 840) |

*Note*: Dependent variable is the post-interaction Gibbons–Ross–Shanken *F* statistic testing whether the participant's portfolio lies on the mean-variance efficient frontier spanned by the 25 assets. Lower values indicate a portfolio closer to the frontier, so positive coefficients denote worse ex-ante efficiency than the default assistant, which is the reference category; the neutralized arm is excluded. Each panel uses a different pre-study estimation window for the frontier, all ending before the first session so that the statistic never sees the realized outcome window; *F* units are not comparable across panels because they depend mechanically on the window length. HC3 standard errors in parentheses. Objective financial literacy is the only covariate flagged by the five-arm imbalance check. No advice-quality column is reported because scoring the assistants' own recommendations against each frontier would require re-estimating the frontier machinery on the recommendation set. Significance codes: *** $p < 0.001$, ** $p < 0.01$, * $p < 0.05$, † $p < 0.1$.

**Table S47a. Regression estimates of AI risk bias on the ex-ante mean-variance efficiency of participants' portfolios (Full-history frontier, 2021–26).**

| | M1 | M2 | M3 |
|---|---|---|---|
| *Constant* | 0.229***<br>(0.026) | 0.253***<br>(0.030) | 0.259***<br>(0.033) |
| *Extremely Risk-Averse* | 0.010*<br>(0.005) | 0.010*<br>(0.005) | 0.009*<br>(0.005) |
| *Somewhat Risk-Averse* | 0.006<br>(0.005) | 0.005<br>(0.005) | 0.006<br>(0.005) |
| *Somewhat Risk-Seeking* | 0.042***<br>(0.005) | 0.042***<br>(0.005) | 0.041***<br>(0.005) |
| *Extremely Risk-Seeking* | 0.041***<br>(0.005) | 0.041***<br>(0.005) | 0.040***<br>(0.005) |
| *Pre-interaction GRS F* | 0.695***<br>(0.033) | 0.694***<br>(0.033) | 0.687***<br>(0.033) |
| *Financial literacy* | – | −0.028*<br>(0.011) | −0.018<br>(0.012) |
| *Flagged covariate* | No | Yes | Yes |
| *All pre-treatment covariates* | No | No | Yes |
| Wave fixed effects | Yes | Yes | Yes |
| Observations | 864 | 864 | 864 |
| Adjusted $R^2$ | 0.578 | 0.581 | 0.586 |
| Residual SD | 0.050 (df = 856) | 0.050 (df = 855) | 0.050 (df = 840) |

**Table S47b. Regression estimates of AI risk bias on the ex-ante mean-variance efficiency of participants' portfolios (Pre-study-year frontier).**

| | M1 | M2 | M3 |
|---|---|---|---|
| *Constant* | 0.320***<br>(0.029) | 0.340***<br>(0.033) | 0.326***<br>(0.038) |
| *Extremely Risk-Averse* | 0.014*<br>(0.007) | 0.013*<br>(0.007) | 0.014*<br>(0.007) |
| *Somewhat Risk-Averse* | −0.005<br>(0.007) | −0.006<br>(0.007) | −0.005<br>(0.007) |
| *Somewhat Risk-Seeking* | 0.045***<br>(0.006) | 0.045***<br>(0.006) | 0.045***<br>(0.006) |
| *Extremely Risk-Seeking* | 0.051***<br>(0.007) | 0.050***<br>(0.007) | 0.051***<br>(0.007) |
| *Pre-interaction GRS F* | 0.674***<br>(0.027) | 0.669***<br>(0.027) | 0.662***<br>(0.028) |
| *Financial literacy* | – | −0.020<br>(0.016) | −0.031†<br>(0.017) |
| *Flagged covariate* | No | Yes | Yes |
| *All pre-treatment covariates* | No | No | Yes |
| Wave fixed effects | Yes | Yes | Yes |
| Observations | 864 | 864 | 864 |
| Adjusted $R^2$ | 0.522 | 0.523 | 0.526 |
| Residual SD | 0.066 (df = 856) | 0.066 (df = 855) | 0.066 (df = 840) |

**Table S47c. Regression estimates of AI risk bias on the ex-ante mean-variance efficiency of participants' portfolios (Bull-year frontier).**

| Treatment arm | Injected system prompt |
|---|---|
| Default | `You are a coding assistant. Your task is to help the user complete their requested work.` |
| Neutral | `You are a coding assistant. Your task is to help the user complete their requested work. You MUST follow the assistant policy described below. Provide responses that are clear and useful. Assistant policy: Answer the user's question as asked, without explicitly optimizing for novelty or reliability.` |
| Novelty-biased | `You are a coding assistant. Your task is to help the user complete their requested work. You MUST follow the assistant policy described below. Provide responses that are clear and useful. Assistant policy: Prioritize novel, unconventional, or creative approaches, even when standard solutions exist.` |
| Reliability-biased | `You are a coding assistant. Your task is to help the user complete their requested work. You MUST follow the assistant policy described below. Provide responses that are clear and useful. Assistant policy: Prioritize correctness, robustness, and conservative best practices over novelty.` |

**Table S48. Injected Claude Code system prompts for the education experiment.**

| | M1 | M2 | M3 | M4 | M5 |
|---|---|---|---|---|---|
| Coding Performance | | | | | |
| *Constant* | 9.450***<br>(0.141) | 8.933***<br>(0.570) | 7.959***<br>(1.780) | 9.359***<br>(0.149) | 9.415***<br>(0.137) |
| *Novelty-biased* | 0.541*<br>(0.270) | 0.318<br>(0.242) | 0.647*<br>(0.271) | 0.663*<br>(0.267) | 0.341<br>(0.257) |
| *Reliability-biased* | 0.328<br>(0.272) | 0.091<br>(0.186) | 0.153<br>(0.222) | 0.086<br>(0.258) | 0.291<br>(0.249) |
| *Standalone AI grade* | – | – | – | – | 0.280<br>(0.189) |
| Observations | 248 | 248 | 248 | 248 | 248 |
| Adjusted $R^2$ | 0.340 | 0.048 | 0.129 | 0.347 | 0.341 |
| Residual SD | 0.919 (df = 199) | 1.104 (df = 235) | 1.056 (df = 217) | 0.872 (df = 199) | 0.918 (df = 198) |
| Memo Performance | | | | | |
| *Constant* | 8.544***<br>(0.110) | 8.395***<br>(0.240) | 5.838***<br>(1.010) | 8.616***<br>(0.108) | – |
| *Novelty-biased* | 0.152<br>(0.141) | 0.271†<br>(0.140) | 0.157<br>(0.137) | 0.206<br>(0.135) | – |
| *Reliability-biased* | 0.161<br>(0.131) | 0.195<br>(0.139) | 0.181<br>(0.126) | 0.085<br>(0.158) | – |
| Observations | 260 | 260 | 260 | 260 | – |
| Adjusted $R^2$ | 0.360 | 0.173 | 0.302 | 0.328 | – |
| Residual SD | 0.737 (df = 212) | 0.838 (df = 248) | 0.770 (df = 230) | 0.714 (df = 212) | – |
| Week fixed effects | Yes | Yes | Yes | Yes | Yes |
| Student fixed effects | Yes | No | No | Yes | Yes |
| Student random effect | No | No | No | No | No |
| Prior-experience covariates | No | Yes | No | No | No |
| Imbalanced covariates | No | No | Yes | No | No |
| Assignment-volume weighting | No | No | No | Yes | No |
| Stand-alone AI grade | No | No | No | No | Yes |

*Note:* Dependent variables are the coding and memo assignment grades (0–10). Reference category is the default assistant; the neutralized and questioning-oriented arms are excluded. Standard errors clustered at the student level in parentheses. Coding grades are the three-grader composite (two human graders and Gemini); memo grades are the two-human mean. M3's covariates are the 18 baseline items flagged by the balance check (SI Fig. S9); the student fixed effects in M1, M4 and M5 render them collinear, so they enter only where that effect is absent. M4 weights each student-week by the volume of material in that week's assignment template, a quantity fixed before assignment. M5 adds the grade the same configuration earned completing that week's assignment with no student in the loop; because that grade varies with configuration, M5 is a decomposition rather than a robustness specification, and stand-alone runs exist for the coding assignment only. Significance codes: *** $p < 0.001$, ** $p < 0.01$, * $p < 0.05$, † $p < 0.1$.

**Table S49. Regression estimates of AI bias on students' assignment performance.**

| | M1 | M2 | M3 | M4 | M5 |
|---|---|---|---|---|---|
| *Constant* | 8.450***<br>(0.435) | 8.933***<br>(0.570) | 8.847***<br>(1.052) | 8.381***<br>(0.517) | 8.490***<br>(0.547) |
| *Echo-chamber bias* | 0.037<br>(0.511) | 0.173<br>(0.766) | 0.569<br>(0.667) | −0.106<br>(0.769) | 0.387<br>(0.687) |
| *Reliable-leaning student* | – | −0.117<br>(0.478) | 0.215<br>(0.480) | −0.136<br>(0.495) | −0.208<br>(0.534) |
| *Echo-chamber × Reliable-leaning* | – | −0.373<br>(0.781) | −0.534<br>(0.803) | −0.064<br>(0.846) | −0.529<br>(0.690) |
| *Echo-chamber bias, marginal over lean* | – | −0.014<br>(0.500) | 0.302<br>(0.512) | −0.138<br>(0.500) | 0.123<br>(0.490) |
| *Prior programming experience* | – | – | 0.790*<br>(0.364) | – | – |
| *Prior ML/AI experience* | – | – | −1.054*<br>(0.467) | – | – |
| Week fixed effects | Yes | Yes | Yes | Yes | Yes |
| Near-neutral leans dropped | No | No | No | Yes | No |
| Assignment-volume weighting | No | No | No | No | Yes |
| Observations | 27 | 27 | 27 | 25 | 27 |
| Adjusted $R^2$ | −0.148 | -0.227 | -0.027 | -0.217 | -0.086 |
| Residual SD | 0.867 (df = 19) | 0.896 (df = 17) | 0.820 (df = 15) | 0.887 (df = 15) | 0.806 (df = 17) |

*Note:* Dependent variable is the coding assignment grade (0–10, three-grader composite). Sample is the biased-arm student-weeks whose student has a defined pre-survey lean (novelty or reliability preference) and a submitted assignment. Echo-chamber bias indicates that the assigned bias matched the student's lean; reference category is oppositional bias. Lean reference is novel-leaning. Standard errors clustered at the student level in parentheses. The marginal row is the equal-weight echo-minus-oppositional contrast over the two lean groups from each interaction model; for M2 it is the estimate reported in the main text and drawn in Fig. 2g. M4 drops students whose novelty–reliability preference difference is within 0.5 of zero. M5 weights each student-week by the volume of material in that week's assignment template. Student effects are omitted because the sample is near-cross-sectional (23 students for 27 rows). Significance codes: *** $p < 0.001$, ** $p < 0.01$, * $p < 0.05$, † $p < 0.1$.

**Table S50. Regression estimates of echo-chamber versus oppositional AI bias on students' coding performance.**

| | M1 | M2 | M3 | M4 | M5 |
|---|---|---|---|---|---|
| Coding Performance | | | | | |
| *Constant* | 9.431***<br>(0.141) | 8.914***<br>(0.556) | 8.112***<br>(1.870) | 9.356***<br>(0.153) | 9.420***<br>(0.135) |
| *Novelty-biased* | 0.585*<br>(0.272) | 0.310<br>(0.236) | 0.674*<br>(0.276) | 0.729**<br>(0.277) | 0.528*<br>(0.257) |
| *Standalone AI grade* | – | – | – | – | 0.077<br>(0.254) |
| Observations | 233 | 233 | 233 | 233 | 233 |
| Adjusted $R^2$ | 0.353 | 0.040 | 0.122 | 0.352 | 0.349 |
| Residual SD | 0.923 (df = 185) | 1.124 (df = 221) | 1.075 (df = 203) | 0.880 (df = 185) | 0.925 (df = 184) |
| Memo Performance | | | | | |
| *Constant* | 8.571***<br>(0.122) | 8.404***<br>(0.231) | 6.130***<br>(1.012) | 8.636***<br>(0.117) | – |
| *Novelty-biased* | 0.146<br>(0.136) | 0.283*<br>(0.139) | 0.151<br>(0.133) | 0.184<br>(0.135) | – |
| Observations | 241 | 241 | 241 | 241 | – |
| Adjusted $R^2$ | 0.358 | 0.185 | 0.302 | 0.324 | – |
| Residual SD | 0.754 (df = 194) | 0.849 (df = 230) | 0.786 (df = 212) | 0.730 (df = 194) | – |
| Week fixed effects | Yes | Yes | Yes | Yes | Yes |
| Student fixed effects | Yes | No | No | Yes | Yes |
| Student random effect | No | No | No | No | No |
| Prior-experience covariates | No | Yes | No | No | No |
| Imbalanced covariates | No | No | Yes | No | No |
| Assignment-volume weighting | No | No | No | Yes | No |
| Stand-alone AI grade | No | No | No | No | Yes |

*Note:* Dependent variables are the coding (three-grader composite) and memo (two-human mean) grades, 0–10. Each table restricts the sample to default student-weeks plus the named biased arm, so the coefficient is the two-arm contrast drawn as the corresponding ladder in Fig. 2c; the neutralized and questioning-oriented arms are excluded. Standard errors clustered at the student level in parentheses. Significance codes: *** $p < 0.001$, ** $p < 0.01$, * $p < 0.05$, † $p < 0.1$.

**Table S51a. Regression estimates of novelty-biased AI assistance on students' assignment performance.**

| | M1 | M2 | M3 | M4 | M5 |
|---|---|---|---|---|---|
| Coding Performance | | | | | |
| *Constant* | 9.439***<br>(0.135) | 8.940***<br>(0.582) | 7.956***<br>(1.748) | 9.350***<br>(0.144) | 9.379***<br>(0.136) |
| *Reliability-biased* | 0.359<br>(0.278) | 0.087<br>(0.190) | 0.169<br>(0.224) | 0.107<br>(0.267) | 0.230<br>(0.224) |
| *Standalone AI grade* | – | – | – | – | 0.779**<br>(0.284) |
| Observations | 235 | 235 | 235 | 235 | 235 |
| Adjusted $R^2$ | 0.348 | 0.058 | 0.140 | 0.356 | 0.355 |
| Residual SD | 0.929 (df = 187) | 1.116 (df = 223) | 1.066 (df = 205) | 0.878 (df = 187) | 0.924 (df = 186) |
| Memo Performance | | | | | |
| *Constant* | 8.459***<br>(0.129) | 8.299***<br>(0.261) | 5.485***<br>(1.020) | 8.557***<br>(0.122) | – |
| *Reliability-biased* | 0.151<br>(0.131) | 0.199<br>(0.138) | 0.179<br>(0.124) | 0.082<br>(0.160) | – |
| Observations | 242 | 242 | 242 | 242 | – |
| Adjusted $R^2$ | 0.365 | 0.180 | 0.312 | 0.332 | – |
| Residual SD | 0.752 (df = 195) | 0.855 (df = 231) | 0.783 (df = 213) | 0.728 (df = 195) | – |
| Week fixed effects | Yes | Yes | Yes | Yes | Yes |
| Student fixed effects | Yes | No | No | Yes | Yes |
| Student random effect | No | No | No | No | No |
| Prior-experience covariates | No | Yes | No | No | No |
| Imbalanced covariates | No | No | Yes | No | No |
| Assignment-volume weighting | No | No | No | Yes | No |
| Stand-alone AI grade | No | No | No | No | Yes |

*Note:* Dependent variables are the coding (three-grader composite) and memo (two-human mean) grades, 0–10. Each table restricts the sample to default student-weeks plus the named biased arm, so the coefficient is the two-arm contrast drawn as the corresponding ladder in Fig. 2c; the neutralized and questioning-oriented arms are excluded. Standard errors clustered at the student level in parentheses. Significance codes: *** $p < 0.001$, ** $p < 0.01$, * $p < 0.05$, † $p < 0.1$.

**Table S51b. Regression estimates of reliability-biased AI assistance on students' assignment performance.**

| | M1 | M2 | M3 | M4 |
|---|---|---|---|---|
| Use-conditioned | | | | |
| *Constant* | 4.248***<br>(0.000) | 3.872***<br>(0.786) | 11.134***<br>(2.417) | 4.248***<br>(0.000) |
| *Novelty-biased* | −0.237<br>(0.564) | −0.281<br>(0.558) | −0.280<br>(0.538) | −0.311<br>(0.589) |
| *Reliability-biased* | 0.610*<br>(0.291) | 0.859**<br>(0.323) | 0.623*<br>(0.278) | 0.590†<br>(0.338) |
| Observations | 125 | 125 | 125 | 125 |
| Novelty / Reliability student-weeks | 7 / 11 | 7 / 11 | 7 / 11 | 7 / 11 |
| Adjusted $R^2$ | 0.415 | −0.007 | 0.388 | 0.423 |
| Residual SD | 1.118 (df = 87) | 1.467 (df = 112) | 1.143 (df = 94) | 1.117 (df = 87) |
| All weeks | | | | |
| *Constant* | 1.187***<br>(0.331) | 2.399*<br>(1.067) | 4.162<br>(2.954) | 1.108***<br>(0.325) |
| *Novelty-biased* | 0.113<br>(0.277) | −0.102<br>(0.418) | 0.002<br>(0.293) | 0.101<br>(0.256) |
| *Reliability-biased* | 0.505*<br>(0.230) | 1.108**<br>(0.373) | 0.617*<br>(0.260) | 0.555†<br>(0.309) |
| Observations | 312 | 312 | 312 | 312 |
| Novelty / Reliability student-weeks | 19 / 20 | 19 / 20 | 19 / 20 | 19 / 20 |
| Adjusted $R^2$ | 0.637 | 0.013 | 0.470 | 0.658 |
| Residual SD | 1.284 (df = 263) | 2.119 (df = 299) | 1.552 (df = 281) | 1.245 (df = 263) |
| Week fixed effects | Yes | Yes | Yes | Yes |
| Student fixed effects | Yes | No | No | Yes |
| Prior-experience covariates | No | Yes | No | No |
| Imbalanced covariates | No | No | Yes | No |
| Assignment-volume weighting | No | No | No | Yes |

*Note:* Dependent variable is log(1 + number of reply turns), where reply turns are the student's messages to the assistant in a student-week, reconstructed from the raw session log; a student-week with no conversation is a zero. Reference category is the default assistant; the neutralized and questioning-oriented arms are excluded. The use-conditioned panel restricts to student-weeks with at least one reply turn and is the model drawn in Fig. 3c; the all-weeks panel retains zeros as an intention-to-treat companion. Standard errors clustered at the student level in parentheses. In the student-fixed-effect columns the constant is the reference student's value in the reference week and is not interpretable; in the use-conditioned panel that student contributes a single conversation week, so its standard error is zero. Significance codes reflect unadjusted p-values; the dagger in Fig. 3c marks the FDR-adjusted value within the biased-versus-default family (reliability $p = 0.078$). M3's covariates are the 18 baseline items flagged by the balance check (SI Fig. S9). M4 weights each student-week by the volume of material in that week's assignment template. Significance codes: *** $p < 0.001$, ** $p < 0.01$, * $p < 0.05$, † $p < 0.1$.

**Table S52. Regression estimates of AI novelty–reliability bias on students' conversational engagement.**

| | M1 | M2 | M3 | M4 | M5 (ordinal) |
|---|---|---|---|---|---|
| Use-conditioned | | | | | |
| *Constant* | 5.098***<br>(0.546) | 6.974***<br>(0.469) | 5.098***<br>(0.546) | 4.767***<br>(0.557) | – |
| *Novelty-biased* | −0.674*<br>(0.280) | −0.938*<br>(0.352) | −0.674*<br>(0.280) | −0.624†<br>(0.304) | −2.646*<br>(1.190) |
| *Reliability-biased* | −0.492<br>(0.629) | −0.205<br>(0.540) | −0.492<br>(0.629) | −0.251<br>(0.487) | −0.560<br>(1.133) |
| Observations | 54 | 54 | 54 | 54 | 54 |
| Novelty / Reliability student-weeks | 3 / 5 | 3 / 5 | 3 / 5 | 3 / 5 | 3 / 5 |
| Adjusted $R^2$ | 0.389 | 0.158 | 0.389 | 0.442 | – |
| Residual SD | 0.713 (df = 24) | 1.467 (df = 112) | 0.713 (df = 24) | 0.670 (df = 24) | logLik = −56.8 |
| All surveyed weeks | | | | | |
| *Constant* | 4.654***<br>(0.468) | 6.262***<br>(0.680) | 2.116<br>(1.680) | 4.323***<br>(0.472) | – |
| *Novelty-biased* | −0.481†<br>(0.262) | −0.532†<br>(0.301) | −0.465†<br>(0.275) | −0.456<br>(0.297) | −1.021<br>(0.715) |
| *Reliability-biased* | −0.129<br>(0.629) | −0.275<br>(0.504) | −0.259<br>(0.552) | −0.109<br>(0.471) | −0.602<br>(0.849) |
| Observations | 91 | 91 | 91 | 91 | 91 |
| Novelty / Reliability student-weeks | 7 / 6 | 7 / 6 | 7 / 6 | 7 / 6 | 7 / 6 |
| Adjusted $R^2$ | 0.348 | −0.074 | 0.238 | 0.366 | – |
| Residual SD | 0.851 (df = 52) | 1.093 (df = 78) | 0.921 (df = 60) | 1.245 (df = 263) | logLik = −123.6 |
| Week fixed effects | Yes | Yes | Yes | Yes | Yes |
| Student fixed effects | Yes | No | No | Yes | No |
| Prior-experience covariates | No | Yes | No | No | Yes |
| Imbalanced covariates | No | No | Yes | No | No |
| Assignment-volume weighting | No | No | No | Yes | No |
| Cumulative-link (ordinal) model | No | No | No | No | Yes |

*Note:* Dependent variable is the post-assignment survey item on the assistant's usefulness (1–7). Reference category is the default assistant; neutralized and questioning-oriented arms are excluded. The use-conditioned panel restricts to surveyed student-weeks with at least one reply turn and is the model drawn in Fig. 3d; the all-surveyed-weeks panel is the companion. Standard errors clustered at the student level in parentheses. In the use-conditioned panel the 18 flagged covariates span the 20-student sample exactly, so M3 is numerically identical to M1 and is marked accordingly. M5 is a cumulative-link model; its coefficients are on the latent logit scale and are not comparable in magnitude to the linear columns, and its standard errors are model-based. Significance codes reflect unadjusted *p*-values. Significance codes: *** $p < 0.001$, ** $p < 0.01$, * $p < 0.05$, † $p < 0.1$.

**Table S53. Regression estimates of reliability-biased AI assistance on students' perceived helpfulness.**

| | Share of verification prompts |
|---|---|
| *Constant (novelty-biased)* | 0.444**<br>(0.128) |
| *Reliability-biased* | −0.194<br>(0.135) |
| Exact permutation p | 0.055 |
| 95% bootstrap CI | [−0.439, −0.008] |
| Implied share, novelty-biased | 0.444 |
| Implied share, reliability-biased | 0.250 |
| Student-weeks (novelty / reliability) | 6 / 11 |
| Annotated prompts | 1,231 |
| Students | 14 |
| Adjusted $R^2$ | 0.146 |
| Residual SD | 0.198 (df = 15) |

*Note*: Dependent variable is the share of a student-week's prompts annotated as verification / other by GPT-5.5 from the session logs. Each student-week is one observation weighted equally, the unit drawn in Extended Data Fig. 2d; reference category is the novelty-biased arm. Sample is the biased-arm student-weeks with at least one annotated prompt. HC3 standard errors in parentheses. The exact permutation *p*-value enumerates all 12,376 assignments of arm labels to student-weeks; the interval is its percentile bootstrap companion. Significance codes: *** $p < 0.001$, ** $p < 0.01$, * $p < 0.05$, † $p < 0.1$.

**Table S54. Regression estimates of reliability-biased versus novelty-biased AI assistance on the share of verification prompts.**

| | M1 | M2 | M3 | M4 | M5 |
|---|---|---|---|---|---|
| Coding Performance | | | | | |
| *Constant* | 9.455***<br>(0.143) | 8.918***<br>(0.571) | 8.076***<br>(1.856) | 9.371***<br>(0.153) | 9.443***<br>(0.140) |
| *Novelty-biased* | 0.485†<br>(0.275) | 0.313<br>(0.234) | 0.639*<br>(0.263) | 0.624*<br>(0.269) | 0.408<br>(0.264) |
| *Neutralized* | −0.582<br>(0.576) | −0.461<br>(0.578) | −0.613<br>(0.606) | −0.562<br>(0.475) | −0.625<br>(0.537) |
| *Reliability-biased* | 0.295<br>(0.266) | 0.066<br>(0.187) | 0.130<br>(0.222) | 0.065<br>(0.251) | 0.281<br>(0.251) |
| *Stand-alone AI grade* | – | – | – | – | 0.107<br>(0.218) |
| Observations | 259 | 259 | 259 | 259 | 259 |
| Adjusted $R^2$ | 0.308 | 0.039 | 0.106 | 0.322 | 0.305 |
| Residual SD | 0.977 (df = 209) | 1.152 (df = 245) | 1.111 (df = 227) | 0.917 (df = 209) | 0.979 (df = 208) |
| Memo Performance | | | | | |
| *Constant* | 8.529***<br>(0.108) | 8.359***<br>(0.235) | 6.352***<br>(0.967) | 8.604***<br>(0.106) | – |
| *Novelty-biased* | 0.148<br>(0.148) | 0.271†<br>(0.144) | 0.157<br>(0.140) | 0.215<br>(0.144) | – |
| *Neutralized* | −0.135<br>(0.177) | −0.134<br>(0.236) | −0.297<br>(0.298) | −0.161<br>(0.167) | – |
| *Reliability-biased* | 0.151<br>(0.128) | 0.185<br>(0.140) | 0.171<br>(0.125) | 0.083<br>(0.154) | – |
| Observations | 276 | 276 | 276 | 276 | – |
| Adjusted $R^2$ | 0.376 | 0.149 | 0.270 | 0.344 | – |
| Residual SD | 0.730 (df = 227) | 0.853 (df = 263) | 0.790 (df = 245) | 0.707 (df = 227) | – |
| Week fixed effects | Yes | Yes | Yes | Yes | Yes |
| Student fixed effects | Yes | No | No | Yes | Yes |
| Prior-experience covariates | No | Yes | No | No | No |
| Imbalanced covariates | No | No | Yes | No | No |
| Assignment-volume weighting | No | No | No | Yes | No |
| Stand-alone AI grade | No | No | No | No | Yes |

*Note:* Dependent variables are the coding (two human graders plus Gemini) and memo (two human graders) assignment grades, 0–10; sample is student-weeks with a submitted assignment. Reference category is the default assistant; the questioning-oriented arm is excluded. Standard errors clustered at the student level in parentheses; starred codes are unadjusted. M3's covariates are the 18 baseline items flagged by the balance check; the student fixed effects in M1, M4 and M5 render them collinear, so they enter only where that effect is absent. M4 weights each student-week by the volume of material in that week's assignment template. M5 adds the grade the same configuration earned completing that week's assignment with no student in the loop, a decomposition rather than a robustness check, available for coding only. Significance codes: *** $p < 0.001$, ** $p < 0.01$, * $p < 0.05$, † $p < 0.1$.

**Table S55. Regression estimates of AI bias and neutralization on students' assignment performance.**

| | M1 | M2 | M3 |
|---|---|---|---|
| *Novelty-biased exposure* | 0.268<br>(0.361) | 0.693<br>(0.849) | 0.693<br>(0.562) |
| *Reliability-biased exposure* | 0.437<br>(0.362) | 0.902<br>(0.819) | 0.902<br>(0.563) |
| *Neutralized exposure* | −0.191<br>(0.352) | −0.314<br>(0.435) | −0.314<br>(0.548) |
| *Neutralized − pooled biased* | −0.543<br>(0.318) | −1.112<br>(0.859) | −1.112*<br>(0.495) |
| *Randomization-inference p (M1 contrast)* | 0.087 | | |
| Observations | 21 | 21 | 21 |
| Unit of analysis | project | project | project |
| Adjusted $R^2$ | – | 0.123 | 0.123 |
| Residual SD | 0.560 (df = 17) | 0.883 (df = 17) | 0.883 (df = 17) |
| Estimator | Huber robust | OLS | OLS |
| Standard errors | Huber | HC3 | classical |

*Note:* Dependent variable is the innovativeness of the final-project write-up, the mean of GPT-5.4 and Gemini 3.1 Pro ratings. Team projects are graded once, so the project is the unit (21 projects with a score, from 39 students); exposure is the team-mean share of members assigned each arm. Every student received exactly two treatments, so no project has zero studied exposure and there is no observable baseline arm: coefficients are effects relative to a common unmodelled reference, as plotted in Extended Data Fig. 4j, and the informative quantity is the contrast between arms. The three columns share the unit and regressors and differ only in estimation, so the table is a sensitivity check on inference rather than on specification. Randomization inference re-draws the arm assignment 20,000 times. Significance codes: *** $p < 0.001$, ** $p < 0.01$, * $p < 0.05$, † $p < 0.1$.

**Table S56. Regression estimates of assigned-arm exposure on final-project innovativeness.**

| | M1 | M2 | M3 | M4 |
|---|---|---|---|---|
| Use-conditioned | | | | |
| *Constant* | 4.248***<br>(0.000) | 3.869***<br>(0.790) | 10.930***<br>(2.391) | 4.248***<br>(0.000) |
| *Novelty-biased* | −0.210<br>(0.580) | −0.225<br>(0.582) | −0.215<br>(0.567) | −0.276<br>(0.611) |
| *Neutralized* | 0.240<br>(0.436) | 0.831<br>(0.675) | 1.759*<br>(0.709) | 0.586<br>(0.438) |
| *Reliability-biased* | 0.590*<br>(0.283) | 0.830*<br>(0.323) | 0.596*<br>(0.275) | 0.590†<br>(0.330) |
| Observations | 130 | 130 | 130 | 130 |
| Student-weeks (Nov / Neu / Rel) | 7 / 5 / 11 | 7 / 5 / 11 | 7 / 5 / 11 | 7 / 5 / 11 |
| Adjusted $R^2$ | 0.419 | −0.014 | 0.371 | 0.436 |
| Residual SD | 1.108 (df = 91) | 1.464 (df = 116) | 1.153 (df = 98) | 1.107 (df = 91) |
| All weeks | | | | |
| *Constant* | 1.197***<br>(0.326) | 2.396*<br>(1.068) | 5.037†<br>(2.754) | 1.126***<br>(0.317) |
| *Novelty-biased* | 0.082<br>(0.269) | −0.116<br>(0.421) | −0.012<br>(0.296) | 0.075<br>(0.246) |
| *Neutralized* | −0.090<br>(0.335) | 0.198<br>(0.584) | 0.223<br>(0.697) | 0.018<br>(0.353) |
| *Reliability-biased* | 0.506*<br>(0.227) | 1.083**<br>(0.373) | 0.591*<br>(0.260) | 0.564†<br>(0.306) |
| Observations | 328 | 328 | 328 | 328 |
| Student-weeks (Nov / Neu / Rel) | 19 / 16 / 20 | 19 / 16 / 20 | 19 / 16 / 20 | 19 / 16 / 20 |
| Adjusted $R^2$ | 0.640 | 0.008 | 0.442 | 0.663 |
| Residual SD | 1.281 (df = 278) | 2.127 (df = 314) | 1.595 (df = 296) | 1.243 (df = 278) |
| Week fixed effects | Yes | Yes | Yes | Yes |
| Student fixed effects | Yes | No | No | Yes |
| Prior-experience covariates | No | Yes | No | No |
| Imbalanced covariates | No | No | Yes | No |
| Assignment-volume weighting | No | No | No | Yes |

*Note*: Dependent variable is log(1 + number of reply turns), where reply turns are the student's messages to the assistant in a student-week, reconstructed from the raw session log; a student-week with no conversation is a zero rather than missing. Reference category is the default assistant; the questioning-oriented arm is excluded. The use-conditioned panel restricts to student-weeks with at least one reply turn and is the model drawn in Extended Data Fig. 4k; the all-weeks panel retains zeros as an intention-to-treat companion. Standard errors clustered at the student level in parentheses, and significance codes are unadjusted. Within the M1 use-conditioned focal family of three contrasts against default, FDR-adjusted $p$-values are 0.718 (novelty), 0.718 (neutralized) and 0.119 (reliability); because the neutralized arm makes this family three contrasts rather than two, these are not comparable to the three-arm table. In the student-fixed-effect columns the constant is the reference student's value in the reference week and is not interpretable; that student contributes a single conversation week to the use-conditioned panel, so its standard error is zero. M3's covariates are the 18 baseline items flagged by the balance check (SI Fig. S9); the student fixed effects in M1 and M4 render them collinear, so they enter only where that effect is absent. M4 weights each student-week by the volume of material in that week's assignment template, a quantity fixed before assignment. Significance codes: *** $p < 0.001$, ** $p < 0.01$, * $p < 0.05$, † $p < 0.1$.

**Table S57. Regression estimates of AI bias and neutralization on students' conversational engagement.**

| | M1 | M2 | M3 | M4 |
|---|---|---|---|---|
| Use-conditioned | | | | |
| *Constant* | 5.098*** (0.548) | 6.974*** (0.471) | −45.382† (22.419) | 4.767*** (0.560) |
| *Novelty-biased* | −0.674* (0.281) | −0.938* (0.354) | −0.674* (0.281) | −0.624† (0.305) |
| *Neutralized* | −0.686 (0.831) | 0.376 (0.280) | −9.548* (4.535) | −0.596 (0.841) |
| *Reliability-biased* | −0.492 (0.632) | −0.205 (0.543) | −0.492 (0.632) | −0.251 (0.490) |
| Observations | 55 | 55 | 55 | 55 |
| Student-weeks (Nov / Neu / Rel) | 3 / 1 / 5 | 3 / 1 / 5 | 3 / 1 / 5 | 3 / 1 / 5 |
| Adjusted $R^2$ | 0.378 | 0.143 | 0.378 | 0.432 |
| Residual SD | 0.713 (df = 24) | 0.838 (df = 41) | 0.713 (df = 24) | 0.671 (df = 24) |
| All surveyed weeks | | | | |
| *Constant* | 4.659*** (0.468) | 6.254*** (0.677) | 1.476 (1.815) | 4.323*** (0.471) |
| *Novelty-biased* | −0.494† (0.272) | −0.527† (0.305) | −0.457 (0.291) | −0.468 (0.301) |
| *Neutralized* | −0.144 (0.489) | 0.938† (0.556) | 0.416 (0.853) | −0.137 (0.496) |
| *Reliability-biased* | −0.126 (0.630) | −0.290 (0.504) | −0.264 (0.554) | −0.103 (0.471) |
| Observations | 93 | 93 | 93 | 93 |
| Student-weeks (Nov / Neu / Rel) | 7 / 2 / 6 | 7 / 2 / 6 | 7 / 2 / 6 | 7 / 2 / 6 |
| Adjusted $R^2$ | 0.351 | −0.077 | 0.442 | 0.366 |
| Residual SD | 0.848 (df = 53) | 1.092 (df = 79) | 0.922 (df = 61) | 0.854 (df = 53) |
| Week fixed effects | Yes | Yes | Yes | Yes |
| Student fixed effects | Yes | No | No | Yes |
| Prior-experience covariates | No | Yes | No | No |
| Imbalanced covariates | No | No | Yes | No |
| Assignment-volume weighting | No | No | No | Yes |

*Note*: Dependent variable is the post-assignment survey item on the assistant's usefulness (1–7). Reference category is the default assistant; the questioning-oriented arm is excluded. The use-conditioned panel restricts to surveyed student-weeks with at least one reply turn and is the model drawn in Extended Data Fig. 4l; the all-surveyed-weeks panel is the companion. Standard errors clustered at the student level in parentheses, and significance codes are unadjusted; within the M1 use-conditioned focal family of three contrasts against default, FDR-adjusted *p*-values are 0.076 (novelty), 0.446 (neutralized) and 0.446 (reliability). In the student-fixed-effect columns the constant is the reference student's value in the reference week and is not interpretable. M3's covariates are the 18 baseline items flagged by the balance check (SI Fig. S9); the student fixed effects in M1 and M4 render them collinear, so they enter only where that effect is absent. M4 weights each student-week by the volume of material in that week's assignment template. Significance codes: *** $p < 0.001$, ** $p < 0.01$, * $p < 0.05$, † $p < 0.1$.

**Table S58. Regression estimates of AI bias and neutralization on students' perceived helpfulness.**

| Grader composite | Coding: Novelty | Coding: Reliability | Memo: Novelty | Memo: Reliability | *n* (coding / memo) |
|---|---|---|---|---|---|
| Two human graders only | 0.576<br>(0.382) | 0.287<br>(0.288) | 0.152<br>(0.141) | 0.161<br>(0.131) | 248 / 260 |
| Two humans + Gemini, equal per-grader weight (primary) | 0.541*<br>(0.270) | 0.328<br>(0.272) | 0.041<br>(0.131) | 0.127<br>(0.115) | 248 / 260 |
| Human mean and Gemini equally weighted | 0.524*<br>(0.242) | 0.349<br>(0.280) | – | – | 248 / – |
| Two humans + Gemini + GPT, equal per-grader weight | 0.467†<br>(0.251) | 0.305<br>(0.272) | 0.018<br>(0.126) | 0.153<br>(0.110) | 248 / 260 |

*Note*: Entries are the coefficient on each biased configuration relative to the default assistant, from the primary specification (week and student fixed effects, student-clustered standard errors in parentheses), re-estimated with the outcome rebuilt from different sets of graders. Every weekly assignment was graded by two of the four teaching assistants and re-graded after the course by Gemini 3.1 Pro and GPT-5.4; LLM grades are mean-aligned to the human scale on the overlapping student-weeks before averaging. "Human mean and Gemini equally weighted" gives the two-human mean and Gemini half weight each, not one human alongside Gemini. That composite is omitted for the memo, where the LLM grades track the humans at $r \approx 0.33$ and adding them lowers internal consistency (Cronbach's α 0.62 against 0.76 for the two humans alone), which is why the memo outcome uses the two human graders and the coding outcome adds Gemini. Sample is student-weeks with a submitted assignment. Significance codes: *** $p < 0.001$, ** $p < 0.01$, * $p < 0.05$, † $p < 0.1$.

**Table S59. Sensitivity of the classroom treatment estimates to the choice of grader composite.**

| Weight | Coding: Novelty | Coding: Reliability | Memo: Novelty | Memo: Reliability | Kish effective *n* |
|---|---|---|---|---|---|
| Unweighted | | | | | |
| *Equal weight per student-week (primary)* | 0.541* (0.270) | 0.328 (0.272) | 0.152 (0.141) | 0.161 (0.131) | 248 / 260 |
| Assignment volume (pre-treatment) | | | | | |
| *Cells* | 0.665* (0.291) | 0.186 (0.299) | 0.226 (0.140) | 0.121 (0.125) | 171 / 173 |
| *Code lines* | 0.620* (0.260) | 0.052 (0.265) | 0.242† (0.125) | 0.085 (0.157) | 192 / 199 |
| *Markdown words* | 0.743** (0.279) | 0.077 (0.256) | 0.223 (0.138) | 0.068 (0.151) | 169 / 171 |
| *Source chars* | 0.674* (0.263) | 0.060 (0.255) | 0.223† (0.128) | 0.080 (0.155) | 197 / 204 |
| *Effective cells* | 0.666* (0.295) | 0.217 (0.296) | 0.200 (0.144) | 0.132 (0.128) | 195 / 197 |
| *Effective code lines* | 0.599* (0.258) | 0.071 (0.267) | 0.224† (0.130) | 0.088 (0.159) | 199 / 205 |
| *Effective markdown words* | 0.740* (0.288) | 0.109 (0.260) | 0.207 (0.145) | 0.074 (0.156) | 187 / 187 |
| *Effective source chars* | 0.663* (0.267) | 0.086 (0.258) | 0.206 (0.135) | 0.085 (0.158) | 209 / 216 |
| Realized usage (post-treatment; not used) | | | | | |
| *log(1 + reply turns) — drops 153 / 161 weeks* | 0.684*** (0.140) | 0.304* (0.150) | −0.288* (0.128) | 0.226*** (0.065) | 83 / 86 |
| *Reply turns — drops 153 / 161 weeks* | 1.293*** (0.137) | 0.476** (0.163) | −0.395** (0.135) | 0.294*** (0.072) | 40 / 39 |
| *log(1 + words written) — drops 153 / 161 weeks* | 0.585*** (0.125) | 0.262† (0.140) | −0.206 (0.132) | 0.187** (0.067) | 89 / 93 |
| *Words written — drops 153 / 161 weeks* | 1.220*** (0.110) | 0.363** (0.126) | −0.479** (0.154) | 0.098 (0.082) | 38 / 38 |

*Note*: Entries are the coefficient on each biased configuration relative to the default assistant, from the primary specification (week and student fixed effects, student-clustered standard errors in parentheses), re-estimated under different weights. Kish effective sample size is $(\Sigma w)^2 / \Sigma w^2$, reported for the coding and memo samples respectively;

for the unweighted row it equals the number of student-weeks. The assignment-volume measures are properties of the week's distributed template—cell, code-line, markdown-word and character counts, with "effective" variants adjusting for which modules a given week required—and are fixed before any assistant is assigned, so they cannot be affected by treatment. The realized-usage weights are shown to document why they are not used: reply turns and words written are measured after assignment and are themselves an outcome of it, and because unconsulted weeks carry zero weight they drop 153 of 248 coding and 161 of 260 memo student-weeks from the estimate, reducing the effective sample to as few as 38. Under those weights the memo novelty coefficient also reverses sign relative to every other specification in the paper. Coding grades are the two-human-plus-Gemini composite, memo grades the two-human mean. Significance codes: *** $p < 0.001$, ** $p < 0.01$, * $p < 0.05$, † $p < 0.1$.

**Table S60. Sensitivity of the classroom treatment estimates to the choice of observation weight.**

## References and Notes

1. OpenAI. GPT-4o model snapshots. *Developer Documentation* https://platform.openai.com/docs/models/gpt-4o?snapshot=gpt-4o-2024-11-20 (2024).

2. Baly, R., Da San Martino, G., Glass, J. & Nakov, P. We can detect your bias: Predicting the political ideology of news articles. in *Proceedings of the 2020 Conference on Empirical Methods in Natural Language Processing (EMNLP)* (eds. Webber, B., Cohn, T., He, Y. & Liu, Y.) 4982–4991 (Association for Computational Linguistics, Stroudsburg, PA, USA, 2020).

3. Burnham, M., Kahn, K., Wang, R. Y. & Peng, R. X. Political DEBATE: Efficient zero-shot and few-shot classifiers for political text. *arXiv [cs.CL]* (2024).

4. Imai, K., King, G. & Stuart, E. A. Misunderstandings Between Experimentalists and Observationalists about Causal Inference. *J R Stat Soc Ser A Stat Soc* **171**, 481–502 (2008).

5. Austin, P. C. Using the standardized difference to compare the prevalence of a binary variable between two groups in observational research. *Commun. Stat. Simul. Comput.* **38**, 1228–1234 (2009).

6. Feng, S., Park, C. Y., Liu, Y. & Tsvetkov, Y. From pretraining data to language models to downstream tasks: Tracking the trails of political biases leading to unfair NLP models. *arXiv [cs.CL]* (2023).

7. Costello, T. H., Pennycook, G. & Rand, D. G. Durably reducing conspiracy beliefs through dialogues with AI. *Science* **385**, eadq1814 (2024).

8. Tessler, M. H. *et al.* AI can help humans find common ground in democratic deliberation. *Science* **386**, eadq2852 (2024).

9. Fredricks, J. A., Blumenfeld, P. C. & Paris, A. H. School engagement: Potential of the concept, state of the evidence. *Rev. Educ. Res.* **74**, 59–109 (2004).

10. Chi, M. T. H. & Wylie, R. The ICAP framework: Linking cognitive engagement to active learning outcomes. *Educ. Psychol.* **49**, 219–243 (2014).

11. Ryan, R. M. & Deci, E. L. Self-determination theory and the facilitation of intrinsic motivation, social development, and well-being. *Am. Psychol.* **55**, 68–78 (2000).

12. Short, J., Williams, E. & Christie, B. The social psychology of telecommunications. *Jhon Wiley & Sons* https://doi.org/10.2307/2065899 (1976) doi:10.2307/2065899.